\documentclass[12pt,german,english,titlepage,twoside]{report}
     \usepackage{babel}
     \usepackage{longtable}
     \usepackage{graphicx}
     \usepackage[latin1]{inputenc}

\newcommand{\dcauthorpre}{Herr Dipl.-Phys. } 
\newcommand{\dcauthorsurname}{J\"uttner } 
\newcommand{\dcauthorname}{Andreas } 
\newcommand{\dcauthoradd}{geboren am 19.08.1976 in N\"urnberg } 
\newcommand{\dctitle}{\hspace{5mm}Precision Lattice Computations in the\newline Heavy Quark Sector} 
\newcommand{\dcsubtitle}{~}  

\newcommand{\dcapprovala}{Dr. Jonathan Flynn} 
\newcommand{\dcapprovalb}{Dr. Rainer Sommer} 
\newcommand{\dcapprovalc}{Prof. Dr. U. Wolff} 

\newcommand{\dcdegree}{doctor rerum naturalium\\ (Dr. rer. nat.)} 
\newcommand{\dcsubject}{Physik} 
\newcommand{\dcfaculty}{Mathematisch-Naturwissenschaftlichen Fakult\"at I}
\newcommand{\dcuniversity}{der Humboldt-Universit\"at zu Berlin}
\newcommand{\dcdean}{Prof. Thomas Buckhout, Ph.D.}
\newcommand{\dcpresident}{Prof. Dr. Jürgen Mlynek}
\newcommand{\dcdatesubmitted}{26. Juli 2004} 
\newcommand{\dcdateexam}{1. Oktober 2004} 

\newcommand{\dckeydea}{HQET}
\newcommand{\dckeydeb}{Gitter QCD}
\newcommand{\dckeydec}{schwere Quarks}
\newcommand{\dckeyded}{sytematisch Fehler}

\newcommand{\dckeywordsde}{\vspace*{2cm} \\{\bf{Schlagw\"orter:}}\\ \dckeydea, \dckeydeb, \dckeydec, \dckeyded \\}

\newcommand{\dckeyena}{HQET}
\newcommand{\dckeyenb}{lattice QCD}
\newcommand{\dckeyenc}{heavy quarks}
\newcommand{\dckeyend}{systematic errors}
\newcommand{\dckeywordsen}{\vspace*{2cm} \\{\bf{Keywords:}}\\ \dckeyena, \dckeyenb, \dckeyenc, \dckeyend \\}

\author{von \\ \dcauthorpre  \dcauthorname  \dcauthorsurname  \\ \dcauthoradd}

\title{
\vspace{-6cm} \begin{flushright}
   \normalsize HU-EP-05/14\\
   \normalsize SFB/CPP-05-09\\
 \end{flushright}
\vspace{1cm}
\dctitle \\ 
\vspace{0.5cm}
\large{\dcsubtitle} \\ 
\vspace{0.5cm} {\Large{D I S S E R T A T I O N }}\\ 
\vspace{0.5cm} \large{zur Erlangung des akademischen Grades \\ 
\dcdegree\\ im Fach \dcsubject \\ 
\vspace{0.5cm} eingereicht an der \\ 
\dcfaculty \\ 
\dcuniversity \\}}
\date{\vspace{0.5cm}
\raggedright{
Pr\"asident der Humboldt-Universit\"at zu Berlin:\\
\dcpresident \vspace{-0.3cm}
}\vspace{0.5cm}\\
\raggedright{
Dekan der \dcfaculty:\\
\dcdean \vspace{-0.3cm}
}\vspace{0.5cm}\\
\raggedright{
Gutachter:
\begin{enumerate} 
\item{\dcapprovala} \vspace{-0.3cm}\\
\item{\dcapprovalb} \vspace{-0.3cm}\\
\item{\dcapprovalc} \vspace{-0.3cm}\\
\end{enumerate}} \vspace{0.5cm}
\raggedright{
\begin{tabular}{lll}
eingereicht am: &  &\dcdatesubmitted\\
Tag der m\"undlichen Pr\"ufung: & & \dcdateexam
\end{tabular}}\\ 
}

\usepackage{exscale}                    
\usepackage{dsfont}                     
\usepackage{rotating}                   
\usepackage{amsfonts}
\usepackage{supertabular}		
\usepackage{textcomp}
\usepackage{bbold}
\usepackage{bbm}
\usepackage{psfrag}
\usepackage{fancyhdr}
\usepackage{amssymb}
\usepackage{subfigure}
\usepackage{multirow}
\usepackage{a4,cite,epsfig}
\usepackage{colortbl}                   

\newcommand{\oh}{{1\over2}}

\newcommand{\be}{\begin{equation}}
\newcommand{\ee}{\end{equation}}
\newcommand{\bea}{\begin{equation}\begin{array}}
\newcommand{\eea}{\end{array}\end{equation}}
\newcommand{\bdm}{\begin{displaymath}}
\newcommand{\edm}{\end{displaymath}}
\newcommand{\ba}{\begin{array}}
\newcommand{\ea}{\end{array}}
\newcommand{\bi}{\begin{itemize}}
\newcommand{\ei}{\end{itemize}}
\newcommand{\bver}{\begin{verbatim}}
\newcommand{\everbatim}{\end{verbatim}}

\newcommand{\rmd}{{\rm{d}}}

\newcommand{\rmX}{{\rm{X}}}

\newcommand{\Tr}{\mathrm{Tr}}

\newcommand{\fds}{{\rm{F}_{\rm{D}_s}}}

\newcommand{\fPS}{{\rm{F}_{\rm PS}}}

\newcommand{\fdsstar}{{\rm{F}_{\rm{D}_s^\ast}}}
\newcommand{\fV}{{\rm{F}_{\rm V}}}
\newcommand{\fbs}{{\rm{F}_{\rm{B}_s}}}
\newcommand{\fbsstar}{{\rm{F}_{\rm{B}_s^\ast}}}

\newcommand{\dsub}{{\rm{D}_s}}
\newcommand{\dsubstar}{{\rm{D}_s^\ast}}
\newcommand{\bsubstar}{{\rm{B}_s^\ast}}
\newcommand{\bsub}{{\rm{B}_s}}
\newcommand{\mPS}{{m_{\rm PS}}}
\newcommand{\mV}{{m_{\rm V}}}
\newcommand{\MSbar}{{\rm \overline{MS}}}

\newcommand{\mbar}{\overline{m}}

\newcommand{\orda}{{O}(a)}

\newcommand{\Dslash}{\ensuremath \raisebox{0.025cm}{\slash}\hspace{-0.32cm} D}

\newcommand{\PP}{\mathbbm{P}}

\makeatletter
\def\overbracket#1{\mathop{\vbox{\ialign{##\crcr\noalign{\kern3\p@}
\downbracketfill\crcr\noalign{\kern3\p@\nointerlineskip}
$\hfil\displaystyle{#1}\hfil$\crcr}}}\limits}
\def\underbracket#1{\mathop{\vtop{\ialign{##\crcr
$\hfil\displaystyle{#1}\hfil$\crcr\noalign{\kern3\p@\nointerlineskip}
\upbracketfill\crcr\noalign{\kern3\p@}}}}\limits}
\def\overparenthesis#1{\mathop{\vbox{\ialign{##\crcr\noalign{\kern3\p@}
\downparenthfill\crcr\noalign{\kern3\p@\nointerlineskip}
$\hfil\displaystyle{#1}\hfil$\crcr}}}\limits}
\def\underparenthesis#1{\mathop{\vtop{\ialign{##\crcr
$\hfil\displaystyle{#1}\hfil$\crcr\noalign{\kern3\p@\nointerlineskip}
\upparenthfill\crcr\noalign{\kern3\p@}}}}\limits}
\def\downparenthfill{$\m@th\braceld\leaders\vrule\hfill\bracerd$}
\def\upparenthfill{$\m@th\bracelu\leaders\vrule\hfill\braceru$}
\def\upbracketfill{$\m@th\makesm@sh{\llap{\vrule\@height3\p@\@width.7\p@}}%
\leaders\vrule\@height.7\p@\hfill
\makesm@sh{\rlap{\vrule\@height3\p@\@width.7\p@}}$}
\def\downbracketfill{$\m@th
\makesm@sh{\llap{\vrule\@height.7\p@\@depth2.3\p@\@width.7\p@}}%
\leaders\vrule\@height.7\p@\hfill
\makesm@sh{\rlap{\vrule\@height.7\p@\@depth2.3\p@\@width.7\p@}}$}
\makeatother

\makeatletter
\newcommand{\fmslash}[2][0mu]{%
  \mathchoice
    {\fmsl@sh\displaystyle{#1}{#2}}%
    {\fmsl@sh\textstyle{#1}{#2}}%
    {\fmsl@sh\scriptstyle{#1}{#2}}%
    {\fmsl@sh\scriptscriptstyle{#1}{#2}}}
\newcommand{\fmsl@sh}[3]{%
  \m@th\ooalign{$\hfil#1\mkern#2/\hfil$\crcr$#1#3$}}
\makeatother

\definecolor{lightblue}{rgb}{.3,.3,1.0}
\definecolor{lightlightblue}{rgb}{.9,.9,1.0}
\definecolor{lightlightgrey}{rgb}{.9,.9,0.8}
\definecolor{lightyellow}{rgb}{.9,.9,.4}
\newcolumntype{A}{>{\columncolor{lightlightblue}}l}
\newcolumntype{B}{>{\columncolor{white}}l}

\newcolumntype{C}{>{\columncolor{lightlightblue}}c}
\newcolumntype{R}{>{\columncolor{lightlightblue}}r}
\newcolumntype{D}{>{\columncolor{lightlightgrey}}c}

\begin{document}
\ifx\href\undefined\else\hypersetup{linktocpage=true}\fi 
\pagenumbering{roman}
\maketitle
\selectlanguage{english}
\abstract

The phenomenology of the pseudo scalar mesons $\dsub$ and $\bsub$ and of the vector mesons $\dsubstar$ and $\bsubstar$, each of which contain a heavy and a light quark, was investigated in simulations of quenched lattice QCD. The work is particularly focused on the minimisation of all systematic errors within this approximation.

The decay constants $\fds$ and $\fdsstar$ and the difference in the masses between the pseudo scalar $\dsub$-meson and the corresponding vector meson $\dsubstar$ were determined from the direct computer simulation of lattice QCD in large physical volume ($L\approx 1.5$ fm). As an aside, the renormalisation group invariant charm quark mass $M_c$ could be obtained from the simulation results.

A platform independent software was developed for the Monte-Carlo simulations of lattice QCD within the Schr\"odinger Functional. A number of simulations at different lattice constants allowed the extrapolation of the results to the continuum.

Since comparable simulations for the $\bsub$- and the $\bsubstar$-meson are not feasible due to the large mass of the $b$-meson, an interpolation in the meson mass to its physical point was carried out for the decay constant and the mass splitting. The interpolation was carried out between the static limit and the range of meson masses of order $m_{\dsub}$. The desired observables were therefore determined and extrapolated to the continuum for altogether six meson masses. 
The functional form of the subsequent interpolation in the meson mass to the static limit was guided by the prediction of the Heavy Quark Effective Theory (HQET). In order to apply it to the results obtained in QCD, a set of conversion functions between HQET and QCD were derived and evaluated numerically with input from results in perturbation theory.

The final results are $\fds=226(7)$MeV, $\fdsstar=239(18)$MeV,  \linebreak$\fbs
=198(9)$MeV, $m_{\dsubstar}-m_{\dsub}=136(9)$MeV, $m_{\bsubstar}-m_{\bsub}=63(6)
$MeV and $M_{\rm c}=1.60(3)$GeV. The result for the renormalisation group invariant charm quark mass is equivalent to $\mbar_{\rm c}^{\MSbar}(\mbar_{\rm c})=1.27(3)$GeV.

The analysis of the interpolation furthermore allowed to estimate, that the lowest order corrections to the static limit in HQET are relatively small. One therefore can expect HQET to offer a good approximation in the range of $B$-physics.
\dckeywordsen
\selectlanguage{german}
\vspace{-1cm}\abstract \setcounter{page}{2}  

Die Ph\"anomenologie der pseudoskalaren Mesonen $\dsub$ und $\bsub$ sowie der Vektormesonen $\dsubstar$ und $\bsubstar$, welche jeweils ein schweres und ein leichtes Quark enthalten, wurde in numerischen Simulationen von Gitter-QCD unter Vernachl\"assigung virtueller Fer\-mion\-schlei\-fen untersucht. Besonderer Wert wurde auf die Kontrolle und Minimierung aller systematischen Fehler innerhalb dieser N\"aherung gelegt. 

Die Zerfallskonstanten $\fds$ und $\fdsstar$ und die Massendifferenz zwischen dem $\dsub$- und dem $\dsubstar$-Meson wurden aus der direkten Computersimulation von Gitter-QCD in gro\ss em physikalischen Volumen ($L\approx 1.5$ fm) bestimmt. Als Nebenprodukt konnte auch ein pr\"aziser Wert der re\-nor\-mierungs\-gruppen-invarianten Charm-Quarkmasse $M_{\rm c}$ ermittelt werden.

F\"ur die Monte-Carlo Simulationen von QCD auf dem Gitter, speziell im hier verwendeten Schr\"odinger Funktional, wurde eine plattformunabh\"angige Software entwickelt. Eine Reihe von Simulationen bei verschiedenen Gitterabst\"anden erlaubte die Extrapolation der Ergebnisse zum Kontinuum.
 
 Da vergleichbare Simulationen f\"ur das $\bsub$- und $\bsubstar$-Meson aufgrund der gro\ss en Masse des enthaltenen $b$-Quarks nicht m\"oglich sind, wurde eine Interpolation in der Mesonmasse zu ihrem experimentell bekannten Punkt f\"ur die Zerfallskonstante und f\"ur den Wert der Massendifferenz durchgef\"uhrt. Interpoliert wurde dazu zwischen dem statischen Limes (unendliche Mesonmasse) und dem Bereich von Mesonmassen in der Gr\"o\ss enordnung von $m_{\dsub}$. F\"ur insgesamt sechs Mesonmassen in diesem Bereich wurden die gew\"unschten Observablen deshalb aus Simulationen von Gitter-QCD in gro\ss em Volumen bestimmt und die Ergebnisse zum Kontinuum extrapoliert. 
 Die Form der anschlie\ss enden Interpolation in der Mesonmasse zum statischen Limes wurde den Vorhersagen der \emph{Heavy Quark Effective Theory} (HQET) entsprechend gew\"ahlt. Um diese auf QCD zu \"ubertragen, wurden Konversionsfunktionen zwischen HQET und QCD hergeleitet und mit Hilfe von Ergebnissen aus der St\"orungstheorie numerisch bestimmt.

\hspace{-1mm}Die Endergebnisse sind $\fds\!\!=\!226(7)$MeV, $\fdsstar\!\!=\!239(18)$MeV, $\fbs\!\!=\!197(9)$MeV, $m_{\dsubstar}-m_{\dsub}=136(9)$MeV, $m_{\bsubstar}-m_{\bsub}=63(7)$MeV und $M_{\rm c}=1.60(3)$GeV. Das Ergebnis f\"ur die Quarkmasse ist \"aquivalent zu $\mbar_{\rm c}^{\MSbar}(\mbar_{\rm c})=1.27(3)$GeV.

Aus der Analyse der so bestimmten Interpolationen lie\ss $\;$ sich au\ss erdem ab\-sch\"at\-zen, da\ss $\;$ die f\"uhrenden Korrekturen zum statischen Limes in der HQET relativ klein sind. Man erwartet deshalb, da\ss $\;$HQET im Bereich der $B$-Physik eine gute N\"aherung darstellt.
\vspace{-1cm}
\dckeywordsde
\newcommand{\tstamp}{\today}   
\renewcommand{\chaptermark}[1]{\markboth{#1}{}}
\renewcommand{\sectionmark}[1]{\markright{#1}}
\lhead[\fancyplain{}{\thepage}]         {\fancyplain{}{\rightmark}}
\chead[\fancyplain{}{}]                 {\fancyplain{}{}}
\rhead[\fancyplain{}{\leftmark}]       {\fancyplain{}{\thepage}}
\lfoot[\fancyplain{}{}]                 {\fancyplain{}{}}
\cfoot[\fancyplain{}{}]         	{\fancyplain{}{}}
\rfoot[\fancyplain{} {}]	  	{\fancyplain{}{}}
\selectlanguage{english}
\setcounter{page}{1}
\tableofcontents
\vspace{3mm}
{\bf Erratum}\newpage
\pagebreak
\newpage
\listoffigures\newpage
\pagebreak
\listoftables\newpage
\pagebreak
\pagestyle{empty}
\pagenumbering{arabic}
\pagestyle{fancy}

\newpage
\newpage
\thispagestyle{empty}
\newpage

\chapter{Introduction}
With the goal of a unification of quantum mechanics and special relativity, P. A. M. Dirac formulated his famous equation of motion for free spin-$\oh$ particles in 1928 \cite{Dirac:1928hu}. 
The Dirac equation has solutions with both positive and negative energy. The latter found an interpretation in terms of anti-particles four years later, when C. D. Anderson discovered the positron in cosmic rays \cite{Anderson:1933mb}.

Today, based on Dirac's findings, elementary particles are described as the quanta of fields in local quantum field theories. The CPT-theorem \cite{Lueders:1952fc,Schwinger:1951xk,Pauli}, according to which the field theory has to be invariant under the combined application of charge conjugation (C), parity (P) and time-reversal (T), postulates an anti-particle to be associated to each particle. Leaving aside gravity due to its weak coupling to elementary particles at the energy scales accessible to experiments today, the electro-magnetic (QED), the weak and the strong (QCD) interactions have been combined in the Standard Model of elementary particles which has an underlying local ${\rm SU(3)_c}\times{\rm SU(2)_L}\times{\rm U(1)_Y}$ gauge-symmetry. The gluons, the gauge bosons in QCD, and the photons, the gauge bosons in QED, couple to both the left-handed and the right-handed fermions. $C$ and $P$ are therefore good quantum numbers in these cases. In contrast, the gauge bosons $W^{\rm \pm}$ and $Z$ of the electro-weak sector only couple to the left handed fermions and therefore parity is violated. But at least the combination $CP$ for the time being seemed to be a symmetry of the electro-weak interactions.

However, the Standard Model for three generations of quarks
\be\ba{ccc}
\left(\!\ba{c} u\\d\ea\!\right),\;
\left(\!\ba{c} s\\c\ea\!\right),\;
\left(\!\ba{c} b\\t\ea\!\right),
\ea
\ee
comprises the possibility of $CP$-violation.
Although not confirmed experimentally, the generation of particle masses in the Standard Model is explained by the Higgs mechanism. The scalar Higgs field interacts with the lepton and the quark fields through Yukawa couplings. By spontaneous electro-weak symmetry breaking, the Higgs field acquires a non-vanishing vacuum expectation value and thereby dynamically generates a mass term for all fields. The physical quark fields $u,\,d,\,s,\,c,\,b,\,t$, are in the mass eigenstate basis, where the associated mass matrix is diagonal. Their relation to the quark states in the weak eigenstate basis is given by a $3\times3$ unitary matrix, the Cabbibo-Kobayashi-Maskawa (CKM) mixing matrix \cite{Kobayashi:1973fv}
\be
U_{\rm CKM}=\left(\ba{ccc}
V_{ud}	&V_{us}	&V_{ub}	\\
V_{cd}	&V_{cs}	&V_{cb}	\\
V_{td}	&V_{ts}	&V_{tb}	
\ea\right).
\ee
It depends on three real angles and six phases. Five phases can be removed due to the freedom to redefine the phases of the quark mass eigenstates, leaving a single physical phase $\delta_{\rm KM}$, the Kobayashi-Maskawa phase. 
In the case of the $CP$-invariance of the electro-weak interactions, the phase has to vanish. However, in 1964, Cronin, Fitch and Christenson \cite{Christenson:1964fg} found experimental evidence for a $CP$-violating $2\pi$-decay of the neutral Kaon.

Since the discovery of $CP$-violation, a lot of effort has been put into precision measurements of the CKM-matrix. 
New experiments, like CLEO-c \cite{Asner:2003uz} and the B\emph{-factories} BaBar \cite{Aubert:2004am} and BELLE \cite{Zhang:2004wz} have been set up for high precision measurements of $CP$-violating effects.

The question arises as to why these measurements are interesting.

Once the parameters of the Standard Model have been fixed by experiment, the consistency of the theory can be checked.  In particular, as $CP$-violation in the Standard Model is induced only by the phase $\delta_{\rm KM}$, its measurement in one process will constrain the $CP$-violation allowed in other processes.  
For example, the $CP$-violation in the decay $B\to\psi K_S$ is related to the $CP$-violation in $K\to\pi\nu\bar\nu$ \cite{Nir:1999mg}. If inconsistencies were discovered by experiment, this would be a sign for physics beyond the Standard Model. Indeed, super-symmetric models predict $CP$-violating effects, that would exceed the magnitude of $CP$-violation allowed by the Standard Model \cite{Grossman:1998pa}.

According to one of Sakharov's three conditions \cite{Sakharov:1967dj}, all of which must be met in order to allow for baryogenesis, $CP$ must be violated in order to favor baryon over anti-baryon production.
Therefore, a deeper understanding of $CP$-violation in connection with electro-weak symmetry breaking in the early universe may help in understanding the observed asymmetry between matter and anti-matter in the observable part of the universe. 

In order to assess possible inconsistencies in the flavor physics of the Standard Model, precision measurements on the one hand and precision predictions from theory on the other hand are required to reduce the error on the elements of the CKM-matrix.
For example, the CLEO-c experiment intends to measure precisely the branching ratio of the leptonic decay of a $\dsub$-meson,
\be\label{branchingratio}
{\rm BR}(\dsub\to l\bar\nu)=\frac{G_{\rm F}^2}{8\pi}\tau_\dsub {\rm F_{D_s}^2} |V_{\rm cs}|^2 m_\dsub m_l^2\left(1-\frac{m_l^2}{m_\dsub^2}\right)^2.
\ee
In this expression, which is correct up to radiative corrections, $G_{\rm F}$ is the Fermi constant, $\fds$, $\tau_\dsub$ and $m_\dsub$ the lifetime, the decay constant and the mass of the $\dsub$-meson respectively and $m_l$ is the mass of the lepton taking part in the decay. In order to extract the value of the matrix element $|V_{\rm cs}|$, theorists have to deliver precise predictions for the decay constant $\fds$. However, the $\dsub$-meson is particular, as $|V_{\rm cs}|$ can also be obtained very precisely from constraints on the CKM-matrix \cite{Soldner-Rembold:2001zk} and therefore the experiment can also measure $\fds$. Hence, its study offers the possibility to test precision predictions from the lattice and to ensure, that the same techniques applied to experimentally less explored meson systems produce reliable results.

In similar ways, from the measurements of the mixing $B_d\leftrightarrow \overline B_d$ and $B_s\leftrightarrow \overline B_s$, the product of CKM-matrix elements $|V_{\rm td}||V_{\rm tb}|$ and the ratio $|V_{\rm td}|/|V_{\rm ts}|$ could be determined. To do this, theorists would have to predict accurately the products of the decay constant and the square root of the bag-parameter, ${\rm F_{B_d}}\sqrt{{\rm B_d}}$ and ${\rm F_{B_s}}\sqrt{{\rm B_s}}$ from the study of semi-leptonic decays \cite{Flynn:1998ca}.

On the theory side, in order to determine these low-energy quantities, QCD sum rules \cite{Penin:2001ux,Narison:2001pu} or relativistic quark models \cite{Ebert:2002qa} may be applied. Lattice QCD however, where a Euclidean space-time lattice is used as the regulator for QCD, appears to be the most promising approach. Physical quantities can be obtained as the expectation values of observables evaluated on an ensemble of field configurations, which have to be computer-generated by means of a Monte-Carlo simulation \cite{Montvay,Rothe:1992nt}. 

For precision lattice phenomenology, a number of systematic uncertainties present in the lattice approach have to be taken into account \cite{Flynn:1998ca}:

\bi
\item \emph{Finite volume effects} - Simulations of lattice QCD are numerically very costly and therefore the size of the lattices which can be simulated is limited. Especially for light quarks, where the Compton wavelength is large, one has to make sure that results are not affected by the presence of a space-time boundary. 
\item \emph{Continuum limit and cutoff effects} - Lattice QCD has to be simulated at various finite lattice spacings in order to allow for a controlled extrapolation to the continuum limit. This has to be done along a line of constant physics, which amounts to renormalizing the theory. Systematic effects can be reduced by renormalizing the theory non-perturbatively \cite{Jansen:1996ck}. Furthermore, the approach to the continuum can be accelerated and systematic effects can be reduced by non-perturbatively improving the theory \cite{Luscher:1996sc}. In this way, lattice artifacts vanish quadratically in the lattice spacing  \cite{Symanzik:1983dc,Symanzik:1983gh} instead of linearly, as it is the case for example for standard Wilson fermions \cite{Wilson:1974sk}.

\item \emph{Extrapolations to physical quark masses} - Lattice computations become very costly for light quarks ($u$ and $d$) and for heavy quarks with masses above the charm quark mass. Light quarks require a large physical volume and at the same time, they are particularly costly with current algorithms due to the required inversions of the badly conditioned Dirac operator. For the heavy quarks, mass dependent cutoff effects at finite lattice spacing are a source of concern due to their short Compton wave length. Both problems can be circumvented by simulating at unphysical but less costly quark masses and then extrapolating the results to the physical point. 
The form of the extrapolation is suggested by effective theories. Chiral perturbation theory is commonly used for extrapolations to light quark masses, whereas heavy quark effective theory (HQET) suggests a polynomial expansion of observables in the inverse quark mass for heavy quarks.
The extrapolations in each case have to be done carefully to avoid uncontrolled systematic errors.

\item \emph{Excited states} - In lattice computations, physical observables are mostly extracted from the time dependence of correlation functions. One does not know exactly how to construct particle wave functions in QCD which would allow for a projection onto particular states of the spectrum. Therefore, contributions of excited states to the desired observables cannot be completely avoided. The magnitude of such contributions can however in some cases be estimated and controlled by an accurate data analysis.

\item \emph{Quenching} - The generation of a representative ensemble of gauge configurations in the lattice Monte-Carlo simulation is very costly in the case of full QCD. Full QCD means that the simulation takes into account both virtual gluon loops and virtual quark loops. In the quenched approximation, one neglects the contributions of virtual quark loops. This reduces the costs immensely, at the expense of significance of the results: Quenched QCD is an uncontrolled approximation to QCD and precise and reliable results for phenomenology cannot be obtained. But still, simulations in the quenched  theory give estimates for phenomenological quantities which in some cases are surprisingly good and are an important tool to assess techniques for later use in the full theory.
\ei

In this work, a feasibility study of precision lattice computations in the heavy-flavor sector of quenched QCD has been accomplished. A particular emphasis was placed on keeping the above sources of systematic errors, apart from quenching, under control. The focus of the study was on the phenomenology of the heavy-light pseudo scalar mesons $\dsub$ and $\bsub$ and the vector mesons $\dsubstar$ and $\bsubstar$. In particular, the simulations aimed at the computation of the leptonic decay constants $\fds$, $\fdsstar$, $\fbs$ and $\fbsstar$, which are important for the determination of CKM-matrix elements. 

Simulations of lattice QCD for heavy-light mesons in large physical volume ($L \gtrsim 1.5$ fm) and with a controlled continuum extrapolation currently can be accomplished at the physical point of charm and strange as the heavy and light quark respectively. 
Thus, the ${\rm D_s}$-meson can be simulated directly and systematic effects due to extrapolations in the quark mass can be avoided.
Since very precise data for the matrix element $|V_{\rm cs}|$ exists, CLEO-c will determine $\fds$ to a precision below 2\% until 2005 and thereby offers a very accurate test of the lattice approach and all the applied techniques.

\begin{figure}
\centering
\psfrag{oom}[t][c][1][0]{\large$1/m_Q$}
\psfrag{oomBs}[l][l][1][0]{$1/m_b$}
\psfrag{oomDs}[l][l][1][0]{$1/m_c$}
\psfrag{0}[t][l][1][0]{\large$0$}
\psfrag{F}[b][c][1][0]{\large some observable $\Phi(m_{Q})$}
\epsfig{scale=.6,file=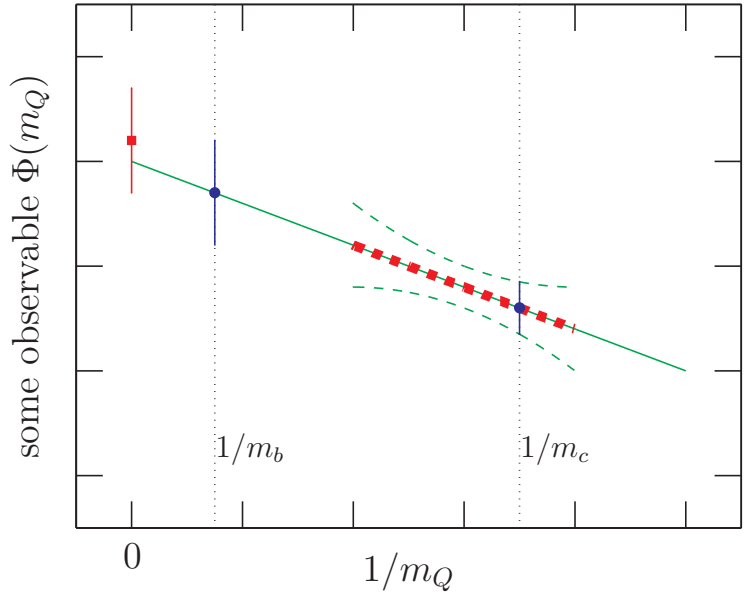}
\caption{Using data from lattice simulations of QCD  with heavy quark masses around the charm quark mass $m_c$ in the continuum limit (bold dashed line with error band), together with predictions or simulation results in the static limit $m_Q\to\infty$, allows for an interpolation to the physical point of the $b$-quark $m_b$.}\label{interpol}
\end{figure}

Once one has gained confidence in the lattice computations, the techniques can be applied to sectors of the Standard Model, like the $\bsub$-mesons, which are not easily accessed through experiments. The large mass of the $b$-quark however does not allow for a direct lattice simulation of the $\bsub$ mesons. 
Instead, the following procedure, which is sketched in figure \ref{interpol}, can be applied. In addition to the simulations at the physical point of the $\dsub$-meson, one simulates for the desired meson observable also at a number of unphysical heavy quark masses around charm (indicated by the dashed bold line). After the continuum limit has been taken, the data would allow for an extrapolation to the physical point of the $b$-quark mass. However, the range of heavy quark masses accessible to relativistic simulations of QCD is limited. Therefore, such an extrapolation has little significance and systematic effects cannot be estimated.

Fortunately, the extrapolation can be constrained further. HQET makes exact predictions for some mesonic observables in the limit of infinite heavy quark mass, the \emph{static} limit. Also, results for the decay constant in the continuum, obtained from lattice simulations in the static approximation with reasonable statistical errors, exist \cite{DellaMorte:2003mn}. Furthermore, HQET predicts the mass dependence of mesonic observables as a polynomial in the inverse heavy quark mass.
Combining predictions from HQET in the static limit with lattice QCD simulations in the charm region allows for a controlled interpolation to the beauty region \cite{Bernard:1988dy,Gavela:1988cf,Wittig:1997tr}, while keeping systematic effects under control.
Furthermore, in assessing the functional form of the resulting interpolation, an estimate of the order of magnitude of the leading order coefficients in the heavy quark expansion is possible and at the same time constitutes a test for HQET.\\

HQET \cite{Eichten:1990vp,Grinstein:1990mj,Falk:1990yz} is an effective theory for QCD with heavy quarks of mass $m_Q\gg\Lambda_{\rm QCD}$. A short motivation and its definition are given in chapter \ref{hqet}. Conversion functions that relate observables of heavy-light mesons in HQET in the continuum to their analog in QCD have been derived, computed and parameterized in terms of the renormalization group invariant heavy quark mass $M_Q$. These conversion functions will finally enable an interpolation between results from QCD in the charm region and the static limit guided by the predictions from HQET. 

Section \ref{technicalissues} establishes the QCD Schr\"odinger Functional as the preferred framework for the lattice computation of decay constants, meson masses and quark masses, for which explicit expressions in terms of correlation functions and finally in terms of quark propagators will be derived. 

A major part of this work consisted in obtaining a platform independent program code that can accomplish all the necessary computations. The program code presented in chapter \ref{chapPCcode} is based on the MILC collaboration's lattice gauge theory code \cite{MILC}. All major changes, improvements and tests of the code will be discussed. 

Chapter \ref{simulations} discusses and tabulates all parameters for the Monte-Carlo simulations.

The analysis of all data is described in chapter \ref{analysis}. After the discussion and estimation of all sources of systematic errors, the continuum extrapolation for the $\dsub$- and the $\dsubstar$-meson are presented. Afterwards, the results from the interpolation to the static limit are discussed. On the one hand, the values of meson observables at the physical point of the $\bsub$ have been obtained and on the other hand an estimate for the order of magnitude of the leading order coefficient of the heavy quark expansion will be given. All results are discussed and compared with other lattice computations and experiment.

The last chapter summarizes all findings of this work and gives an outlook.

\chapter{Non-perturbative test of HQET with QCD}\label{hqet}
After a short motivation of HQET in section one, section two describes the computation of conversion functions, which are necessary to relate matrix elements in QCD to those in HQET. 
They will allow for an interpolation in the heavy quark mass between the region of the charm quark mass and the static limit, guided by predictions from HQET.
The necessary relations are given in section three. 
\section{Heavy quark effective theory}\label{HQETeins}
\begin{figure}\centering
\epsfig{scale=.7,file=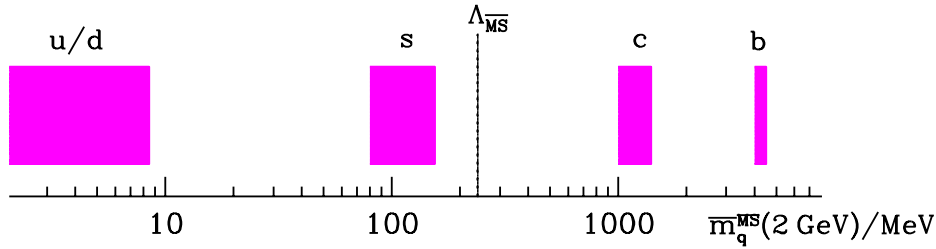}
\caption{Quark mass ranges in the $\overline{\rm MS}$-scheme of dimensional regularization \cite{PDBook}.}\label{quarkmasseseps}
\end{figure}

The typical energy carried by the light constituents in mesons ($u$-, $d$-, $s$-quarks and anti quarks and gluons) is of order $\Lambda_{\rm QCD}\approx200$ MeV.
The phenomenology of mesons containing a light quark $q$ and a heavy quark $Q$ with $m_Q\gg\Lambda_{\rm QCD}$\footnote{A particular choice for the quark mass definition will be done in section \ref{CompCX}. At this point $m_{\tiny\rm \emph{Q}}$ may for example be the heavy quark's pole mass.}
(cf. figure \ref{quarkmasseseps}) is therefore governed by the two different energy scales $m_Q$ and $\Lambda_{\rm QCD}$. With the heavy quark's Compton wavelength being $\lambda_Q\sim 1/m_Q$, the gluons cannot resolve the heavy quark's quantum numbers - the light degrees of freedom are blind to spin and flavor (mass) of the heavy quark, leading to heavy quark spin and flavor symmetry. For example the experimentally determined spin splittings \cite{PDBook}
\be\ba{rcl}
m^2_{{\rm B}^\ast}-m^2_{\rm B}&\approx&0.49\,{\rm GeV}^2,\\
m^2_{{\rm D}^\ast}-m^2_{\rm D}&\approx&0.55\,{\rm GeV}^2,
\ea
\ee
and the mass splittings \cite{PDBook}
\be\ba{rcl}
m_{{\rm B}_s}-m_{{\rm B}_d}&=&(90\pm3)\,{\rm MeV},\\
m_{{\rm D}_s}-m_{{\rm D}_d}&=&(99\pm1)\,{\rm MeV},
\ea
\ee
for different heavy-light mesons are approximately the same.
One expects these symmetries to be exact for heavy-light mesons with one infinitely heavy, or \emph{static}, quark. The symmetry breaking which can be observed experimentally at a finite but large heavy quark mass can be interpreted as the consequence of small perturbations to the theory with a static quark due to the interaction with the chromo-magnetic and chromo-electric fields mediated by soft gluons. This idea has been formulated in terms of an effective theory, the heavy quark effective theory (HQET) \cite{Eichten:1990zv,Politzer:1988bs,Georgi:1990um} which will be derived briefly in the following. For an extensive derivation the reader may refer to the reviews \cite{Neubert:1996wg,Neubert:1994mb} and \cite{Mannel:1996rg}.

Restricting the study on heavy-light mesons with momentum $p$, containing one flavor of heavy quarks $Q(x)$ and one flavor of light quarks $q(x)$, the starting point is the QCD path integral
\be\ba{rcl}
\mathcal{Z}_{\rm QCD}&=&\int_{\bar q,q, \bar Q,Q,U}e^{-i\int d^4x\left\{\mathcal{L}^{\rm YM}[U(x)]+\mathcal{L}^{\rm Q}[\bar q(x),q(x),U(x)]+\mathcal{L}^{\rm Q}[\bar Q(x),Q(x),U(x)]\right\}}.\\
\ea
\ee
$\mathcal{L}^{\rm YM}$ is the SU(3) Yang-Mills Lagrangian, and $\mathcal{L}^{\rm Q}$ is the QCD Lagrangian for quark fields coupled to the gauge field $U$ in the adjoint representation,
\be\ba{rcl}
\mathcal{L}^{\rm Q}[\bar \psi(x),\psi(x),U(x)]&=&\bar\psi(x)(i\,\Dslash  +m)\psi(x),
\ea\ee
with the Dirac operator \,$\Dslash$. 
The focus will now be on the heavy quark Lagrangian. The heavy quark in the meson is approximately on shell and therefore behaves like a free particle moving at four-velocity $v$. Removing the space time dependence of a solution of the free Dirac equation, the four-component Dirac field $Q(x)$ can be rewritten in terms of the large and small component fields\footnote{This nomenclature stems from the free Dirac theory, where in the non-relativistic limit $E\to mc^2$, the upper components of the Dirac spinor remain of $O(1)$ while the lower components vanish. One therefore refers to the upper components as the ``large components'' and to the lower components as the ``small components''.} $h_v(x)$ and $H_v(x)$ by
\be\ba{rclrcl}
h_v(x)&=&e^{im_Q v\cdot x} P_+^v Q(x)\;{\rm and}\;H_v(x)&=&e^{im_Q v\cdot x} P_-^v Q(x).
\ea
\ee
$P^v_+$ and $P^v_-$ are the projection operators
\be
P_\pm^v={1\pm \fmslash{v}\over 2}.
\ee
The time dependence of the fields $h(x)$ and $H(x)$ is then expected to be determined by the residual momentum $k=p-m_Qv$ which is of order $\Lambda_{\rm QCD}$. The heavy quark will only be considered in its rest frame throughout this work and therefore $v_\mu=(1,0,0,0)$. In this case, $h(x)\equiv h_v(x)$ corresponds to the upper components of $Q(x)$ and $H(x)\equiv H_v(x)$ to the lower components. 

\begin{figure}
\centering
\psfrag{1}[c][c][1][0]{$x_0$}
\epsfig{file=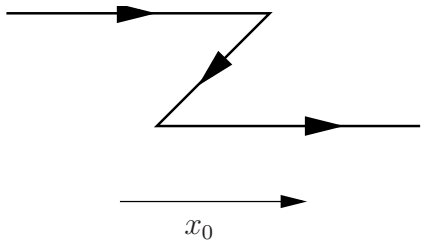}
\caption{Virtual fluctuation of a heavy quark.}\label{feyn1}
\end{figure}

The small components $H(x)$ of the heavy quark field $Q(x)$ only become relevant at high energies and are the origin of the short distance effects - for example, effects involving pair creation of heavy quarks or the zig-zag depicted in figure \ref{feyn1}, where the intermediate state has an energy that differs from the initial one by at least $2m_Q$ and therefore propagates only over a short distance. 

In terms of the fields $H(x)$ and $h(x)$, the heavy quark Lagrangian in the rest frame can be rewritten as
\be\ba{rcl}\label{HQETrestframe}
\mathcal{L}[\bar Q(x),Q(x),U(x)]&=& \bar h(x)i D_0h(x)-\overline H(x) (iD_0+2m_Q) H(x)\\
\\
&+&\bar h(x) i\;\Dslash_\perp H(x)+\overline H(x) i\;\Dslash_\perp h(x).\\
\ea
\ee
with $D_\perp=(0,\vec D)$.
The large component fields $h(x)$ do no longer have a mass term, whereas the small component fields $H(x)$ appear with a mass term with twice the heavy quark mass. It is this term which will be eliminated in the construction of the effective theory.

By Gaussian integration, which in this case is equivalent to applying the classical equation of motion 
\be
(iD_0+2m_Q)H(x)=\Dslash_\perp h(x),
\ee
the small component fields can be eliminated and one arrives at the non-local effective Lagrangian
\be\ba{rcl}\label{HQETrestframe}
\mathcal{L^{\rm eff}}[\bar h(x),h(x),U(x)]&=& \bar h(x)i D_0h(x)+\overline h(x)i\;\Dslash_\perp{1\over 2m_Q(1+{iD_0\over 2m_Q})}i\;\Dslash_\perp h(x).\\
\ea
\ee
The second term in this Lagrangian represents the virtual processes suppressed by at least $1/2m_Q$. In momentum space the operator that acts on $h(x)$ corresponds to powers of the momentum. As the residual momenta of the heavy quark field $h(x)$ are small with respect to the heavy quark's mass, the quotient in the second term can be expanded in $D_0/m_Q$ 
by means of a derivative expansion which results in a an effective Lagrangian, in which the operators are ordered in powers of $1/m_Q$. This is the HQET Lagrangian. Up to the 1st order in $1/m_Q$ it reads\footnote{Higher order terms will not be considered in this thesis.}
\be\ba{rcl}\label{HQETeff1}
\mathcal{L}^{\rm HQET}[\bar h(x), h(x), U(x)]&=& \mathcal{L}^{\rm stat}[\bar h(x), h(x), U(x)]\\
\\
&+&{1\over 2m_Q}\mathcal{L}_{1/m_Q}[\bar h(x), h(x), U(x)]+{O}(1/m_Q^2)\\
\ea
\ee
with\\
\be\ba{rcccc}
\mathcal{L}^{\rm stat}[\bar h(x), h(x), U(x)]&\hspace{-.3cm}=&\hspace{-.3cm}{O}_{\rm stat}[\bar h(x), h(x), U(x)]&\hspace{-.3cm}=&\hspace{-.3cm}\overline h(x)\,iD_0h(x),\\
\\
\mathcal{L}_{1/m_Q }[\bar h(x), h(x), U(x)]&\hspace{-.3cm}=&\hspace{-.3cm}{O}_{1/m_Q}^{\rm kin}[\bar h(x), h(x), U(x)]&\hspace{-.3cm}+&\hspace{-.3cm}{O}_{1/m_Q}^{\rm spin}[\bar h(x), h(x), U(x)]\\
\\
&\hspace{-.3cm}=&\hspace{-.3cm}\bar h(x)i \vec{D}^2 h(x)&\hspace{-.3cm}+&\hspace{-.3cm}\bar h(x)i\vec{S}\cdot \vec{B}(x) h(x).\\
\\
\ea\ee
The $S^i$ are the generators of spin ${\rm SU(2)}$ rotations and can be chosen as
\be
S^i=\oh\left(\ba{cc}\sigma^i&0\\0&\sigma^i\ea\right),\;\;[S^i,S^j]=i\epsilon^{ijk}S^k,
\ee
where the $\sigma_i$ are the Pauli matrices (cf. appendix \ref{notation}).
$B^i(x)=-\oh \epsilon^{ijk}G^{ij}(x)$ is the chromo-magnetic gluon field where $[iD^\alpha(x),iD^\beta(x)]=ig G^{\alpha\beta}(x)$ is the gluon field strength tensor.
The term with ${O}_{1/m_Q}^{\rm kin}$ is responsible for fluctuations of order $\Lambda_{\rm QCD}$ in the heavy quark's motion and ${O}_{1/m_Q}^{\rm spin}$ describes the coupling of the heavy quark's spin to the chromo-magnetic field. Both terms introduce the leading order flavor and spin symmetry breaking interactions at finite heavy quark mass, which were mentioned at the beginning of this chapter.

The theory with the Lagrangian (\ref{HQETeff1}) is not renormalizable by a finite number of counter terms. 
Due to the presence of couplings with negative mass dimension, terms of a given order in $1/m_Q$ may mix with terms of higher order under renormalization \cite{Collins} and an infinite number of counter terms would be necessary. 

Thus, one expands the Boltzmann-factor in the corresponding path integral in the heavy quark mass $1/m_Q$,
\be\ba{rcl}\label{HQETpartfun}
\mathcal{Z}_{\rm HQET}[U]&=&\int_{\bar h, h}e^{-i\int d^4x\mathcal{L}^{\rm HQET}[\bar h(x), h(x), U(x)]}\\
\\
&=&\int_{\bar h, h}e^{-i\int d^4x\mathcal{L}^{\rm stat}[\bar h(x), h(x), U(x)]}\\
\\
&\times&\qquad\left\{1+{1\over 2m_Q}\int d^4x\mathcal{L}_{1/m_Q}[\bar h(x), h(x), U(x)]+{O}(1/m_Q^2)\right\}.
\\
\ea
\ee
From power counting one concludes, that the static theory defined by $\mathcal{L}^{\rm stat}$ is renormalizable with a finite number of parameters.

In the same way as for the derivation of the HQET Lagrangian, an operator $O^{\rm X}(x)$ containing heavy quark degrees of freedom, at tree-level can be expanded in a power series in $1/m_Q$,
\be
O^{\rm X}(x)=O^{\rm X}_0(x)+{1\over 2m_Q}O^{\rm X}_1(x)+{O}(1/m_Q^2).
\ee
This may for example be done for the heavy-light axial vector current $A_\mu(x)=\bar q(x)\gamma_\mu\gamma_5 Q(x)$ (${\rm X=PS}$\footnote{This common notation refers to the transformation properties of $A_\mu(x)$ under parity (odd), which are the same as for a pseudo scalar.}) and the vector current $V_\mu(x)=\bar q(x)\gamma_\mu Q(x)$ (${\rm X=V}$) which then at leading order are defined as
\be\ba{c}
O^{{\rm PS}}_{\mu}(x)=A_\mu(x)=\bar q(x)\gamma_\mu\gamma_5 h(x),\\
\\
O^{{\rm V}}_{\mu}(x)=V_\mu(x)=\bar q(x)\gamma_\mu h(x).
\ea
\ee
Unlike the analog weak current operators in QCD, $O^{{\rm PS}}_{0,\mu}(x)$ and $O^{{\rm V}}_{0,\mu}(x)$ become scale dependent under renormalization. Also the chromo-electric moment $O^{\rm spin}_{1/m_Q}$ receives a scale dependence. In contrast, $O^{\rm kin}_{1/m_Q}(x)$ stays scale independent due to re-parameterization invariance \cite{Luke:1992cs,Chen:1993sx}.

For the cases X=PS, V and spin one then writes 
\be
O^{\rm X}_{\rm R}(x,\mu)=Z^{\rm X}(\mu)O^{\rm X}(x),
\ee
with the renormalization constant $Z^{\rm X}(\mu)$ whose scale dependence is determined by the renormalization group equation
\be\label{matchingADs}
\gamma^{\rm X,\MSbar}(\bar g(\mu))=\mu\frac{d \log Z^{\rm X}(\mu)}{d\mu}.
\ee
The renormalized coupling $\bar g(\mu)$ is the one in the $\MSbar$-scheme of dimensional regularization and the anomalous dimension $\gamma^{\rm X,\MSbar}(g)$ has the generic perturbative expansion
\be\label{RGE}
\gamma^{\rm X,\MSbar}(g)=-\gamma_0^{\MSbar}g^2-\gamma_1^{\MSbar}g^4-\gamma_2^{\MSbar}g^6+\dots\;.
\ee
It is equivalent for X=PS and V and
has been determined in the $\MSbar$-scheme of dimensional regularization at one-loop 
in \cite{Kilian:1994mw,Georgi:1990um}, at two-loop in
\cite{Golden:1991dx,Neubert:1994za} and at three-loop precision
in \cite{Chetyrkin:2003vi}.
For ${\rm X=spin}$, the one-loop anomalous dimension is given
in \cite{Eichten:1990vp,Falk:1991pz} and at two-loop
in \cite{Amoros:1997rx,Kilian:1994mw}.
The corresponding coefficients are given in table \ref{anomalous dimension}.
\begin{table}
\centering
\begin{tabular}{ccc}
\hline
\hline\\[-2ex]
&X=PS,V&X=spin\\[.5ex]
\hline\\[-2ex]
$\gamma^{\rmX,\MSbar}_0$&$-{1\over(4\pi^2)}$			 &${6\over(4\pi)^2}$\\[2.0ex]
$\gamma^{\rmX,\MSbar}_1$&$-\left({254\over9}+ {56\over27\pi^2}\right){1\over(4\pi)^4}$    &${68\over (4\pi)^4}$\\[2ex]
$\gamma^{\rmX,\MSbar}_2$&$-{12.941\over(4\pi^2)^3}$\\[2.0ex]
\hline\hline
\end{tabular}
\caption{Coefficients for the 3- resp. 2-loop anomalous dimension for renormalized heavy-light quark currents (axial vector and vector current) and the chromo-magnetic moment of a heavy quark. }
\label{anomalous dimension}
\end{table}
The vacuum expectation value of an operator $ O^{\rm X}_{\rm R}(x,\mu)$ in HQET then takes the form
\be\ba{rcl}\label{thatsHQET}
\Big<O^{\rm X}_{\rm R}(x,\mu)\Big>_{}\hspace{-.3cm}
&=&\hspace{-.2cm}\Big< ({O}^{\rm X}_0)_{\rm R}(x,\mu)\Big>_{}+\hspace{-.0cm}{1\over 2m_Q}\Big< ({O}^{\rm X}_1)_{\rm R}(x,\mu)\Big>_{}\hspace{-.2cm}\\
\\
&+&{1\over 2m_Q}\Big< ({O}^{\rm X}_0)_{\rm R}(x,\mu)\int d^4y\left({}({O}_{1/m_Q}^{\rm kin})_{\rm R}(y,\mu)\hspace{-0.3mm}+\hspace{-0.3mm}{}({O}_{1/m_Q}^{\rm spin})_{\rm R}(y,\mu)\right) \Big>_{}\\
\\
&+&\hspace{-.2cm}{O}({1/ m_Q^2}),\\
\ea
\ee
where the operator expectation values have to be understood in the theory defined by the path integral
\be\ba{rcl}
\mathcal{Z}&=&\int_{\bar q,q, \bar h,h,U}e^{-i\int d^4x\left\{\mathcal{L}^{\rm YM}[U(x)]+\mathcal{L}^{\rm Q}[\bar q(x),q(x),U(x)]+\mathcal{L}^{\rm stat}[\bar h(x),h(x),U(x)]\right\}}.\\
\ea
\ee

\section{Matching the effective theory to QCD}\label{secHQET}
By explicitly integrating out the short distance physics associated with the heavy quark in the last section, an effective theory for heavy quarks has been derived which one expects to correctly describe the long-distance physics of QCD. 

It is known from QCD, that quarks couple to gluons which can have virtual momenta as high as the quark mass. In HQET, when taking the limit $m_Q\to \infty$, this introduces logarithmic divergences for example in weak matrix elements. Those matrix elements therefore have to be renormalized. 

The matching of the effective theory to QCD amounts to reintroduce the high energy behavior of matrix elements in HQET in terms of Wilson coefficients. They allow to define conversion functions $C$, which relate QCD matrix elements for heavy quarks of mass $m_Q$ to the corresponding renormalization group invariant matrix elements in HQET.
\subsection{The conversion functions $C_{\rm X}(m_Q)$ for \\$\rm X=PS,\,V,\,PS/V,\,spin$}\label{matchcoeffs}
The Wilson coefficients are defined by the relation between the matrix element of the corresponding operator $O_{\rm R}(x,m_Q)$ in QCD, containing heavy degrees of freedom of mass $m_Q$, and the operator in the effective theory, renormalized at the scale $\mu$,
\be\ba{rcl}\label{opexp}
\Big<O^{\rm X}_{\rm R}(x,m_Q)\Big>_{\rm QCD}&\hspace{-.4cm}= &C_{\rm X}(m_Q,\mu)\Big< ({O}^{\rm X}_0)_{\rm R}(x,\mu)\Big>_{\rm stat}\\
\\
&&\hspace{-4cm}+\;{B_{\rm X}(m_Q,\mu)\over 2m_Q}\Bigg\{\Big< ({O}^{\rm X}_1)_{\rm R}(x,\mu)\Big>_{\rm stat}+\;\Big< ({O}^{\rm X}_0)_{\rm R}(x,\mu)\int d^4y\left(C_{\rm kin}(\mu,\mu)({O}_{1/m_Q}^{\rm kin})_{\rm R}(y,\mu)\right.\\
\\
&&\hspace{-4cm}+\;\left.C_{\rm spin}(\mu,\mu)({O}_{1/m_Q}^{\rm spin})_{\rm R}(y,\mu)\right) \Big>_{\rm stat}\Bigg\}+\;{O}({1/ m_Q^2}).\\
\ea
\ee
The coefficient $B_{\rm X}(m_Q,\mu)$ is mentioned for completeness but will be of no relevance for this work.
In practice, one determines the Wilson coefficients $C_{\rm X}(m_Q,\mu)$ in perturbation theory at the scale $\mu=m_Q$ from a comparison or {matching} of suitable matrix elements in the full and in the effective theory\footnote{In \cite{Heitger:2003nj}, a method, how to do the matching non-perturbatively has been suggested.}. Here, $m_Q$ is the heavy quark's pole mass. As the pole mass does not have a well defined perturbative expansion \cite{Beneke:1998ui}, it will be replaced by the renormalization scheme independent renormalization group invariant quark mass $M_Q$ in the next section. 

$C_{\rm X}(m_Q,m_Q)$ depends on the particular Dirac structure of the operator\linebreak ${O^{\rm X}_{\rm R}}(x,\mu)$ and has been determined in perturbation theory for a number of heavy-light current matrix elements.
The coefficients have an expansion in a power series in the renormalized coupling
\be\label{expandedmatching}
C_{\rm X}(m_Q,m_Q)=1+c^\rmX_1 \bar g^2(m_Q)+ c^\rmX_2 \bar g^4(m_Q)+\dots\;.
\ee
For the axial vector and the vector current, the one-loop computation has been accomplished in \cite{Eichten:1990zv} and at two-loop precision it is given in \cite{Broadhurst:1995se}. 
In the case of the kinetic term, $C_{\rm kin}(\mu,\mu^\prime)=1$ holds due to re-parameterization invariance \cite{Luke:1992cs,Chen:1993sx}. For $C_{\rm spin}(m_Q,m_Q)$ only the one-loop coefficient is known \cite{Eichten:1990vp}. 
The factors $c_1^{\rm X}$ and $c_2^{\rm X}$ for the quenched theory are collected in table \ref{rhoX} for the phenomenologically important cases X=PS, V and spin.

The scale dependence of the Wilson coefficients derives from the renormalization of the associated heavy quark current (cf. section \ref{secHQET}). After integrating the renormalization group equation (\ref{matchingADs}) one gets the relation
\be\label{erratum1}
C_{\rm X}(m_Q,\mu)\hspace{-.1cm}=\hspace{-.1cm}C_{\rm X}(m_Q,m_Q)\exp\hspace{-1mm}\left\{\int\limits_{\bar g(m_Q)}^{\bar g(\mu)}\hspace{-.3cm}dg{\gamma^{\rm X,\MSbar}(g)\over\beta(g)}\right\}\hspace{-1mm}.
\ee 
Here, $\gamma^{\rm X,\MSbar}(g)$ is the anomalous dimension introduced in the last section and $\beta(g)$ is the anomalous dimension of the renormalized coupling $\bar g(\mu)$ in the $\MSbar$-scheme of dimensional regularization which is known at 4-loop accuracy \cite{vanRitbergen:1997va},
\be\label{betafunction}
\beta(g)=-b_0g^3-b_1g^5-b_2g^7-b_3g^9-\dots\,.
\ee
The leading coefficients are $b_0=11/(4\pi)^2$ and $b_1=102/(4\pi)^4$ and the higher order coefficients are collected in appendix \ref{poleRGI}. 

To eliminate any dependence on the renormalization scale in the relation between matrix elements in QCD and in HQET, it is convenient to take the limit $\mu\to \infty$ in the above expressions. The Wilson coefficients then relate matrix elements in QCD to the renormalization group invariants 
\be
O_{\rm RGI}^{\rm X}(x)=\lim\limits_{\mu\to\infty}\left\{[2b_0\bar g^2(\mu)]^{-\gamma_0^{\rm X,\MSbar}/(2b_0)}O_{\rm R}^{\rm X}(x,\mu)\right\}
\ee
in HQET and one can write
\be
C_{\rm X}(m_Q,\mu)O^{\rm X}_{\rm R}(x,\mu)\to C_{\rm X}(m_Q)O^{\rm X}_{\rm RGI}(x).
\ee

\subsection{Computation of $C_{\rm X}(M_Q/\Lambda_{\rm QCD})$}\label{CompCX}
\begin{table}
\centering
\begin{tabular}{ccccccccc}
\hline\hline \\[-2.0ex]
   X    &matrix element& & & $c_1^\rmX$ && $c_2^\rmX$ \\[1.0ex]
\hline\hline \\[-1.0ex]
   PS   &\footnotesize$\Phi_{\rm PS}=\langle 0|\bar q \gamma_0 \gamma_5 Q|{\rm PS}\rangle$	& & & $-{2\over 3}{1\over 4\pi^2}$ 
&& $-4.2\,{1\over (4\pi^2)^2}$  \\\\
   V    &\footnotesize$\Phi_{\rm V}=\langle 0|\bar q \gamma_0 Q|{\rm V}\rangle	$		& & & $-{4\over 3}{1\over 4\pi^2}$ 
&& $-11.5\,{1\over (4\pi^2)^2}$ \\[2ex]
\multirow{2}{*}{spin} &\footnotesize$\Phi^{\rm spin}_{\rm PS}=\langle {\rm PS}|\bar Q \vec{S}\cdot\vec{B}Q|{\rm PS}\rangle$&&&\multirow{2}{*}{ ${13\over 6}{1\over 4\pi^2}$} && \multirow{2}{*}{---} \\
&\footnotesize$\Phi^{\rm spin}_{\rm V}=\langle {\rm V}|\bar Q \vec{S}\cdot\vec{B}Q|{\rm V}\rangle$\\[2.0ex]
\hline\hline 
\end{tabular}
\caption{
Coefficients for the matching factors $C_{\rm X}(m_Q,m_Q)$ in the quenched theory.
}\label{rhoX}
\end{table}
Since the pole mass $m_Q$ has a badly behaved perturbative expansion due to non-perturbative infrared effects \cite{Beneke:1998ui}, it will now be eliminated in favor of the dimensionless ratio between the renormalization group invariant quark mass $M_Q$ and the $\Lambda_{\rm QCD}$-parameter as the new argument of the conversion functions\footnote{Since only the case $N_f=0$ was considered in this thesis, the non-perturbatively determined value $\Lambda_{\rm QCD}=\Lambda_{\MSbar}=238(19)$ MeV (quenched) \cite{Capitani:1998mq} is used.}. $M_Q$ is scale- and scheme-independent. It is defined via the limiting behavior of any renormalized mass $\mbar (\mu)$,
\be
M_Q=\lim\limits_{\mu\to\infty}\left\{[2b_0 \bar g^2(\mu)]^{-d_0/(2b_0)}\mbar(\mu)\right\},
\ee
where $d_0=8/(4\pi)^2$ is the universal leading order coefficient of any quark mass anomalous dimension.
How $m_Q$ and $M_Q$ are related to each other in detail is explained in the appendix \ref{poleRGI}.

As an intermediate step, in the computation of the coefficients $C_{\rm X}(M_Q/\Lambda_{\rm QCD})$, one defines the conversion functions parameterized with the 
renormalized mass  $\mbar_\ast=\mbar(\mbar_\ast)$ in the $\overline{\rm MS}$-scheme,
\be
C_{\rm X}(\mbar_\ast)=
\left[\,2b_0\bar g^2(\mbar_\ast)\,\right]^{\gamma_0^{\rm X}/(2b_0)}
\exp\left\{\int\limits_0^{\bar g(\mbar_\ast)} \rmd g 
\left[\,\frac{\gamma^{\rm X}(g)}{\beta(g)}-\frac{\gamma_0^{\rm X}}{b_0g}
\,\right]\right\}.
\label{Chat}
\ee
The anomalous dimension $\beta(g)$ and the the anomalous dimension for ${\rm X=PS}$, V,
\be\label{gammamatch}
\gamma^{\rm X}(g) = 
-\gamma_0^{\rm X}g^2-\gamma_1^{\rm X}g^4 -\gamma_2^{\rm X}g^6-\dots\,
\ee
will always be taken at 4- respectively 3-loop precision. The difference to taking the 3-loop $\beta$-function instead, turned out to be tiny. The perturbative error introduced by $\gamma^{\rm X}(g)$ was estimated with half the difference between the values for $C_{\rm X}$ obtained with the 2-loop and the 3-loop expression.
For X=PS, V, the $\gamma_i$ are defined as
\be\ba{lcl}\label{truegammamatch}
\gamma_0^{\rmX}&=&\gamma_0^{\rm X,\MSbar },\\[1ex]
\gamma_1^{\rmX}&=&\gamma_1^{\rm X,\MSbar }+2b_0c_1^{\rmX},\\[1ex]
\gamma_2^{\rmX}&=&\gamma_2^{\rm X,\MSbar }+4b_0(c_2^{\rm X}+\gamma_0^{\rm X}k)+2b_1c_1^{\rm X}-2b_0[c_1^{\rm X}]^2.\\
\ea
\ee
All the coefficients are collected in the tables \ref{anomalous dimension} and \ref{rhoX}.
$\gamma^\rmX(g)$ contains a contribution which has been derived from the 
matching (\ref{expandedmatching}) of the HQET operators 
and a contribution that originates from a 
re-parameterization: The matching was originally done at the matching scale 
given in terms of the heavy quark's pole mass $m_Q$. 
Using the ratio $m_Q/\mbar_\ast$, which is known at three-loop 
precision \cite{Melnikov:2000qh,Gray:1990yh,Broadhurst:1991fy} (cf. appendix \ref{poleRGI}), the pole mass can be replaced by $\mbar_\ast$. 
However, given the anomalous dimensions of the currents
to three-loop order, only the one-loop term actually contributes to 
$\gamma^{\rm X}(g)$ (appearing as the piece proportional to $k=-1/(3\pi^2)$ 
in equation (\ref{truegammamatch})). 

The chromo-magnetic operator $\bar Q(x)\vec{S}\cdot\vec B(x)Q(x)$ in the heavy quark expansion is multiplied by the inverse pole mass. 
Since the preferred expansion parameter for HQET in this thesis is $M_Q$ rather than $m_Q$, the corresponding conversion function $C_{\rm spin}(M_Q/\Lambda_{\rm QCD})$ must also include the factors $\mbar_\ast/m_Q$ and $M_Q/\mbar_\ast$ in order to cancel the factor $1/m_Q$ in favor of $1/M_Q$. Using the relation (cf. appendix \ref{poleRGI})
\be\label{Mombar}
M_Q/\mbar_\ast= 
\left[\,2b_0\bar g^2(\mbar_\ast)\,\right]^{-d_0/(2b_0)}
\exp\left\{-\int\limits_0^{\bar g(\mbar_\ast)} \rmd g 
\left[\,\frac{\tau^{\MSbar}(g)}{\beta(g)}-\frac{d_0}{b_0g}\,\right]\right\}\,,
\label{Movermbar}
\ee
where 
\be
\tau^\MSbar(g)=-g^2 d_0-g^4d_1+\dots
\ee
 denotes the quark mass anomalous dimension in the 
$\MSbar$ scheme in QCD known up to four-loop 
precision \cite{Chetyrkin:1997dh,Vermaseren:1997fq}, one then obtains for X=spin
\be\ba{lcl}
\gamma_0^{\rm spin}&=&\gamma_0^{\rm spin,\MSbar }-d_0,\\[1ex]
\gamma_1^{\rm spin}&=&\gamma_1^{\rm spin,\MSbar }-d_1+2b_0(c_1^{\rm spin}+k),\\[1ex]
\ea
\ee
where $d_1=404/(3(4\pi)^4)$.
For the case ${\rm X=PS/V}$, all but the contributions from the matching 
cancel and one gets 
\be
 C_{\rm PS/V}(\mbar_\ast)=
\exp\left\{\int\limits_0^{\bar g(\mbar_\ast)} \rmd g\,
\frac{\gamma^{\rm PS}(g)-\gamma^{\rm V}(g)}{\beta(g)}\right\}\,.
\ee
Using (\ref{Mombar}), one finally changes the argument of the various ${C}_{\rm X}(\mbar_\ast)$ to the
renormalization group invariant ratio $M_Q/\Lambda_{\rm QCD}$
and arrives at expressions for the 
conversion functions
\be
C_{\rm X}(M_Q/\Lambda_{\rm QCD})
\quad{\rm with}\quad{\rm X=PS,V,PS/V\, and\,spin}.
\ee

\begin{figure}
\hspace{-1.cm}
\begin{tabular}{cc}
\psfrag{1t}[c][c][1][0]{$\Lambda/M_Q$}
\psfrag{2t}[c][c][1][0]{$C_{\rm PS}$}
\psfrag{1-loop loop}[c][c][1][0]{$1-{\rm loop}\, \gamma$}
\psfrag{2-loop loop}[c][c][1][0]{$2-{\rm loop}\, \gamma$}
\psfrag{3-loop loop}[c][c][1][0]{$3-{\rm loop}\, \gamma$}
\epsfig{scale=.5,file=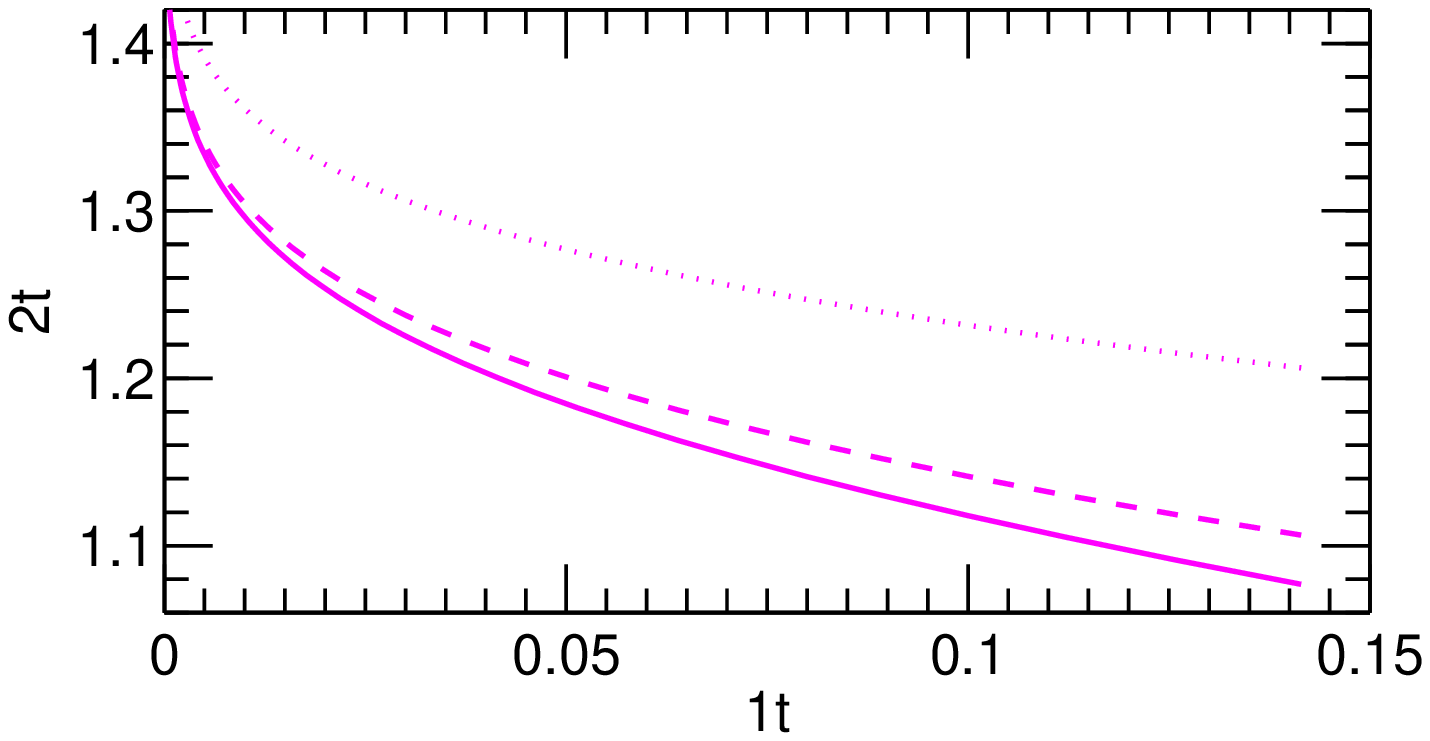}
&
\psfrag{1t}[c][c][1][0]{$\Lambda/M_Q$}
\psfrag{2t}[c][c][1][0]{$C_{\rm V}$}
\psfrag{1-loop loop}[c][c][1][0]{$1-{\rm loop}\, \gamma$}
\psfrag{2-loop loop}[c][c][1][0]{$2-{\rm loop}\, \gamma$}
\psfrag{3-loop loop}[c][c][1][0]{$3-{\rm loop}\, \gamma$}
\epsfig{scale=.5,file=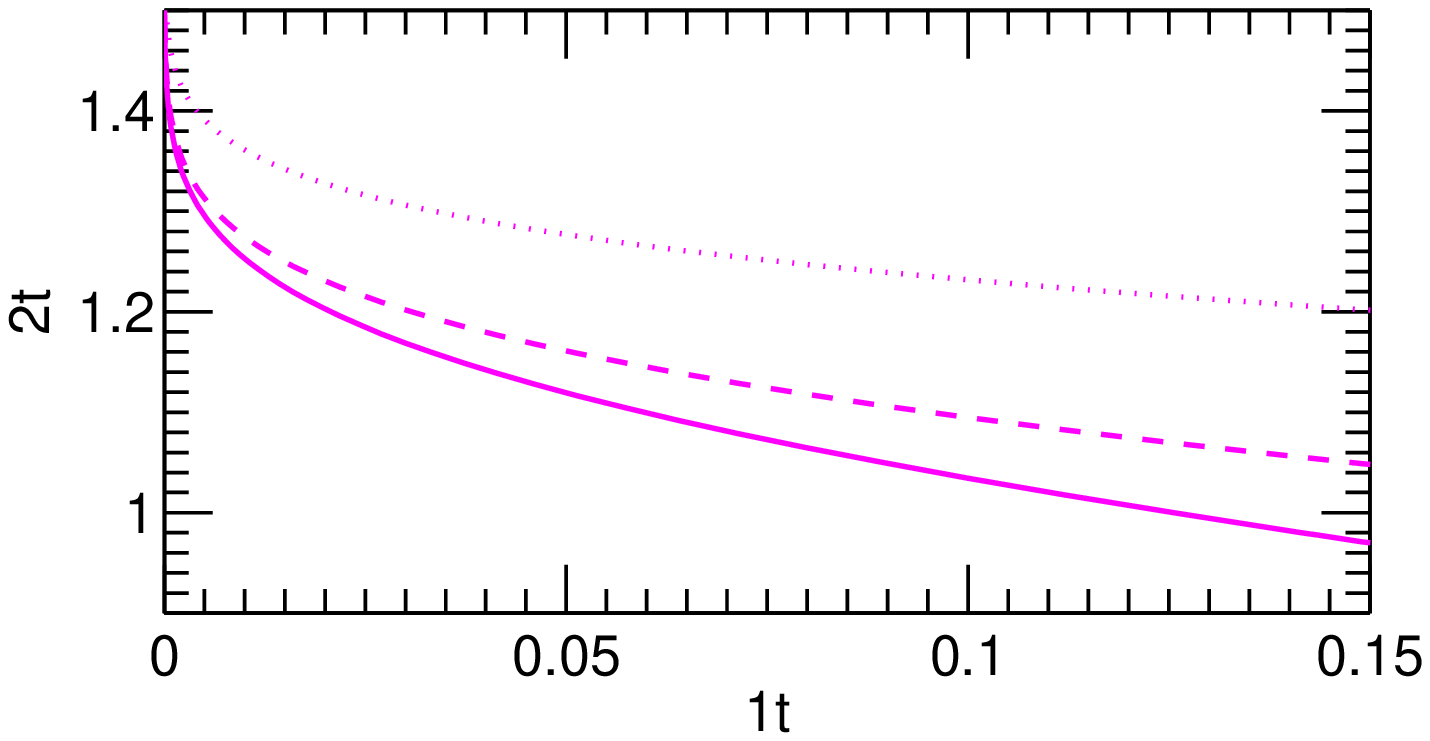}
\\
\psfrag{1t}[c][c][1][0]{$\Lambda/M_Q$}
\psfrag{2t}[c][c][1][0]{$C_{\rm PS/V}$}
\psfrag{1-loop loop}[c][c][1][0]{$1-{\rm loop}\, \gamma$}
\psfrag{2-loop loop}[c][c][1][0]{$2-{\rm loop}\, \gamma$}
\psfrag{3-loop loop}[c][c][1][0]{$3-{\rm loop}\, \gamma$}
\epsfig{scale=.5,file=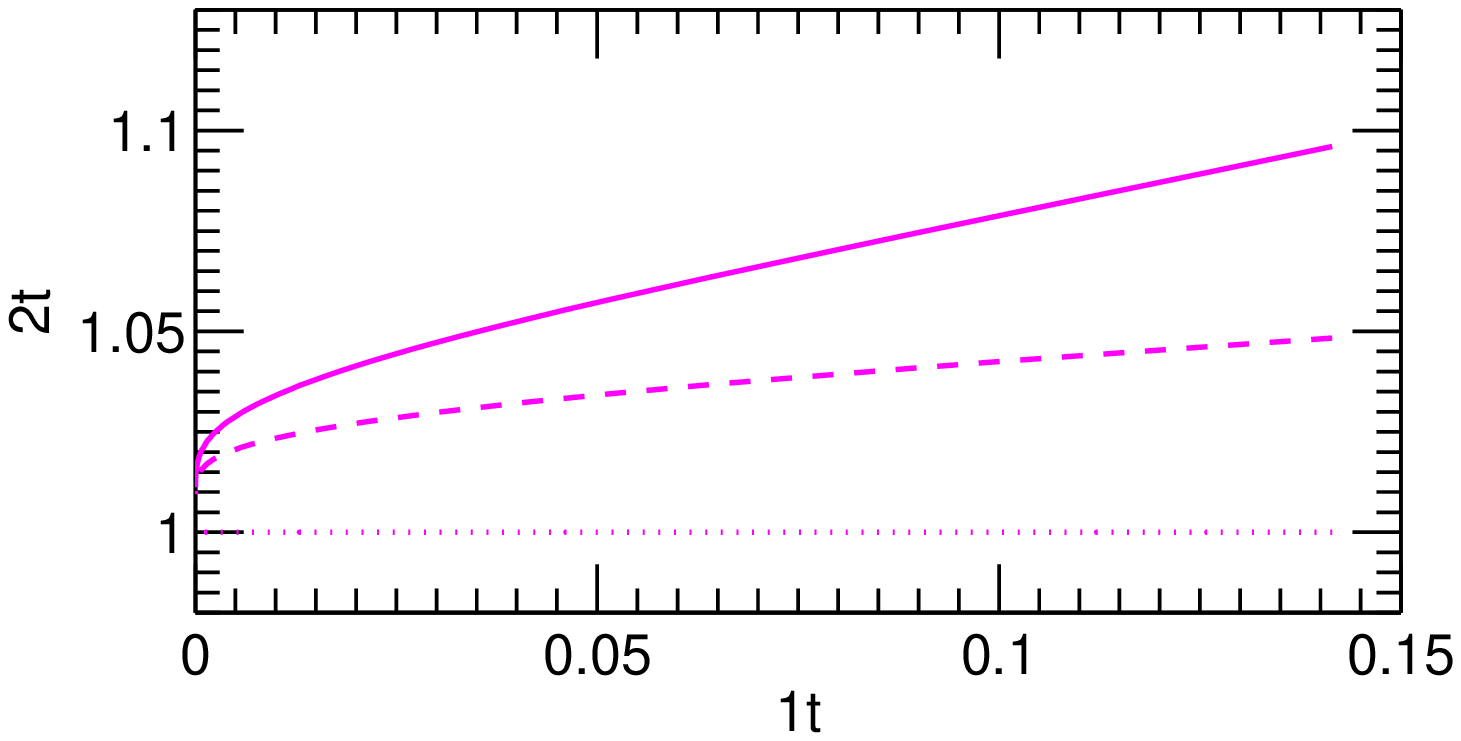}
&
\psfrag{1t}[c][c][1][0]{$\Lambda/M_Q$}
\psfrag{2t}[c][c][1][0]{$C_{\rm spin}$}
\psfrag{1-loop loop}[c][c][1][0]{$1-{\rm loop}\, \gamma$}
\psfrag{2-loop loop}[c][c][1][0]{$2-{\rm loop}\, \gamma$}
\psfrag{3-loop loop}[c][c][1][0]{$3-{\rm loop}\, \gamma$}
\epsfig{scale=.5,file=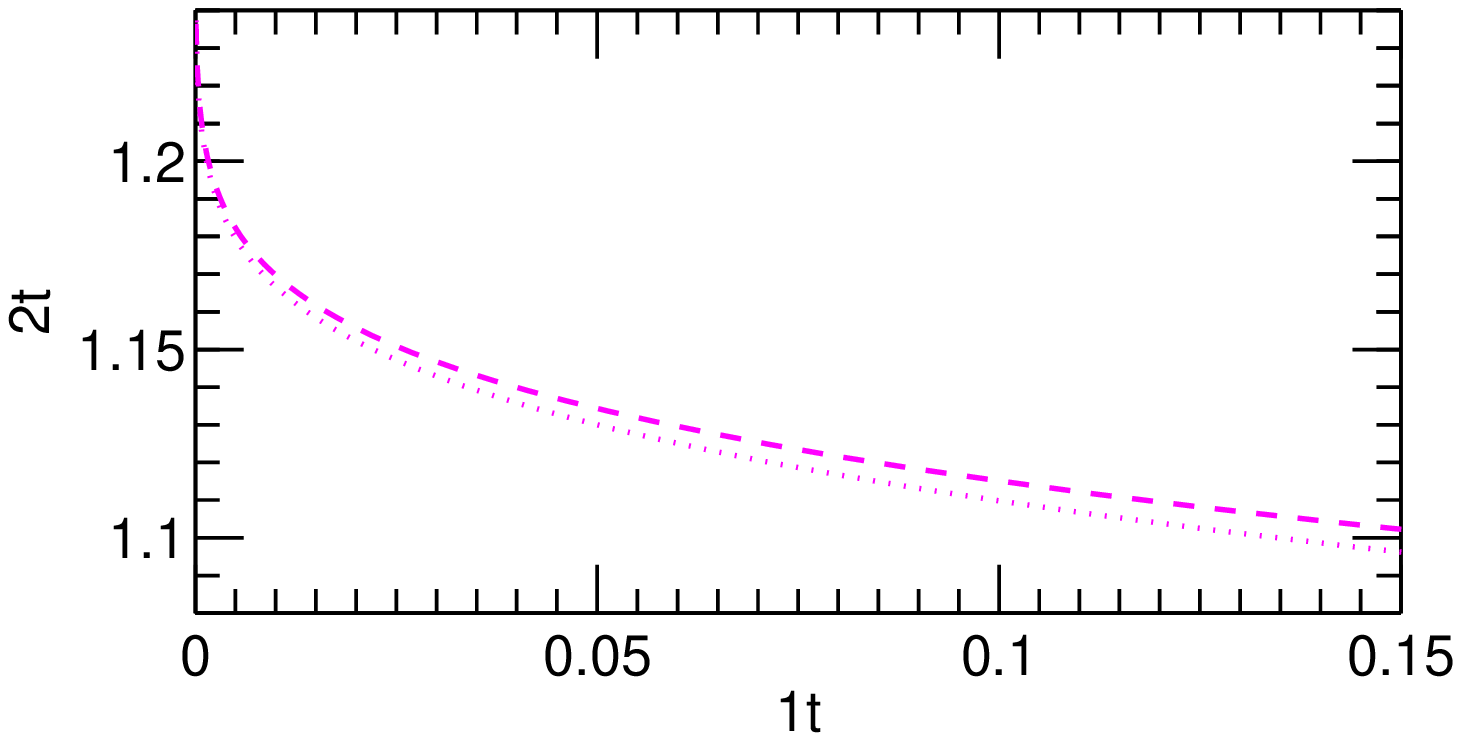}

\end{tabular}
\caption{Plots of the perturbative matching coefficients $C_{\rm X}$, X=PS, V, PS/V and spin for 1-loop, 2-loop and 3-loop $\gamma$-function (dotted, dashed and solid line respectively). }\label{matchingplots}
\end{figure}

For practical purposes, such as repeated use in the fits of the heavy 
quark mass dependence of QCD observables that will be considered, 
a parameterization of all conversion functions in terms of the variable 
\be
x\equiv
\frac{1}{\ln\left(M_Q/\Lambda_{\rm QCD}\right)}
\ee
was determined from a numerical evaluation. This parameterization is suggested by the asymptotic behavior of the conversion functions
\be
C_{\rm X}(M_Q/\Lambda_{\rm QCD})\stackrel{M_Q\to\infty}{\sim}\left(\ln(M_Q/\Lambda_{\rm QCD})\right)^{-\gamma_0^{\rm X}/(2b_0)}\left\{1+O\left(\frac{\ln[\ln(M_Q/\Lambda_{\rm QCD})]}{\ln(M_Q/\Lambda_{\rm QCD})}\right)\right\}
\ee
for X=PS, V and spin. 
The functions decompose into a prefactor encoding the leading
asymptotics as $x\rightarrow0$, multiplied by a polynomial of appropriate
order in $x$. The results are given in table \ref{parametrizations} and plotted in figure \ref{matchingplots}.

The parameterizations of the matching factors deviate little from the numerical data and the error from perturbation theory is not too large (cf. table \ref{parametrizations}). Taking as an estimate for it half the difference between the numerical data based on the $n$-loop $\gamma$-function and the $(n-1)$-loop $\gamma$-function, the functions $C_{\rm PS}$, $C_{\rm V}$, $C_{\rm PS/V}$ have an error in the range of $1\%-5.3\%$. 

The results for $C_{\rm spin}$ with the 1-loop and the 2-loop anomalous dimension differ only little. As this may be accidental, the size of the three-loop term in $C_{\rm PS}$ has been taken as the uncertainty in order to arrive at a conservative estimate for the error. A better error estimate would require the knowledge of $\gamma_2^{\rm spin}$. 

\begin{sidewaystable}
\centering
\small
\begin{tabular}{lclccccc}
\hline
\hline
\\
\multicolumn{3}{c}{parameterization of matching factors}&precision&$\Delta^{\rm max}_{\rm num.}$			&$\Delta^{\rm max}_{\rm pert.}$ 	&$\Delta_{\overline{m}_{b,\overline{\rm MS}}}$&$ \Delta_{\overline{m}_{c,\overline{\rm MS}}}$\\[2ex]
\hline
\\
$ C_{\rm PS}(x)   $&\hspace{-.2cm}$=$&\hspace{-.2cm}$ x^{\gamma^{\rm PS}_0/(2b_0)}(1-0.068x-0.087x^2+0.0779x^3)$ 		&3-loop $\gamma^{\rm PS}$&$10^{-2}\%$  &$1.8\%$  &$0.7\%$  &$1.1\%$\\
\\
$ C_{\rm PS}(x)   $&\hspace{-.2cm}$=$&\hspace{-.2cm}$ x^{\gamma^{\rm PS}_0/(2b_0)}(1-0.065x+0.048x^2)$ 		&2-loop $\gamma^{\rm PS}$\\
\\                                                      
$ C_{\rm V}(x)    $&\hspace{-.2cm}$=$&\hspace{-.2cm}$ x^{\gamma^{\rm V}_0/(2b_0)}(1- 0.196x-0.222 x^2 + 0.193x^3)$	&3-loop $\gamma^{\rm V}$&$10^{-2}\%$  &$5.3\%$  &$1.9\%$  &$3.0\%$\\
\\
$ C_{\rm V}(x)    $&\hspace{-.2cm}$=$&\hspace{-.2cm}$ x^{\gamma^{\rm V}_0/(2b_0)}(1- 0.180x+0.099 x^2)$			&2-loop $\gamma^{\rm V}$\\
\\                                                      
$ C_{\rm PS/V}(x) $&\hspace{-.2cm}$=$&\hspace{-.2cm}$ 1 + 0.124x   + 0.187  x^2 - 0.102 x^3$				&2-loop matching&$10^{-2}\%$  &$2.9\%$  &$1.1\%$  &$1.8\%$\\
\\
$ C_{\rm PS/V}(x) $&\hspace{-.2cm}$=$&\hspace{-.2cm}$ 1 + 0.117x   - 0.043  x^2 $						&1-loop matching\\
\\                                                      
$ C_{\rm spin}(x) $&\hspace{-.2cm}$=$&\hspace{-.2cm}$ x^{\gamma^{\rm spin}_0/(2b_0)}(1+ 0.087x -0.021x^2)$		&2-loop $\gamma^{\rm spin}$&$0.3\%$  &-  &$0.2\%$  &$0.3\%$\\
\\
$ C_{\rm spin}(x) $&\hspace{-.2cm}$=$&\hspace{-.2cm}$ x^{\gamma^{\rm spin}_0/(2b_0)}(1- 0.066x )$		&1-loop $\gamma^{\rm spin}$\\
\\
\hline
\vspace{-.4cm}&&&&&&&\\
Errors:&\multicolumn{7}{l}{The last four columns give the relative error for the parameterizations:}\\
&\multicolumn{7}{l}{\textbullet$\,\Delta^{\rm max}_{\rm num.}$ is the maximal deviation of the parameterization from the numerical data for the conversion functions }\\
&\multicolumn{7}{l}{\hspace{.25cm}at 3-loop precision,}\\
&\multicolumn{7}{l}{\textbullet$\,\Delta^{\rm max}_{\rm pert.}$ is half the maximal difference between the numerical data at $n$ and $(n-1)$-loop order,}\\
&\multicolumn{7}{l}{\textbullet$\,\Delta_{\overline{m}_{b,\overline{\rm MS}}}=\Delta_{\rm pert.}(\overline{m}_{\ast}=4.25{\rm GeV})$,}\\
&\multicolumn{7}{l}{\textbullet$\, \Delta_{\overline{m}_{c,\overline{\rm MS}}}=\Delta_{\rm pert.}(\overline{m}_{ \ast }=1.2{\rm GeV})$.}\\[1.5ex]
\hline
\hline
\end{tabular}\caption{Parameterization of matching factors. $x=1/\ln(M_Q/\Lambda_{\rm QCD})<0.6$.}\label{parametrizations}
\end{sidewaystable}

\section{Combining QCD and HQET}\label{asymptotics}
The decay constants for pseudo scalar mesons ${\rm F_{PS}}(m_{\rm PS})$ and  vector mesons ${\rm F_{V}}(m_{\rm V})$ are defined as
\be\ba{rcl}\label{erratum2}
\langle 0|A_\mu(0)|{\rm P}\rangle&=&ip_\mu{\rm F_{PS}},\\[1ex]
\langle 0|V_i(0)|{\rm V(\lambda)}\rangle&=&i\epsilon^\lambda_im_{\rm V}{\rm F_{V}},
\ea
\ee
where $|{\rm P}\rangle$ and $|{\rm V}\rangle$ are zero-momentum states with the quantum numbers of a pseudo scalar and a vector meson, respectively, and the $|0\rangle$ represents the ground state of the gluonic vacuum. In the case of the vector meson decay constant, $\epsilon^\lambda_\mu$ is a polarization vector. $m_{\rm PS}$ and $m_{\rm V}$ are the associated meson masses.
With the matching coefficients of the last section one obtains the following relations between meson decay constants in the full theory (l.h.s.) and in HQET (r.h.s.):
\be\ba{rcl}
{Y_{\rm PS}\over C_{\rm PS}}\equiv\frac{{\rm F}_{\rm PS}(m_{\rm PS})\sqrt{m_{\rm PS}}}{{C}_{\rm PS}(M_Q/\Lambda_{\rm QCD})}&=&\Phi^{\rm stat}_{\rm PS,\,RGI}+O(1/M_Q), \\
\ea\ee
\be\ba{rcl}
{Y_{\rm V}\over C_{\rm V}}\equiv\frac{{\rm F}_{\rm V}(m_{\rm V})\sqrt{m_{\rm V}}}{{C}_{\rm V}(M_Q/\Lambda_{\rm QCD})}&=&\Phi^{\rm stat}_{\rm V,\, RGI}+O(1/M_Q). \\
\ea\ee
In the static limit no interactions of the gluon field with the heavy quark's spin survive and therefore 
\be
{\Phi^{\rm stat}_{\rm PS,\,RGI}\over \Phi^{\rm stat}_{\rm V,\,RGI}}=1
\ee
holds. $\Phi^{\rm stat}_{\rm PS,\,RGI}$, which is the renormalization group invariant of the matrix element defined in table \ref{rhoX}, has been computed non-perturbatively in \cite{DellaMorte:2003mn} in the static approximation. 

For the ratios of decay constants one expects a behavior like
\be\ba{rcl}
{R\over C_{\rm PS/V}}\equiv\frac{{\rm F}_{\rm PS}(m_{\rm PS})}{{\rm F}_{\rm V}(m_{\rm V})}\frac{\sqrt{m_{\rm PS}}}{\sqrt{m_{\rm V}}}{1\over {C}_{\rm PS/V}(M_Q/\Lambda_{\rm QCD})}&=&1+O(1/M_Q).\\
\ea
\ee

Furthermore, HQET predicts the relation
\be\label{massformula}
m_{\rm X} = m_Q + \bar\Lambda + {1 \over 2m_Q}\Delta m^2 + O(1/m_Q^2)\;{\rm for\; X=PS,\,V}.
\ee
between the heavy-light meson mass $m_{\rm X}$ and the heavy quark mass \cite{Falk:1993wt}, where
\be
\Delta m^2 = -\lambda_1 + 2\left[J(J+1)-{3\over 2}\right] \lambda_2.
\ee
$J$ is the total spin of the meson, i.e. $J=0$ for pseudo scalar mesons and  $J=1$ for vector mesons.
$\bar\Lambda$ is a parameter, that describes the properties of the light degrees of freedom in the background of the static color source provided by the heavy quark and and $\lambda_1\propto \langle {\rm X}|-\bar Q (i\vec{D})^2 Q|{\rm X}\rangle$ and $\lambda_2 \propto \Phi^{\rm spin}_{\rm X}$. 
The mass splitting 
\be
\ba{rcl}
{\Delta m\over C_{\rm spin}}\equiv{m_{\rm V}-m_{\rm PS}\over {C}_{\rm spin}(M_Q/\Lambda_{\rm QCD})}
&=& O(1/M_Q),\\
\ea\ee
where the lowest order contribution comes from $\lambda_2$,
is therefore expected to vanishes in the limit $M_Q\to \infty$.
 
By producing data for the l.h.s. of these equations from relativistic lattice QCD for the range of heavy-light meson masses accessible to current lattice simulations, it is the scope of this work to try to  
\begin{itemize}
\item estimate, down to which heavy quark mass in heavy-light meson systems observables scale with $1/M_Q$ without sizeable contributions from the higher orders
\item check the compatibility of the relativistic simulations with the static approximation from an interpolation in $1/M_Q$
\item obtain a value for the decay constant and the mass splitting for a meson containing a $b$-quark from an interpolation in the mass, i.e. including the prediction from HQET in the static limit.
\item estimate the order of the $1/M_Q$ corrections to the static limit from a fit-ansatz \'a la
\be
{{\rm F}_{\rm PS,V}\over C_{\rm PS,V}}=a_0+{a_1\over M_Q}+\dots\;.
\ee
\end{itemize}

\chapter{Masses and meson decay constants on the lattice}\label{technicalissues}
It has been shown in the previous chapter that the validity of
HQET can be tested by exploring the mass dependence of meson observables in QCD.
At their physical point such observables are also an important input for
the phenomenology of the Standard Model.

A particularly suitable framework in which a non-perturbative determination of mesonic observables is possible is the Euclidean QCD Schr\"odinger Functional on the lattice \cite{Luscher:1992an,Sint:1994un}. It has been demonstrated, that its Monte-Carlo simulation allows for the determination of mesonic observables with smaller statistical fluctuations than with conventional methods like lattice QCD on the torus. In addition, systematic errors introduced by excited states can be estimated more reliably \cite{Guagnelli:1999zf}.

First, the $O(a)$-improved QCD Schr\"odinger Functional will be introduced in this chapter. Then, the meson mass and the decay constant of pseudo scalar mesons and vector mesons, and also the quark mass will be expressed in terms of renormalized and improved quark bilinear currents, whose expectation values can be evaluated in a Monte-Carlo simulation of the Schr\"odinger Functional. Finally, expressions of these currents in terms of quark propagators will be derived for the direct implementation in a computer program. 
\section{The Schr\"odinger Functional - geometry and fields}
The QCD Schr\"odinger Functional on the lattice is the partition function
\be\label{SFdef}
\mathcal{Z}[C,C^\prime,\bar\rho,\rho,\bar\rho^\prime,\rho^\prime]=\int_{\bar\psi,\psi,U}e^{-S[U,\bar\psi,\psi]},
\ee
discretized on a hyper-cubic 4-dimensional Euclidean space-time cylinder
\be
\Gamma_{\rm E}=\left\{x|\;{x/ a}\in \mathbbm{Z};\,0\le x_0\le T;\,0\le x_k<L;\,k=1,2,3\right\}
\ee
with boundaries in the time direction.
Here, $S[U,\bar \psi,\psi]$ is the QCD action discretized on $\Gamma_{\rm E}$ and will be specified in the next section. The integration in (\ref{SFdef}) is the short form of
\be
\int_{\bar\psi,\psi,U}=\int\prod\limits_x d\bar\psi(x) d\psi(x) dU(x).
\ee
The quark fields $\psi_{a,A,\alpha}(x)$ are assignments of Grassman numbers to each lattice site $x\in\Gamma_{\rm E}$ and carry the flavor-, Dirac- and color-indices $a,A,\alpha$ respectively. The gauge fields $U(x,\mu)\in {\rm SU(3)}$ are associated to the links between two adjacent sites $(x,x+\hat{\mu})$, $\hat{\mu}$ being a unit vector in the $\mu$-direction.

In the spatial directions, the gauge fields $U(x,\mu)$ obey periodic boundary conditions 
\be
U(x+L\hat{k},k)=U(x,k),\;\; k=1,2,3. \\
\ee
while the fermion fields $\psi(x)$ are $\theta$-periodic 
\be\label{fermbound}
\psi(x+L\hat{k})=e^{i\theta_k}\psi(x)\;{\rm and}\;\bar\psi(x+L\hat{k})=e^{-i\theta_k}\bar\psi(x),\; k=1,2,3.  \\
\ee

The functional explicitly depends on Dirichlet conditions at the boundaries in the time direction. In particular, the gauge fields on the time slices $x_0=0$ and $x_0=T$ are set to identity matrices\footnote{Other choices of the boundary conditions have been used to define the running coupling constant in the Schr\"odinger Functional scheme, that scales with the size of the lattice volume like $\mu=1/L$ \cite{Luscher:1991wu,Luscher:1992an}.}
\be
U(x,k)_{|x_0=0}=\exp\{C_k\}=U(x,k)_{|x_0=T}=\exp\{C^\prime_k\}=\mathbbm{1}_{3\times3}\;{\rm for}\,k=1,2,3.
\ee
As the Dirac equation is a first order differential equation, only two of the four components of the Dirac spinors $\psi(x)$ on the boundaries have to be prescribed \cite{Sint:1994un}. With $P_\pm={1\over 2}(1\pm\gamma_0)$, the boundary conditions are 
\be
\ba{rclrcl}
P_+\psi(x)_{|x_0=0}&=&\rho(\vec{x}),&P_-\psi(x)_{|x_0=T}&=&\rho^\prime(\vec{x}),\\
\\
\overline{\psi}(x)P_{-|x_0=0}&=&\overline{\rho}(\vec{x}),&\overline{\psi}(x)P_{+|x_0=T}&=&\overline{\rho}^\prime(\vec{x}).
\ea
\ee

\section{Lattice action}
As in the continuum, a generic lattice action for QCD consists of a gauge action $S_{\rm G}[U]$, and a quark action $S_{\rm F}[U,\bar\psi,\psi]$, describing the dynamics of massive quarks coupled to the gauge field,
\be
S[U,\bar\psi,\psi]=S_{\rm G}[U]+S_{\rm F}[U,\bar\psi,\psi].
\ee
Wilson suggested the plaquette action, which in the version adapted for the Schr\"odinger Functional reads 
\be\label{WilsonGauge}
S_{\rm G}[U]={1 \over g_0^2}\sum\limits_p w(p)\Tr\left\{1-U(p)\right\}.
\ee
The sum runs over all oriented plaquettes $p$. A plaquette is the smallest closed loop of link variables $U(p)=U(x,\mu)U(x+\hat{\mu},\nu)U^{-1}(x+\hat{\nu},\mu)U^{-1}(x,\nu)$. The coefficient $w(p)$ will be specified in the next section.

The quark action is
\be
S_{\rm F}[U,\overline\psi,\psi]=a^4\sum\limits_x \overline\psi(x)\left(D+m_0\right)\psi(x).
\ee
Here, $D$ is the Wilson Dirac operator 
\be\label{WilsonDO}
D={1\over 2}\left\{\gamma_\mu\left(\nabla_\mu^\ast+\nabla_\mu)-a\nabla^\ast_\mu\nabla_\mu\right)\right\}
\ee
with the covariant lattice forward and backward derivatives 
\be
\ba{rcl}
\nabla_\mu\psi(x)&=&{1\over a}\left\{\lambda_\mu U(x,\mu)\psi(x+a\hat{\mu})-\psi(x)\right\},\\
\\
\nabla^\ast_\mu\psi(x)&=&{1\over a}\left\{\psi(x)-\lambda^\ast_\mu U^{-1}(x-a\hat{\mu},\mu)\psi(x-a\hat{\mu})\right\}.\\
\ea
\ee
$\lambda_\mu=e^{ia\theta_\mu/L}$, with $\theta_0=0$ and $-\pi<\theta_k\le\pi$ ($k=1,2,3$), is a phase factor that incorporates the spatial boundary conditions (\ref{fermbound}) for the fermion fields.
Also the covariant lattice derivatives acting to the left side,
\be
\ba{rcl}
\bar\psi(x)\stackrel{\leftarrow}{\nabla}_\mu&=&{1\over a}\left\{\lambda^\ast_\mu \bar\psi(x+a\hat\mu)U^{-1}(x,\mu)-\bar\psi(x)\right\},\\
\\
\bar\psi(x)\stackrel{\leftarrow}{\nabla}^{\ast}_\mu&=&{1\over a}\left\{\bar\psi(x)-\lambda_\mu\bar\psi(x-a\hat\mu) U(x-a\hat{\mu},\mu)\right\},\\
\ea
\ee
can be defined. They will be useful at a later point.

With the hopping parameter $\kappa=(8+2am_0)^{-1}$, the fermion fields can be rescaled like $\psi(x)\to\sqrt{2\kappa}\psi(x)$ and $\bar\psi(x)\to\sqrt{2\kappa}\bar\psi(x)$.
The fermionic part of the action can then be rewritten as
\be
S_F[U,\overline\psi,\psi]=a^4\sum\limits_{x,y} \overline\psi(x)M\psi(x),
\ee
with
\be\ba{rcl}
M\psi(x)& =& \psi(x)-\kappa\sum\limits_{\mu=0}^3\left\{\lambda_\mu U(x,\mu)(1-\gamma_\mu)\psi(x+a\hat\mu)\right.\\
&+&\left.\lambda^\ast_\mu U^{-1}(x-a\hat\mu,\mu)(1+\gamma_\mu)\psi(x-a\hat\mu)\right\}.
\ea
\ee
\section{$O( a )$-improvement }
In Wilson's original formulation of lattice fermions \cite{Wilson:1974sk}, discretization effects due to the finite lattice cut-off $\pi/a$ turn up at leading order in $a$. At the same time, in the Schr\"odinger Functional, also the Wilson gauge action is affected by ${O}(a)$ cut-off effects due to the presence of the boundary. Using an $O(a)$-improved action and $O(a)$-improved expressions for observables, lattice-discretization effects can be reduced (ideally removed) at leading order in $a$ and the approach to the continuum limit is accelerated.

The basic idea was first formulated by Symanzik \cite{Symanzik:1983dc,Symanzik:1983gh}, who showed for the $\phi^4$-theory, that the corresponding lattice field theory can be described by a local effective continuum field theory with the action
\be
S_{\rm eff}=\int dx^4\left\{\mathcal{L}_0+a\mathcal{L}_1+a^2\mathcal{L}_2+\dots\right\}.
\ee
$\mathcal{L}_0$ is the generic continuum Lagrangian. The terms $\mathcal{L}_k$ ($k=1,2,3,\dots$) are combinations of local operators of dimension $4+k$, all of them respecting the exact discrete symmetries of the lattice theory. In the effective continuum theory, the lattice fields are represented by the renormalized effective fields 
\be
\phi_{\rm eff}(x)=\phi_0(x)+a\phi_1(x)+a^2\phi_2(x)+\dots.
\ee
The fields $\phi_k(x)$ ($k=0,1,2,3,\dots$) also must have the appropriate dimension and respect the same lattice symmetries. 

The $O(a)$-improved theory can now be obtained by adding suitable counter terms with properly tuned improvement coefficients to the lattice action and to the lattice fields, such that the ${O}(a)$-terms in the effective continuum action and the effective continuum fields are canceled. In the cases of interest for this work, only observables on the mass shell will be considered. Thus, \emph{on-shell improvement} can be applied, where the number of counter-terms, that are necessary to improve the action and the fields at $\orda$ can be reduced by exploiting the classical equations of motion. 

$\orda$-improvement has shown to be an efficient tool for lattice-field theory \cite{Jansen:1996ck}. The improvement coefficients have been determined non-perturbatively \cite{Luscher:1997jn,Bhattacharya:2001ks} in most of the cases, or otherwise are available from perturbation theory \cite{Luscher:1997jn,Sint:1997jx,Bode:1999sm}. 

One thing to mention here is, that Symanzik's proof, that $\orda$-improvement works, relies on perturbation theory. On the lattice one therefore always assumes, that $\orda$-improvement is applicable beyond perturbation theory. Experience from lattice simulations of QCD however support, that this assumption is justified \cite{Jansen:1996ck,Heitger:1999dw,Gockeler:1997jx,Wittig:1998bk,Edwards:1998nh}.
\subsection{Improved action}
To cancel the $\orda$-effects in the Wilson ${\rm SU(3)}$ gauge action in the Schr\"odinger Functional, one has to choose the weight $w(p)$ in (\ref{WilsonGauge}) as \cite{Sint:1997jx,Bode:1999sm}
\be
w(p)=\left\{\ba{ll}
{1\over 2}c_s(g_0)&{\rm if}\; p\; \textrm{ is a spatial plaquette at}\; x_0=0\; {\rm or}\; x_0=T,\\
c_t(g_0)&{\rm if}\; p\; \textrm{ is a time-like plaquette attached to a boundary plane},\\
1&{\rm elsewhere}.
\ea
\right.
\ee
$c_s(g_0)$ can be dropped for simulations with vanishing background field, as it will be the case throughout this thesis, while $c_t(g_0)$ has been determined in perturbation theory at 2-loop order \cite{Bode:1999sm} and is given in appendix \ref{parametrizationtables}.

For the $\orda$-improvement of the fermionic part of the QCD Schr\"odinger Functional, volume and boundary counter terms $\delta S_v[U,\bar\psi,\psi]$ and $\delta S_b[U,\bar\psi,\psi]$ have to be added to the action which then reads 
\be
S^I_{\rm F}[U,\bar\psi,\psi]=S_{\rm F}[U,\bar\psi,\psi]+\delta S_b[U,\bar\psi,\psi]+\delta S_v[U,\bar\psi,\psi].
\ee
The superscript $I$ from now on indicates ${O}(a)$-improvement.
The boundary improvement term is
\be\ba{rcl}\label{boundaryimprovement}
\delta S_{b}[U,\bar\psi,\psi]&=&a^4\sum\limits_{\vec{x}}\left\{(\tilde{c}_s(g_0)-1)\left[\hat\mathcal{O}_s(x)+\hat\mathcal{O}_s^\prime(x)\right]\right.\\
\\
&+&\qquad\;\;\;\;\left.(\tilde{c}_t(g_0)-1)\left[\hat\mathcal{O}_t(x)-\hat\mathcal{O}_t^\prime(x)\right]\right\},
\ea
\ee
where
\begin{eqnarray}
\hat\mathcal{O}_s(x)&=&\bar\rho(\vec{x})\oh \gamma_k\left(\nabla^\ast_k+\nabla_k\right)\rho(\vec{x}),\\
\hat\mathcal{O}_s^\prime(x)&=&\bar\rho^\prime(\vec{x})\oh \gamma_k\left(\nabla^\ast_k+\nabla_k\right)\rho^\prime(\vec{x}),\\
\hat\mathcal{O}_t(x)&=&\left\{\bar\psi(x)P_+\nabla_0^\ast\psi(x)+\bar\psi(x)\stackrel{\leftarrow}{\nabla}^\ast_0P_-\psi(x)\right\}_{|x_0=a}\;{\rm and}\\
\hat\mathcal{O}_t^\prime(x)&=&\left\{\bar\psi(x)P_-\nabla_0\psi(x)+\bar\psi(x)\stackrel{\leftarrow}{\nabla}_0P_+\psi(x)\right\}_{|x_0=T-a}.\\
\end{eqnarray}

The volume- or Sheikholeslami-Wohlert-term \cite{Sheikholeslami:1985ij}, has the form
\be
\delta S_{v}[U,\bar\psi,\psi]=a^5\sum\limits_{x_0=a}^{T-a}\sum\limits_{\vec{x}}\bar\psi(x)\, c_{\rm SW}(g_0){i \over 4}\sigma_{\mu\nu}\cdot \hat{F}_{\mu\nu}(x)\psi(x).
\ee
$\sigma_{\mu\nu}={i\over 2}[\gamma_\mu,\gamma_\nu]$
and
$\hat{F}_{\mu\nu}$ is the symmetric lattice field strength tensor 
\be
\hat{F}_{\mu\nu}={1\over {8a^2}}\left\{Q_{\mu\nu}(x)-Q_{\nu\mu}(x)\right\}
\ee
with 
\be\ba{rcl}
Q_{\mu\nu}(x)&=&U(x,\mu)U(x+a\hat{\mu},\nu)U(x+a\hat{\nu},\mu)^{-1}U(x,\nu)^{-1}\\
\\
&&+U(x,\nu)U(x-a\hat{\mu}+a\hat{\nu},\mu)^{-1}U(x-a\hat{\mu},\nu)^{-1}U(x-a\hat{\mu},\mu)\\
\\
&&+U(x-a\hat{\mu},\mu)^{-1}U(x-a\hat{\mu}-a\hat{\nu},\nu)^{-1}U(x-a\hat{\mu}-a\hat{\nu},\mu)U(x-a\hat{\nu},\nu)\\
\\
&&+U(x-a\hat{\nu},\nu)^{-1}U(x-a\hat{\nu},\mu)U(x+a\hat{\mu}-a\hat{\nu},\nu)U(x,\mu)^{-1}.
\ea\ee
$Q_{\mu\nu}(x)$ is the sum over four plaquettes in a hyper plane, all having the site $x$ in common and therefore resembling a clover leaf.

The improved fermion action can equivalently be expressed in terms of an action with the improved Dirac operator 
\be
D^I=D+\delta D_v + \delta D_b,
\ee
where 
\be
\delta D_v\psi(x)=a c_{\rm SW}(g_0){i \over 4}\sigma_{\mu\nu}\cdot \hat{F}_{\mu\nu}(x)\psi(x)
\ee
and
\be\ba{rcl}
\delta D_b\psi(x)& =& \left(\tilde{c}_t(g_0)-1\right){1\over a}\bigg\{\delta_{x_0,a}\left[\psi(x)-U(x-a\hat{0})^{-1}P_+\psi(x-a\hat{0})\right]\\
\\
&&\hspace{.95cm}+\,\delta_{x_0=T-a}\left[\psi(x)-U(x,0)P_-\psi(x+a\hat{0})\right]\bigg\}.
\ea
\ee
As for the unimproved Dirac operator, a parameterization in terms of the hopping parameter $\kappa$ is possible, namely
\be
\left(D^I+m_0\right)\psi(x)={1\over 2\kappa}\left(M+\delta M\right)\psi(x)={1\over 2\kappa}M^I\psi(x)
\ee
with $\delta M=2\kappa\left(\delta D_v + \delta D_b\right)$.

The improvement constant $\tilde{c}_t(g_0)$ has been determined in perturbation theory at 1-loop order \cite{Sint:1997jx} and $c_{\rm SW}(g_0)$ has been determined non-perturbatively \cite{Luscher:1997ug}. Both constants are given in appendix \ref{parametrizationtables}.
\subsection{Improved fields}
As it will be explained in section \ref{SFobservables}, all phenomenological observables that have been determined in this work, can be expressed in terms of quark bilinear currents. Of particular interest are the iso-vector axial current $A_\mu^a(x)=\bar\psi(x)\gamma_\mu\gamma_5{\tau^a\over2}\psi(x)$, the iso-vector vector current $V_\mu^a(x)=$ $\bar\psi(x)\gamma_\mu{\tau^a\over 2}\psi(x)$ and the iso-vector pseudo scalar density $P^a(x)=\bar\psi(x)\gamma_5 {\tau^a\over 2}\psi(x)$.
Except for the pseudo scalar density, these currents get contributions from dimension five operators under improvement \cite{Luscher:1996sc}. The improved axial current is 
\be\label{imprpscurr}
(A_\mu^a)^I(x)=A_\mu^a(x)+c_A(g_0) a \tilde\partial_\mu P^a(x),
\ee
where 
\be\label{symderiv}
\tilde\partial_\mu f(x)={f(x+a\hat\mu)-f(x-a\hat\mu)\over 2a}
\ee
is the symmetric lattice derivative. For the vector cur\-rent $V_\mu^a(x)=$ $\bar\psi(x)\gamma_\mu{\tau^a\over 2}\psi(x)$, one adds the symmetric derivative of the tensor current
\be
T_{\mu\nu}^a=\bar\psi(x)\sigma_{\mu\nu}{\tau^a\over 2}\psi(x),
\ee
\be\label{imprvcurr}
(V_\mu^a)^I(x)=V_\mu^a(x)+{i\over 2}c_V(g_0)a\tilde\partial_\nu T_{\mu\nu}^a
\ee
$c_A(g_0)$ and $c_V(g_0)$ have been determined non-perturbatively in \cite{Luscher:1997ug} and \cite{Guagnelli:1998db,Bhattacharya:2001ks}, and their parameterizations in terms of the bare coupling\footnote{In the case of $c_V(g_0)$, the parameterization derived in \cite{Harada:2002jh} has been used here} are given in appendix \ref{parametrizationtables}. 
\section{Fermionic observables in the Schr\"o\-ding\-er\newline Func\-tio\-nal}\label{fermobs}
Summarizing the last section, the ${O}(a)$-improved Euclidean QCD Schr\"odinger Functional with the lattice action 
\be
S^I[U,\bar\psi,\psi]=S^I_{\rm G}[U]+S^I_{\rm F}[U,\bar\psi,\psi].
\ee
and vanishing background field is
\be
\mathcal{Z}[\rho,\bar{\rho},\rho^\prime,\bar{\rho}^\prime]=\int_{\bar{\psi},\psi,U}e^{-S^I[U,\bar\psi,\psi]}.
\ee
$S_{\rm G}^I$ and $S_{\rm F}^I$ are the improved Wilson gauge and fermion action introduced before. A source term
\be
S_{\rm S}[\bar{\psi},\psi,\bar\eta,\eta]=a^4\sum\limits_{0<x_0<T}\sum\limits_{\vec{x}}\left\{\bar\eta(x)\psi(x)+\bar\psi(x)\eta(x)\right\}
\ee
will now be added to the action.  
One can then identify
\be\ba{rclrcl}\label{funcderiv}
\psi(x)&=&\frac{\delta}{\delta\bar\eta(x)},&\bar\psi(x)&=&-\frac{\delta}{\delta\eta(x)},\\
\\
\zeta(\vec{x})&=&\frac{\delta}{\delta\bar\rho(\vec{x})},&\bar\zeta(\vec{x})&=&-\frac{\delta}{\delta\rho(\vec{x})},\\
\\
\zeta^\prime(\vec{x})&=&\frac{\delta}{\delta\bar\rho^\prime(\vec{x})},&\bar\zeta^\prime(\vec{x})&=&-\frac{\delta}{\delta\rho^\prime(\vec{x})},\\
\ea\ee
where $\zeta(\vec{x})$, $\bar\zeta(\vec{x})$, $\zeta^\prime(\vec{x})$ and $\bar\zeta^\prime(\vec{x})$ can be interpreted as {boundary fields}. Expectation values of operators that are polynomials $\mathcal{O}$ in these fields are defined as
\be\ba{rcl}\label{genfunc}
\langle\mathcal{O}\rangle&=&\left\{{1\over \mathcal{Z}}\int_{\bar{\psi},{\psi},U}\mathcal{O}\,e^{-S^I[U,\bar\psi,\psi]-S_{\rm S}[\bar{\psi},\psi,\bar\eta,\eta]}\right\}_{|\rho=\bar\rho=\rho^\prime=\bar\rho^\prime=\bar\eta=\eta=0}.\\
\ea
\ee
With $\tilde\mathcal{O}$ being the polynomial $\mathcal{O}$, where all fields replaced by the associated functional derivative, the expectation value can be obtained as
\be\ba{rcl}\label{genfunc}
\langle\mathcal{O}\rangle&=&\left\{\tilde\mathcal{O}\ln \mathcal{Z}[\rho,\bar{\rho},\rho^\prime,\bar{\rho}^\prime,\bar\eta,\eta]\right\}_{|\rho=\dots=\eta=0}.
\ea
\ee

For the following, only the fermionic contribution to $\langle\mathcal{O}\rangle$ on a given gauge background, 
\be
[\mathcal{O}]_{\rm F}[U]={1\over \mathcal{Z_{\rm \rm F}}}\int_{\bar\psi,\psi} \mathcal{O}\, e^{-S^I_{\rm F}[U,\bar\psi,\psi]},
\ee
with 
\be\label{funcintF}
\mathcal{Z}_{\rm F}[U,\rho,\bar{\rho},\rho^\prime,\bar{\rho}^\prime,\bar\eta,\eta]=\int_{\bar\psi,\psi} e^{-S^I_{\rm F}[U,\bar\psi,\psi]-S_{\rm S}[\bar\psi,\psi,\bar\eta,\eta]}
\ee
will be considered. The relation to the full expectation value is given by
\be
\langle\mathcal{O}\rangle=\langle[\mathcal{O}]_{\rm F}\rangle_{\rm G},
\ee
where the subscript G indicates, that the expectation value has to be evaluated with respect to the path integral of the quenched theory.

\section{Renormalization}\label{Renormalization}
Usually, a mass independent renormalization scheme is employed  \cite{Luscher:1996sc} in which the renormalization constants are independent of the quark mass and therefore the renormalization group equations, which describe the scale dependence of renormalized quantities, take a particularly simple form. 
In this scheme, the theory is parameterized around a critical line $\kappa_{crit}(g_0)$ for which the subtracted bare quark mass \be\label{hoppingparam}
am_q=\oh\left({1\over\kappa}-{1\over\kappa_{crit}(g_0)}\right),
\ee
and also the renormalized quark mass is zero. 

All the bare parameters and fields receive a multiplicative renormalization factor. However, when taking the continuum limit, uncanceled ${O}(am_q)$-cutoff effects arise \cite{Luscher:1996sc}. In the improved mass independent renormalization scheme, one subtracts these effects by a multiplicative counter term of the form $1+b(g_0)am_q$ and defines \cite{Luscher:1996sc} the improved renormalized subtracted quark mass (with flavor index $i=s,c,b$)
\be\label{rensub}
m_{{\rm R},i}=Z_m(g_0)\tilde m_{q,i}=Z_m(g_0)m_{q,i}(1+b_m(g_0^2)am_{q,i}).
\ee

The renormalization of the heavy-light quark currents can be discussed conveniently after introducing the off-diagonal iso-matrices
\be\label{offdiag}
\tau^\pm=\tau^1\pm i\tau^2.
\ee
Then, the improved renormalized axial current 
\be\label{imprpscurr}
(A_{\mu}^\pm)^I_{\rm R}(x)=Z_A(g_0)(1+b_A(g_0) \oh(am_{q,h}+am_{q,l}))(A_{\mu}^\pm)^I(x),
\ee
the vector current
\be\label{imprvcurr}
(V_{\mu}^\pm)^I_{\rm R}(x)=Z_V(g_0)(1+b_V(g_0)\oh( am_{q,h}+am_{q,l}))(V_{\mu}^\pm)^I(x),
\ee
and the renormalized pseudo scalar density
\be\label{imprpdens}
(P^\pm)_{\rm R}(x,\mu)=Z_P(g_0,\mu)(1+b_P(g_0)\oh( am_{q,h}+am_{q,l}))P^\pm(x)
\ee
can be defined, where $h$ and $l$ indicate a heavy and light flavor respectively.
The mass renormalization constant $Z_m(g_0)$ and the scale dependent renormalization constant $Z_P(g_0,\mu)$ will only be needed in terms of the composite renormalization factor $Z(g_0)=Z_m(g_0)Z_A(g_0)Z_P^{-1}(g_0)$ (cf. section \ref{quarkmasses}). $Z(g_0)$ has been determined non-perturbatively in \cite{Guagnelli:2000jw} and $Z_A(g_0)$ and $Z_V(g_0)$ in \cite{Luscher:1997jn} and \cite{Capitani:1998mq}. Their parameterizations in terms of the bare coupling constant $g_0$ have been summarized in table \ref{renormalizationtab} in appendix \ref{parametrizationtables}.
The improvement constants $b_A(g_0)$, $b_V(g_0)$, $b_P(g_0)$ and $b_m(g_0)$ have to be tuned in order to compensate for the ${O}(am_q)$-terms. $b_A(g_0)$ and $b_V(g_0)$ have been determined non-perturbatively in \cite{Luscher:1997ug} and $b_m(g_0)$ in \cite{Guagnelli:2000jw}. $b_P(g_0)$ will only be needed in the difference $b_A(g_0)-b_P(g_0)$, which also has been determined in \cite{Guagnelli:2000jw}. All the $b$-factors are given in appendix \ref{parametrizationtables} in table \ref{improvementtab}.
\section{Meson masses, decay constants and quark masses in the Schr\"o\-din\-ger Functional}\label{SFobservables}
In this section, expressions for the pseudo scalar (X=PS) and the vector meson (X=V) decay constants ${\rm F}_{{\rm X}}$ and the corresponding meson masses $m_{{\rm X}}$ will be derived. In the the Schr\"odinger Functional, this can be done in terms of correlation functions of quark-bilinear currents $O$ carrying the appropriate quantum numbers,
\be\ba{lrcl}\label{unimpcurr}
{\rm X=PS}:&\; O&\equiv&{A^\pm_0}(x)=\bar\psi(x)\gamma_0\gamma_5{\tau^\pm\over 2}\psi(x)\;{\rm and} \\
\\
{\rm X=V }:&\; O&\equiv&{V^\pm_i}(x)=\bar\psi(x)\gamma_i{\tau^\pm\over 2}\psi(x). \\
\ea\ee
 Furthermore, expressions for renormalized quark masses will be given in this framework. 

The derivation is based on the transfer matrix formalism which has been formulated for the QCD Schr\"odinger Functional in \cite{Sint:1994un}. Although the derivation does not carry over rigorously to the improved theory, universality implies that the construction is valid there as well and that one can replace the currents (\ref{unimpcurr}) by the improved currents (\ref{imprpscurr}) and (\ref{imprvcurr})
\cite{Guagnelli:1999zf}.

\subsection{Correlation functions in the Schr\"odinger Functional}
The lattice Schr\"odinger Functional for QCD (\ref{SFdef}) \cite{Sint:1994un} can be represented as the QCD transition matrix element\footnote{The boundary fields $C$ and $C^\prime$ are mentioned for completeness here and will again be set to $0$ in the following.}
\be
\mathcal{Z}[C, C^\prime,\bar\rho,\rho,\bar\rho^\prime,\rho^\prime]=\langle i^\prime|\left(e^{-\mathbbm{H}}\right)^T\mathbbm{P}|i\rangle
\ee
between an initial state $|i\rangle$ and a final state $\langle i^\prime|$ (at times $x_0=0$ and $x_0=T$ respectively). The exponential $e^{-a\mathbbm{H}}$, with the QCD-Hamiltonian $\mathbbm{H}$, can be interpreted as the transfer matrix connecting two adjacent time slices.
$\mathbbm{P}$ is a projector constraining the dynamics to the gauge invariant subspace of the whole Hilbert space. 

Relevant for this work are the eigenstates $|n,q\rangle$ of the Hamiltonian which are, next to the energy quantum number $n$, specified by the set of quantum numbers $q=(J,C,P,m_h,m_l)$ (total angular momentum, parity, mass of the heavy and light quark respectively).

The pseudo scalar ${\rm D_s}$-meson for example can be specified by the quantum numbers $q=(0,\pm,-,m_c,$$m_s)$, the one of the corresponding vector meson ${\rm D_s^\ast}$ by $q=(1,\pm,-,m_c,m_s)$\footnote{As electromagnetic effects only play a minor role here (cf. \cite{Rolf:2002gu}), the associated charge will be neglected.}.

The $|n,q\rangle$ build a complete set of energy eigenstates of the mesonic sector of the Hamiltonian $\mathbbm{H}$ with
\be
|n,q\rangle,\;\;\;n=0,1,2,\dots,
\ee
\be
\mathbbm{H}|n,q\rangle=E_n^q|n,q\rangle
\ee
and the normalization
\be
\langle n^\prime,q^\prime|n,q\rangle=\delta_{n,n^\prime}\delta_{q,q^\prime}.
\ee

In the case of a Schr\"odinger Functional with vanishing fermion and gluon boundary fields, 
\be
|i^\prime\rangle=|i\rangle=|i_0\rangle
\ee
holds, with $|i_0\rangle$ carrying the quantum numbers of the vacuum.


\begin{figure}[t]
\centering
\hspace{1.0cm}
\begin{minipage}{.3\linewidth}
\vspace{3mm}
\psfrag{1}[c][c][1][0]{$\bar\zeta$}
\psfrag{2}[c][c][1][0]{$\zeta$}
\psfrag{3}[c][c][1][0]{$\bar\psi(x)\Gamma_{O}\psi(x)$}
\psfrag{4}[r][t][1][0]{$x_0=0$}
\psfrag{5}[r][r][1][0]{$x_0=T$}
\psfrag{7}[c][c][1][0]{\large$f_{O}$}
\psfrag{8}[c][c][1][0]{$\Gamma_{ O}^\prime$}
\epsfig{scale=.45,file=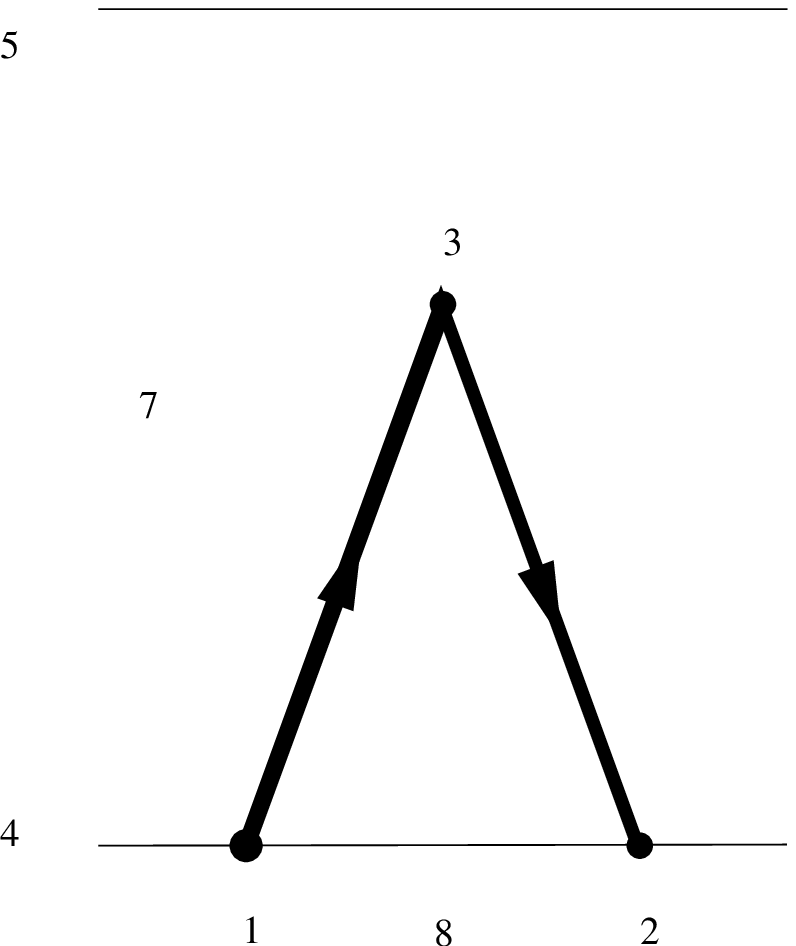}
\end{minipage}
\begin{minipage}{.3\linewidth}
\vspace{-1.15mm}
\psfrag{1}[c][c][1][0]{$\bar\zeta$}
\psfrag{2}[c][c][1][0]{$\zeta$}
\psfrag{3}[c][c][1][0]{$\bar\zeta^\prime$}
\psfrag{6}[c][c][1][0]{$\zeta^\prime$}
\psfrag{4}[r][t][1][0]{}
\psfrag{5}[r][r][1][0]{}
\psfrag{7}[c][c][1][0]{\large$f_{ O}^T$}
\psfrag{8}[c][c][1][0]{$\Gamma_{O}^\prime$}
\psfrag{9}[c][c][1][0]{$\Gamma_{O}^\prime$}
\epsfig{scale=.45,file=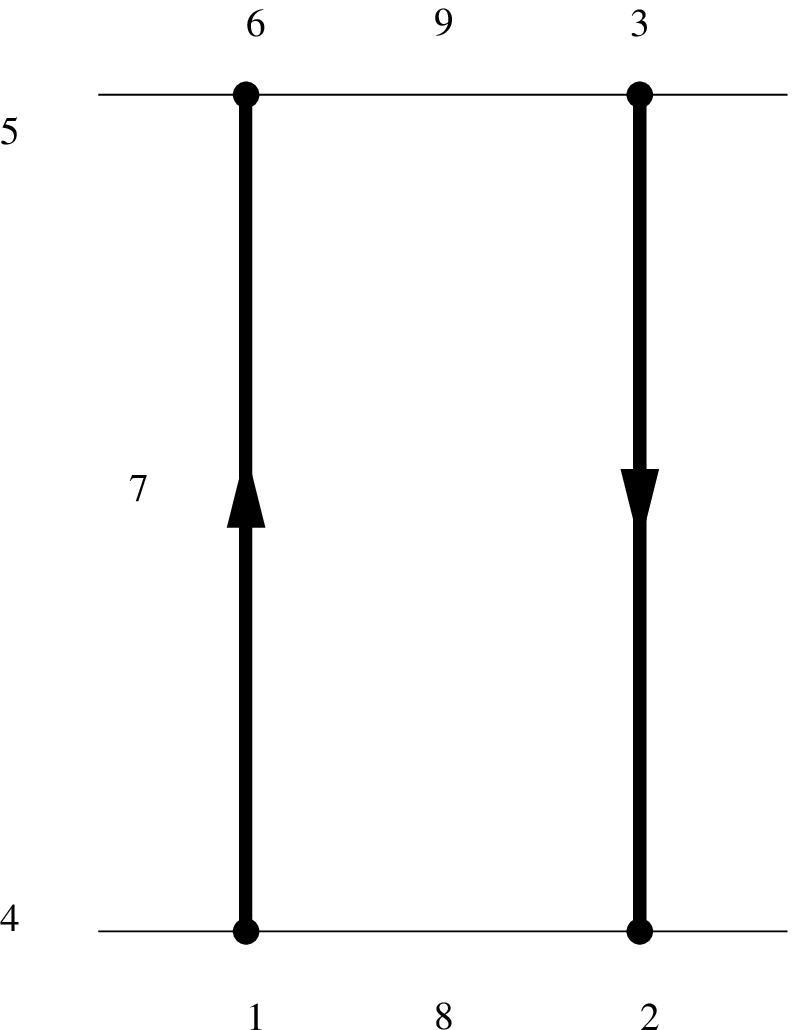}
\end{minipage}\caption{The Schr\"odinger Functional correlation functions $f_{O}$ and $f_{O}^T$.}\label{fig_f_Xandf_1}
\end{figure}


In the following, the focus will be on two types of matrix elements. The first matrix element is 
\be\label{fX}
f_{O}(x_0)=-\mathcal{Z}^{-1}{1\over 2}\langle i_0|e^{-(T-x_0)\mathbbm{H}}\mathbbm{P}\mathbbm{O}(\vec{x})e^{-x_0\mathbbm{H}}\mathbbm{P} |i_{\rm X}\rangle.
\ee
It measures the correlation between the operator $\mathbbm{O}({\vec x})$ at time $x_0$ with the initial and final states $|i_{\rm X}\rangle$ and  $\langle i_0|$. The physical picture is that of a meson with quantum numbers $q_{\rm X}$ being created from fermionic boundary sources at time $x_0=0$ and being annihilated at some later time $x_0$ by the operator $\mathbbm{O}(\vec x)$ (cf. figure \ref{fig_f_Xandf_1}). The second matrix element is\footnote{In the literature on the Schr\"odinger Functional, these matrix elements are also often refered to as $f_1=f_P^T$ and $k_1=f_V^T$.}
\be\label{f1}
f_{O}^T=-\mathcal{Z}^{-1}{1\over 2}\langle i^\prime_{\rm X}|  e^{-T\mathbbm{H}}\mathbbm{P} |i_{\rm X}\rangle.
\ee
It measures the correlation between the initial state $|i_{\rm X}\rangle$ and the final state $\langle i^\prime_{\rm X} |$ after time evolution over the time extent $T$ and thereby represents a meson state, traveling through the space-time from the boundary at $x_0=0$ to the boundary at $x_0=T$.  
In practice, the states $|i_{\rm X}\rangle$ and $\langle i^\prime_{\rm X}|$ are created by dimensionless mesonic boundary operators in the zero-momentum projection,
\be\ba{cc}
\mathcal{O}^{a}_{O}={a^6\over L^3}\sum\limits_{\vec{x},\vec{y}}\bar\zeta(\vec{x})\Gamma_{O}^\prime{\tau^a\over 2}\zeta(\vec{y})|_{x_0=0},& \mathcal{O}^{a\prime}_{O}={a^6\over L^3}\sum\limits_{\vec{x},\vec{y}}\bar\zeta^{\prime}(\vec{x})\Gamma_{O}^\prime{\tau^a \over 2}\zeta^{\prime}(\vec{y})|_{x_0=T}.
\ea
\ee
and the operator $\mathbbm{O}(\vec x)$ typically represents a bilinear quark current of the form
\be
{O}^a(x)=\bar\psi(x)\Gamma_{O}{\tau^a\over 2}\psi(x).
\ee

\begin{table}
\centering
\begin{tabular}{ccccc}
\hline \hline\\[-2ex]
$O$		&$A$			&$V$		&$P$	&$T$\\[1ex]
\hline\\[-2ex]
$\Gamma_{O}$&$\gamma_0\gamma_5$	&$\gamma_i$	&$\gamma_5$&$\sigma_{ij}$\\[1ex]
$\Gamma^\prime_{O}$	&$\gamma_5$		&$-{1\over{3}}\gamma_i$	&$\gamma_5$&$-{1\over 3}\gamma_i$	\\[1ex]
\hline
\hline
\end{tabular}
\caption{Dirac structure for the currents ${O}(x)$ ($\sigma_{ij}={i\over 2}[\gamma_i,\gamma_j]$).}\label{Gammatab}
\end{table}
The Dirac matrices $\Gamma_{O}^{(\prime)}$ are collected in table \ref{Gammatab} for the axial current ($A$), the vector current ($V$), the pseudo scalar density ($P$) and the tensor current ($T$). They determine the desired quantum numbers of angular momentum and parity. The matrix $\Gamma_{A}\equiv\gamma_0\gamma_5$ for example corresponds to $(J,P)=(0,-)$, while the matrix $\Gamma_{ V}\equiv{1 \over \sqrt{3}}\gamma_\mu$ corresponds to $(J,P)=(1,-)$. The ${\rm SU(2)}$ flavor structure of the current is given by the matrices $\tau^a$ ($a=1,2,3$, cf. appendix \ref{notation}).

The matrix elements $f_{O}(x_0)$ and $f_{O}^T$ can then be identified with the expectation values
\be
f_{O}(x_0)=-{L^3 \over 2}\langle {O}^+(x_0)\mathcal{O}^{-}_{O}\rangle\;{\rm for\;} O=A,P,V,T\\
\ee
and
\be\ba{rcll}
f_{ O}^T&=&-{L^3 \over 2}\langle \mathcal{O}^{+ \prime}_{ O} \mathcal{O}^-_{ O}\rangle& \textrm{for}\;X=A, P.\\
\ea
\ee
where again the flavor off-diagonal fields (cf. (\ref{offdiag})) have been used.
In order to arrive at expressions for the meson mass $m_{{\rm X}}$ and the meson decay constant ${\rm F}_{{\rm X}}$ in terms of these correlation functions one first inserts twice the identity $\mathbbm{1}=\sum\limits_{n,q}|n,q\rangle\langle n,q|$ into (\ref{fX}). This yields
\be\ba{ll}
f_{{O}}(x_0)&=-\mathcal{Z}^{-1}{L^3\over 2}
\hspace{-.2cm}
{\sum\limits_{\tiny\ba{c}q,q^\prime\\ n,n^\prime\ea}}
\hspace{-.1cm}
\langle i_0|e^{-(T-x_0)\mathbbm{H}}\mathbbm{P}|n,q\rangle\langle n,q|\mathbbm{O}|n^\prime,q^\prime\rangle\langle n^\prime,q^\prime|e^{-x_0\mathbbm{H}}\mathbbm{P}|i_{\rm X}\rangle\vspace{.2cm}\\

&\approx- \mathcal{Z}^{-1}{L^3\over 2}
\bigg\{
\langle i_0|e^{-(T-x_0)\mathbbm{H}}\mathbbm{P} \big(|0,0\rangle\langle 0,0|+|0,q_{\rm X}\rangle\langle 0,q_{\rm X}|\\
\\
&+\qquad|1,0\rangle\langle 1,0|+|1,q_{\rm X}\rangle\langle 1,q_{\rm X}|\big)
\;\mathbbm{O}\; \big(|0,0\rangle\langle 0,0|+|0,q_{\rm X}\rangle\langle 0,q_{\rm X}|\\
\\
&+\qquad|1,0\rangle\langle 1,0|+|1,q_{\rm X}\rangle\langle 1,q_{\rm X}|\big) \;e^{-x_0\mathbbm{H}}\mathbbm{P}|i_{\rm X}\rangle\bigg\}.\\
\ea\ee
In the second step, all but the ground state, the first excited state $n^{(\prime)}=0,1$ and $q=0,q_{{\rm X}}$ have been neglected, since their contribution will be suppressed exponentially. With
\be
\mathcal{Z}\approx\langle i_0|\mathbbm{P}|0,0\rangle\langle 0,0|\mathbbm{P}|i_0\rangle e^{-E_0^{(0)}T},
\ee 
one arrives at
\be\ba{lll}
f_{{O}}(x_0)&\approx&-{L^3\over 2}
\frac{1}{\langle i_0|\mathbbm{P}|0,0\rangle\langle 0,0|\mathbbm{P}|i_0\rangle}\; {e^{E_0^{(0)}T}}\\
\\
&\times&\hspace{-2.5mm}\bigg\{
\langle i_0|\mathbbm{P}|0,0\rangle\langle0,0|\mathbbm{O}|0,q_{\rm X}\rangle\langle0,q_{\rm X}|\mathbbm{P}|i_{\rm X}\rangle e^{-E_0^{(0)}(T-x_0)}e^{-E_0^{q_{\rm X}}x_0}\\
&&\\
&+&\langle i_0|\mathbbm{P}|0,0\rangle\langle0,0|\mathbbm{O}|1,q_{\rm X}\rangle\langle1,q_{\rm X}|\mathbbm{P}|i_{\rm X}\rangle e^{-E_0^{(0)}(T-x_0)}e^{-E_1^{q_{\rm X}}x_0}\\
&&\\
&+&\langle i_0|\mathbbm{P}|1,0\rangle\langle1,0|\mathbbm{O}|0,q_{\rm X}\rangle\langle0,q_{\rm X}|\mathbbm{P}|i_{\rm X}\rangle e^{-E_1^{(0)}(T-x_0)}e^{-E_0^{q_{\rm X}}x_0}\\
&&\\
&+&\langle i_0|\mathbbm{P}|1,0\rangle\langle1,0|\mathbbm{O}|1,q_{\rm X}\rangle\langle1,q_{\rm X}|\mathbbm{P}|i_{\rm X,}\rangle e^{-E_1^{(0)}(T-x_0)}e^{-E_1^{q_{\rm X}}x_0}\bigg\}.\\
&&\\
\ea\ee
In order to write this in a more compact form, the following abbreviations can be introduced. First, $m_{q_{\rm X}}=E_0^{q_{{\rm X}}}-E_0^0$ and $m_{\rm G }=E_1^0-E_0^0$ are the mass of the ground state meson and the glueball respectively, $\Delta=E_1^{q_{x}}-E_0^{q_{x}}$ is the gap energy in the meson sector. Furthermore one defines the matrix elements
\begin{eqnarray}
\rho&=&\frac{\langle0,q_{\rm X}|\PP|i_{{\rm X}}\rangle}{\langle0,0|\PP|i_0\rangle},\\
\eta_{{{\rm X}}}^{x}&=&\frac{\langle0,0|\mathbbm{O}|1,q_{\rm X}\rangle\langle1,q_{\rm X}|\PP|i_{{\rm X}}\rangle}{\langle 0,0|\mathbbm{O}|0,q_{\rm X}\rangle\langle0,q_{\rm X}|\PP|i_{{\rm X}}\rangle}\qquad{\rm and}\\
\eta_{ {{\rm X}}}^{0}&=&\frac{\langle i_0|\PP|1,0\rangle\langle1,0|\mathbbm{O}|0,q_{\rm X}\rangle}{\langle i_0|\mathbbm{P}|0,0\rangle\langle0,0|\mathbbm{O}|0,q_{\rm X}\rangle}.\\
\end{eqnarray}
$f_{{O}}$ then takes the simple form \cite{Guagnelli:1999zf}
\be\ba{rcl}\label{FX}
f_{{{O}}}(x_0)&\approx&-{L^3\over 2}\rho\;\langle0,0|\mathbbm{O}|0,q_{\rm X}\rangle \;e^{-x_0m_{{{\rm X}}}}\left\{1+\eta_{{{\rm X}}}^{q_{\rm X}}e^{-x_0\Delta}   +\eta_{{{\rm X}}}^{0}e^{-(T-x_0)m_{\rm G}}\right\}, \\
\ea
\ee
which decays with the mass of the meson.
For $f_{O}^T$ one derives
\be
f_{O}^T\approx{1\over 2}\frac{|\langle i_{{\rm X}}|\mathbbm{P}|0,q_{\rm X}\rangle|^2}{|\langle i_0|\mathbbm{P}|0,0\rangle|^2}\;e^{-T(E_0^{q_{\rm X}} -E_0^0)}={1\over 2}\rho^2\,e^{-Tm_{q_{\rm X}}}.
\ee
The correlation function $f_{O}(x_0)$ contains contributions from higher excited states for small $x_0$, which decay  exponentially with the gap energy $\Delta$. For large $x_0$, bound states of gluons, the so called glueballs contribute exponentially enhanced with their mass $m_{\rm G}$. 

For the numerical implementation it is very convenient to also define the {backward correlation functions}
\be\ba{rcll}\label{backwardcorr}
g_{{O}}(T-x_0)&=&{L^3\over 2}\langle \mathcal{O}^{+\prime}_{O}{{O}}^{-}(x_0)\rangle&{\rm for\;}O=A,\,V,\\
\\
g_{{O}}(T-x_0)&=&-{L^3\over 2}\langle \mathcal{O}^{+\prime}_{O}{{O}}^{-}(x_0)\rangle&{\rm for}\;O=P,\,T.\\
\ea
\ee
(next to the {forward correlation functions} $f_{O}$). 
They are constructed in the same way as $f_{O}(x_0)$ but with the meson sources at the time slice at $x_0=T$ and serve the same purpose, provided that the background field is zero. 
In a Monte-Carlo simulation, both the forward correlation functions and the backward correlation functions will be evaluated on the same sample of gauge configurations. Although the resulting data will be correlated, the statistics can be increased in this way. 

The improvement of the axial current (\ref{imprpscurr}) and the vector current (\ref{imprvcurr}) is taken over from \cite{Guagnelli:1999zf} to define the improved correlation functions
\be\label{FPSimpr}
f_{{ A}}(x_0)\to f_{ A}^I(x_0)=f_{ A}(x_0)+ac_A(g_0)\tilde\partial_0 f_{ P}(x_0) 
\ee
and 
\be\label{FVimpr}
f_{{V}}(x_0)\to f_{V}^I(x_0)=f_{V}(x_0)+{i\over 2}ac_V(g_0)\vec{\tilde\partial} \vec{f_{ T}}(x_0) 
\ee
and analogously for the correlation functions $g_{O}(T-x_0)$.

\subsection{The meson mass and the meson decay constant}\label{meffandfds}
The meson mass $m_{{\rm X}}$ can be extracted as the effective mass,
\be\ba{rcl}\label{effectivemass}
m_{\rm eff}(x_0+{a\over 2})&=&{1\over a}\ln\left(\frac{f^I_{{O}}(x_0)}{f^I_{{O}}(x_0+{a})}\right)\\
\\
&\approx& m_{{{\rm X}}} \left\{1+\frac{2 \sinh(a\Delta/2)}{am_{{\rm X}}}\eta^{q_{{\rm X}}}_{{\rm X}}e^{-x_0\Delta}-\frac{2 \sinh(a m_{\rm G}/2)}{am_{{{\rm X}}}}\eta^{0}_{{\rm X}}e^{-(T-x_0)m_{\rm G}}\right\}.\\
\ea
\ee
The connection to the meson decay constant, which is defined as \cite{Guagnelli:1999zf}
\be\label{decaymatelement}
Z_{{O}}\langle 0,0|\mathbbm{O}|0,q_{{\rm X}}\rangle = {\rm F}_{{\rm X}}m_{{\rm X}}(2m_{{\rm X}}L^3)^{-1/2}
\ee
is then given by 
\be\ba{rcl}\label{fdsffinal}
\rm{F}_{{\rm X}}&\approx&-{2}Z_{{O}}(m_{{\rm X}}L^3)^{-1/2}\,e^{(x_0-T/2)m_{{\rm X}}}\frac{f^I_{{O}}(x_0)}{\sqrt{f_{O}^T}}\\
\\
&\times&\qquad\qquad\left\{1-\eta_{{{\rm X}}}^{{q_{\rm X}}}\,e^{-x_0\Delta}   -\eta_{{{\rm X}}}^{0}\,e^{-(T-x_0)m_{\rm G}}\right\}.\\
\ea
\ee
Here, $Z_{O}$ is the renormalization constant of the current ${O}$ and $(2m_{{\rm X}}L^3)^{-1/2}$ in (\ref{decaymatelement}) is the conventional normalization of one-particle states.

In terms of the backward correlation functions, the effective mass and the decay constant can be expressed by replacing $f^I_{{O}}(x_0)$ with $g^I_{{O}}(T-x_0)$ in (\ref{effectivemass}) and (\ref{fdsffinal}). 
\subsection{Quark masses}\label{quarkmasses}
On the lattice, there exist various ways to define quark masses, which differ at finite lattice spacing but converge in the continuum limit. One definition has already been introduced in (\ref{rensub}), where the renormalized valence quark mass is given in terms of the hopping parameter. Another way to obtain the quark mass is guided by the PCAC relation in the continuum
\be
\partial_\mu A_\mu^a=2mP^a,
\ee
where $m$ is the quark mass. 
The average bare current quark mass in the Schr\"odinger Functional can be written as \cite{Luscher:1996sc}
\be\label{barePCACmass}
m_{hl} = \oh \left[\oh(\partial^\ast_0+\partial_0)f_A(x_0)+c_A a \partial_0^\ast\partial_0 f_P(x_0)\right]/f_P(x_0).
\ee
The sum of the renormalized valence quark masses can then be obtained by multiplicative renormalization of the axial current (\ref{imprpscurr}) and the pseudo scalar density (\ref{imprpdens}) \cite{Guagnelli:2000jw},
\be\ba{rcl}\label{PCACmass}
m_{{\rm R},h}+m_{{\rm R},l}&=&2\frac{Z_A(1+b_A(g_0)\oh(am_{q,h}+am_{q,l}))}{Z_P(1+b_P(g_0)\oh(am_{q,h}+am_{q,l}))}m_{hl} +O(a^2)\\
\\
&=&2{Z_AZ_P^{-1}(1+(b_A(g_0)-b_P(g_0))\oh(am_{q,h}+am_{q,l}))}m_{hl} +O(a^2).
\ea
\ee

\subsection{Correlation functions, propagators and symmetries}\label{correlationfunctions}
For their implementation in a Monte-Carlo simulation program, the correlation functions $f_{{O}}(x_0)$, $g_{{O}}(T-x_0)$ and $f^T_{O}$ must be expressed in terms of contractions of quark propagators which will be derived in the following.

Writing down the flavor indices explicitly $f_{{O}}(x_0)$ reads
\be\ba{rcl}
f_{{O}}(x_0)&=&-{a^6L^3\over 2}\sum\limits_{\vec{y},\vec{z}}\,\langle\bar\psi_h(x)\Gamma_O\psi_l(x)\,\bar\zeta_l(\vec{y})\Gamma_O^\prime\zeta_h(\vec{z})\rangle.
\ea\ee
for $O=A,P,V,T$. 
Therefore one can write $f_{{O}}$ as
\be\ba{rl}\label{fXcontracted}
f_{{O}}(x_0)&={a^6 L^3\over 2}\sum\limits_{\vec{y},\vec{z}}\left<{\rm{Tr}}\left\{\left[\zeta_h({\vec{z}})\bar\psi_h(x)\right]_{\rm{F}}\Gamma_{O}\left[\psi_l(x)\bar\zeta_l({\vec{y}})\right]_{\rm{F}}\Gamma^\prime_{O}\right\}\right>_{\rm G},
\ea
\ee
where the trace is over Dirac and color indices. Analogously one obtains
\be\ba{rl}\label{gXcontracted}
g_{{O}}(T-x_0)&=\pm{a^6L^3\over 2}\sum\limits_{\vec{y},\vec{z}}\left<{\rm{Tr}}\left\{\left[\psi_h(x)\bar\zeta_h^\prime(\vec{y})\right]_{\rm{F}}\Gamma^\prime_{O}\left[\zeta_l^\prime(\vec{z})\bar\psi_l(x)\right]_{\rm{F}}\Gamma_{O}\right\}\right>_{\rm G},
\ea
\ee
where the sign has to be chosen according to (\ref{backwardcorr}),
and
\be\label{f1contracted}
f_{O}^T={a^{12} L^3\over 2}\sum\limits_{\vec{v},\vec{w},\vec{y},\vec{z}}\left<{\rm{Tr}}\left\{\left[\zeta_h({\vec{z}})\bar\zeta_h^\prime(\vec{v})\right]_{\rm{F}}\Gamma^\prime_O\left[\zeta_l^\prime(\vec{w})\bar\zeta_l({\vec{y}})\right]_{\rm{F}}\Gamma_O^\prime\right\}\right>_{\rm G}.
\ee
Applying the functional derivatives (\ref{funcderiv}) to the fermionic generating functional $\ln\mathcal{Z}_{\rm F}[U,\rho,\bar{\rho},\rho^\prime,\bar{\rho}^\prime,\bar\eta,\eta]$, the propagators $[\;\cdot\;]_{\rm F}$ in (\ref{fXcontracted}), (\ref{gXcontracted}) and (\ref{f1contracted}) can now be written down explicitly: 
\be\ba{lcl}
\left[\zeta({\vec{z}})\bar\psi(x)\right]_{\rm{F}}&\stackrel{(\ref{funcderiv})}{=}&-\left\{\frac{\delta}{\delta\bar\rho(\vec{z})}\frac{\delta}{\delta\eta(x)}\ln\mathcal{Z}_{\rm F}[U,\rho,\bar\rho,\rho^\prime,\bar\rho^\prime,\eta,\bar\eta]\right\}_{|\rho=\dots=\bar\eta=0}\\
\\
&=&\tilde{c}_tP_-U(z,0)S(z,x)_{|z_0=a},
\ea
\ee
\be\ba{lcl}
\left[\psi(x)\bar\zeta(\vec{y})\right]_{\rm{F}}&{=}&\tilde{c}_tS(x,y)U^{-1}(y,0)P_{+\;|y_0=a},\\
\\
\left[\psi(x)\bar\zeta^\prime(\vec{y})\right]_{\rm{F}}&{=}&\tilde{c}_tS(x,y)U(y,0)P_{-\;|y_0=T-a},\\
\\
\left[\zeta^\prime(\vec{z})\bar\psi(x)\right]_{\rm{F}}&{=}&\tilde{c}_tP_+U^{-1}(z,0)S(z,x)_{|z_0=T-a},\\
\\
\left[\zeta(\vec{z})\bar\zeta^\prime(\vec{v}))\right]_{\rm{F}}&{=}&\tilde{c}_t^2P_-U(z,0)S(z,v)U(v,0)P_{-\,|z_0=a,\,v_0=T-a},\\
\\
\left[\zeta^\prime(\vec{w})\bar\zeta(\vec{y}))\right]_{\rm{F}}&{=}&\tilde{c}_t^2P_+U^{-1}(w,0)S(w,y)U^{-1}(y,0)P_{+\,|w_0=T-a,\,y_0=a}.\\
\ea
\ee
Inserting these expressions into $f_{{O}}(x_0)$ one gets, denoting the quark flavor by the subscripts $h$ and $l$ respectively,
\be\ba{ll}
f_{{O}}(x_0)&={\tilde{c}^2a^6L^3\over 2}\sum\limits_{\vec{y},\vec{z}}\left<{\rm{Tr}}\left\{P_- U({\vec{z}},0)S_h(x,z)\Gamma_O^\prime S_l(x,y) U^\dagger({\vec{y}},0)P_+\Gamma_O\right\}\right>_{{\rm G}|_{y_0=z_0=a}}\\
\\
&={\tilde{c}^2a^6L^3\over 2}\sum\limits_{\vec{y},\vec{z}}\left<{\rm{Tr}}\left\{P_+ U({\vec{z}},0)S_h^\dagger(x,z)\gamma_5\Gamma_O^\prime S_l(x,y) U^\dagger({\vec{y}},0)P_+\Gamma_O\gamma_5\right\}\right>_{{\rm G}|_{y_0=z_0=a}}\\
\\
&={L^3\over 2}\left<{\rm{Tr}}\left(\bar S_h^\dagger(x)\gamma_5\Gamma_O^\prime\bar S_l(x)\Gamma_O\gamma_5\right)\right>_{\rm G}.\\
\ea
\ee
Here, the summation over the boundary fields has been included into the propagator
\be\label{Sbar}
\bar S(x)=\tilde{c}_ta^3\sum\limits_{\vec{y}}S(x,y) U^{-1}({\vec{y}},0)P_{+\,|_{y_0=a}},
\ee
which is the propagator of a zero-momentum quark state at $y_0=0$ to a space-time point $x$ in the interior of the Schr\"odinger Functional.
In the same way, $g_{{O}}(T-x_0)$ is
\be\ba{ll}
g_{{O}}(T-x_0)&=\pm{L^3\over 2}\left<{\rm{Tr}}\left(\bar R_h^\dagger(x)\gamma_5\Gamma_O^\prime\bar R_l(x)\Gamma_O\gamma_5\right)\right>_{\rm G}.\\
\ea
\ee
with 
\be\label{Rbar}
\bar R(x)=\tilde{c}_ta^3\sum\limits_{\vec{y}}S(x,y)U(y,0)P_{-\,|y_0=T-a},
\ee
being the propagator of a zero-momentum quark state at $y_0=T$ to the space-time point $x$.
The boundary-to-boundary correlation function $f_{O}^T$ is
\be
f_{O}^T={\tilde{c}^2a^6L^3\over2}\sum\limits_{\vec{v},\vec{w}}\left<{\rm{Tr}}\left\{\bar{S}_h^\dagger(v)U(\vec{v},0)P_+\gamma_5\Gamma_O^\prime P_+U^{-1}(w,0)\bar{S}_l(w)\Gamma_O^\prime\right\}\right>_{\rm G|v_0=T-a,\,w_0=a}.
\ee
Introducing the boundary-to-boundary quark propagator
\be\label{SbarT}
\bar{S}_T=\tilde{c}_t\sum\limits_{\vec{x}}P_+U^{-1}(x,0)\bar{S}(x)_{|x_0=T-a}
\ee
it takes the form
\be
f_{O}^T={L^3\over2}\left<{\rm{Tr}}\left\{\bar{S}^\dagger_{T,h}\gamma_5\Gamma_O^\prime \bar{S}_{T,l}\Gamma_O^\prime\right\}\right>_{\rm G}.
\ee
Observing that
\be
\bar{S}(x)P_-=\bar{R}(x)P_+=0
\ee
in the chiral basis for the gamma matrices, the relations
\be\ba{cc}\label{propsymmetries}
\bar{S}_{A1}(x)+\bar{S}_{A3}(x)=0,&\bar{S}_{A2}(x)+\bar{S}_{A4}(x)=0,\\
\bar{R}_{A1}(x)-\bar{R}_{A3}(x)=0,&\bar{R}_{A2}(x)-\bar{R}_{A4}(x)=0,\\
\ea
\ee
hold, where $A$ is a Dirac index. Thus, all correlation functions introduced so far are completely determined if only half of the Dirac-components of the forward- and backward propagators $\bar{S}(x)$ and $\bar{R}(x)$ are known. 

In the numerical simulation, these components of $\bar S(x)$ and $\bar R(x)$ have to be determined on each gauge background of the Monte-Carlo history. The equations
\be\ba{rcl}
\left(D^I+m\right)\bar S(x)&=&\tilde{c}_t a^3\sum\limits_{\vec{y}}\left(D^I+m\right)S(x,y)U^{-1}(\vec{y},0)P_{+\,|y_0=a}\\
\\
&=&\tilde{c}_t a^{-1}\delta_{x_0,a}U^{-1}(\vec{x},0)P_+\\
\ea
\ee
and 
\be\ba{rcl}
\left(D^I+m\right)\bar R(x)&=&\tilde{c}_t a^{-1}\delta_{x_0,a}U(\vec{x},0)P_-,\\
\ea
\ee
or equivalently
\be\ba{rcl}
M^I\bar S(x)&=&2\kappa\tilde{c}_t a^{-1}\delta_{x_0,a}U^{-1}(\vec{x},0)P_+,\\
\\
M^I\bar R(x)&=&2\kappa\tilde{c}_t a^{-1}\delta_{x_0,a}U(\vec{x},0)P_-\\
\ea\ee
therefore have to be solved for $\bar S(x)$ and $\bar R(x)$, using a matrix inversion algorithm, the stabilized biconjugate gradient \cite{Frommer:1994vn} for example.

\chapter{The PC-Code}\label{chapPCcode}
For a major part of the simulation parameters, the production runs could be accomplished on the APEMille supercomputers at the John von Neumann Institute for Computing \cite{NIC}. In particular, the simulations for the scaling study with the bare couplings $\beta=6/g_0^2=6.0,6.1,6.2,6.45\;$\footnote{Here, $g_0^2$ denotes the bare gauge coupling.} were all done within this framework using the ALPHA-collaboration's approved program code for the simulation of quenched QCD in the Schr\"odinger Functional. In simulations for the lattice with the finest resolution, i.e. at $\beta=6.7859$, the number of lattice points $L/a$ and therefore the amount of memory space that needs to be allocated to store the associated fields gets very large and the available memory on the APE-computer is not sufficient any more. 

An alternative supercomputer with the appropriate specifications was the newly installed system of 26 IBM pSeries 690 \emph{eservers} at the \emph{Norddeutscher Verbund f\"ur Hoch- und H\"ochst\-leistungs\-rechnen} (HLRN). Each of these servers has 32 IBM-Power4 processors (with a peak performance of $5.2$ GFlop/s each) that share between 64 to 256 GByte of memory. 

As the lattice-gauge code on the APE-computer is written in a proprietary language (TAO), a new simulation code had to be obtained. To this end, the lattice code by the MIMD\footnote{The aforementioned IBM-computer has a  MIMD architecture (\emph{Multiple Instructions, Multiple Data}), as opposed to a computer architecture like the one of the APEMille which is a SIMD (\emph{Single Instruction, Multiple Data}) computer.} Lattice Computation (MILC) Collaboration \cite{MILC}, published under the GNU General Public License \cite{GNU}, seemed to be an appropriate choice.
It consists of a set of routines written in C for doing simulations of four dimensional SU(3) lattice gauge theory. The version which the collaboration shares on the Web, already comprises many of the features that are necessary for the realization of this project. Among those are
\bi
\item Schr\"odinger Functional boundary conditions,
\item Wilson pure gauge code,
\item Wilson Dirac operator,
\item $O(a)$-improvement,
\item stabilized biconjugate gradient inverter,
\item platform-independence,
\item MPI-based parallelism (MIMD).
\ei

This chapter discusses the MILC Code, all changes and improvements added to it and the extensive testing previous to the data production. Other topics are the resource allocation for the production runs and the corresponding applications for CPU-time at the HLRN. 
\section{The MILC Code}

The two building blocks in the lattice simulation of gauge theories are the generation of a Monte-Carlo series of gauge-configurations and the subsequent evaluation of observables on the given gauge background.

The gauge updates in the MILC Code are done with a quasi-heatbath gauge update by three SU(2) subgroups \`a la Kennedy-Pendleton \cite{Kennedy:1985nu} and Cabibbo-Marinari \cite{Cabibbo:1982zn}. Also micro canonical over relaxation steps by doing SU(2) gauge hits have been implemented. The pseudo random number generator is the ex\-clu\-sive-OR of a 127 bit feedback shift register and a 32 bit integer congruence generator. It runs with a different seed on each of the parallel processors to avoid correlations.

Based on the even-odd preconditioned Dirac operator \cite{DeGrand:1990dk}, the quark propagators on a given gauge background are evaluated using the stabilized biconjugate gradient algorithm \cite{Frommer:1994vn}.

Up to the random number generator, the implementation of the algorithms resemble very much the ones in the ALPHA Collaboration's lattice code. However, some peculiarities should be mentioned here. The MILC code uses the Weyl-basis ($\gamma_5$ diagonal) Dirac matrices. In contrast to the standard convention, the projectors $P_\pm={1\over 2}(1\pm \gamma_0)$ are implemented with opposite sign, that means, the projectors in the MILC code $P_\pm^{\rm MILC}$ are
\be
P_\pm^{\rm MILC}=P_\mp
\ee

The MILC code uses MPI-based (MPI stands for Message Passing Interface \cite{MPIhttp} ) parallelization. The code has been designed in such a way, that the user in most cases does not have to bother about the parallelization, e.g. when implementing a set of correlation functions. Only at a deeper level, for example for the implementation of a new action, these issues have to be considered. 
Assuming, that the number of lattice sites is divisible by the number of nodes, which is a power of two, the lattice volume is divided by factors of two in any of the four directions. Dividing the direction of the largest lattice extent is favored in order to keep the area of the surface minimal. Similarly, dividing directions, which have already been divided is preferred, thereby keeping the number of off-node directions minimal. 

The root project directory has the following listing:
\begin{verbatim}
generic/
generic_clover/
generic_pg/
generic_schroed/
generic_wilson/
include/
libraries/
schroed_pg/
f_A/
\end{verbatim}
The two main directories are \verb|schroed_pg/| and \verb|f_A/|. They contain the main program file for the gauge-update and the code for the evaluation of observables on a given gauge background.

The \verb|generic/| directory contains high level routines that are more or less independent of the physics. Examples of generic code are the communication routines (MPI), random number routines, routines to evaluate the plaquette, etc. A set of slightly more specific directories is \verb|generic_clover/|, \verb|generic_wilson/|, and \verb|generic_schroed/|. They contain the implementation of the Dirac operator, the clover term, the matrix inversion routines or routines for setting up the lattice topology. All applications share the \verb|include/| directory, containing most of the header files and the \verb|libraries/| directory containing low-level routines, mostly linear algebra routines.

Detailed instructions on how to use the code can be found in appendix \ref{runit}. Further information about the MILC code in general can be found on the project web site \cite{MILC}
\section{Changes}
The MILC code as it is available on the web had to be customized in a couple of places in order to conform to the requirements demanded by the project.\\
\subsection{The main program}
Next to setting up the lattice including the fermionic sources, initializing MPI-based inter-node communication and loading the gauge configuration, the main program \verb|control_cl.c| contains a nested loop in which the inverter is called for a given set of indices like
\begin{verbatim}
for(color = 0; color < 3; color++){
  for(spin = 0; spin < 2; spin++){
    for(kappa = 0; kappa < num_kap; k++){
      *** CALL BiCG (color, spin, kappa)***
    } /* kkappa */
    for(kkappa = 0; kkappa < num_kap; k++){
      for(lkappa = k; lkappa < num_kap; l++){
      *** EVAL CORRELATION FUNCTIONS (color, spin, kappa) ***
      } /* lkappa */
    } /* kkappa */
  } /* spin */
} /* color */
\end{verbatim}
In this loop-structure, the symmetries of the propagators pointed out at the end of section \ref{correlationfunctions} have already been taken into account. The loop over the spin components only runs from 1 to 2. Only one color component of the propagator has to be stored in the memory at a time. Whether the code evaluates the forward- or backward-propagators has to be decided at compilation time and just affects, the way the fermion sources are set up at the boundaries.

The contraction of the propagators that give the correlation functions given in \ref{correlationfunctions} have been implemented in the routines \verb|f_A.c|, \verb|f_P.c|, \verb|f_1.c|, \verb|f_V.c|, \verb|k_T.c|, \verb|k_1.c|.
\subsection{Double precision arithmetics}\label{doublearithmetics}
The official MILC code with Schr\"odinger Functional boundary conditions is based on single precision arithmetics\footnote{According to the IEEE-standard, a single precision number consists of a sign bit, eight exponent bits and 23 bits for the mantissa, while a double precision number consists of one sign bit, 11 exponent bits and 52 bits for the mantissa.}. For simulations of large lattices, this may become a source of concern, especially when evaluating sums over fluctuating numbers over the whole lattice. To be on the safe side, double precision arithmetics have been implemented throughout the  code.

A compiler flag now allows for changing back to single precision arithmetics at compilation time (defined in \verb|include/config.h|). 
Thus, allowing at the price of reduced precision, to allocate only roughly half the memory necessary for double precision arithmetics.

A better number precision results in a better precision of the inversion algorithm. This has been investigated with the following test. First, the Dirac operator applied to a quark source $\phi_{\rm in}$ was inverted on a gauge background ($L/a = 48^3\times96$, $\beta=6.7859$, $\kappa=0.117625$). Then, the Dirac operator was again applied to the result, i.e.,
\be
\phi_{\rm out} = \hat{M}(\hat{M}^{-1}\phi_{\rm in})_{\rm BiCGstab}.
\ee

The inversion algorithm monitors the convergence with an iterated solver residual $\epsilon$ \cite{Frommer:1994vn}. After every iteration, it can be constructed from the change of the solution vector with respect to the previous iteration. Compared to the exact residuum
\be\label{residual} \frac{||\phi_{\rm in}-\phi_{\rm out}||}{||\phi_{\rm in}||}=\sqrt{\frac{\sum\limits_x(\phi_{x,\rm in}-\phi_{x,\rm out})^2}{\sum\limits_x{\phi_{x,\rm in}}^2}},
\ee
one thereby saves one time consuming application of the Dirac operator in each iteration step.

The solution $\phi_{\rm out}$ is accepted as soon as the iterated residual is smaller than the stopping criterion $\epsilon$. This computation has been repeated for a number of residuals $\epsilon = 10^{-4},$$10^{-6},$$\dots,$$10^{-20}$. 
The expectation is, that for single precision arithmetics, a residual of $\approx 10^{-8}$ can be reached and $\approx 10^{-16}$ in the case of double precision.
\begin{figure}
\centering
\psfrag{1}[c][c][1][0]{$\epsilon$}
\psfrag{new}[c][c][1][0]{double}
\psfrag{old}[c][c][1][0]{single}
\psfrag{2}[c][c][1][0]{$||\phi_{\rm in}-\phi_{\rm out}||/||\phi_{\rm in}||$}
\epsfig{scale=.7,file=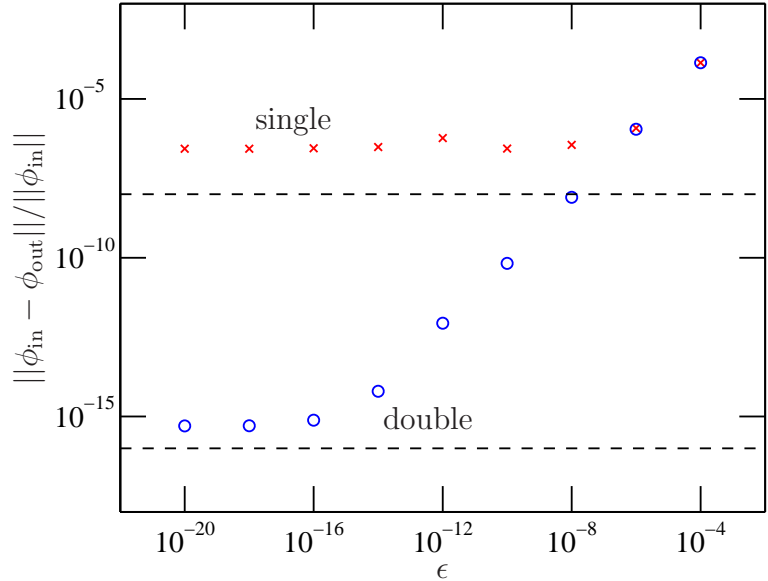}
\caption{Exact residuum $||\phi_{\rm in}-\phi_{\rm out}||/||\phi_{\rm in}||$ plotted against the iterated residuum using single precision arithmetics (crosses) and the newly implemented double precision arithmetics (circles). The dashed lines indicate a residuum of $10^{-8}$ and $10^{-16}$.}\label{singledouble}
\end{figure}

The results shown in figure \ref{singledouble} confirm, that the implementation works in the desired way: The exact residual 
decreases with $\epsilon$ and then saturates at about the corresponding arithmetic precision. 
\subsection{Performance}\label{chapterperformance}
A well-known disadvantage of the MILC code is the way the allocation of memory is organized \cite{Gottlieb:2001ce}. All variables associated to a lattice site are stored in a local structure called \emph{site} - metaphorically speaking this is a container for indices, gauge fields, auxiliary fields etc., defined at each lattice site. It has the form  
\begin{verbatim}
typedef struct {
        /* coordinates of this site */
        short x,y,z,t;
        /* is it even or odd? */
        char parity;
        /* index of the site in the lattice array */
        int index;

       /* Physical fields, application dependent. 
        add or delete whatever is needed.*/
        /* gauge field */
        su3_matrix link[4];
        /* spatial boundary links */
        su3_matrix boundary[3]; 
        /* temporary link variable for Field Major /*
        su3_matrix link_tmp[4];
        /* wilson complex vectors */
        wilson_vector psi;      /* solution vector */
        wilson_vector chi;      /* source vector */
	.
	.
	.
} site; 
\end{verbatim}
(c.f. \verb|lattice.h| in the project directory).

The sites on each node of the parallel machine are the elements of a lo\-cal array which occupies a linear space in the node's memory. One often refers to this type of memory arrangement as \emph{site major} because one first has to point to a particular site before accessing the appendant variables. 

On the one hand \emph{site major} is concise and user-friendly as it allows to easily add new variables, auxiliary fields e.g., that may become necessary when creating new projects. On the other hand, it is very ineffective with respect to the communication between the main memory and the computer's cache.

One reason for this is, that during a lexicographic sweep through the lattice, one often asks for the value of a certain variable at the current site. As detailed before, not each variable on its own, but the array of sites occupies a connected address space in the memory. In each single step of the sweep, the address pointer therefore has to be translated by the length of one {site} in address space, causing sizeable latency.

Another reason is intimately connected with hardware prefetching, a common feature of modern processor architectures (e.g. IBM Power4), where the processor guesses which address space has to be prefetched into the cache for processing during the next clock steps. This feature only works properly if the variables to be processed in a loop are allocated in a reasonably homogeneous way.

A third point is, that some processors allow for multiple data {streams} (e.g. IBM p690 up to 8 streams). That means that the processor can handle a number of simultaneous data interchanges - {streams} - between the main memory and the cache. The processor can identify streams only, if the data is contained in a connected region of the address space.

An important example, where all these issues play a role is the application of the Dirac operator onto a source vector in a lexicographic sweep over all lattice sites on a node. This is the most time consuming operation during the computation of propagators. At each site, the source vector, the link matrices and a destination vector are needed. In the case of site major they are not contained in a connected piece of address space. The pointer to the data of interest therefore has to be moved very often and only a single {stream} is recognized.

\begin{figure}
\centering
\psfrag{1}[c][c][1][0]{ cache}
\psfrag{2}[c][c][1][90]{\small stream 1}
\psfrag{3}[c][c][1][90]{\small stream 2}
\psfrag{4}[c][c][1][90]{\small stream 3}
\psfrag{11}[c][c][1][0]{site 4}
\psfrag{5}[c][c][1][90]{\small stream}
\psfrag{6}[c][c][1][0]{site 3}
\psfrag{7}[c][c][1][0]{site 2}
\psfrag{8}[c][c][1][0]{site 1}
\psfrag{9}[c][c][1][0]{\bf site major}
\psfrag{10}[c][c][1][0]{\bf field major}
\psfrag{12}[c][c][1][90]{unused stream}
\epsfig{scale=.7,file=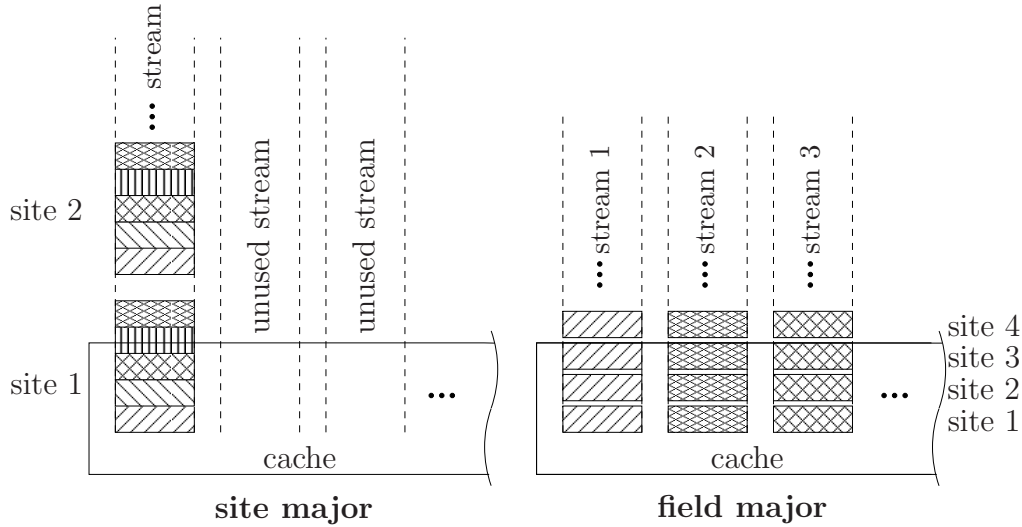}
\caption{Sketch of site major and field major cache processing}\label{field_site_major}
\end{figure}

This situation can be improved by copying the necessary fields in lexicographic order into temporarily allocated variables at the beginning of the inversion. These temporary fields are then used throughout the inversion. Each temporary variable now occupies a connected piece of address space which the processor can recognize as {streams}. This philosophy of memory allocation is usually referred to as \emph{field major} (cf. figure \ref{field_site_major}). In order to maintain the user-friendliness of the \emph{site major} approach, the fields are copied back to the {site} structure ordering after the inversion has stopped. This can be done in a negligible amount of time.

These ideas have been implemented in the MILC code for the stabilized biconjugate gradient algorithm and the Wilson Dirac operator (\verb|dslash_tmp| in \verb|dslash_lean.c| and \verb|bicgilu_cl| in \verb|d_bicgilu_cl_cttilde_lean.c|).  

The measured performance improvement on a single IBM p690 CPU for different lattice sizes has been summarized in table \ref{impresults}. While the achieved speedup first rises with the lattice volume and then saturates, the overall performance reduces with increasing lattice volume. For small lattices, a larger part of the fields permanently resides in the cache, which improves performance in general. At this point the difference between site major and field major is already sizeable. The larger the part of the lattice data residing outside the cache becomes, the more the effects of a sensible memory allocation manifests itself. 
The numbers quoted for the speedup are quite impressive and illustrate that it is of utmost importance to take into account the target computer architecture at the time of the creation of the production code.

Table \ref{impresults} also contains numbers for the efficiency of the code assuming a peak performance of $5.2$ GFlop/s for a single IBM processor \cite{HLRN:2002}. The performance numbers quoted there are disappointing. On custom designed computers for lattice gauge theory, like the APEMille computer \cite{Panizzi:1997jv,Bartoloni:2001he} for example, efficencies of up to 50\% are not unusual. When the experiments with the MILC-code on the IBM-computers at the HLRN code started, no experience with codes for lattice gauge theory, in particular for the application of the Dirac operator with a much better performance existed \cite{HinnerkPC}. 
In the meantime, other groups have achieved better performance numbers. For example the NIC/DESY-Zeuthen group \cite{CarstenPC} have measured the performance of their implementation of the SU(3)-Dirac operator (Clover improved Wilson action) which uses MPI-based parallelism as well. The test was done with a $16^4$-lattice. Although not yet published, they claim to get a single-processor performance of $802$ MFlop/s on the same computer which is by a factor of 2.3 better than the improved version of the MILC code. It would be interesting to learn more about their program code and to transfer their ideas to the MILC code.

\begin{table}                                                
\begin{center}                                               
\begin{tabular}{ccccc}                                     
\hline                                                       
\hline  
\vspace{-.4cm}&&&&\\                                                     
        &\emph{site major}&\emph{field major}    &	&field major\\                  
$L/a$   &performance    &performance    &speedup&efficiency\\           
        &[MFlop/s]      &[MFlop/s]      &	&\\           [.5ex]       
\hline                                                       
\vspace{-.4cm}&&&&\\
6       &350            &510            &1.5	&9.8\% \\               
8       &246            &467            &1.9	&9.0\% \\               
12      &187            &349            &1.9	&6.7\% \\               
14      &173            &355            &2.1	&6.8\% \\               
16      &93             &345            &3.7	&6.6\% \\[.5ex]
\hline                                                       
\hline                                                       

\end{tabular}
\caption{Performance improvement for different lattice sizes after migrating from \emph{site major} to \emph{field major}, measured on a single IBM p690-CPU with a peak performance of 5.2 GFlop/s.}\label{impresults}
\end{center}
\end{table}
\subsection{Miscellaneous other changes}
\subsubsection{O(a)-improvement at the boundary}
The O(a)-improvement term proportional to $(\tilde c_t(g_0)-1)$ (cf. (\ref{boundaryimprovement})) is not implemented in the version of the code that the MILC collaboration shares on the web. It is a term that adds to the diagonal of the Dirac matrix at the boundary and has been implemented in the routine \verb|make_clov|. The value of $\tilde c_t(g_0)$ (cf. table \ref{improvementtab}) has to be specified in the input file.

\subsubsection{Large file support for gauge configurations and propagators}

A gauge configuration consists of 
\be
[(L/a)^3\times T/a]_V\times[3\times3]_{\rm SU(3)}\times [4]_{\hat\mu}\times[2]_{\mathbb{C}}
\ee
real numbers. In the case of double precision arithmetics, a lattice of size $(L/a)^3\times T/a=48^3\times 96$ needs roughly 6.1 GB of memory space. 
A propagator on the other hand consists of
\be
[ (L/a)^3\times T/a]_V\times [3\times3]_{\rm SU(3)}\times [4\times 2]_{\rm Dirac}\times[2]_{\mathbb{C}}
\ee
real numbers which, in double precision, for the above lattice size corresponds to 12.2 GB. Here, the symmetries (\ref{propsymmetries}) have already been exploited.

To be able to allocate and store such large address spaces of memory, all I/O-routines had to be rewritten to work with a 64Bit address space. 

\section{Testing the code}
Previous to the production runs, the adopted MILC code had to pass through a number of tests which are described in the following. All the tests were done on the IBM computer.
In particular, the ALPHA-Collaboration's lattice gauge theory code for the QCD Schr\"odinger Functional was taken as a reference to test all routines of the MILC-code that are involved in the production runs.

\subsection{Testing the plaquette}
Up to the pseudo random number generator, the MILC code is based on the same gauge update algorithm as the one that is implemented in the ALPHA code. The average plaquette value was taken as a test observable to evaluate the functioning of the MILC code gauge-update algorithm.

One test consisted of determining the normalized average plaquette value
\be\label{avplaqu}
\overline{U}_p={1\over 9L^3(2T-1)}\sum\limits_{p}\langle U(p)\rangle
\ee
with both, the ALPHA and the MILC code. The results are compiled in the following table.
\begin{center}
\begin{tabular}{cccccc}
\hline
\hline
\vspace{-0.4cm}\\
$L^3\times T$	&statistic	&$\overline U_p^{\rm MILC}$	&$\overline U_p^{\rm ALPHA}$	&$\tau_{\rm int}^{\rm MILC}$ 	&$\tau_{\rm int}^{\rm ALPHA}$\\[.5ex]
\hline
\vspace{-.4cm}\\
$4^3\times 4$	&1000		&0.62871(37)			&0.62850(32)			&0.051(9)			&0.041(6)	\\
$16^3\times 32$	&100		&0.59319(3)			&0.59323(3)			&0.004(1)			&0.003(1)	\\[.5ex]
\hline
\hline
\end{tabular}
\end{center}
In the case of the smaller lattice, the measurements were separated by one heat bath and 20 over relaxation steps and in the case of the larger lattice they were separated by 100 heat bath steps, each followed by 8 over relaxation steps. The integrated auto correlation time $\tau_{int}$ is quoted in units of sweeps, not distinguishing between over relaxation and heat bath sweeps.

As it will be detailed in chapter \ref{analysis}, the production runs with the MILC code were done with a geometry of $(L/a)^3\times T/a=48^3\times96$ at $\beta=6.7859$. The lattice cutoff is already large for this $\beta$ ($a/r_0\approx0.0625\to 6.3$GeV)\footnote{The Sommer scale $r_0=0.5$ fm was employed to convert to physical units (cf. section \ref{settingthescale}).} and perturbation theory for the value of the plaquette should give a rough estimate.

\begin{figure}
\begin{center}
\psfrag{NSPT}[l][l][1][0]{NSPT}
\psfrag{Monte Carlo data Mon}[l][l][1][0]{Monte Carlo data}
\psfrag{av. Plaquette}[b][c][1][0]{$\overline{U}_p$}
\psfrag{g0sq}[c][c][1][0]{$g_0^2$}
\epsfig{scale=.6,file=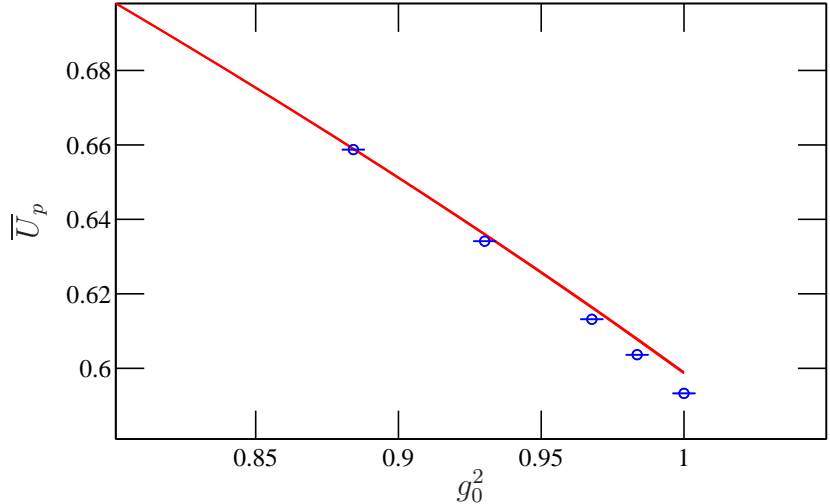}
\caption{Average Plaquette $\overline{U}(p)$ from Numerical Stochastic Petrurbation Theory (line with error band) and Monte Carlo simulation (circles). }\label{NSPT}
\end{center}
\end{figure}

To evaluate whether the MILC-code produces sensible gauge configurations, a comparison of the average value of the plaquette to computations in Numeric Stochastic Perturbation Theory (NSPT) \cite{DiRenzo:2000ua} has been carried out. The result is depicted in figure \ref{NSPT}. The circles correspond to Monte-Carlo data, generated with  the MILC code. The corresponding parameters are (100 measurements at each value of $\beta$):
\begin{center}
\begin{tabular}{cccccc}
\hline
\hline
\vspace{-.4cm}\\
$\beta$& 6.0& 6.1& 6.2& 6.45& 6.7859\\[.5ex]
\hline
\vspace{-.4cm}\\
$L^3\times T$ &$16^3\times32$ &$24^3\times 40$&$24^3\times48$&$32^3\times 64$&$48^3\times 96$\\[.5ex]
\hline
\hline
\end{tabular}
\end{center}
The solid line (where the error band is barely visible) corresponds to the pla\-quette obtained at 10th order in NSPT. Although the non-perturbative data tends to agree with NSPT for larger values of $\beta$, there remains a discrepancy at $\beta=6.7859$. The MILC code gives $\overline{U}_p=0.658732(3)$, while NSPT yields $\overline{U}_p=0.65896(15)$. 

Still, the findings indicate, that the MILC code produces sensible gauge configurations, since the discrepancy between NSPT and the non-per\-tur\-ba\-tive result for the plaquette can be expected to vanish for larger values of $\beta$.
\subsection{Testing the implementation of the correlation functions}
All correlation functions introduced in section \ref{SFobservables} were implemented in the MILC-code and tested against the ALPHA-code. For this test, a PERL-script was written (\verb|alpha2milc|), that converts gauge-configurations produced by the ALPHA-colla\-bo\-ra\-tion's lattice gauge code into a format, which can be imported by the MILC-code. In this way, the correlation functions could be evaluated on the same gauge background with both codes. The test was done on a gauge configuration ($16^3\times 32$-geometry), generated by the ALPHA-code at $\beta=6.0$, and with seven different values of the hopping parameter with a solver stopping criterion of $\epsilon = 10^{-7}$ (The $\kappa$-values are the ones that are tabulated in table \ref{hoppingparams} for $\beta=6.0$). All the correlation functions $f_A,f_P,f_V,f_T$ and $f_P^T$ and $f_V^T$ were evaluated for the combinations of hopping parameters $\kappa_1-\kappa_1,\kappa_1-\kappa_2,\dots,\kappa_1-\kappa_7$. 
Averaged over the correlation functions at all times $x_0$, there was a mean deviation of $0.03\%$. A maximum deviation of $0.5\%$ was observed for the value of the boundary-to-boundary correlation function $k_{\rm V}^T$ at the largest value of the mass. 

As communicated with the authors of \cite{Rolf:2002gu,Juttner:2003ns}, where roundoff errors in the same correlation functions where studied, the deviations of the magnitude observed here lead to systematic errors in the final results, e.g. for the decay constant, that are much smaller than the expected statistical error.

\subsection{Testing the whole setup: A comparative test-run}\label{testrun}
This test was carried out in order to simulate the conditions for a production run with the MILC code, while having results from the ALPHA code with the same simulation parameters as a reference.

The reference from the ALPHA code was given in terms of 200 measurements of the correlation functions necessary to construct the decay constant ${\rm F_{PS}}$, i.e. $f_A^I(x_0)$ and $f_P^T$. The data was taken from the production runs for \cite{Juttner:2003ns} at $\beta=6.0$. 
As a test observable, the pseudo scalar meson decay constant $\rm F_{\rm PS}$ as a function of the inverse pseudo scalar mass $1/r_0 m_{\rm PS}$ was studied. The above data contains all the correlation functions for seven meson masses. While the light quark mass was fixed to the strange quark mass, altogether six heavy quark masses around the charm quark mass were simulated for (The hopping parameters are those summarized in table \ref{hoppingparams}).

\begin{figure}
\centering
\psfrag{oor0mPS}[t][c][1][0]{$1/r_0m_{\rm PS}$}
\psfrag{r0FPS}[c][c][1][0]{$r_0{\rm F_{PS}}$}
\epsfig{scale=.7,file=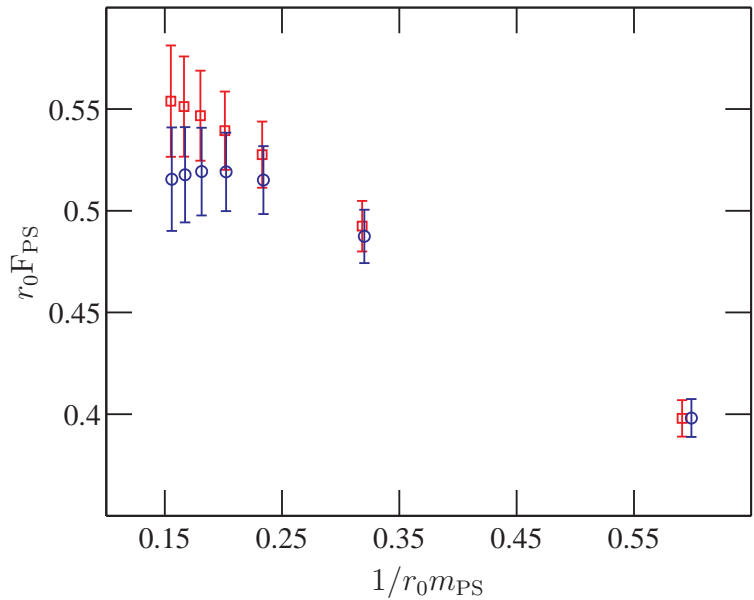}
\caption{The meson mass dependence of the decay constant from a comparative test-run between the ALPHA-code (squares) and the MILC-code (circles) at $\beta=6.0$. $N_{\rm meas.}=200$.}\label{b6p0-comparative}
\end{figure}

The same number of measurements were then carried out with the MILC code, employing double precision arithmetics. The data analysis was done with the procedure that will be explained in detail in chapter \ref{analysis}. The dependence of the decay constant on the meson mass is plotted in figure \ref{b6p0-comparative} which shows, that the data agrees within errors.

Both codes produce compatible results under production conditions.

\chapter{Simulations in quenched QCD}\label{simulations}
The lattice simulations in quenched QCD were done for heavy-light mesons containing a strange quark as the light quark and a quark with a mass around the charm quark mass as the heavy quark. The correlation functions for altogether six different heavy quark masses were simulated at five different values of the lattice spacing in order to allow for a reliable continuum extrapolation. This chapter presents the choice of simulation parameters.
\section{Simulation parameters}
\subsection{Setting the scale}\label{settingthescale}
Dimensionful parameters are often given in terms of the Sommer scale $r_0$ \cite{Guagnelli:1998ud}. This is convenient in so far as the attribution of physical units to results in the quenched approximation is ambiguous.
For example, lattice studies of the hadronic spectrum in the quenched theory have revealed inconsistencies in the comparison with the experiment of up to 10\% \cite{Aoki:2002fd}. Turning this around, one expects different hadronic input to influence the results for observables.

The Sommer scale has two advantages. On the one hand, $[r_0/a](\beta)$ has been determined precisely over the range of cutoff values \cite{Guagnelli:1998ud,Edwards:1998xf,Necco:2001xg} which have been used in the scaling study in this thesis. Furthermore, in the absence of dynamical quarks, $r_0$ is only affected by cutoff effects of order $a^2$ \cite{Necco:2001xg}.

When using physical input from experiments, the conversion will be done by setting $r_0=0.5$ fm. For this value of the reference scale, the decay constant of the $K$-meson takes its physical value $r_0{\rm F_K}=0.405(5)$ in quenched lattice simulations \cite{Garden:1999fg}. One therefore often refers to this choice as \emph{setting the scale with the Kaon decay constant}.
Setting the scale with the nucleon mass instead, the reference scale would have to be $r_0^\prime=0.55$ fm to match the experimental value. This corresponds to a 10\% shift in the scale.
An estimate for the size of the ambiguity for the observables determined in this work will be given in chapter \ref{QSA}.

A parameterization of the ratio $a/r_0$ has been obtained as a function of the bare coupling $\beta=6/g_0^2$ in \cite{Necco:2001xg},
\be\ba{rcl}\label{Sommerscale}
\left[{a\over r_0}\right](\beta)& =& \exp\left\{ -1.6804 - 1.7331(\beta-6) + 0.7849(\beta-6)^2-0.4428(\beta-6)^3\right\},\\
\\
&&{\rm for}\; 5.7\le\beta\le 6.92.
\ea
\ee
The accuracy of this formula ranges from $0.5\%$ at low values of $\beta$ to $1\%$ at large values of $\beta$.

\subsection{Parameters for the scaling study}
For the scaling study, simulations were done for five different lattices of approximately constant physical size $L/r_0\approx 3$, but decreasing lattice spacing. The values $\beta_1,\dots,\beta_4$ were taken over from previous work within the ALPHA-collaboration \cite{Juttner:2003ns}. The value of $\beta_5$ was obtained with (\ref{Sommerscale}). The corresponding lattice geometries are given in table \ref{scalingparams}. 

As the time-extent of the lattice, $T\approx 2L$ was chosen. The experience from earlier simulations with the Schr\"odinger Functional, e.g. in \cite{Garden:1999fg,Rolf:2002gu} showed, that with this asymmetric geometry, the plateaus of the effective masses (\ref{effectivemass}) and decay constants (\ref{fdsffinal}) are well pronounced.
\begin{table}
\begin{center}
\begin{tabular}[h]{lccccccc}
\hline
\hline
\vspace{-.4cm}\\
	&$L/a$	&$T/a$	&$L/r_0$	&$a[fm]$&$n_{\rm meas}$&$n_{\rm update}$&$n_{\rm OR}$\\[.5ex]
\hline\\[-2ex]
$\beta_1=6.0$ 	&$16$	&$32$	&$2.98$ 	&$0.093$&380&100&8\\
$\beta_2=6.1$ 	&$24$ 	&$40$	&$3.79$ 	&$0.079$&201&100&12\\
$\beta_3=6.2$	&$24$ 	&$48$	&$3.26$ 	&$0.068$&251&100&12\\
$\beta_4=6.45$ 	&$32$	&$64$	&$3.06$ 	&$0.048$&289&100&12\\
$\beta_5=6.7859$ 	&$48$	&$96$	&$3.00$	&$0.031$&150&50 &24\\	[.5ex]
\hline
\hline
\end{tabular}
\end{center}
\caption{Lattice geometries and simulation parameters used in the simulations for the scaling study.}\label{scalingparams}
\end{table}
A numerical check for pseudo scalar mesons in \cite{Garden:1999fg} revealed, that with a lattice extent of $L/r_0\approx 3$ fm, $T=2L$, finite volume effects can be neglected when simulating for mesons containing a strange quark as the light quark. The corresponding vector mesons are heavier and therefore have a shorter Compton wavelength. It is concluded, that finite volume effects for this channel are also negligible.

A peculiar choice for the lattice geometry had to be made at $\beta_2=6.1$, where the hardware and memory configuration of the APEMille computers at NIC/DESY-Zeuthen \cite{NIC} did not allow to follow the relation $T=2L$.
\subsection{Improvement and renormalization constants}\label{imprandrenormconstants}
Most of the improvement and renormalization constants needed for the derivation of simulation parameters and for the data analysis have already been introduced in chapter \ref{technicalissues}. Where available, only non-perturbatively determined improvement and renormalization constants were used. 
The parameterizations of all constants in terms of the bare coupling $g_0$ or $\beta=6/g_0^2$ are given in appendix \ref{parametrizationtables}, in the tables \ref{renormalizationtab} and \ref{improvementtab}. 

Two additional renormalization constants, namely $Z$ \cite{Guagnelli:2000jw} and $Z_M$ \cite{Capitani:1998mq} will be needed. On the one hand,
\be
Z={m_i\over m_{q,i}}
\ee
relates the bare current quark mass $m_i$ (PCAC) of some flavor $i$ to the subtracted bare quark mass $m_{q,i}$, while
\be\label{ZMzerlegt}
Z_M(g_0)=\frac{M_i}{\overline{m}_i(\mu)}\frac{Z_A(g_0)}{Z_P(g_0,\mu)}
\ee
on the other hand that relates the bare current quark mass to the renormalization group invariant mass $M_i$. 

In \cite{Capitani:1998mq}, the non-perturbative parameterization of $Z_A(\beta)$ \cite{Luscher:1997jn} together with simulations for the ratio $M_i/\overline{m}_i(\mu=(2L_{\rm max})^{-1})=1.157(12)$ and for $Z_P(g_0,\mu=(2L_{\rm max})^{-1})$ at five values of $\beta$ were combined to give a parameterization for $Z_M(g_0)$ in the range $6.0\le \beta =6/g_0^2\le6.5$. The scale dependence of the parameters was computed in the mass independent Schr\"odinger Functional scheme, where the renormalization scale is given by the finite box size, $L=1/\mu$. In this particular case, $\mu=(2L_{\rm max})^{-1}\approx(1.436r_0)^{-1}$ was used.

With further data from \cite{Capitani:1998mq}, it is possible to derive a parameterization of $Z_M(g_0)$ for a larger range of $\beta$ which will be needed in the continuum extrapolation for the renormalization group invariant charm quark mass $M_c$ (cf. table \ref{scalingparams}). 

First, one has to find an extended parameterization of $Z_P(g_0,(2L_{\rm max})^{-1})$. This factor has been determined for several values of $\beta$, once along lines of constant physics defined by the fixed volume $L_{\rm max}$ and once along an equivalent (in the continuum) condition  defined by the renormalized coupling in the Schr\"odinger Functional \cite{Luscher:1994gh}, $Z_P(g_0,(2L)^{-1})_{|\bar g^2(1/L)=3.48}$ \cite{Capitani:1998mq}. The corresponding data is summarized in table \ref{ZPdet}. The two definitions of lines of constant physics differ by $O(a^2)$, since the simulations have been done in the improved theory.

\begin{figure}
\begin{center}
\psfrag{beta}[c][c][1][0]{$\beta$}
\psfrag{ZP}[c][c][1][0]{$Z_P(\beta,\mu\approx1/436r_0)$}
\psfrag{val}[lb][l][1][0]{\small$\kappa_{{\rm crit}}(6.7859)=0.13020(5)$}
\epsfig{scale=.8,file=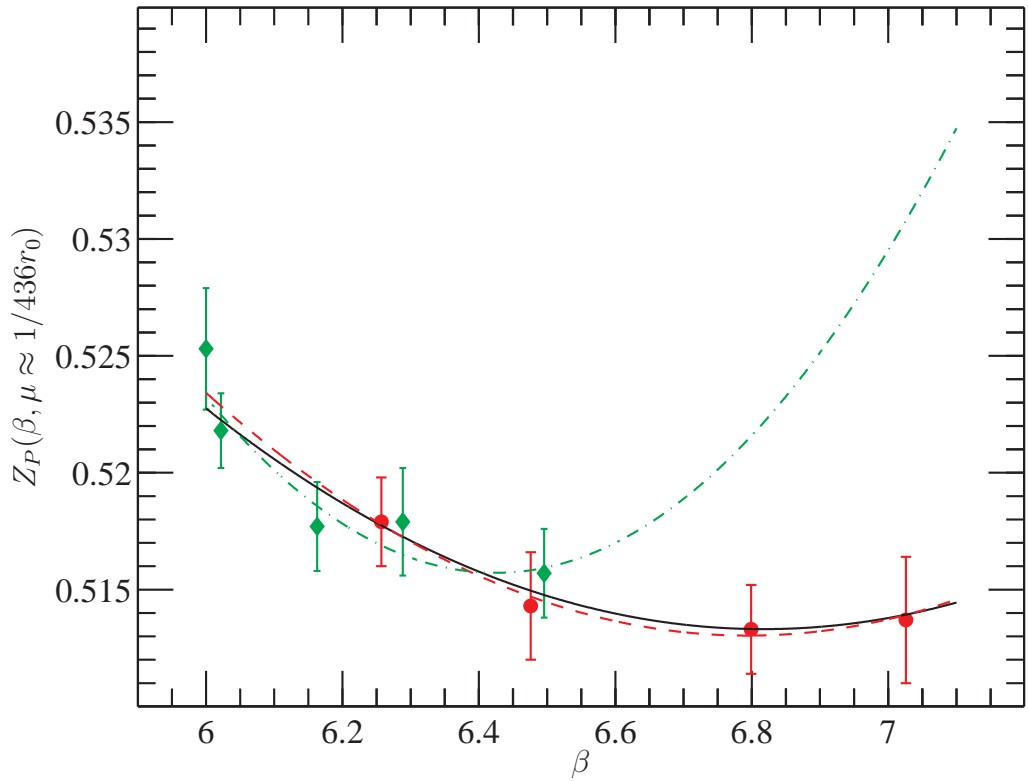}
\end{center}
\caption{The renormalization factor $Z_P(g_0,1/2L_{\rm max})$. The diamonds correspond to non-perturbative data at fixed $2L_{\rm max}=1.436r_0$ and the circles correspond to non-perturbative data obtained from the simulations at fixed $\bar g^2(1/L)=3.48$.}\label{ZPplot}
\end{figure}

The two data sets are compatible, as can be seen in figure \ref{ZPplot}. The diamonds represent the data along constant volume and the circles along constant renormalized coupling. The parameterizations for $Z_P(g_0,(2L_{\rm max})^{-1})$ \cite{Capitani:1998mq} and a new parameterization for the data of $Z_P(g_0,(2L)^{-1})_{|\bar g^2(1/L)=3.48}$ have been added to the plot (dash-dotted and dashed line). The latter parameterization describes both the data sets very well in the range $6.0\le \beta\le 7.0$ and the combination of both data sets into one parameterization is thus justified. For a polynomial ansatz one obtains
\be\label{erratum3}
Z_P(\beta,(2L_{\rm max})^{-1})= 0.5228  -0.0231(\beta-6)+ 0.0142(\beta-6)^2.
\ee
It describes the data with a maximal deviation of 0.5\% and is plotted in figure \ref{ZPplot} as the solid line. 

In a second step, the data for $Z_P(g_0,(2L_{\rm max})^{-1})$ can be combined with the ratio $M_i/\overline{m}_i((2L_{\rm max})^{-1})=1.157(12)$ and with the non-perturbatively determined values of $Z_A(\beta)$ to give $Z_M(g_0)$ which then can be parameterized as
\be\label{ZMnew}
Z_M(\beta)=    1.754+  0.27(\beta-6) -0.10(\beta-6)^2.
\ee
This parameterization describes the data within 0.5\% accuracy. The error of $Z_A(g_0)$ and of the ratio $M_i/\overline{m}_i((2L_{\rm max})^{-1})$ was taken into account in the error analysis.
\begin{table}
\centering
\begin{tabular}{ccccccc}
\hline\hline&&&&&&\\[-2ex]
\multicolumn{7}{c}{fixed $2L_{\rm max}\approx1.436r_0$}\\[1ex]
\hline\\[-2ex]
$\beta$		&6.0 &6.0219 &6.1628 &6.2885 &6.4956\\
$Z_P(g_0,(2L_{\rm max})^{-1})$	&0.5253(26) &0.5218(16)&0.5177(19)&0.5179(23)&0.5157(19)\\[1ex]
\hline\\[-2ex]
\multicolumn{7}{c}{fixed $\bar g^2(1/L)=3.480(13)$}\\[1ex]
\hline\\[-2ex]
$\beta$		&6.257 &6.476 &6.799 &7.026\\
$Z_P(g_0,(2L)^{-1})$  &0.5179(19) &0.5143(23) &0.5133(19) &0.5137(27)\\[1ex]
\hline\hline
\end{tabular}\caption{Non-perturbative data for $Z_P(g_0,(2L_{\rm max})^{-1})$.}\label{ZPdet}
\end{table}

\subsection{Hopping parameters and stopping criteria}\label{Hoppsandstops}
In order to obtain correlation functions for mesons containing a strange quark and for various heavy quark masses in the range of charm, propagators for seven different hopping parameters were determined in the Monte-Carlo simulation. They were chosen such, that the lightest quark mass corresponded to the strange quark mass. A second hopping parameter was tuned to simulate at the charm quark mass and the remaining five parameters were distributed homogeneously over the mass range between the strange quark mass and the lattice cut-off.\\
\subsubsection{$\kappa_{{\rm crit}}$ - the critical hopping parameter}
\begin{figure}
\begin{center}
\psfrag{beta}[c][c][1][0]{\large$\beta$}
\psfrag{kappac}[c][c][1][0]{\large$\kappa_{{\rm crit}}$}
\psfrag{val}[lb][l][1][0]{$\kappa_{{\rm crit}}(6.7859)=0.135120(5)$}
\epsfig{scale=.7,file=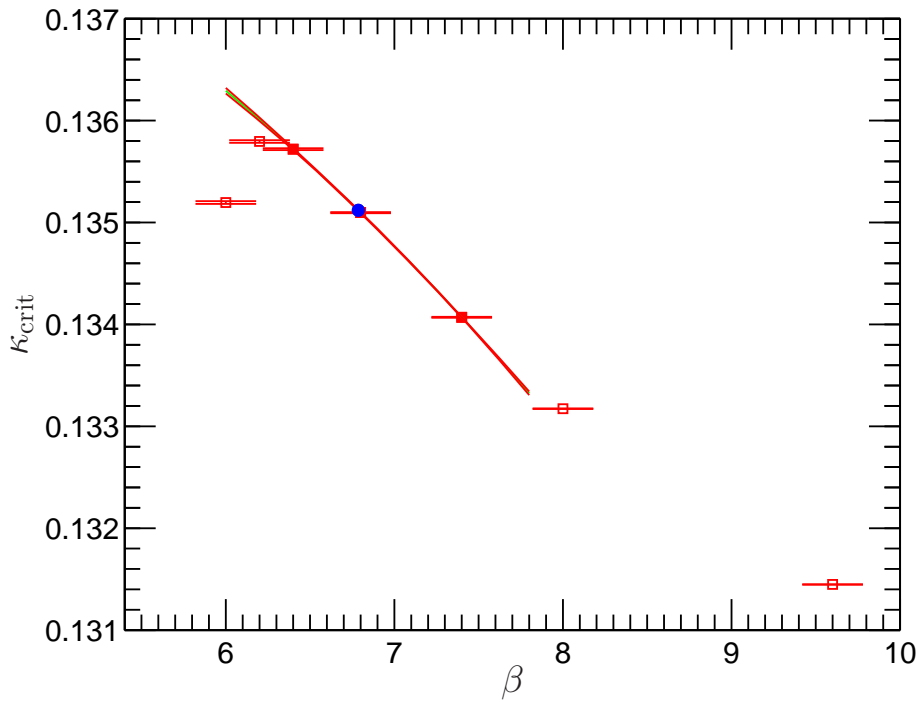}
\caption{Quadratic interpolation for $\kappa_{{\rm crit}}(\beta)$. Only the data at the filled squares was included into the fit.}\label{kappacdetermination}
\end{center}
\end{figure}

$\kappa_{{\rm crit}}$ has been determined in \cite{Luscher:1997ug} in quenched QCD for a large range of $\beta$-values (c.f. table 1 therein). For $\beta=6.0$ and $6.2$, the values could be taken over directly. In collaboration with the authors of \cite{Luscher:1997ug}, the critical hopping parameters for $\beta=6.1,6.45,$ and $6.7859$ were determined from quadratic interpolations in $\beta$, of which the case $\beta=6.7859$ has been illustrated representatively in figure \ref{kappacdetermination}.
\subsubsection{$\kappa_s$ - the hopping parameter corresponding to the strange quark}
\begin{figure}%
\centering
\psfrag{aor0sq}[t][t][1][0]{$({a/ r_0})^2$}
\psfrag{ZmRor0}[b][t][1][0]{$Z_{\rm M}{{\rm F}_{\rm K}\over G_{\rm K}}$}
\psfrag{val}[lb][l][1][0]{$Z_{\rm M}{{\rm F}_{\rm K}\over G_{\rm K}}(\beta=6.7859)=0.2268(57)$}
\epsfig{scale=.8,file=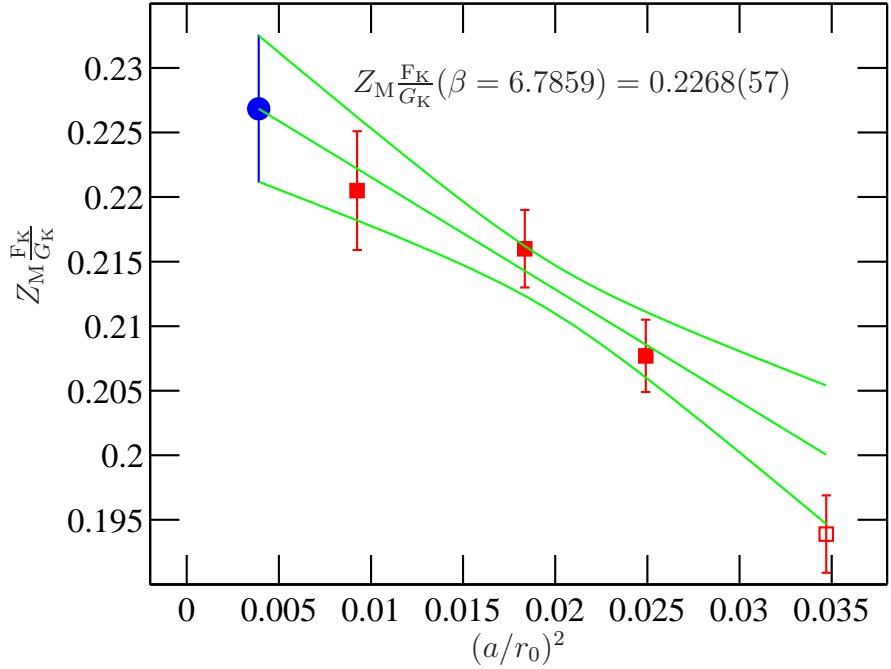}
\caption{Linear extrapolation in $({a/ r_0})^2$ for $Z_{\rm M}{{\rm F}_{\rm K}\over G_{\rm K}}$ with $1\sigma$-error-band. Only the data at the filled squares was included into the fit. The circle represents the extrapolated value at $\beta=6.7859$.}\label{extrapols2}
\end{figure}

The hopping parameters for the strange quark for $\beta_1-\beta_5$ were obtained, using previous work by the ALPHA- and UKQCD-collaboration \cite{Garden:1999fg}. There, the sum of the renormalization group invariant strange and light quark mass $M_s+\hat M$ with $\hat M = \oh (M_u+M_d)$ has been determined in a quenched simulation with $\orda$-improved Wilson fermions. The basic idea in this work was to exploit the PCAC-relation
\be\label{M_m}
M_s+\hat M=Z_{\rm M}{{\rm F}_{\rm K}\over G_{\rm K}}m_{\rm K}^2
\ee
between the renormalization group invariant quark masses of the strange quark, $M_s$, the average light quark mass $\hat M=\oh(M_u+M_d)$ and the mass of the corresponding meson, $m_{\rm K}$. $Z_{\rm M}$, as detailed in the previous section, relates the bare current quark mass to the renormalization group invariant quark mass. ${\rm F}_{\rm K}$ is the Kaon decay constant and $G_{\rm K}$ denotes the vacuum-to-$K$ matrix element of the pseudo scalar density, which was determined in that work \cite{Garden:1999fg}.\\
The computation used $r_0m_{\rm K}=1.5736$ \cite{PDBook} as experimental input. The following table summarizes the results for $Z_{\rm M}{{\rm F}_{\rm K}\over G_{\rm K}}$ at $\beta=6,6.1,6.2,6.45$ obtained with $O(a)$-improved Wilson fermions. The value at $\beta=6.7859$ has been extrapolated linearly in $(a/r_0)^2$ based on the numerical data at $\beta = 6.1, 6.2, 6.45$, as is depicted in figure (\ref{extrapols2}).
\begin{center}
\begin{tabular}{cccccc}
\hline\hline\\[-2ex]
$\beta$		&$6$		&$6.1$		&$6.2$		&$6.45$		&$6.7859$\\[1ex]
${Z_{\rm M}{\rm F}_{\rm K}\over G_{\rm K}r_0}$
		&$0.1939(30)$ 	&$0.2077(28)$ 	&$0.2160(30)$ 	&$0.2205(46)$	&$0.2268(57)$\\[1ex]
\hline\hline
\end{tabular}
\end{center}
$M_s$ can be extracted from (\ref{M_m}) by using the ratio $M_s/\hat M=24.4\pm1.5$ from chiral perturbation theory \cite{Leutwyler:1996qg}. 
Together with the definition of the subtracted bare quark mass (\ref{hoppingparam}), this defines the hopping parameter $\kappa_s$ of the strange quark as the solution of the quadratic equation in $am_{q,s}=\oh\left({\kappa_s^{-1}}-{\kappa_{\rm crit}^{-1}(g_0)}\right)$,
\be\label{RGI_curr}
r_0(M_s+\hat M)=r_0M_s(1+\frac{\hat M}{M_s})=\left[\frac{r_0}{a}\right]Z_M Z  a m_{q,s}(1+b_mam_{q,s}).
\ee
\begin{figure}%
\centering
\psfrag{aor0sq}[t][t][1][0]{$({a/ r_0})^2$}
\psfrag{r0Mc}[t][t][1][0]{$r_0M_c$}
\psfrag{val}[lb][l][1][0]{$r_0M_c(\beta=6.7859)=4.277(85)$}
\epsfig{scale=.8,file=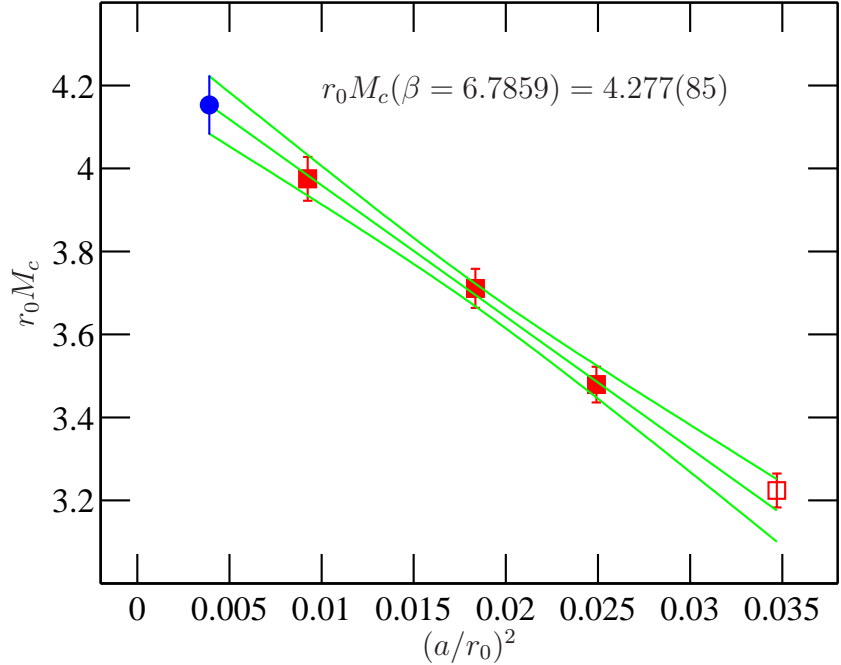}
\caption{Linear extrapolation in $({a/ r_0})^2$ for $r_0M_c$. Only the filled squares entered the fit. The circle represents the extrapolated value at $\beta=6.7859$.}\label{extrapols}
\end{figure}
\subsubsection{$\kappa_c$ - the hopping parameter corresponding to the charm quark}
The hopping parameters at $\beta_1\dots\beta_4$ for the charm quark have been determined in the computation of the renormalization group invariant mass $M_c$ \cite{Rolf:2002gu}.  
With the values for $M_c$ at finite lattice spacing ($\beta_1-\beta_4$) listed in the following table, one can extrapolate linearly in $(a/r_0)^2$ to obtain a first guess at $\beta_5$. The extrapolation is illustrated in  (c.f. figure \ref{extrapols}). 
\begin{center}
\begin{tabular}{cccccc}
\hline\hline\\[-2ex]
$\beta$		&$6$		&$6.1$		&$6.2$		&$6.45$		&$6.7859$\\[1ex]
$r_0M_c$		&$3.224(41)$ 	&$3.479(43)$ 	&$3.711(47)$ 	&$3.975(53)$	&$4.277(85)$\\[1ex]
\hline\hline\\
\end{tabular}
\end{center}
As in the case of the strange quark hopping parameter, $\kappa_c$ can again be obtained by solving the quadratic equation (\ref{RGI_curr}).

\subsubsection{Five additional hopping parameters }
Five additional hopping parameters were guessed such that a roughly uniform distribution of quark masses between the strange quark and half the $b$-quark mass for $\beta_1,\dots,\beta_4$ and 4.5 GeV in the case $\beta_5$ was achieved.
In particular, the parameters were determined from a linear inter- resp. extrapolation with respect to the pseudo scalar masses $r_0m_{\rm K}=1.5735$ and $r_0m_{{\rm D}_s}=4.988$ \cite{PDBook}. 
\begin{table}\begin{center}
\begin{tabular}{c|ccccc}
$\beta$	&6		&6.1		&6.2		&6.45		&6.7859\\
\hline\\[-2ex]
$\kappa_{\rm crit}$
	 &0.135196	& 0.135496	& 0.135795	& 0.135701	& 0.135120\\[1ex]
$\kappa_1$&\multicolumn{1}{|C}{0.134108}	&\multicolumn{1}{C}{0.134548}	&\multicolumn{1}{C}{0.134959}	&\multicolumn{1}{C}{0.135124}	&\multicolumn{1}{C}{0.134739}\\
$\kappa_2$&0.128790	&0.130750	&0.131510	&0.132690	&0.132440\\	
$\kappa_3$&0.123010	&0.125870	&{0.127470}	&0.130030	&\multicolumn{1}{C}{0.130253}\\
$\kappa_4$&\multicolumn{1}{|C}{0.119053}	&\multicolumn{1}{C}{0.122490}	&\multicolumn{1}{C}{0.124637}	&\multicolumn{1}{C}{0.128131}	&0.128439\\
$\kappa_5$&0.115440	&0.119370	&0.122000	&0.126330	&0.126774\\
$\kappa_6$&0.112320	&0.116640	&0.119680	&0.124730	&0.123571\\
$\kappa_7$&0.109270	&0.113960	&0.117370	&0.123120	&0.117625\\
\end{tabular}

\caption{Summary of all hopping parameters. The shaded fields indicate the value of the hopping parameters corresponding approximately to the strange quark and to the charm quark.}\label{hoppingparams}
\end{center}
\end{table}
\subsubsection{The solver stopping criterion $\epsilon$}\label{stoppingcrit}
Heavy-light correlation functions decay exponentially with the meson mass in time. Thus, large components contribute in the heavy quark propagator for initial times and small components for large times $x_0$. 
If one takes for example the heavy-light correlation function $f_A(x_0)$, that decays by 28 orders of magnitude between $x_0=a$ and $x_0=T-a$ for the cases with a very large heavy quark mass ($(L/a)^3\times (T/a)=48^3\times 96$, $\beta=6.7859$, $\kappa_1$ and $\kappa_7$ as in table \ref{hoppingparams} ). In its computation, sums over numbers of different order of magnitude have to be evaluated, thereby possibly introducing roundoff errors if the number precision is not sufficient. 
\begin{figure}
\centering
\begin{minipage}{\linewidth}
\centering
\psfrag{fa}[c][c][1][0]{$f_A$}
\psfrag{fp}[c][c][1][0]{$f_P$}
\psfrag{kv}[c][c][1][0]{$k_V$}
\psfrag{kt}[c][c][1][0]{$k_T$}
\psfrag{prozent}[b][c][1][0]{Relative error in $\%$}
\psfrag{kombination}[bc][c][1][0]{for $\kappa_1-\kappa_3$}
\psfrag{exp}[bc][c][1][0]{\tiny $\times 10^{-4}$}
\epsfig{scale=.8,file=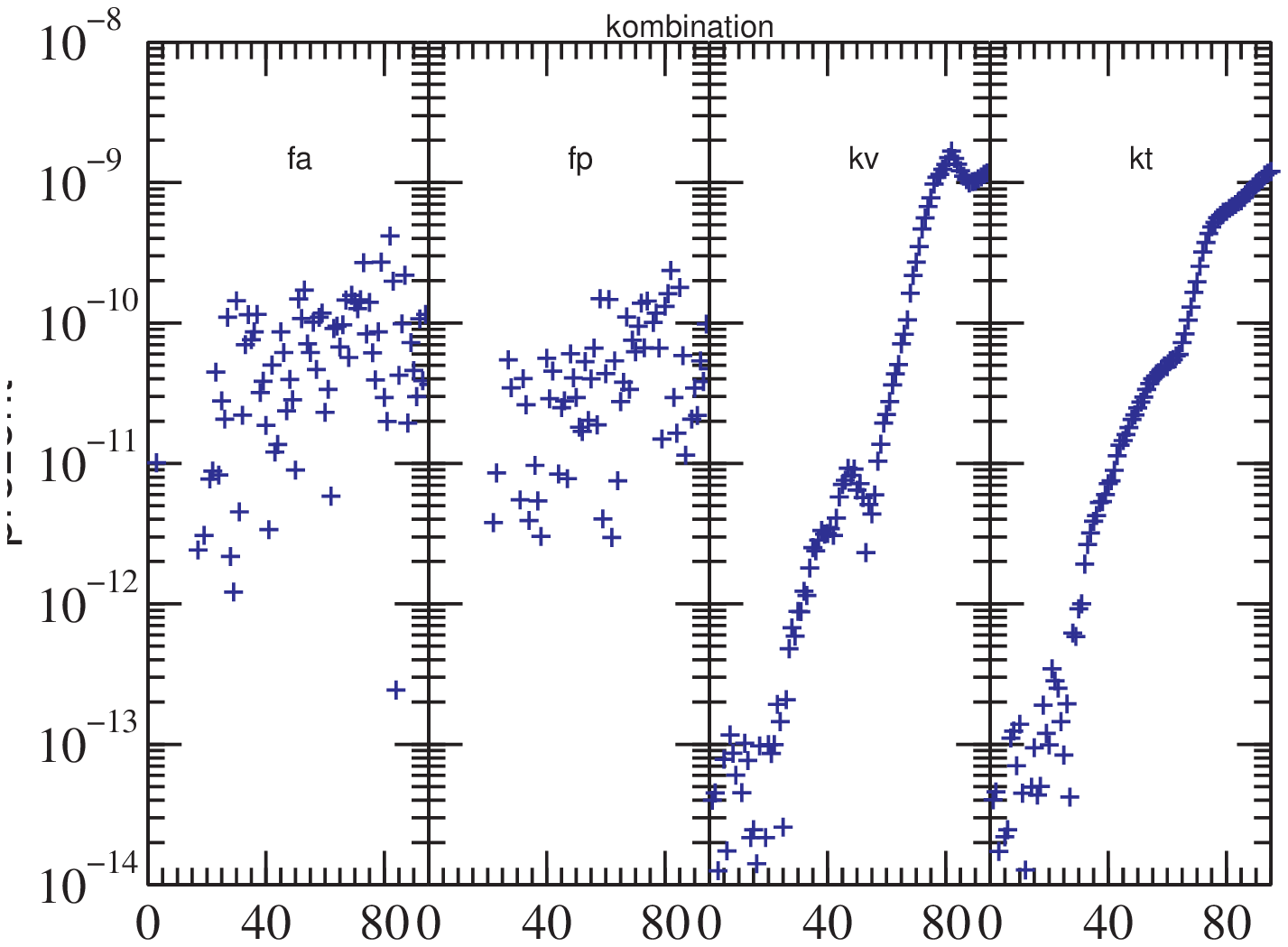}\label{Obsrelerr}
\end{minipage}
\begin{minipage}{\linewidth}
\centering
\vspace{.3cm}
\psfrag{fa}[t][c][1][0]{$f_A$}
\psfrag{fp}[t][c][1][0]{$f_P$}
\psfrag{kv}[t][c][1][0]{$k_V$}
\psfrag{kt}[t][c][1][0]{$k_T$}
\psfrag{prozent}[b][c][1][0]{Relative error in $\%$}
\psfrag{kombination}[bc][c][1][0]{for $\kappa_1-\kappa_5$}
\epsfig{scale=.8,file=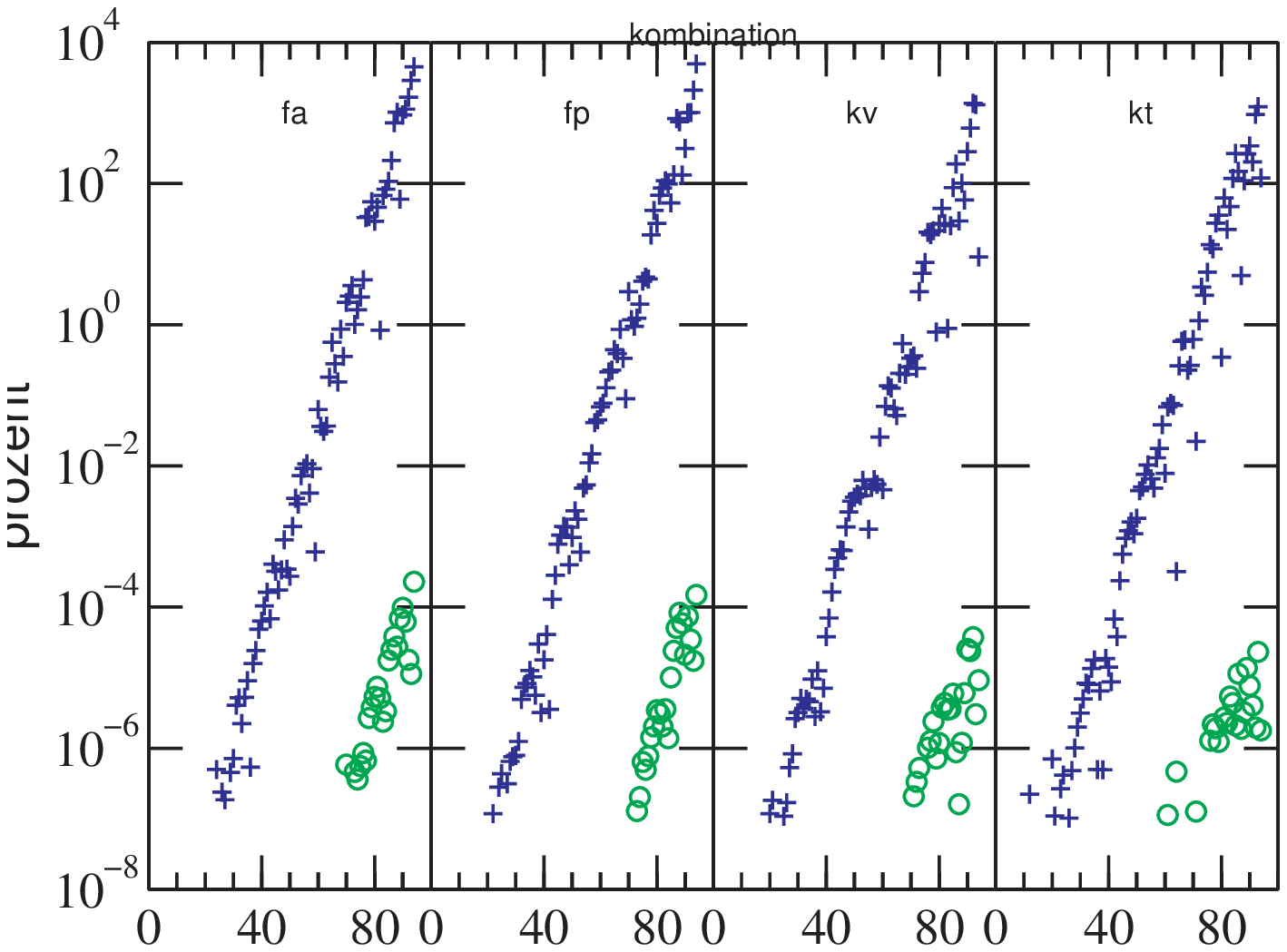}\label{Obsrelerr}
\end{minipage}
\caption{Upper plot: Relative deviation between correlation functions evaluated once with a stopping criterion of $\epsilon = 10^{-7}$ and once with $\epsilon = 10^{-16}$. Lower plot: Relative deviation between $\epsilon = 10^{-12}$ and $\epsilon = 10^{-13}$ (plus signs) and between $\epsilon = 10^{-13}$ and $\epsilon = 10^{-14}$ (circles).
Each plot shows the relative deviation for the correlation functions $f_A(x_0)$, $f_P(x_0)$, $k_V(x_0)$ and $k_T(x_0)$ for $x_0/a=1\dots95$. Missing markers indicate vanishing relative deviation.}\label{relativedeviation} 
\end{figure}

Adjusting the stopping criterion of the inverter is a crucial task in the preparation of the production runs. A stricter value for $\epsilon$ causes the inverter to iterate longer. But the stopping criterion is also connected with the precision to which the observables constructed from quark propagators are determined. The runs at $\beta=6.0,6.1,6.2$ and 6.45 were all done in single precision and the value of $\epsilon=10^{-7}$ was taken for all hopping parameters. The run with the MILC code at $\beta=6.7859$ used double precision arithmetics.  
In order to find out the optimal stopping criteria for this production run, the quark propagators for all seven values of the hopping parameter were computed on a gauge background, once with $\epsilon = 10^{-7}$ and once with $\epsilon = 10^{-16}$. All meson correlation functions defined in chapter \ref{fermobs}, were then computed for all combinations of the hopping parameters $\kappa_1-\kappa_2,\kappa_1-\kappa_3,\dots,\kappa_1-\kappa_7$. Then, the relative deviations of correlation functions were computed. It turned out, that especially the light quark propagators are not sensible to the changed stopping criterion (cf. figure \ref{relativedeviation}). The maximum deviation for the combinations of hopping parameters $\kappa_1 - \kappa_2,\dots,\kappa_1-\kappa_4$ was less than 0.01\% and the mean deviation of all correlation functions at all values of $x_0$ was better than 0.002\%. In contrast, for the combinations $\kappa_1 - \kappa_5,\kappa_1-\kappa_6$ and $\kappa_1-\kappa_7$ the correlation functions for the two different stopping criteria differed by orders of magnitude for times $x_0/a>50$. 
In a second test, the meson correlation functions for the combination of hopping parameters $\kappa_1-\kappa_5$ were computed with stopping criteria $\epsilon=10^{-13},10^{-14}$ and $10^{-15}$. The lower plot in figure \ref{relativedeviation} shows the result. While the correlation functions deviate sizably when changing the solver precision from $\epsilon=10^{-13}$ to $\epsilon=10^{-14}$ (crosses), the change is small between the two solver precisions $\epsilon=10^{-14}$ and $\epsilon=10^{-15}$ (circles). Thus, a conservative solver residual of $\epsilon=10^{-16}$ for the large quark mass associated to $\kappa_5$ was used. The same stopping criterion was applied to $\kappa_6$ and $\kappa_7$. However, roundoff errors could not be ruled out in these cases.

\section{The simulations}
The simulations at $\beta=6.0,\,6.1,\,6.2$ and $6.45$ were all done on the APEMille computer at NIC/DESY-Zeuthen \cite{NIC} with the ALPHA-collaboration's TAO-code, while the simulation at $\beta=6.7859$ was done with the MILC-code at the HLRN \cite{HLRN:2002}. The data of the former runs were already available when this project started and could be taken over. 

Beginning with a cold gauge configuration, O(100) update sweeps were done at each value of $\beta$ to guarantee thermalization. The average plaquette value (\ref{avplaqu}) was also monitored but took a stable mean earlier at every $\beta$.
The runs on the APEMille computer were not so time consuming and a large number of intermediate updates, followed by $O(L/2a)$ over relaxation steps was done at all values of $\beta$. The number of intermediate updates was reduced for the run at $\beta=6.7859$, but no correlation effects in the data were observed. 

After the thermalization phase, the configurations for the production were generated.\\ 

\noindent{\bf The runs on the IBM computers at the HLRN}\\[2ex]
The generation of gauge configurations was carried out with a parallelization over 64 CPUs where the 50 update steps take 3.4 hours. 

To save CPU-time, the generation of the field configurations was done using the code with single precision arithmetics.The gauge configurations obtained in this way were then converted to double precision arithmetics using a PERL-script.

During the computation of the correlation functions with double precision arithmetics, a total of 114 GB of memory had to be allocated: 6 GB for the gauge field, 56 GB for the storage of one color and two spin components of the seven quark propagators and the rest for auxiliary fields, necessary for the biconjugate gradient routine. This large amount of memory assigned to the auxiliary fields was due to the re-shuffling which was necessary for changing from \emph{site-major} to \emph{field-major} (cf. \ref{chapterperformance}).

As the IBM servers at the HLRN are divided into units of 8 CPUs (LPAR) with a shared memory of at least 64 GByte, it was expected, that the code performs properly with a parallelization over 64 CPUs. In this case, the program allocates 57GB of RAM on each LPAR. However, the code then ran very instable. Fluctuations of the total runtime by a factor of two were observed. The project consultants at the HLRN found out, that memory intensive system software  (e.g. the operating system itself and software connected to the GPFS file system management) consumed a considerable part of the shared memory. This caused the production code to swap, i.e., parts of the allocated memory temporarily had to be stored to disk in order to free dynamic memory, thereby slowing down the program considerably. In order to avoid these losings in performance, the production was reorganized to use 128 CPUs instead. This unfortunately prolonged waiting times in the Queue, because larger numbers of CPUs were free less frequently.

With a parallelization with 128 CPUs, the computation of the propagators for the seven hopping parameters given in table \ref{hoppingparams} took 9.5 hours. Loading the gauge configuration in the beginning, contracting the propagators to compute the correlation functions and saving propagators took a negligible amount of time.\\

\noindent{\bf Application for CPU-time}\\[2ex]
In the beginning, the plan was to produce a statistic of 200 measurements of all the forward and backward correlation functions at $\beta_5$.

The thermalization and most of the initial tests could be done at the HLRN before the official accounting started and was for free. For the production runs however, applications for CPU-time had to be written.

At the HLRN, the CPU-time is measured in NPL (\emph{Norddeutsche Parallel\-rechner-Leistungseinheit}). 1 NPL is defined as 1 hour wall-clock time on 32 IBM p690 CPU's.

At the time of writing of the first application, some of the improvements of the code were not yet implemented and one point in the Monte-Carlo-History was estimated to cost 128 NPL. On this basis, an application for 25600 NPL for one year was written, of which 10000 NPL were  granted  for the period 04-2003 until 03-2004. Another 10000 NPL were granted after an application for an upgrade in 09-2003.

The progress during the first six months of the production runs for $\beta=6.7859$ was very slow because the computing facilities did not provide the expected services. On the one hand, the computer at the HLRN had a lot of down times and the code did not perform very well, as detailed above. Furthermore, a fair queueing system, that prefers massively parallel applications from jobs with a small number of CPUs was introduced only after the first six months of the production. 

Most of the problems were finally resolved after nine months when the project was granted an extension of another three months. At this point, the generation of one gauge configuration cost 7.8 NPL and the measurement of the corresponding forward and backward correlation functions roughly 80 NPL. 

The project finished with a statistic of 151 measurements at $\beta=6.7859$.
The forward propagators and the backward propagators for the hopping parameters $\kappa_1$ and $\kappa_3$, which approximately correspond to the mass of the strange quark and the charm quark respectively, were stored on tape for 80 measurements. They can be restored for further usage.

\chapter{Data analysis}\label{analysis}
This chapter addresses the analysis of the data which were obtained from the Monte-Carlo simulations of lattice QCD in the quenched approximation. The continuum limit was taken for the decay constant and the mass splitting of the ${\rm {D_s^{(\ast)}}}$-meson and the renormalization group invariant charm quark mass. The combined analysis of the simulation results for a number of heavy quark masses around the charm quark mass, together with predictions and simulation results from HQET, allowed to determine the decay constant of the ${\rm B_s^{(\ast)}}$-meson and the corresponding mass splitting at the physical point from an interpolation in the meson mass. Furthermore, estimates for the magnitude of the first spin- and flavor-symmetry breaking terms in the $1/M_Q$-expansion in HQET could be given.
\section{Data analysis - general remarks}\label{dataanalysis}
The results and errors from the simulations at all values of $\beta$  have been obtained from statistical samples 
of the primary quantities
\be\ba{rl}
f_{ O}(x_0), g_{ O}(T-x_0)&{\rm for}\; O=A,\,V,\,P,\,T\; {\rm and}\\
\\
f_{ O}^T&{\rm for}\; O=P,\,V,
\ea
\ee
over the whole range of $x_0\in[a,T-a]$, using the jackknife method. The correlation functions for the forward and the backward direction $f_{ O}(x_0)$ and $g_{O}(T-x_0)$ have been averaged at each step of the Monte-Carlo history\footnote{This is possible only for vanishing background field in the Schr\"odinger Functional.},
\be
f_{ O}(x_0)\equiv \oh(f_{O}(x_0)+g_{ O}(T-x_0)).
\ee
Using the improvement and renormalization constants introduced in section \ref{imprandrenormconstants}, the secondary quantities
\be\ba{c}\label{2ndaryquant}
\fPS,\; \fV,\; \fPS/\fV,\; m_{\rm PS},\; m_{\rm V},\; \mPS-\mV \;{\rm and}\\
\\
{{Y_{\rm PS}}\over{C_{\rm PS}}},\;{Y_{\rm V}\over C_{\rm V}},\; {R\over C_{\rm PS/V}},\;\frac{\Delta m}{{C_{\rm spin}}}\\
\ea\ee
were constructed from them following the definitions in section \ref{meffandfds} and section \ref{asymptotics}. No autocorrelation in the data was observed.

The renormalization and improvement coefficients are only known up to statistical and systematic errors from their determination in lattice simulations. 
In order to take this error properly into account, a statistical sample of all constants with a Gaussian distribution around their mean and with the width defined by their error was generated, using a pseudo random number generator (\verb|randn| in MATLAB). These samples were handed over to the jackknife routine and then treated in the same way as the Monte-Carlo data of the primary observables. 

The error due to perturbation theory in the conversion functions $C_{\rm X}(M_Q/\Lambda_{\MSbar})$ as listed in table \ref{parametrizations} has been taken into account by error propagation.

\subsection{Plateaus}
First, jackknife results were determined for the effective mass
\be\ba{rcl}
m^{\rm X}_{\rm eff}(x_0+{a\over 2})&=&{1\over a}\ln\left(\frac{f^I_{ O}(x_0)}{f^I_{ O}(x_0+{a})}\right).\\
\ea\ee 
\begin{figure}
\begin{minipage}{.49\linewidth}
\centering
\psfrag{x0}[t][c][1][0]{\large $x_0/a$}
\psfrag{r0mPS}[c][c][1][0]{\large$r_0m_{\rm PS}^{\rm eff}(x_0)$}
\epsfig{scale = .5,file=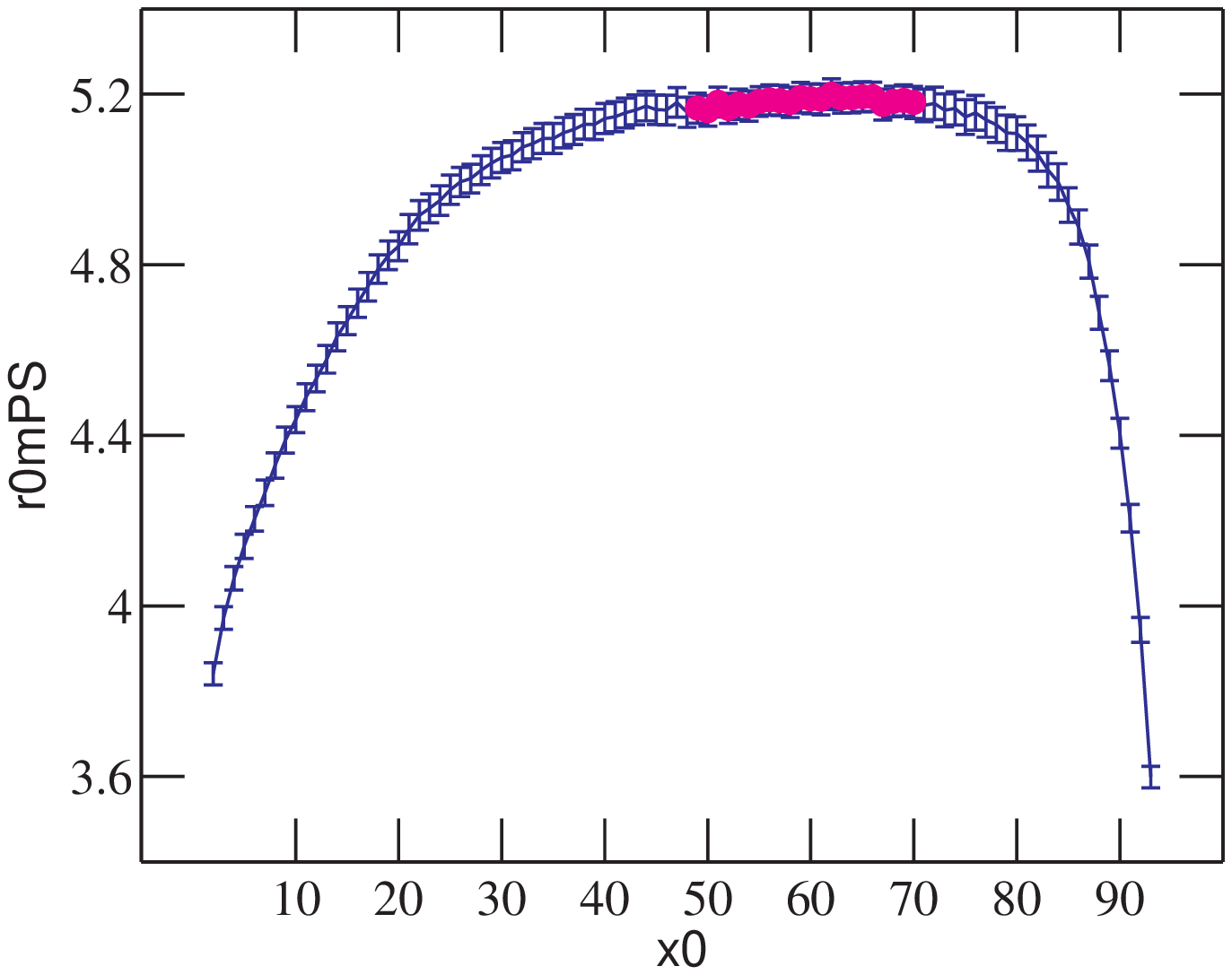}\\[3ex]
\psfrag{x0}[t][c][1][0]{\large $x_0/a$}
\psfrag{r0mPS}[c][c][1][0]{\large$r_0{\rm F_{PS}}(x_0)$}
\epsfig{scale = .5,file=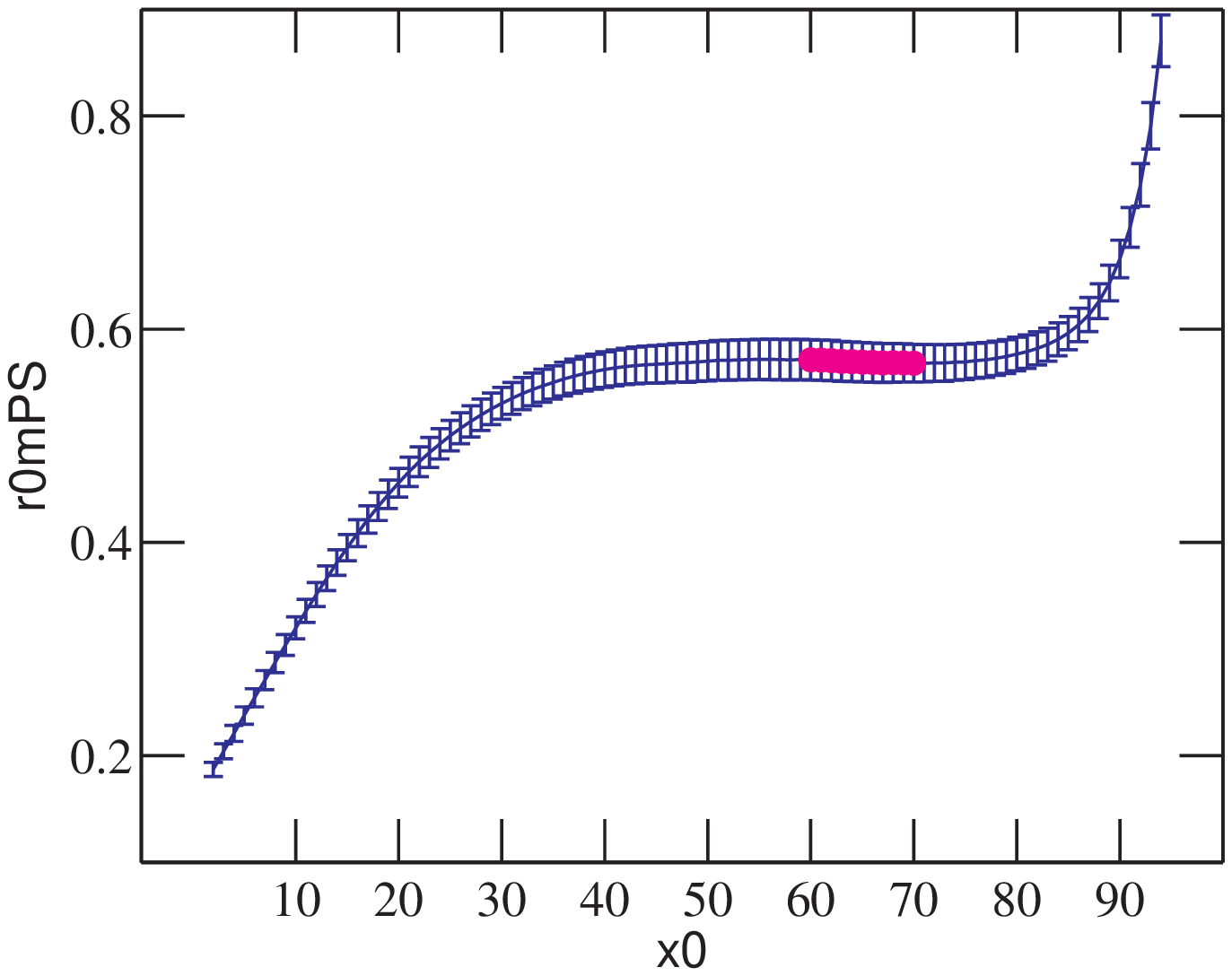}
\end{minipage}
\begin{minipage}{.49\linewidth}
\centering
\psfrag{x0}[t][c][1][0]{\large $x_0/a$}
\psfrag{r0mPS}[c][c][1][0]{\large$r_0m_{\rm V}^{\rm eff}(x_0)$}
\epsfig{scale = .5,file=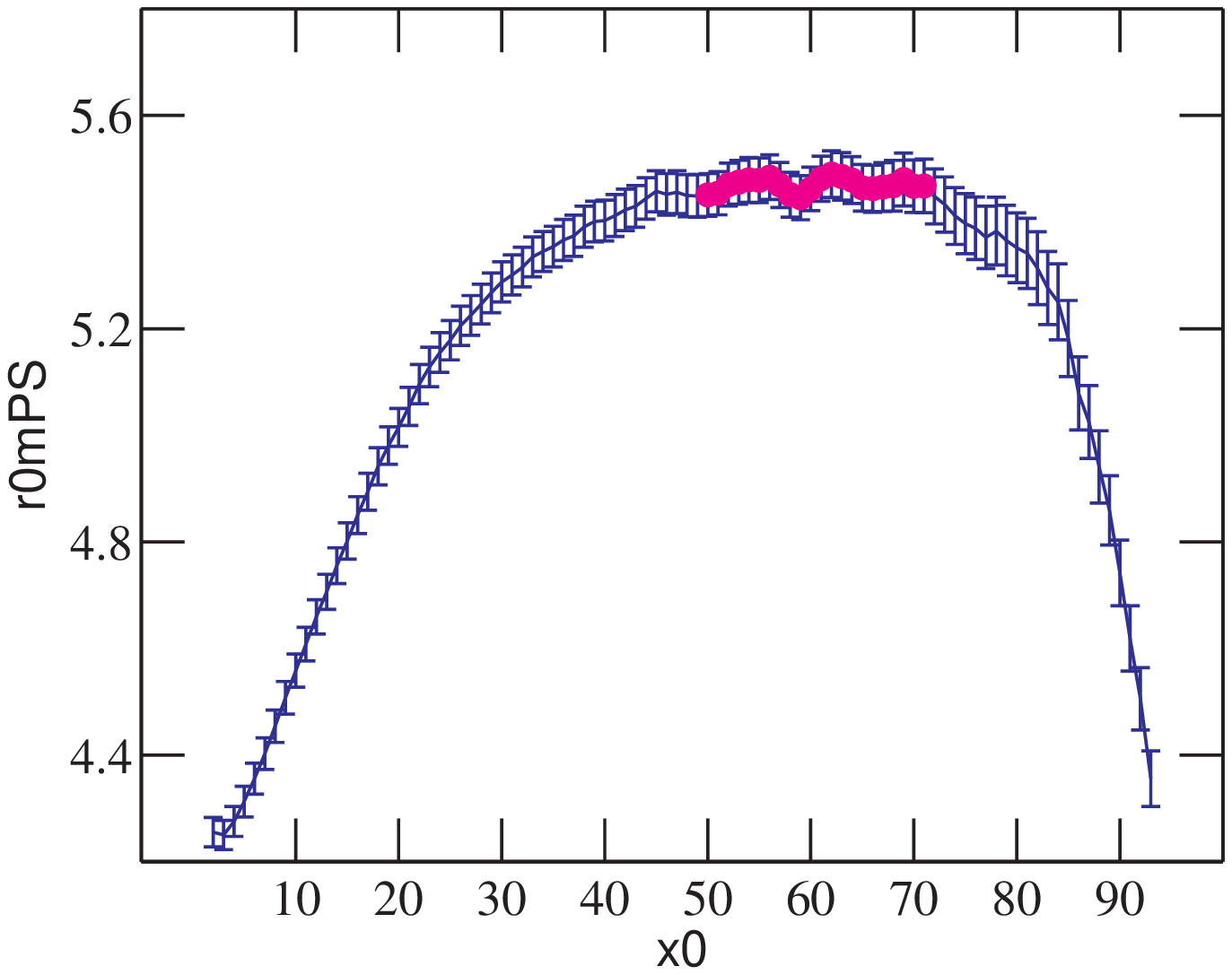}\\[3ex]
\psfrag{x0}[t][c][1][0]{\large $x_0/a$}
\psfrag{r0mPS}[c][c][1][0]{\large$r_0{\rm F_V}(x_0)$}
\epsfig{scale = .5,file=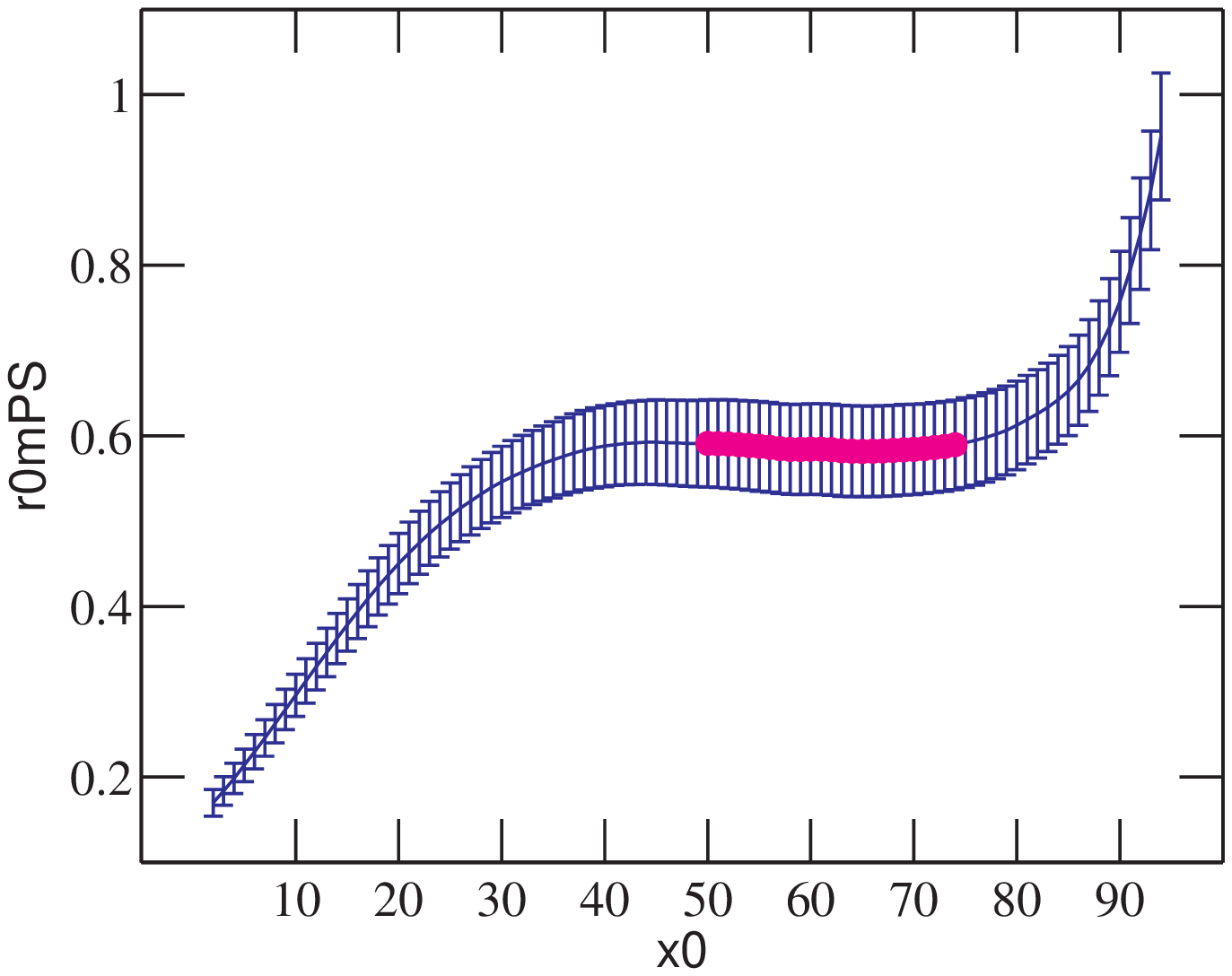}
\end{minipage}\\[1ex]
\caption{Plot of the effective mass $m_{\rm X}^{\rm eff}$ and the decay constant ${\rm F_X}(x_0)$ at $\beta=6.7859$ for the combination of hopping parameters $\kappa_1$ and $\kappa_3$.}\label{repeffmass}
\end{figure}
Figure \ref{repeffmass} shows the results for the combination $\kappa_1-\kappa_3$ of hopping parameters  at $\beta_5=6.7859$.
As expected from (\ref{effectivemass}), the effective mass exhibits a plateau for intermediate times, and contributions from excited states of mass $\Delta$ and glueballs of mass $m_{\rm G}$ for small and large times $x_0$, respectively. 

The meson mass $m_{\rm X}$ can be extracted as the average over the plateau.
In order to keep the systematic errors in $m_{\rm X}$ due to contaminations by excited states under control, the time interval, where their relative contributions are below a chosen threshold was determined. The thresholds that have been used are given in table \ref{thresholdvals}. Thus, the statistical error will always exceeds the systematic error. The time interval, or plateau range, can be determined by means of the following iterative procedure:

One first subjectively chooses a sensible plateau range and subtracts the average over the plateau from the data.

\begin{table}
\centering
\begin{tabular}{cccc}
\hline\hline\\[-2ex]
$r_0m_{\rm PS}$	&$r_0m_{\rm V}$	&$r_0{\rm F_{PS}}$  	&$r_0{\rm F_{V}}$\\
\\
$0.5\%$		&$0.7\%$	&$0.5\%$		&$0.7\%$\\[1ex]
\hline\hline
\end{tabular}
\caption{Thresholds for the accepted contribution of excited states to the plateau range.}\label{thresholdvals}
\end{table}

\begin{figure}
\centering
\begin{minipage}{.49\linewidth}
\centering
\psfrag{x0oa}[t][c][1][0]{\large $x_0/a$}
\psfrag{logr0mPSfAI513mplateau}[c][c][1][0]{\footnotesize$\log|r_0(m^{\rm eff}_{\rm PS}(x_0)-m_{\rm PS}^{\rm plateau})|$}
\epsfig{scale = .5,file=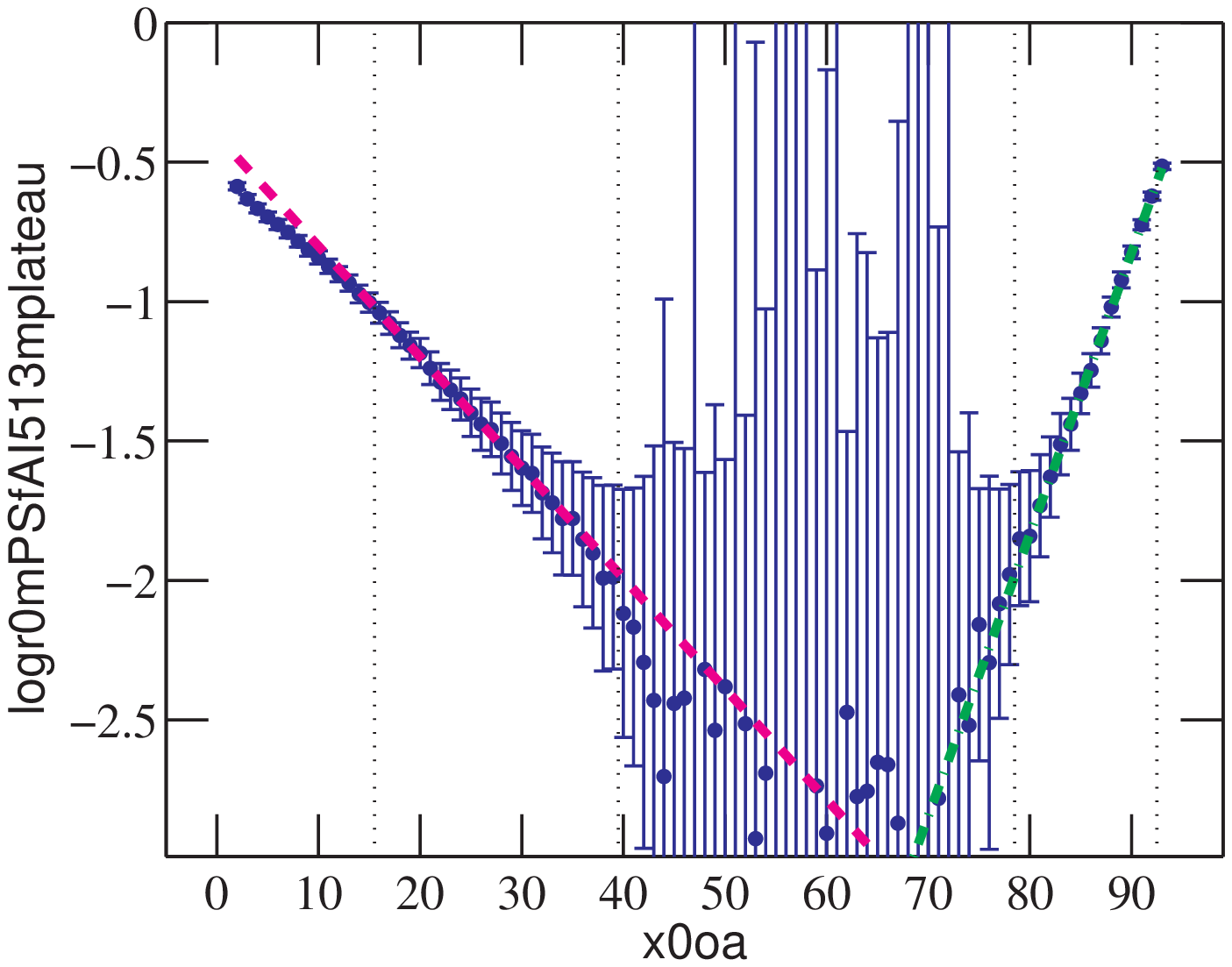}\\[3ex]
\psfrag{x0oa}[t][c][1][0]{\large $x_0/a$}
\psfrag{logr0fds513mplateau}[c][c][1][0]{\footnotesize$\log|r_0({\rm F_{PS}}(x_0)-{\rm F}_{\rm PS}^{\rm plateau})|$}
\epsfig{scale = .5,file=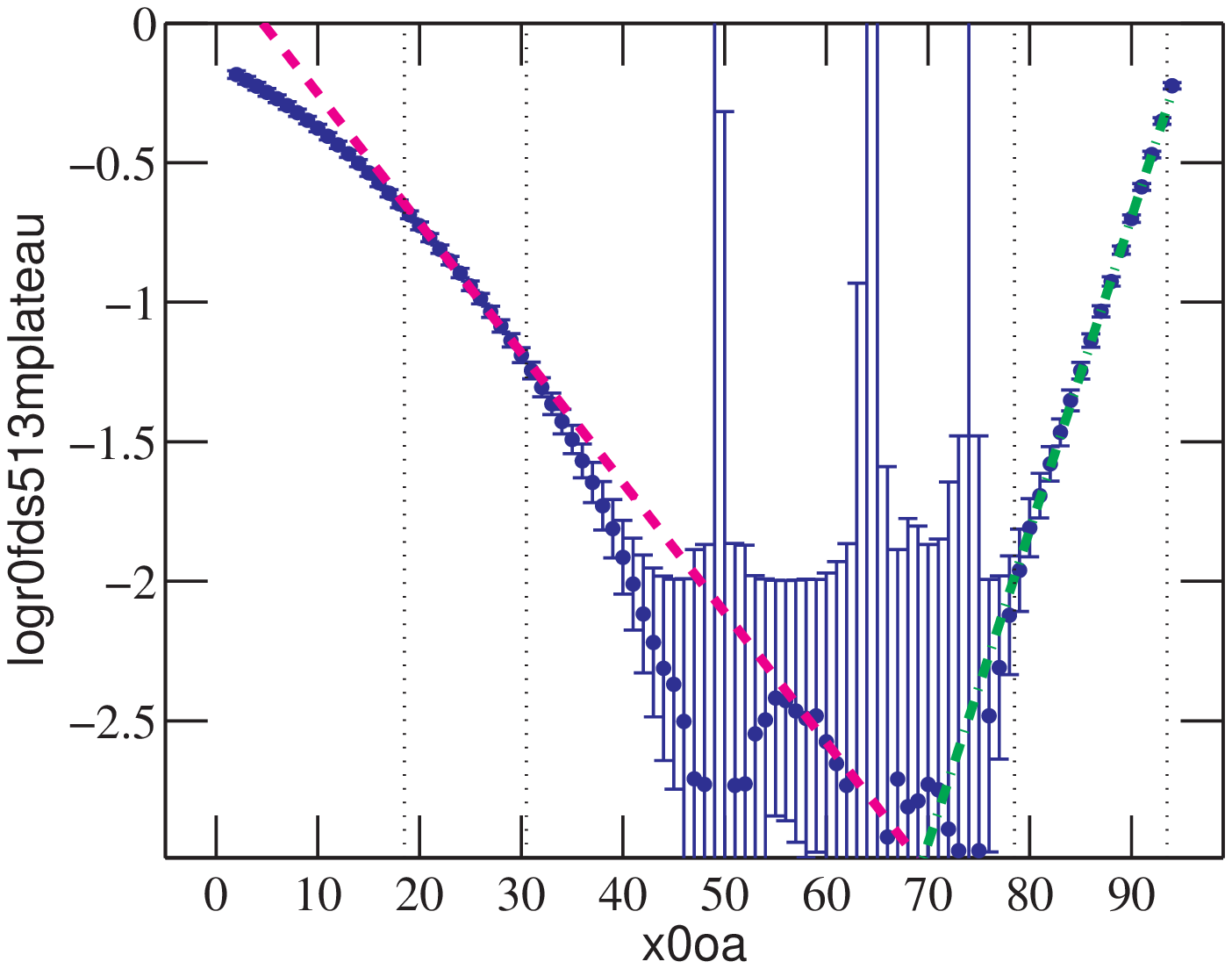}
\end{minipage}
\begin{minipage}{.49\linewidth}
\centering
\psfrag{x0oa}[t][c][1][0]{\large $x_0/a$}
\psfrag{logr0mVkVI513mplateau}[c][c][1][0]{\footnotesize$\log|r_0(m^{\rm eff}_{\rm V}-m_{\rm V}^{\rm plateau})|$}
\epsfig{scale = .5,file=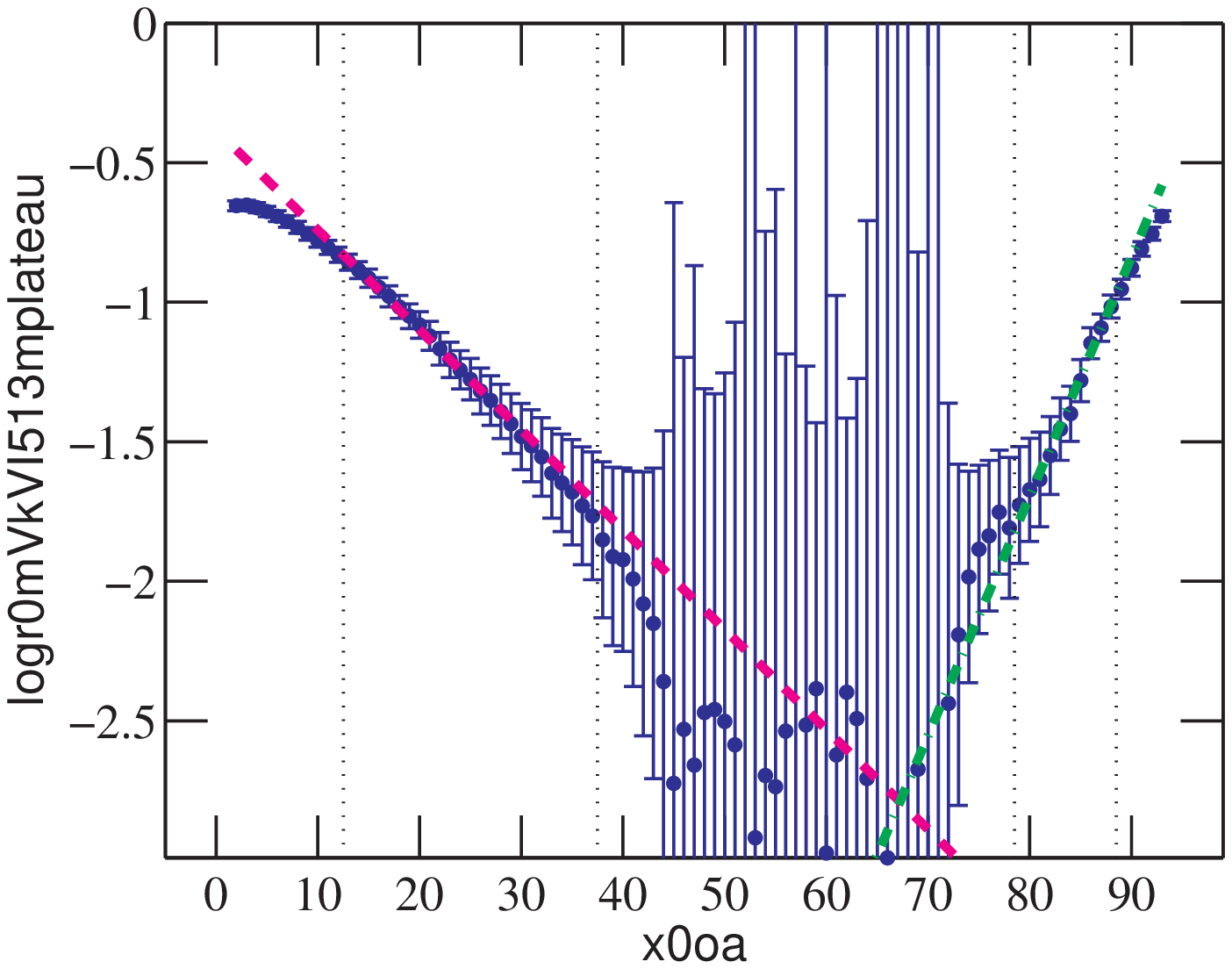}\\[3ex]
\psfrag{x0oa}[t][c][1][0]{\large $x_0/a$}
\psfrag{logr0fdsstar513mplateau}[c][c][1][0]{\footnotesize$\log|r_0({\rm F_{V}}(x_0)-{\rm F}_{\rm V}^{\rm plateau})|$}
\epsfig{scale = .5,file=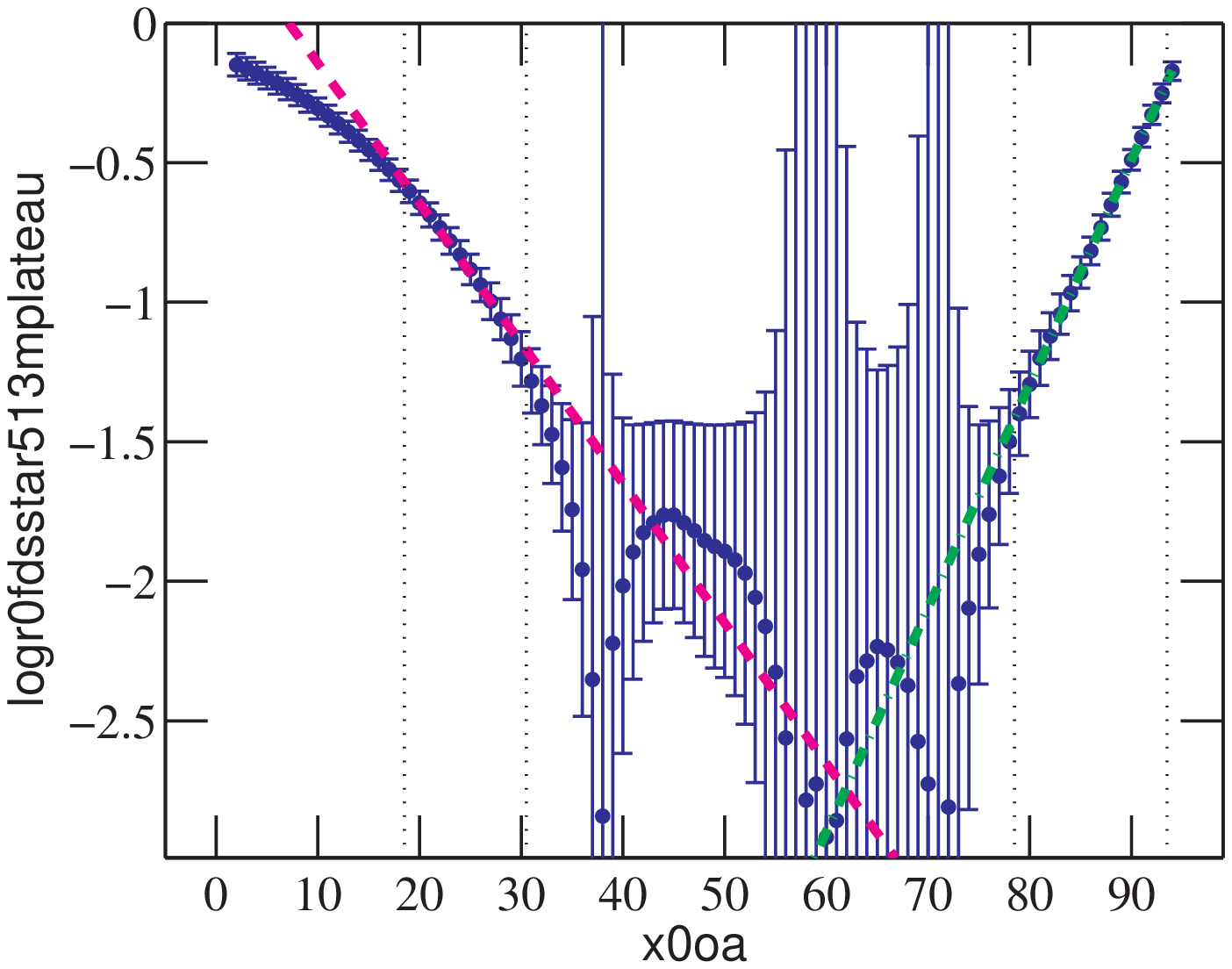}
\end{minipage}\\[1ex]
\caption{Fits to the logarithm of the subtracted effective mass and the subtracted decay constant. The slope of the linear fits give an estimate for the mass gap $\Delta$ (dashed line) and the glueball mass $m_{\rm G}$ (dash-dotted line). The dotted lines indicate the corresponding fit range in each case.}\label{thresholds}
\end{figure}

\begin{figure}
\begin{minipage}{.49\linewidth}
\centering
\psfrag{x0oa}[t][c][1][0]{\large $x_0/a$}
\psfrag{relerr}[c][c][1][0]{\footnotesize relative error in $m_{\rm PS}^{\rm eff}(x_0)$}
\epsfig{scale = .5,file=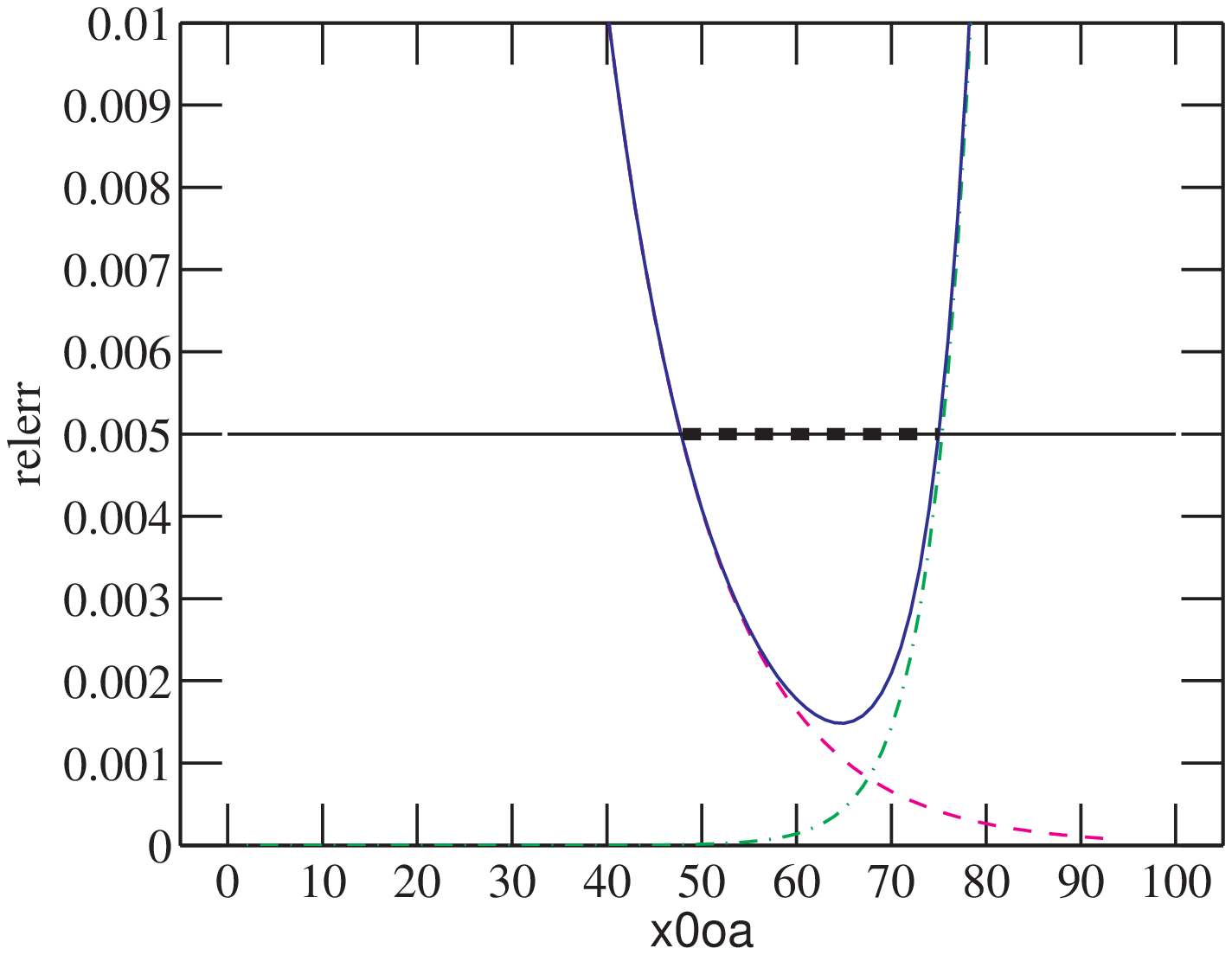}\\[3ex]
\psfrag{x0oa}[t][c][1][0]{\large $x_0/a$}
\psfrag{relerr}[c][c][1][0]{\footnotesize relative error in ${\rm F_{PS}}(x_0)$}
\epsfig{scale = .5,file=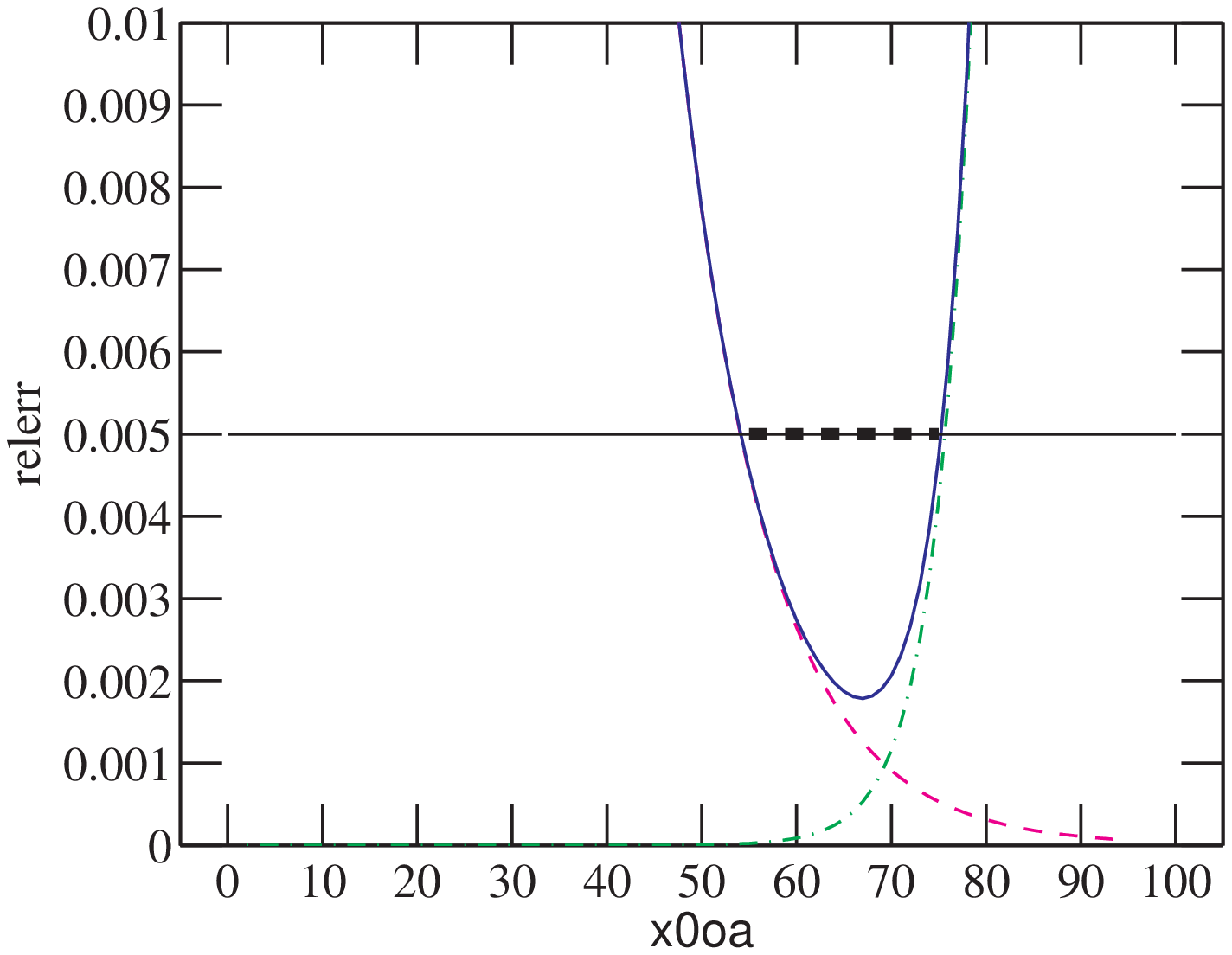}
\end{minipage}
\hspace{.2cm}
\begin{minipage}{.49\linewidth}
\centering
\psfrag{x0oa}[t][c][1][0]{\large $x_0/a$}
\psfrag{relerr}[c][c][1][0]{\footnotesize relative error in $m_{\rm V}^{\rm eff}(x_0)$}
\epsfig{scale = .5,file=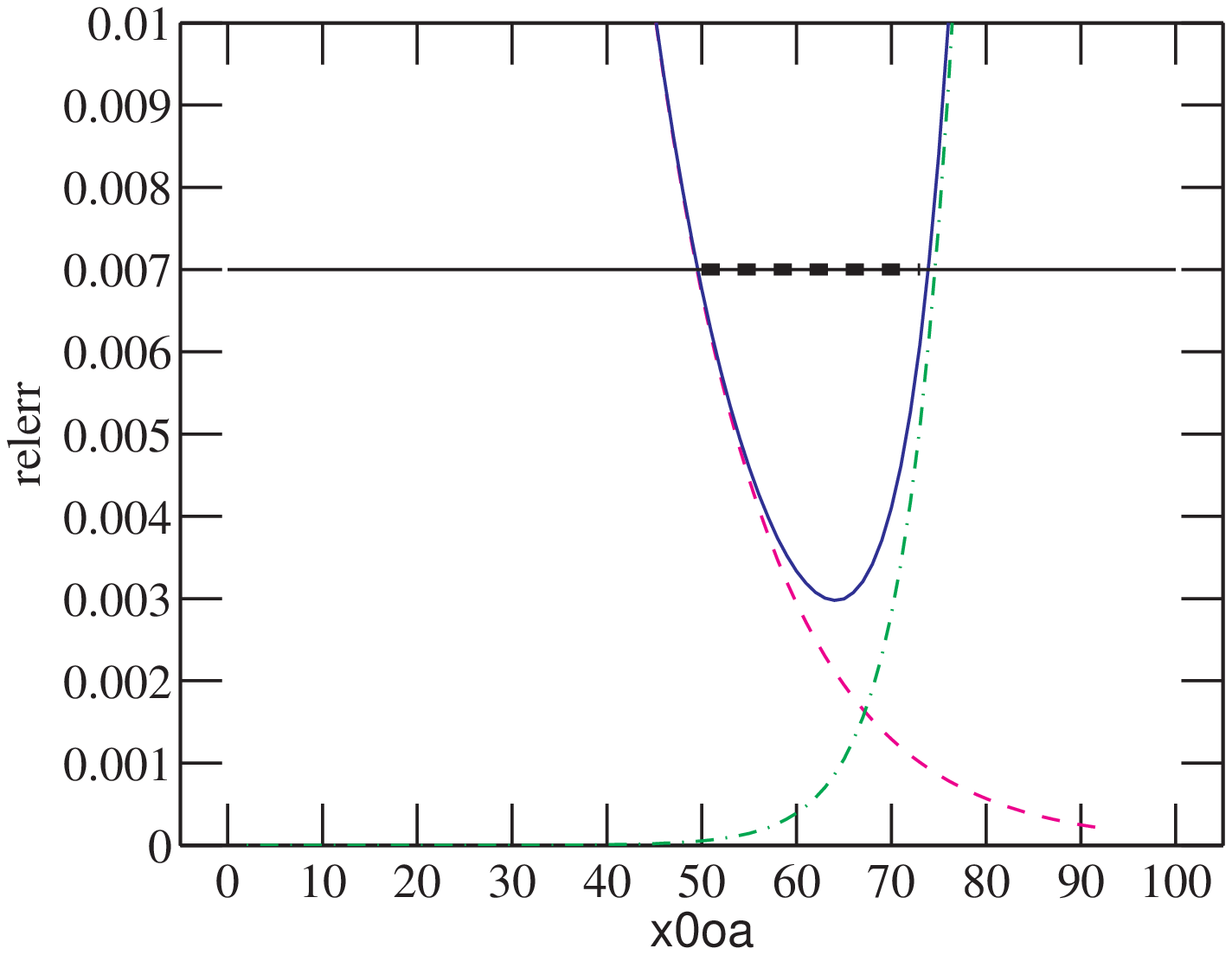}\\[3ex]
\psfrag{x0oa}[t][c][1][0]{\large $x_0/a$}
\psfrag{relerr}[c][c][1][0]{\footnotesize relative error in ${\rm F_{V}}(x_0)$}
\epsfig{scale = .5,file=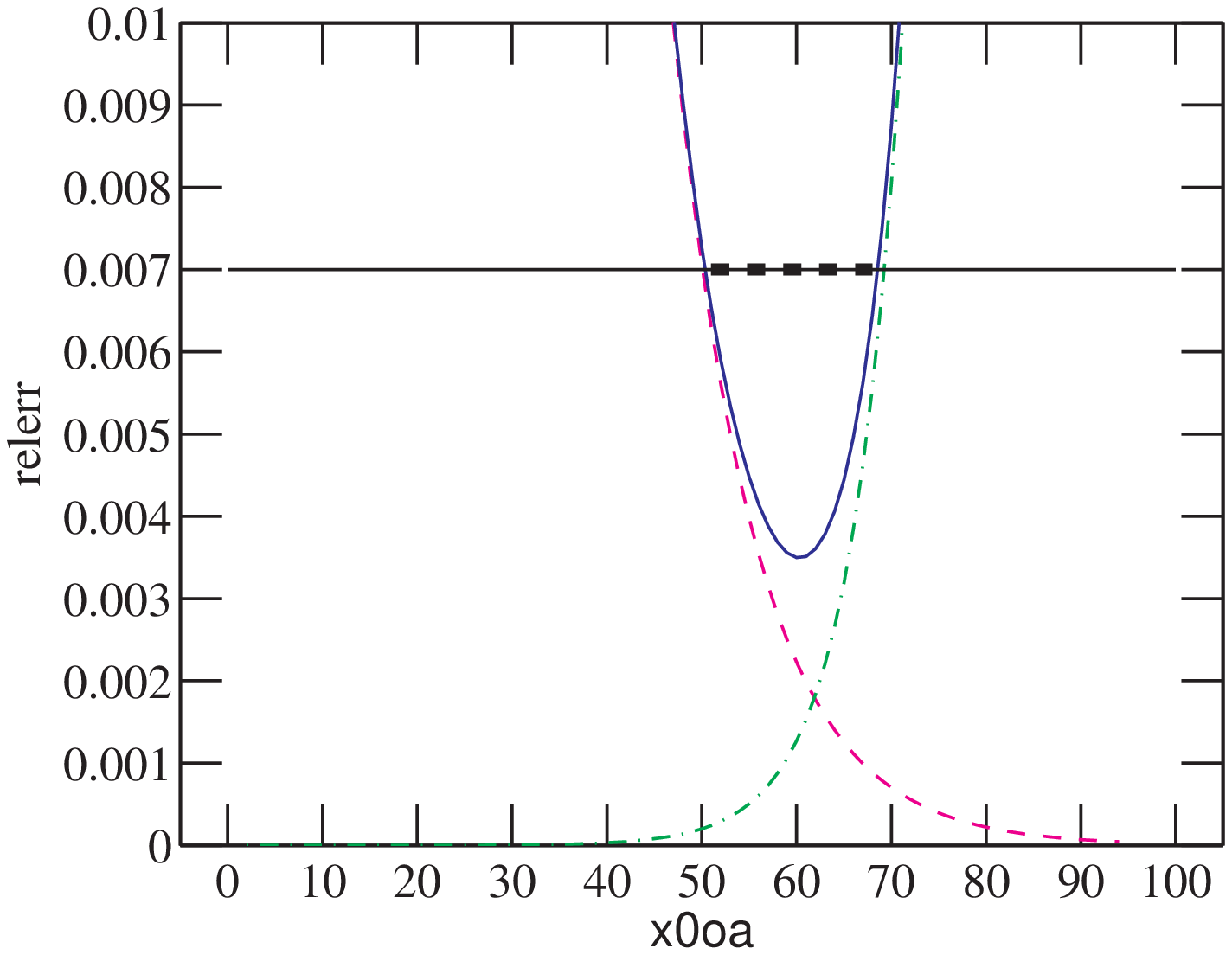}
\end{minipage}\\[1ex]
\caption{Estimated systematic contribution to the plateau due to excited meson states (dashed line), due to glueballs (dash-dotted line) and the sum of both contributions (solid line). The plateau average is taken over the range, where both contributions are below the threshold (indicated by the bold dashed line).}\label{estimateplots}
\end{figure}

\bi
\item[1.)] 
From the logarithm of this data, estimates for the contributions (cf. (\ref{effectivemass}))
\be\ba{c}\label{exccontrib}
{2 \sinh(a\Delta/2)}\eta^{q_{{\rm X}}}_{{\rm X}}e^{-x_0\Delta},\;\;{2 \sinh(a m_{\rm G}/2)}\eta^{0}_{{\rm X}}e^{-(T-x_0)m_{\rm G}}\\
\ea
\ee
to the effective mass can be obtained in terms of linear fits to the time dependence for small and large times. The fit ranges have to be chosen subjectively. This procedure is illustrated in figure \ref{thresholds}, again for $\kappa_1-\kappa_3$ at $\beta=6.7859$. The data sometimes does not exhibit a clear linear behavior and the fit range cannot be chosen without ambiguities. Therefore, the extracted glueball masses and mass gaps can only be interpreted as estimates, which however suffices for the purposes here.
\item[2.)] 
The sum of the relative contributions of excited meson states and glueballs to the plateau as a function of the time $x_0$ is shown in figure \ref{estimateplots}. The new plateau range is defined as the time interval, where this contribution is below the threshold given in table \ref{thresholdvals}. 
\item[3.)] 
The procedure can be repeated with the newly defined plateau average, until the plateau range is stable, which usually occurs after one or two iterations. 
\ei

One finally obtains the meson mass $m_{\rm X}$ as the average over all the plateau ranges listed in table \ref{plateauranges} in the appendix. They typically extend over $1r_0$ to $1.5r_0$ and their position approximately scales with $\beta$. 
The estimates for the glueball mass have been summarized in table \ref{mG}. The values given there for each value of $\beta$ are the averages over the fits obtained from the six hopping parameter combinations $\kappa_1-\kappa_2\dots\kappa_1-\kappa_7$. The estimates for the gap energy are tabulated in table \ref{gaps}.

With the plateau ranges of $m_{\rm X}$ at hand, one now has to repeat the procedure for the decay constant 
\be\ba{rcl}
\rm{F}_{\rm X}(x_0)&=&-2Z_{O}(m_{\rm X}L^3)^{-1/2}\,e^{(x_0-T/2)m_{\rm X}}\frac{f_{ O}(x_0)}{\sqrt{f_O^T}}.\\
\ea\ee
Here, the contributions of the excited states to the plateau are of the form 
\be\ba{c}\label{exccontrib}
\eta_{{{\rm X}}}^{{q_{\rm X}}}\,e^{-x_0\Delta},\;   \;\eta_{{{\rm X}}}^{0}\,e^{-(T-x_0)m_{\rm G}}. \\
\ea\ee
Representative plots of the procedure are shown next to the ones for the effective mass in the figures \ref{repeffmass} to \ref{estimateplots}.

	Although feasible in principle, a continuum limit for the glueball mass has not been taken because of the uncertainties in the determination of the corresponding fit ranges. But even at finite lattice spacing, the data is mostly compatible with the glueball mass of the $0^{++}$-glueball as obtained in the dedicated lattice simulations and summarized in table \ref{mGlit}.

\begin{table}
\centering
\begin{tabular}{crrrr}
\hline\hline\\[-2ex]
&$r_0m_{\rm G,F_{PS}}$&$r_0m_{\rm G,F_{V}}$&$r_0m_{\rm G,m^{\rm eff}_{PS}}$&$r_0m_{\rm G,m^{\rm eff}_{V}}$\\
\\[-2ex]
$\beta_1$&5.9&-&5&4\\
$\beta_2$&5.3&4.9&5&4\\
$\beta_3$&4.5&4.5&4.1&3.7\\
$\beta_4$&4.7&4.4&4.0&3.3\\
$\beta_5$&4.38&3.4&3.5&3.5\\
 [1ex]
\hline\hline
\end{tabular}
\caption{Estimates for the glueball mass. In the case $r_0m_{\rm G,F_{V}}$ at $\beta_1$, the data was too noisy to allow for a sensible fit. }\label{mG}
\end{table}
\begin{table}
\centering
\begin{tabular}{lcl}
\hline\hline\\[-2ex]
Collab. &&$r_0m_{0^{++}}$ \\[1ex]
Morningstar \emph{et. al.} 	&\cite{Morningstar:1999rf}	&4.21(11)(4) \\
Bali \emph{et. al.} 		&\cite{Vaccarino:1999ku} 	&4.33(10)\\
Teper 				&\cite{Teper:1998kw} 		&4.35(11) \\
UKQCD 				&\cite{Bali:1993fb} 		&4.05(16) \\
Niedermayer \emph{et. al.}      &\cite{Niedermayer:2000yx} 		&4.12(21)\\[1ex]
\hline\hline\\
\end{tabular}\caption{Lattice data for the lowest glueball mass ($0^{++}$) in the continuum with (combined) statistical and systematic errors.}\label{mGlit}
\end{table}

For the determination of the charm quark mass using the PCAC-relation, no plateau average over the involved correlation functions has been taken, as the statistical errors turned out to be very small. Instead, the value at $x_0=2/3T$ was used, where the contamination by excited states was reasonably small. 

\subsection{Interpolation in the meson mass}\label{cutoff}
The observables at different quark masses but at constant $\beta$ have been evaluated on the same set of gauge configurations and are therefore correlated. Indeed, the corresponding normalized covariance matrix of the observables which was determined in the jackknife procedure has large off-diagonal elements. It was checked, that this does not affect the error estimation in the data interpolation.

\subsubsection{Interpolation to lines of constant physics}\label{interpolation}
The pseudo scalar and vector meson masses computed at each lattice spacing do not coincide exactly (cf. figure \ref{interpolpoints} and table \ref{tabplatav1} in the appendix). 
Thus, for a continuum limit along a line of constant physics, the secondary quantities (\ref{2ndaryquant}) were interpolated to common values. For the decay constant and the mass splitting, this was done linearly in the inverse meson mass $m_{\rm PS}$ or $m_{\rm V}$, whereas in the case of the renormalization group invariant charm quark mass, linearly in the meson mass. These parameterizations are not only suggested by HQET, but are also supported by visual examination of the mass dependence of all secondary quantities.

Figure \ref{interpolpoints} shows the choice of meson masses given in table \ref{exactintmasses} for the pseudo scalar and the vector meson channel, to which the secondary observables were interpolated. Most of the masses were chosen such as to lie in the vicinity of the simulated ones in order to reduce the error introduced by the interpolation. In one case, the interpolation was performed to the experimentally known masses \cite{PDBook} of the $\rm D_s$ and $\rm D_s^\ast$ meson respectively.

The data obtained for $\kappa_6$ and $\kappa_7$ at $\beta_5=6.7859$ were discarded for the whole analysis, since roundoff effects were observed for the heavy quark masses simulated for, which could be ruled out reliably only for $\kappa_5$ (cf. \ref{stoppingcrit}). 
\begin{table}
\centering
\begin{tabular}{cccCccc}
\hline\hline
&&&&\\[-2ex]
i			&1	&2	&3		&4		&5		&6\\[1ex]
\hline
&&&&&\\[-2ex]
$r_0m^i_{\rm PS}$	&3.768	&4.327	&4.987  	& 5.653 	& 6.211 	& 6.560\\[1ex]
$aM_Q<0.64$	&$\beta_2-\beta_5$&$\beta_2-\beta_5$&$\beta_3-\beta_5$&$\beta_3-\beta_5$&$\beta_3-\beta_5$&$\beta_3-\beta_5$\\[1ex]
\hline&&&&&\\[-2ex]
$r_0m^i_{\rm V}$	&4.210&4.660    &5.363  &5.920  &6.280  &6.550\\[1ex]
$aM_Q<0.64$	&$\beta_2-\beta_5$&$\beta_2-\beta_5$&$\beta_3-\beta_5$&$\beta_3-\beta_5$&$\beta_3-\beta_5$&$\beta_3-\beta_5$\\[1ex]
\hline\hline\\
\end{tabular}\caption{Meson masses to which all observables were interpolated before taking the continuum limit. The masses in the third column ($i=3$) correspond to the experimentally known values of the $\rm D_s$ and $\rm D_s^\ast$ meson respectively \cite{PDBook}. In addition, the lattices which will be included into the continuum extrapolation are given in terms of the corresponding coupling constant (cf. section \ref{cutoff}).}\label{exactintmasses}
\end{table}
\begin{figure}
\centering
\psfrag{oormPS1}[c][c][1][0]{ $1/(r_0m_{\rm PS})$}
\psfrag{oormV2}[c][c][1][0]{ $1/(r_0m_{\rm V})$}
\psfrag{b11}[c][c][1][0]{ $\beta_1$}
\psfrag{b12}[c][c][1][0]{ $\beta_2$}
\psfrag{b13}[c][c][1][0]{ $\beta_3$}
\psfrag{b14}[c][c][1][0]{ $\beta_4$}
\psfrag{b15}[c][c][1][0]{ $\beta_5$}
\epsfig{scale=.8,file=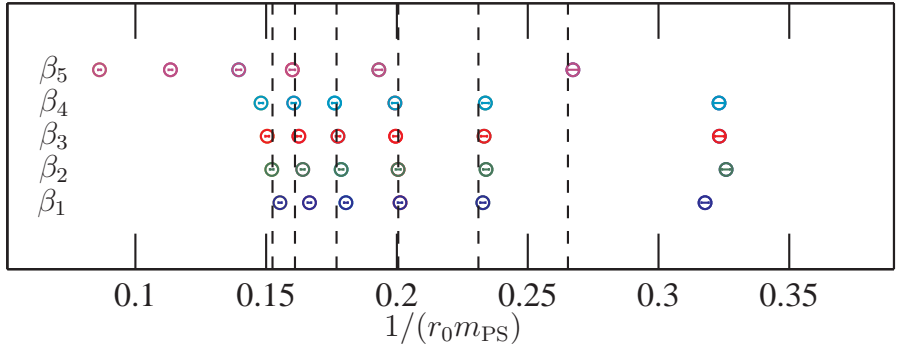}
\caption{Simulated inverse pseudo scalar masses for the combinations of hopping parameters $\kappa_1-\kappa_2,\dots,\kappa_1-\kappa_7$ (circles). The dashed lines indicate the inverse masses to which the observables have been interpolated at each value of $\beta$ before the continuum limit has been taken.}\label{interpolpoints}
\end{figure}
\subsubsection{Interpolation to the $b$-quark}
Prior to the interpolation between the region of the charm quark mass and the static limit, the continuum extrapolation for the desired observables at the masses given in table \ref{exactintmasses} was carried out.
This ordering of the procedure is suggested by the expectation, that discretization errors are quark mass dependent. Since all the observables have been determined in the $O(a)$-improved theory, one would naively assume, that they approach the continuum limit approximately linearly in $a^2$.
In \cite{Kurth:2001yr}, deviations from this scaling behavior for heavy quarks were observed.
A comparison of lattice results in the finite volume $\orda$-improved Schr\"odinger Functional at non-vanishing lattice spacing with results from perturbation theory at zero lattice spacing revealed, that the expected scaling behavior breaks down for heavy quark with $(am_Q^{\MSbar})^2\gtrsim 0.2$. Thus, the bound
\be\label{cutoffcrit}
aM_Q\lesssim 0.64. 
\ee
has to be fulfilled by all the data that enter the analysis.
Table \ref{exactintmasses} shows the meson masses together with the range of $\beta$-values that are compatible with this bound.

The interpolations were carried in the meson mass. This allowed to localize the physical point of the $\bsub$-meson exactly, because its mass is known very precisely from experiment ($m_{\bsub}=5.36966(24)$ GeV \cite{PDBook}). In units of the Sommer-scale, the mass translates to $r_0m_{\bsub}=13.6056(6)$.

In HQET, one expects that the heavy quark mass differs from the heavy-light meson mass only by the binding energy $\bar\Lambda$ and terms of $O(1/M_Q)$ (cf. (\ref{massformula})). Thus, the functional dependence of the observables on the meson mass is compatible with the fit-ansatz
\be
a_0+{a_1\over r_0m_{\rm PS}}+{a_2\over r_0m_{\rm PS}^2}+\dots\;.
\ee
The coefficients $a_i$ can be taken as estimates for the magnitude of the contributions to heavy-light observables from higher orders in the HQET expansion. 

\section{The ${\rm D_s}$- and the $\dsubstar$-meson and the $c$-quark mass}
In this section, a precise determination of the observables
\be
\fds,\,\fdsstar,\,\fds/\fdsstar\,{\rm and}\,r_0(m_{\dsubstar}-m_{\dsub})
\ee
is presented. 

Simulations with a continuum extrapolation for the decay constant from rather coarse lattice resolutions have been done in \cite{Alexandrou:1994mr,Gupta:1996zd,Aoki:1997xe,Bernard:1997by} and more recent studies with an extended approach to the continuum limit, $O(a)$-improved Wilson fermions and also partly with dynamical quarks have been carried out in \cite{El-Khadra:1998hq,Bernard:1998xi,AliKhan:2000eg,Becirevic:1998ua,Bowler:2000xw,deDivitiis:2003wy,Juttner:2003ns,Bernard:2002pc,Maynard:2001zd}.
A recent summary of lattice data for the $\dsub$-system can be found for example in \cite{Ryan:2001ej}. Some representative results are summarized in table \ref{otherfds}.

\subsection{The decay constants $\fds$ and $\fdsstar$ and the ratio $\fds/\fdsstar$}
\begin{figure}
\centering
\psfrag{aor0sq}[t][c][1][0]{\large$(a/r_0)^2$}
\psfrag{fds}[b][c][1][0]{\large$r_0 \rm F_{D_s}$}
\psfrag{val}[c][c][1][0]{\large$r_0\rm F_{D_s}=$\input{./tables/fdscont.txt}}
\epsfig{scale=.8,file=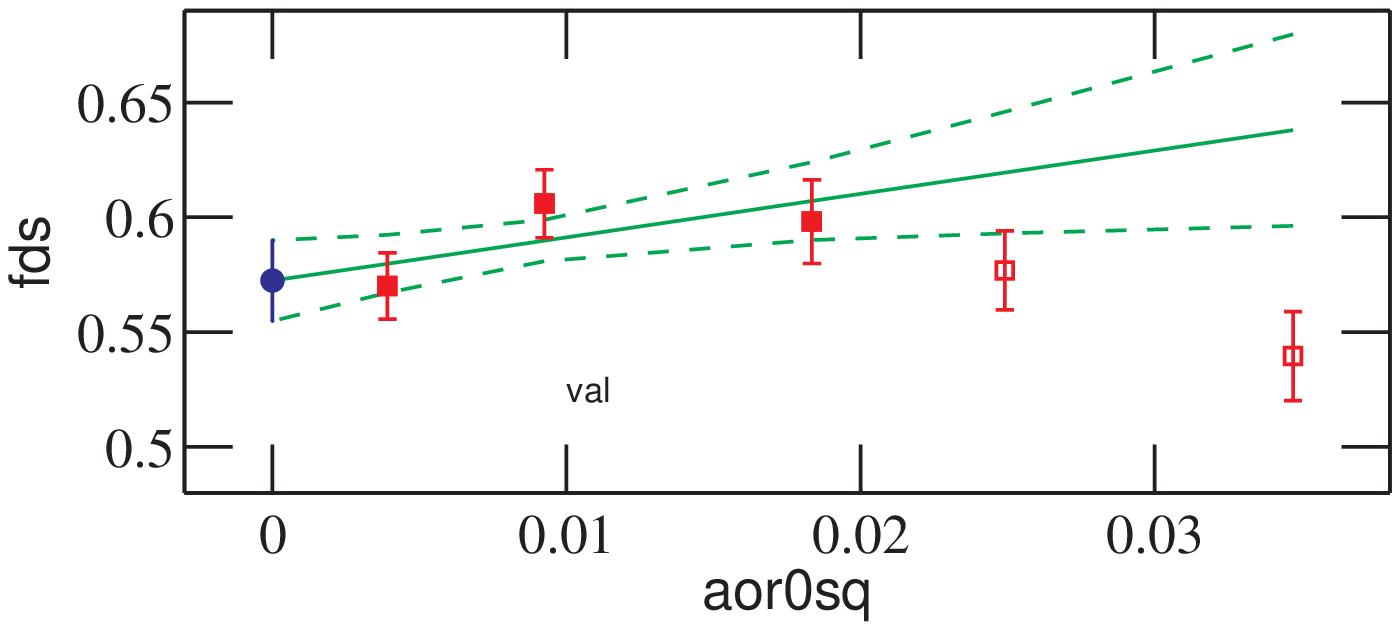}\\\vspace{1cm}
\psfrag{aor0sq}[t][c][1][0]{\large $(a/r_0)^2$}
\psfrag{fds}[b][c][1][0]{\large$\fds/ \fdsstar $}
\psfrag{valstar}[c][c][1][0]{\large$\fds/ \fdsstar=$\input{./tables/FoFstarcont.txt}}
\epsfig{scale=.8,file=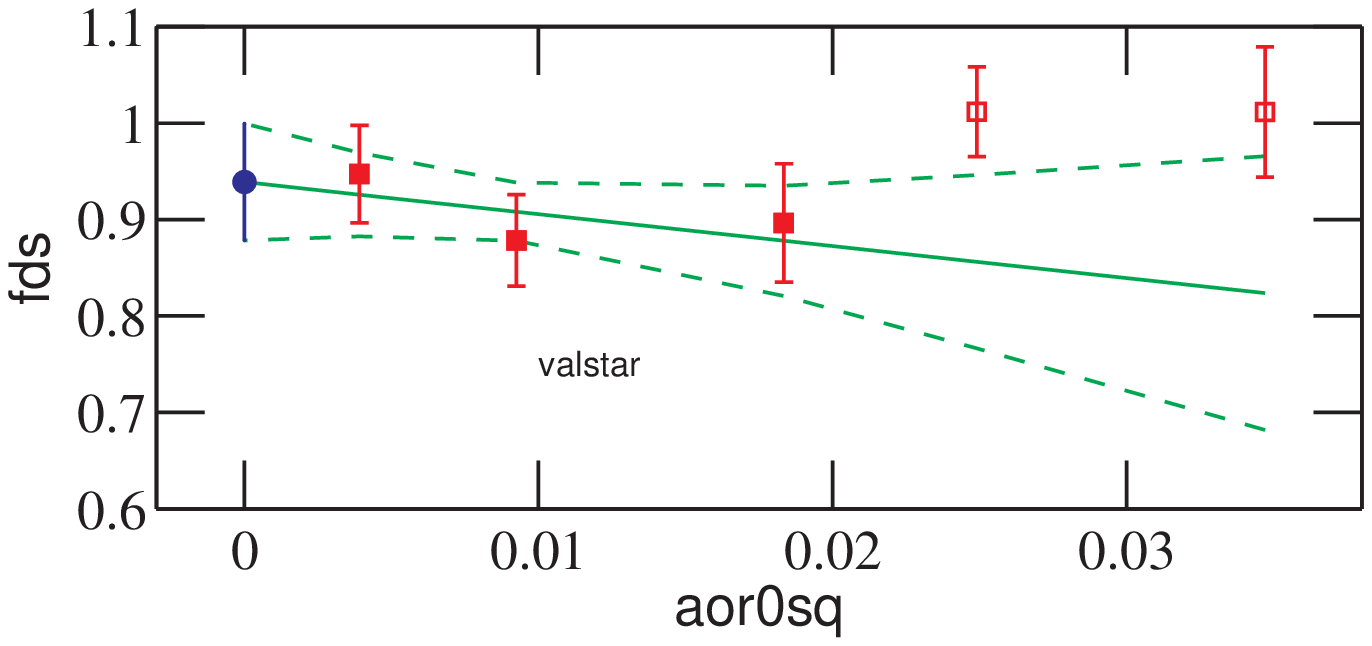}
\caption{Continuum limit for the decay constant ${\rm F_{D_s}}$ and the ratio $\fds/{\rm F_{D_s^\ast}}$.}\label{fdscont}
\end{figure}

After interpolating the data for the decay constants $\fds$ and $\fdsstar$ at each finite lattice spacing (cf. table \ref{tabplatav1}) to the physical point, given in terms of the mass $r_0m_{\dsub}=4.987$ \cite{PDBook}, one obtains the corresponding values in the continuum from a linear extrapolation in $(a/r_0)^2$. The extrapolations for $\fds$ and $\fds/\fdsstar$ are shown in figure \ref{fdscont}. For $\fds$, the fit has $\chi^2/{\rm d.o.f.}=1.9$, for $\fdsstar$ $\chi^2/{\rm d.o.f.}=1.6$ and for $\fds/\fdsstar$ a value of $\chi^2/{\rm d.o.f.}=0.7$. The final results in the continuum are
\begin{center}
\begin{tabular}{lcl}
$r_0\fds$&=&0.572(18),\\
$r_0\fdsstar$&=&0.605(47),\\
$\fds/\fdsstar$&=&0.939(61).\\
\end{tabular}
\end{center}
Using $r_0=0.5$ fm, the results for the decay constants translate to \linebreak$\fds=226(7)$MeV and $\fdsstar=239(18)$MeV. The error is the combination of statistical and systematic contributions within the quenched approximation. 

Following the arguments of section \ref{cutoff}, the data from the simulations at $\beta_1=6.0$ and $\beta_2=6.1$ has not been included into the fits.  
However, the stability of the extrapolation has been tested by studying the change in observables under the inclusion of $\beta_2=6.1$. This re\-sul\-ted in $r_0\fds$=0.585(15) with $\chi^2/{\rm d.o.f.}=1.9$, $r_0\fdsstar$=0.658(38) with $\chi^2=2.4$ and $\fds/\fdsstar$=0.885(49) with $\chi^2/{\rm d.o.f.}=1.5$. Thus, the observables change by  2\%, 8\% and 6\% respectively, and the results from both fit ranges agree within errors.

The data at $\beta_2-\beta_4$ has previously been used to determine the pseudo scalar decay constant in the continuum with a final value of $\fds=252(9)$MeV \cite{Juttner:2003ns}, where the scale was also set with the Kaon decay constant. This value differs from the one obtained here by 26 MeV and the two results are not compatible within errors.

Table \ref{otherfds} summarizes some recent values for the decay constants from quenched as well as from dynamical simulations by other groups.
There are indications, that the effect of unquenching is a shift in the decay constant to about 10\% higher values \cite{Ryan:2001ej,Bernard:2002pc}. However, results with dynamical quarks still suffer from large statistical and systematic uncertainties so that a reliable estimate of the quenching error is not yet possible. 
\begin{table}\centering
\begin{tabular}{l@{\hspace{0.5mm}}c@{\hspace{2mm}}l@{\hspace{2.5mm}}rl@{\hspace{3.5mm}}lc}
\hline\hline\\[-2ex]
\multicolumn{2}{c}{Reference	}			&\multicolumn{1}{c}{${\rm F}^{N_f=0}_{\rm D_s}/$MeV}		&$N_f$&\multicolumn{1}{c}{$\fds/$MeV}&\multicolumn{1}{c}{${\rm F}_{\rm D_s^\ast}^{N_f=0}/$MeV}&{\begin{tabular}{c}scale\\setting\end{tabular}}\\[1ex]
\hline\\[-2ex]
ALPHA &\cite{Juttner:2003ns}	&$252(9)$  &&&&$r_0$\\[.3ex]
Becirevic \emph{et. al.}&\cite{Becirevic:1998ua}	&$231(12)(^{+6}_{-0})$ 	&&&$272(16)(^{+0}_{-20})$ &$m_\rho$\\[.3ex]
Bowler \emph{et. al.}&\cite{Bowler:2000xw}	&$229(3)(^{+23}_{-12})$ &&&$264(10)(^{+15}_{-20})$ &$f_\pi$\\[.3ex]
CP-PACS&\cite{AliKhan:2000eg}		&$250(1)(^{+24}_{-18})$ 	&2&$267(13)(^{+27}_{-17})$ &&$m_\rho$\\[.3ex]
de Divitiis&\cite{deDivitiis:2003wy}	&240(5)(5) &&&&$r_0$\\[.3ex]

MILC&\cite{Bernard:2002pc}		&$223(5)(^{+18}_{-17})$ 	&2&$241(5)(^{+  29}_{-26})$ &&$f_\pi$\\[.3ex]
UKQCD&\cite{Maynard:2001zd}		&$229(3)(^{+23}_{-12})$ &&&&$f_\pi$\\[.3ex]

Wingate \emph{et. al.}&\cite{Wingate:2003gm}	&		&3&290(7)(41) &&$\Upsilon$\\[.3ex]
Ryan (\emph{world av.})&\cite{Ryan:2001ej}	&$230(15)$ 	&&$250(30) $ \\[1ex]
\hline\hline\\
\end{tabular}\caption{Results for the decay constant $\fds$ and $\fdsstar$ from other groups. The errors are statistical and systematic.}\label{otherfds}
\end{table}

Only the decay constant for the $\dsub$-meson has been determined in an experiment. Two recent measurements quote ${\rm F}_{\rm D_s}^{\rm exp}=285(19)(40)$MeV \cite{Heister:2002fp} and ${\rm F_{\rm D_s}^{\rm exp}}=280(19)  (28) (34)$MeV \cite{Chadha:1998zh}, where the first error is statistical and the second systematic. In the latter case, the third error is also systematic due to an uncertainty in branching ratios. The present world average from experimental determinations is ${\rm F}_{\rm D_s}^{\rm exp}=267(33)$MeV \cite{PDBook}.

Also QCD sum rules have been used to determine $\fds$. In \cite{Narison:2001pu} a value of $\fds=235(24)$MeV has been suggested.

\subsection{The mass splitting $r_0(m_{\rm D_s^\ast}-m_{\rm D_s})$}
\begin{figure}
\centering
\psfrag{aor0sq}[t][c][1][0]{\large $(a/r_0)^2$}
\psfrag{r0mV}[b][c][1][0]{\large$r_0(m_{\rm D_s^\ast}-m_{\rm D_s})$}
\psfrag{val}[c][c][1][0]{$r_0(m_{\rm D_s^\ast}-m_{\rm D_s})=$\input{./tables/r0mV_r0mPSmPScont.txt}}
\epsfig{scale=.8,file=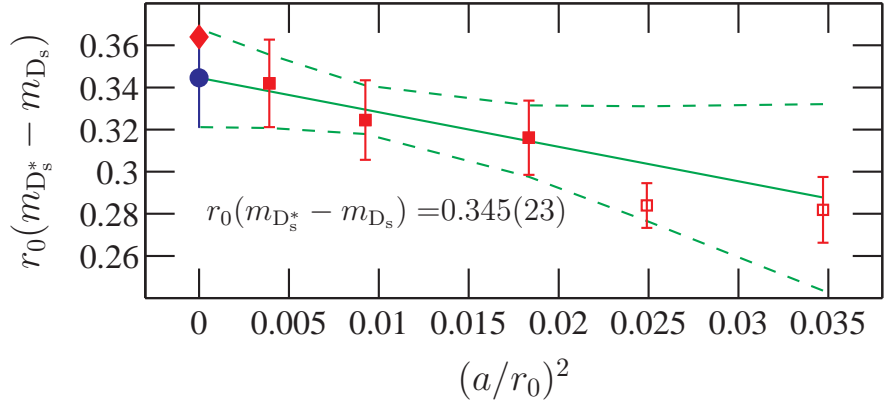}
\caption{Continuum limit for the mass splitting. The diamond represents the experimental value. Only the data at the filled squares entered the fit.}\label{clms1}
\end{figure}

In the same way as for the decay constant, after interpolating the data to the physical point given by the mass of the $\dsub$-meson, the continuum extrapolation can be carried out for the mass splitting $r_0(m_{\rm D_s^\ast}-m_{\rm D_s})$
(cf. figure \ref{clms1}). The resulting value is
\begin{center}
\begin{tabular}{rcl}
$r_0(m_{\rm \dsubstar}-m_{\rm \dsub})$&=&0.345(23)\hspace{-1mm},\\
\end{tabular}
\end{center}
The linear fit has $\chi^2/{\rm d.o.f.}=0.1$.
Converted to physical units with $r_0=0.5\,$ fm, the value is $(m_{\rm \dsubstar}-m_{\rm \dsub})=\,$136.0(92)MeV.
The result is compatible with the current experimental data taken from \cite{PDBook}, $(m_{\rm \dsubstar}-m_{\rm \dsub})^{\rm exp}=143.8(4)$MeV.

In a previous lattice study, where the same techniques were applied but the continuum extrapolation was made from coarser lattices \cite{Rolf:2002gu}, a value of $(m_{\rm \dsubstar}-m_{\rm \dsub})=122(14)$MeV was quoted. Another computation \cite{Becirevic:1998ua} gave $(m_{\rm \dsubstar}-m_{\rm \dsub})=97(12)$MeV as the final result. This result was obtained on rather coarse lattices and the discrepancy with respect to experiment was ascribed to cutoff effects.

\subsection{The renormalization group invariant charm quark mass $M_c$}
The renormalization factor $Z_M(g_0)$, which has been introduced in section \ref{imprandrenormconstants}, can be used to relate the two definitions of renormalized quark masses (\ref{rensub}) and (\ref{PCACmass}) to the renormalization group invariant quark masses.

On the one hand, employing the definition of the PCAC-mass (\ref{barePCACmass}), the charm quark mass can be extracted from the correlation functions containing non-degenerate quarks,
\be\ba{rcl}
  r_0M_{c{|m_{sc}}} & = & Z_M \Bigl\{2r_0 m_{sc}
  \left[ 1+(b_A-b_P){1\over 2} (am_{\rm q,c}+am_{\rm q,s})\right]\\
  &-&r_0m_{\rm s}\left[1+(b_A-b_P)am_{\rm q,s}\right]\Bigr\}.
\ea\ee
Here, the flavor indices have been replaced by the particular quark flavor ($c$ for charm and $s$ for strange).
The relation can also be applied to the mass degenerate case, where one obtains \be
  r_0M_{c{|m_c}} =  Z_M r_0 m_c \left[ 1+(b_A-b_P)am_{\rm q,c}\right].
\ee
On the other hand, it can be extracted directly from the corresponding bare subtracted quark mass via the relation 
\be
   r_0M_{c{|m_{\rm q,c}}} = Z_M Z r_0 m_{\rm q,c}
   \left[ 1+b_m am_{\rm q,c}\right].
   \label{RGIs}
\ee

\begin{figure}
\centering
\psfrag{aor0sq}[t][c][1][0]{\large$(a/r_0)^2$}
\psfrag{r0Mc}[b][c][1][0]{\large$r_0M_c$}
\psfrag{M1}[c][c][1][0]{\large$r_0M_{c{|m_{cc}}}$}
\psfrag{M2}[c][c][1][0]{\large$r_0M_{c{|m_{sc}}}$}
\psfrag{M3}[c][c][1][0]{\large$r_0M_{c{|m_{q,c}}}$}
\psfrag{val}[c][c][1][0]{\large$r_0M_{c{|m_{sc}}}$=\input{./tables/r0Mccont.txt}}
\epsfig{scale=.8,file=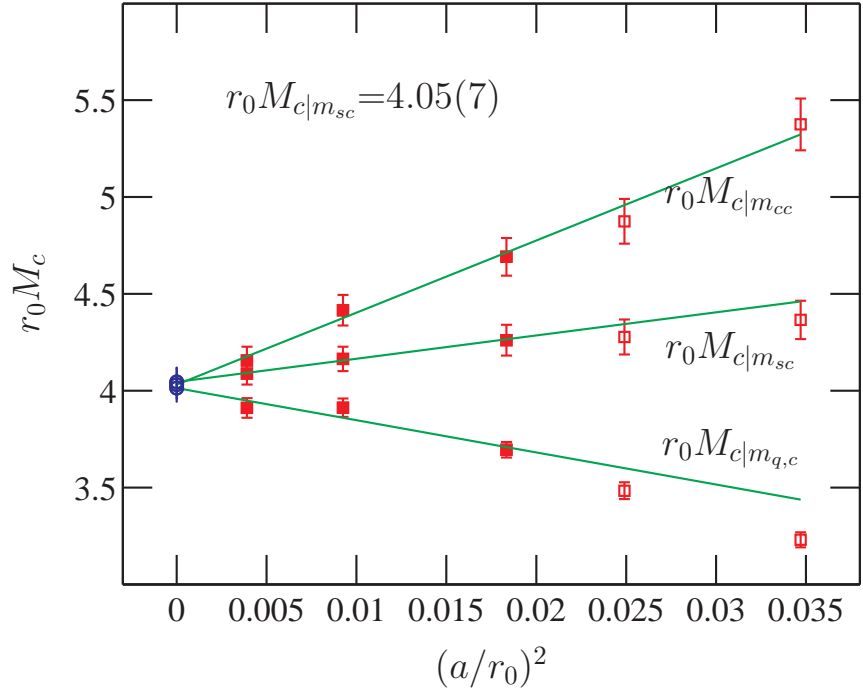}
\caption{Continuum extrapolation for the renormalization group invariant charm quark mass $M_c$. Only the data at the filled squares entered the respective fit.}\label{Mccont}
\end{figure}

The continuum extrapolations of all the three definitions of the renormalization group invariant charm quark mass $M_c$ are illustrated in figure \ref{Mccont}. Within the errors, they all converge to the same continuum result.  The final result was taken from the continuum limit of $r_0M_{c{|m_{sc}}}$, since the smallest cutoff effects were observed in this case. The corresponding linear fit has  $\chi^2/{\rm d.o.f.}=0.32$ and the extrapolated value at $a=0$ is
\begin{center}
\begin{tabular}{rcl}
$r_0M_c$&=&4.05(7).
\end{tabular}
\end{center}
Using $r_0=0.5$ fm, this translates into $M_c=$1.597(28)GeV. By inverting equation (\ref{Mombar}) with MAPLE, this value can be translated to the $\MSbar$-scheme and one gets $\mbar_c(\mbar_c)=1.27(3)$ GeV. In this step, the 4-loop anomalous dimensions $\gamma^{\MSbar}$ and $\beta$ have been employed and the error of $\Lambda_{\rm QCD}=238(19)$ MeV \cite{Capitani:1998mq} has been taken into account.

The Particle Data group \cite{PDBook} gives $\mbar_c(\mbar_c)\approx 1.15 - 1.35$ GeV as a range for the charm quark mass, excluding current data from the lattice. In addition, they also quote $\mbar_c(\mbar_c)=1.26(13)(20)$GeV as the world average from the lattice. Here, the second error is an estimated uncertainty of 15\% due to the quenched approximation.
 
The determination of the charm quark mass in lattice QCD has a long history \cite{Allton:1994ae,Bochkarev:1996ai,Kronfeld:1998zc,Gimenez:1998uv,Rolf:2002gu,Becirevic:2001yh,deDivitiis:2003iy}.
Other recent measurements in quenched lattice QCD include \cite{Becirevic:2001yh} with $\mbar_c(\mbar_c)=1.26(3)(12)$GeV and \cite{Kronfeld:1998zc} with $\mbar_c(\mbar_c)=1.33(8)$GeV, with statistical and systematic error in the first case and combined statistical and systematic error in the second case. The first result has been obtained in the $O(a)$-improved theory at the rather coarse lattice spacing $a\approx0.07$ fm and the latter has been obtained from a continuum extrapolation. In \cite{deDivitiis:2003iy}, the continuum limit for the charm quark mass was taken by using a step scaling method. The final result quoted there is $\mbar_c(\mbar_c)=1.319(28)$ GeV. A result for the charm quark mass which was obtained in the same way as here, but without a simulation at $\beta_5$ has been carried out in \cite{Rolf:2002gu}, with a final value of $\mbar_c(\mbar_c)=1.301(34)$ GeV. 

An unquenched computation has been carried out by UKQCD \cite{Wingate:2003gm} with $N_f=3$ flavors of dynamical light quarks. Although only preliminary and at very large lattice spacing ($a\approx 1.1$ fm), no effects of sea quarks compared to the quenched results \cite{Rolf:2002gu,Becirevic:2001yh} were observed.

The value for the charm quark mass obtained here is compatible within errors with all previous determinations. 
\subsection{Quenched scale ambiguity}\label{QSA}
In order to assess the quenched scale ambiguity (cf. section \ref{settingthescale}), the data has been analyzed again, with the line of constant physics for the continuum extrapolation defined at the physical meson masses $r_0^\prime m_{\dsub}=5.486$, to which the data has been interpolated. The results and the ambiguity with respect to the case $r_0=0.5$ fm are the following:
\begin{center}
\begin{tabular}{lcrcr}
\hline\hline\\[-2ex]
&&&&ambiguity\\[1ex]
\hline\\[-2ex]
$\fds_{|r_0^{\prime}}  $&=&$205.8(70)$ MeV&$\to$&$-9\%$\\[1ex]
$\fdsstar_{|r_0^{\prime}}  $&=&$ 213(17)$ MeV&$\to$&$-11\%$\\
\\
$ (m_{\dsubstar}-m_{\dsub})_{|r_0^{\prime}}  $&=&$ 111.9(66)$ MeV&$\to$&$-18\%$\\
\\
$M_{c|r_0^{\prime} } $&=&$ 1.690(33)$ GeV&$\to$&$+6\%$\\
\\
$\mbar_c(\mbar_c)$&$=$&1.32(3) GeV&$\to$&+4\%\\[1ex]
\hline\hline\\
\end{tabular}
\end{center}
The magnitude of the ambiguities agree with equivalent estimates in \cite{Juttner:2003ns} for the decay constant and in \cite{Rolf:2002gu} for the charm quark mass.
It should be mentioned here, that the {quenched scale ambiguity} only gives an estimate for the ambiguity inherent to the quenched approximation. The true quenching error can only be determined by direct comparison to results in the full theory. 

\subsection{Discussion}
\begin{table}
\centering
\begin{tabular}{lrrc}
\hline\hline\\[-2ex]
observable		&experiment\cite{PDBook}&\multicolumn{1}{c}{lattice}&precision\\
			&			&	&(quenched)\\[1ex]
\hline\\[-2ex]
$\fds$			&267(33) MeV	&226(7)MeV&3\%\\
$\fdsstar$		&		&239(18)MeV&8\%\\
$\fds/\fdsstar$		&		&0.939(61)&7\%\\[.5ex]

$m_{\rm D_s}$		&1.9683(5) GeV	&\multicolumn{1}{l}{input}\\
$m_{\dsubstar}-m_{\dsub}$ &143.8(4) MeV	&136.0(92)MeV&7\%\\[1ex]
$M_c$		&		&1.597(28)GeV&2\%\\
$\mbar_c(\mbar_c)$&		&1.27(3) GeV&2\%\\[1ex]
\hline\hline\\
\end{tabular}\caption{Summary of results for the $\dsub$- and the $\dsubstar$-meson and the $c$-quark mass.}\label{DSres}
\end{table}
All results in this section have been summarized in table \ref{DSres} together with the final error and the corresponding experimental values (where available).
Most of the observables determined here, are compatible within errors with the previous lattice simulations and with QCD sum rules. Especially in the case of the pseudo scalar decay constant it has been demonstrated, that it is possible to produce lattice data, which matches the precision of experiments, i.e. CLEO-c with a predicted error for $\fds$ of 2\% \cite{CLEO-c}. 

In the case of the meson mass splitting and the renormalization group invariant charm quark mass, the data obtained here is nicely compatible with the expected linear scaling in $(a/r_0)^2$, which is confirmed by the small $\chi^2$ of the corresponding fits. 

For the decay constant $\fds$, although still acceptable, the relatively large value $\chi^2/{\rm d.o.f.}=1.9$ indicates, that the data is not quite compatible with the expected linear scaling behavior. 
In contrast to the mass splitting and the quark mass, the boundary-to-boundary correlation functions $f_A^T$ and $f_V^T$ enter the definition of the pseudo scalar and the vector meson decay constant, respectively. Both quantities are known to fluctuate strongly in the Monte-Carlo history. The situation might therefore improve with larger statistics at $\beta_5$. 

The simulation results from the coarser lattices, which have been excluded from the continuum extrapolation in order to avoid mass dependent cutoff effects, deviate considerably from a linear scaling in $(a/r_0)^2$. This confirms the findings in \cite{Kurth:2001yr} and the upper bound $aM_Q\le 0.64$ for the heavy quark mass at finite lattice spacing, suggested there.

Despite working in the quenched approximation, most of the results presented in this section are compatible within errors with the corresponding values from experiment. A slight discrepancy was discovered for the decay constant. However, unquenching in this case is expected to increase the value by about 10\%. 

Although all sources of systematic error have been taken into account during the data analysis, it could be shown, that lattice results can meet the accuracy of precision experiments.

\section{The $\rm B_s$- and the  $\bsubstar$-meson and HQET}
This section presents the combined analysis of predictions from HQET together with the relativistic data for heavy quarks with masses around the charm quark mass. 

\subsection{Incorporation of  results from the static limit}    
Similar to the observables for the $\dsub$-meson, the quantities
\be\label{intobs}
r_0^{3/2}{{Y_{\rm  PS}}\over C_{\rm PS}},\;{R\over C_{\rm PS/V}}\,{\rm and}\,r_0{{\Delta m}\over C_{\rm spin}},\\
\ee
which are defined in section \ref{asymptotics}, have been extrapolated to the continuum at the pseudo scalar masses which are collected in table \ref{exactintmasses} together with the associated fit range. 

\begin{figure}
\centering
\psfrag{mass}[c][c][1][0]{\large$r_0m_{\rm PS}=\input{./tables/PSmass6.txt},\,{\rm c.l.}=\input{./tables/clPS6.txt}$}
\psfrag{aor0sq}[t][c][1][0]{\large$(a/r_0)^2$}
\psfrag{fsqrtmC}[b][c][1][0]{\large$r_0^{3/2}{\rm F_{ PS}\sqrt{m_{\rm PS}}\over C_{\rm PS}(M_Q/\Lambda_{\rm QCD})}$}
\epsfig{scale=.8,file=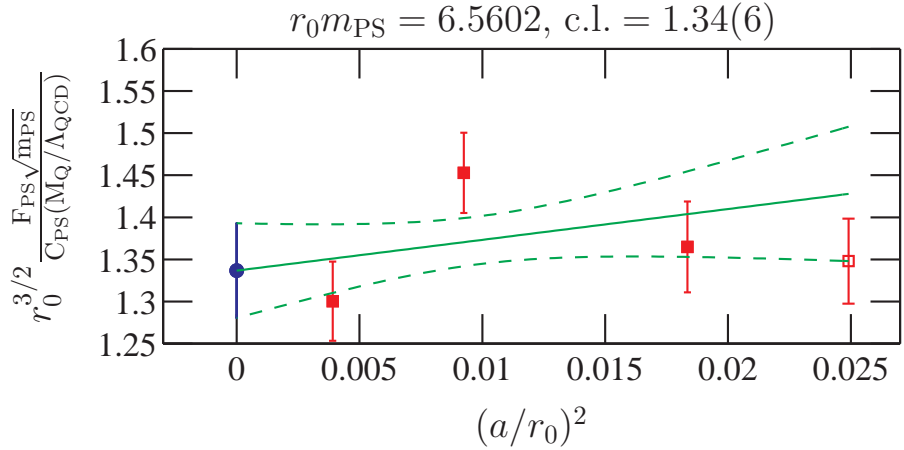}
\caption{Continuum limit for $r_0^{3/2}{\rm F_{ PS}}\sqrt{m_{\rm PS}}\over C_{\rm PS}$ at the  largest pseudo scalar mass $m_{\rm PS}^6$ with $\chi^2/{\rm d.o.f.}=2.5$. Only the data at the filled symbols entered the fit.}\label{worstcase}
\end{figure}

In the case of $r_0^{3/2}{{Y_{\rm  PS}}\over C_{\rm PS}}$ at the largest mass $m_{\rm PS}^6$, the linear continuum extrapolation of three data points as depicted in figure \ref{worstcase} has a large $\chi^2/{\rm d.o.f.}=2.5$. Although $P(\chi^2/{\rm d.o.f.}=2.5)\lesssim 10\%$, the continuum value has been included into the further data analysis, since the extrapolations to $a=0$ at all other pseudo scalar masses behaved much better.

\begin{figure}\begin{minipage}[c]{\linewidth}
\centering

\psfrag{1or0mPS}[t][c][1][0]{\large $1/r_0m_{\rm PS}$}
\psfrag{mDs}[b][l][1][0]{\large $1/r_0m_{\rm D_s}$}
\psfrag{mBs}[b][l][1][0]{\large $1/r_0m_{\rm B_s}$}
\psfrag{FmC}[c][c][1][0]{\large$r_0^{3/2}{{Y_{\rm PS}}\over C_{\rm PS}}$}
\epsfig{scale=.8,file=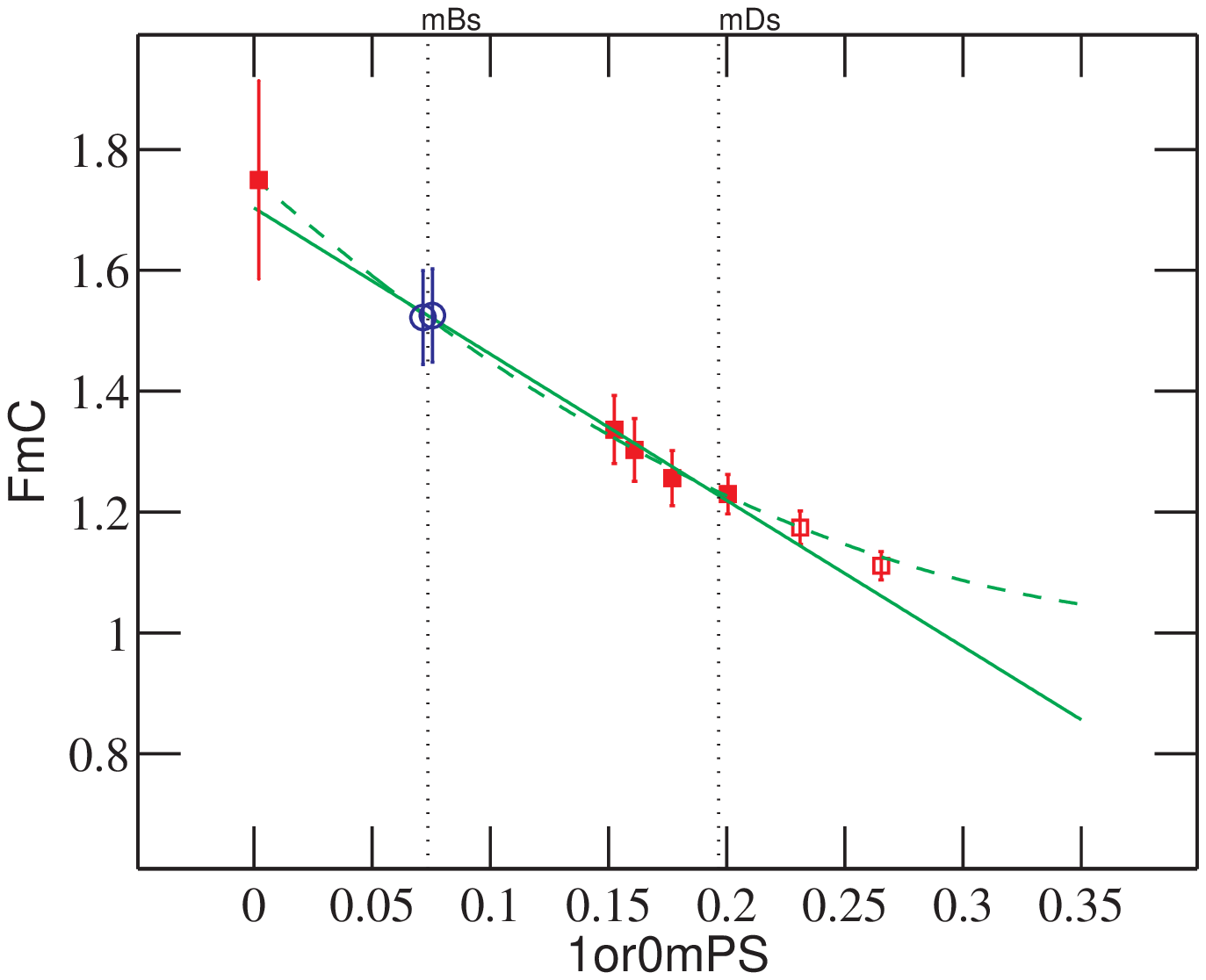}\label{FBs36}\\
\vspace{1cm}
\psfrag{1or0mPS}[t][c][1][0]{\large$1/r_0m_{\rm PS}$}
\psfrag{mDs}[b][l][1][0]{\large $1/r_0m_{\rm D_s}$}
\psfrag{mBs}[b][l][1][0]{\large $1/r_0m_{\rm B_s}$}
\psfrag{FmC}[c][c][1][0]{\large$r_0^{3/2}{Y_{ PS}}\over C_{\rm PS}$}
\epsfig{scale=.8,file=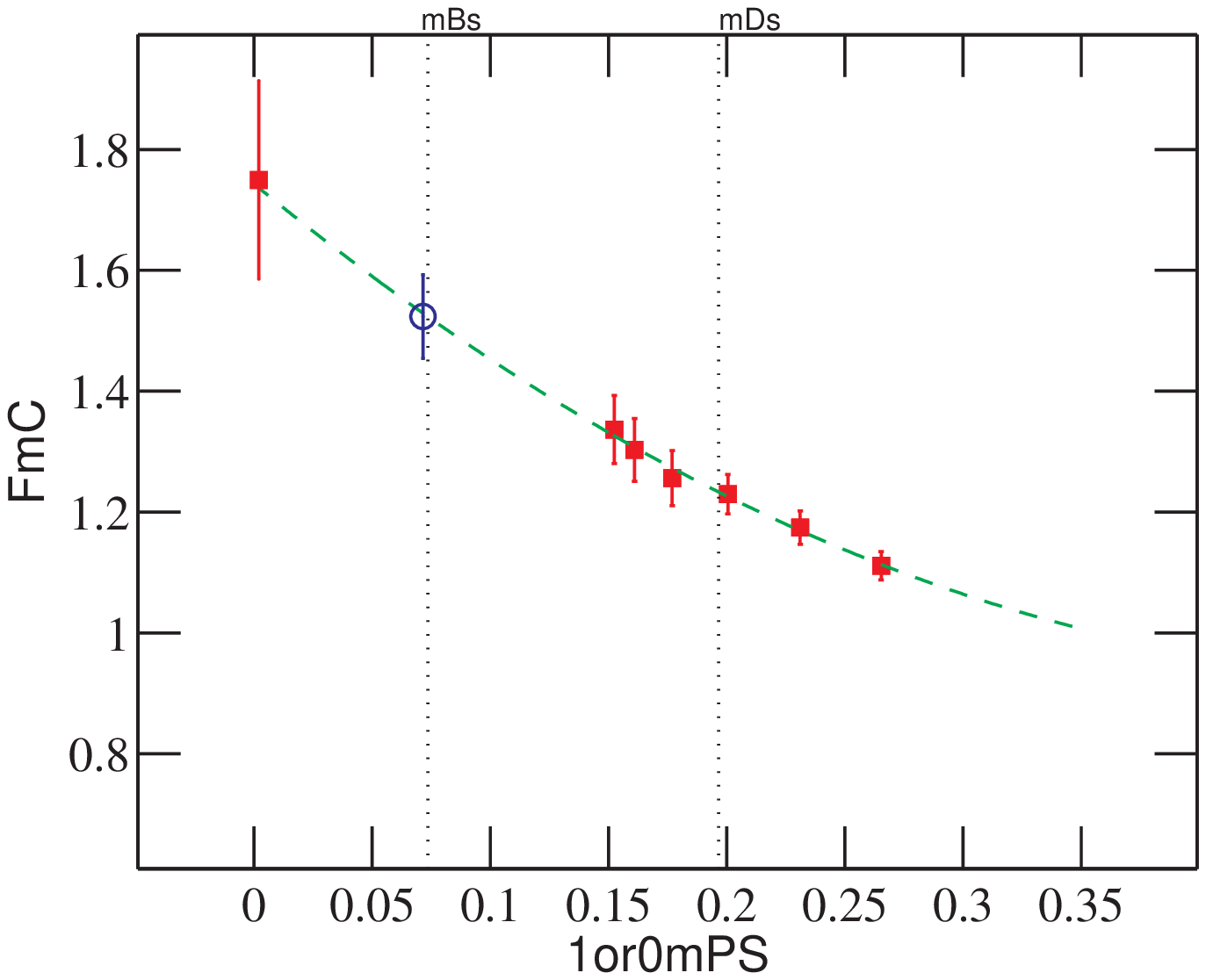}
\end{minipage}
\caption{Interpolation for the decay constant of the pseudo scalar meson with a  linear and a quadratic fit ansatz (solid and dashed line resp.). Only the data at the filled squares was included into the fit. The circles represent the values at the physical point of the $\bsub$-meson.}\label{interF}
\end{figure}

\begin{figure}
\begin{minipage}{\linewidth}
\centering
\psfrag{oor0mPS}[t][c][1][0]{\large$1/r_0m_{\rm PS}$}
\psfrag{mDs}[b][l][1][0]{\large $1/r_0m_{\rm D_s}$}
\psfrag{mBs}[b][l][1][0]{\large $1/r_0m_{\rm B_s}$}
\psfrag{fof}[b][c][1][0]{\large${{ R_{PS}}\over C_{\rm PS/V}}$}
\epsfig{scale=.8,file=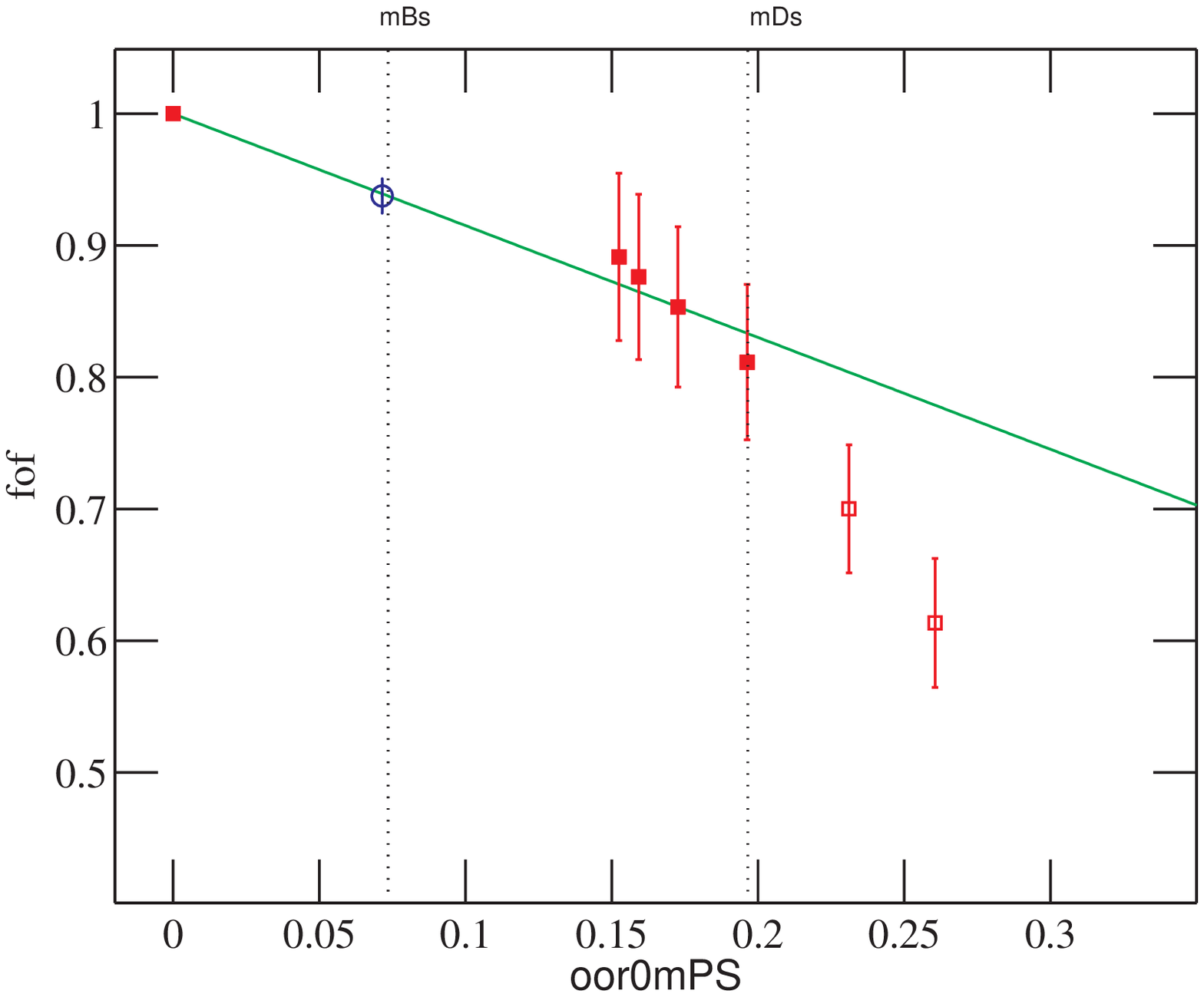}\\
\vspace{1cm}
\psfrag{oor0mPS}[t][c][1][0]{\large$1/r_0m_{\rm PS}$}
\psfrag{mDs}[b][l][1][0]{\large $1/r_0m_{\rm D_s}$}
\psfrag{mBs}[b][l][1][0]{\large $1/r_0m_{\rm B_s}$}
\psfrag{fof}[b][c][1][0]{\large${{R_{PS}}\over C_{\rm PS/V}}$}
\epsfig{scale=.8,file=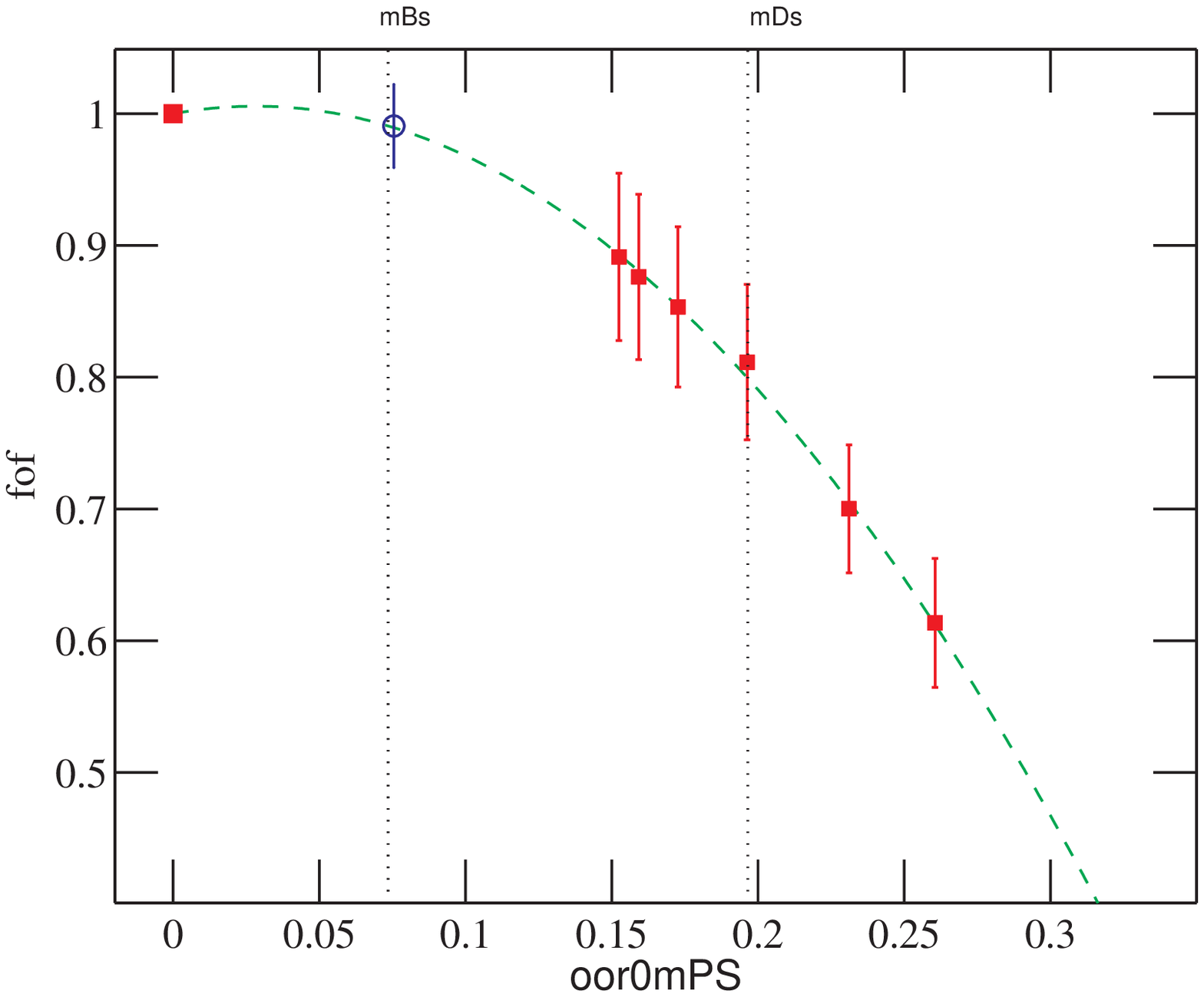}
\end{minipage}
\caption{Interpolation for the ratio of the pseudo scalar and the vector meson decay constant with a linear and a quadratic fit ansatz (solid and dashed line resp.). Only the the data at the filled squares was included into the fit. The circles represent the values at the physical point of the $\bsub$-meson.}\label{interfof}
\end{figure}

\begin{figure}
\begin{minipage}{\linewidth}
\centering
\psfrag{oor0mPS}[t][c][1][0]{\large$1/r_0m_{\rm PS}$}
\psfrag{mDs}[b][l][1][0]{\large $1/r_0m_{\rm D_s}$}
\psfrag{mBs}[b][l][1][0]{\large $1/r_0m_{\rm B_s}$}
\psfrag{r0mVmPSoCspin}[b][c][1][0]{\large${r_0\Delta m\over C_{\rm spin}}$}
\epsfig{scale=.8,file=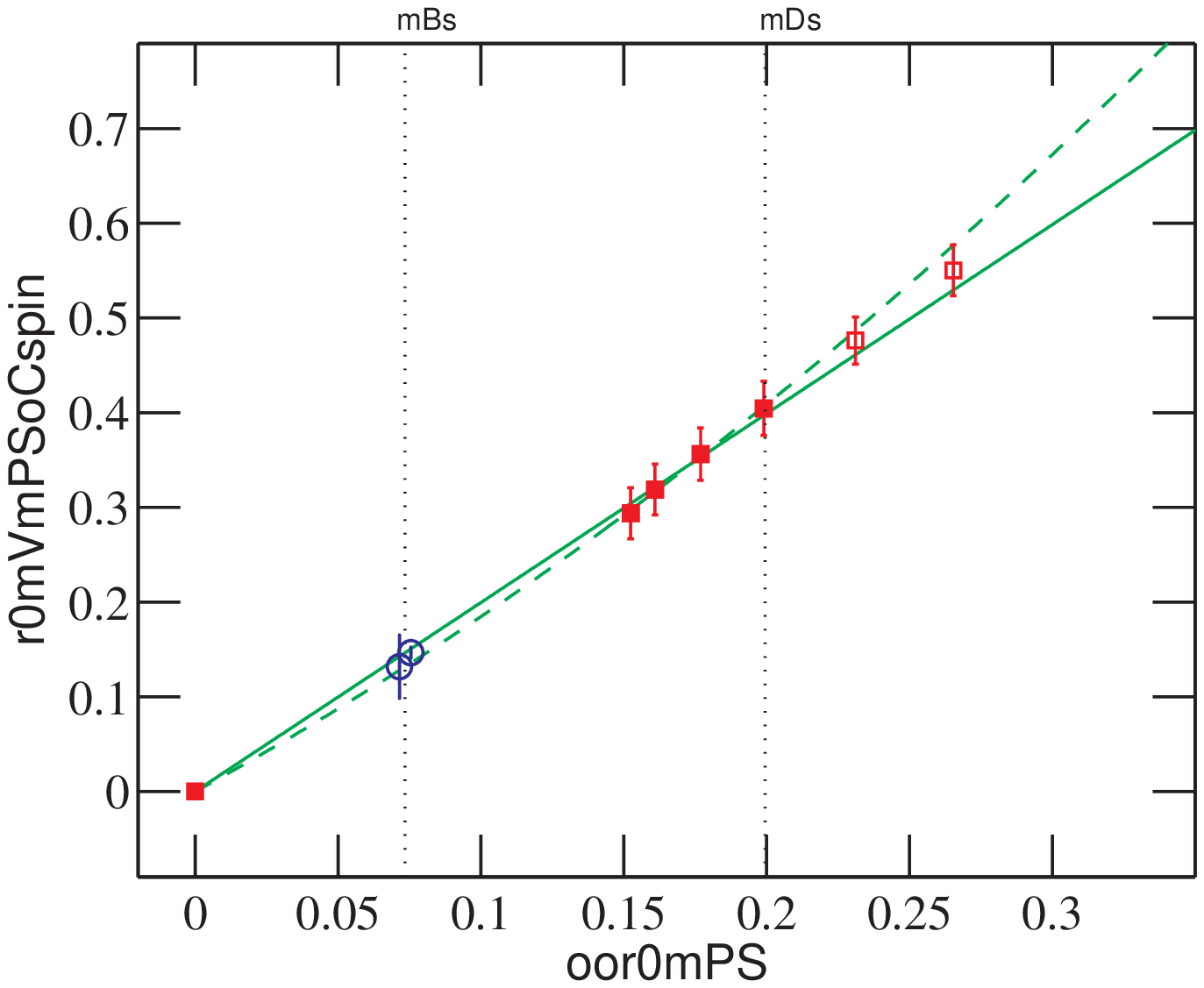}\\
\vspace{1cm}
\psfrag{oor0mPS}[t][c][1][0]{\large$1/r_0m_{\rm PS}$}
\psfrag{mDs}[b][l][1][0]{\large $1/r_0m_{\rm D_s}$}
\psfrag{mBs}[b][l][1][0]{\large $1/r_0m_{\rm B_s}$}
\psfrag{r0mVmPSoCspin}[b][c][1][0]{\large${r_0\Delta m\over C_{\rm spin}}$}
\epsfig{scale=.8,file=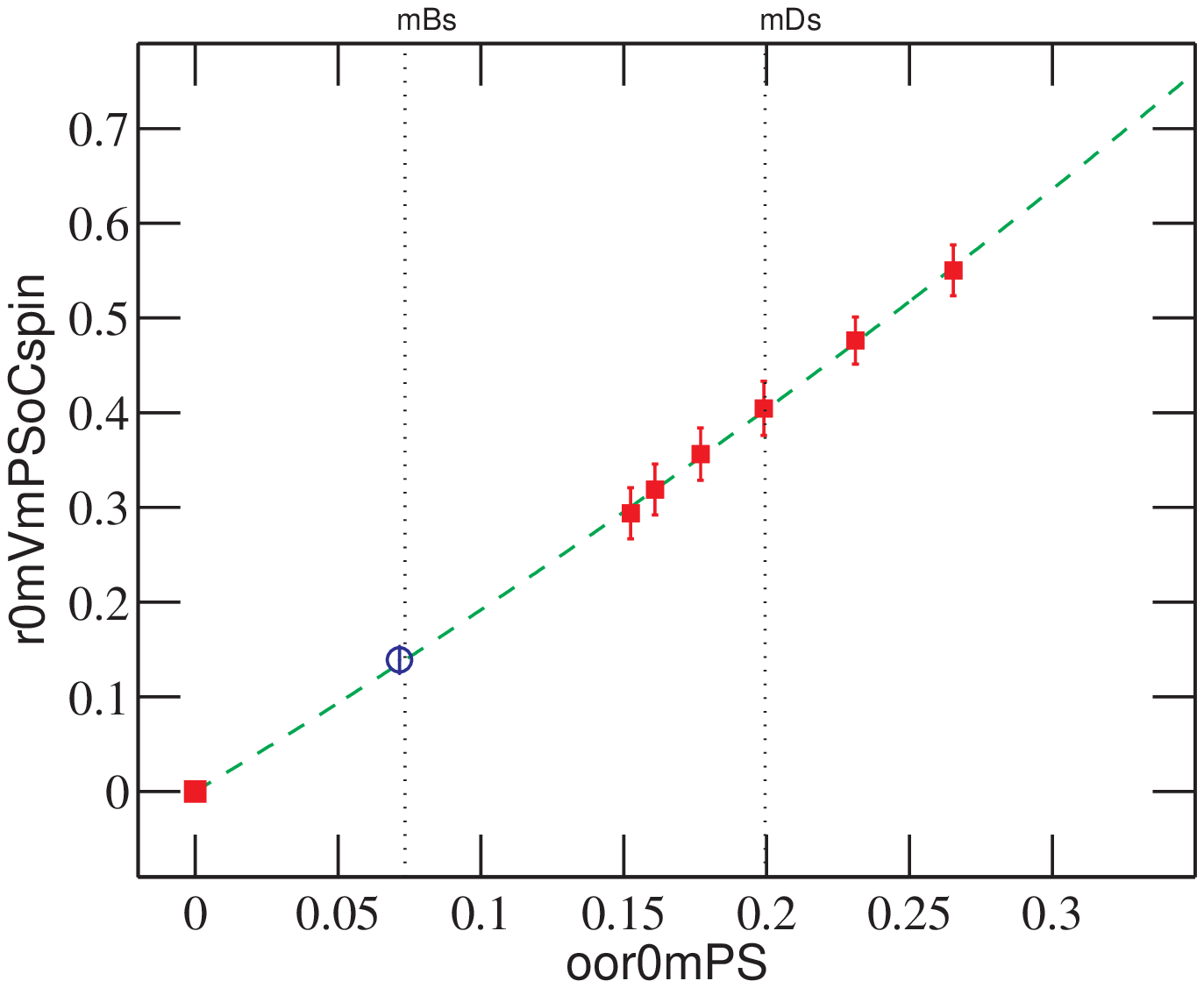}
\end{minipage}
\caption{Interpolation for the mass splitting with a linear and a quadratic fit ansatz (solid and dashed line resp.). Only the data at the filled squares was included into the fit. The circles represent the values at the physical point of the $\bsub$-meson.}\label{intersplit}
\end{figure}

For $r_0^{3/2}{{Y_{\rm  PS}}/ C_{\rm PS}}$, the non-perturbatively determined value in the static approximation \cite{DellaMorte:2003mn}
\be
r_0^{3/2}\Phi_{\rm PS,RGI}^{\rm stat}=1.74(13)
\ee
has been included into the fit as a constraint in the static limit. For the ratio ${R/ C_{\rm PS/V}}$ and the mass splitting $r_0{{\Delta m}/ C_{\rm spin}}$,
the asymptotics as detailed in section \ref{asymptotics} have been included to constrain the interpolation.
Figure \ref{interF}, \ref{interfof} and \ref{intersplit} show the results. The interpolation has been carried out, once including the data in the range of pseudo scalar masses from $r_0m_{\rm PS}^1 - r_0m_{\rm PS}^6$ and once with the masses $r_0m_{\rm PS}^3 - r_0m_{\rm PS}^6$. For the larger mass range, a second order polynomial was used as the fit-ansatz, while for the smaller mass range also a linear fit was applied. The physical point of the $\bsub$-meson and the corresponding results from the interpolations are indicated by the circles. Table \ref{HQETcoeffs} shows the parameters of the fit polynomial  
\be
a_0+{a_1\over r_0m_{\rm PS}}+{a_2\over (r_0m_{\rm PS})^2}+\dots\;.
\ee

The errors associated to the fit-parameter $a_1$ in the case of linear interpolations is not too large. Thus, $a_1$ can be taken as a rough estimate of the magnitude of the first-order correction. 

A fit of ${R/ C_{\rm PS/V}}$ over the mass range $r_0m_{\rm PS}^3 - r_0m_{\rm PS}^6$ with a 2nd order polynomial is not feasible due to the large error associated with the data. In the other cases, 2nd order polynomials fit the data well for both mass ranges. However, the corresponding coefficients can only be determined very inaccurately and significant statements about higher order contributions cannot be made.

\begin{table}\centering
\begin{tabular}{c@{\hspace{1cm}}cc@{\hspace{1.5cm}}ccc}
\hline\hline\\[-2ex]
observable	&\multicolumn{2}{c}{\hspace{-1.5cm}linear}	&\multicolumn{3}{c}{quadratic}\\
		&$a_0$	&$a_1$			&$a_0$  &$a_1$	&$a_2$\\[1ex]
\hline\hline\\[-2ex]
\multicolumn{6}{c}{mass range $m_1\dots m_6$}\\
\\[-2ex]
\hline\\
$r_0^{3/2}{Y_{\rm  PS}}\over C_{\rm PS}$
		&-&-&1.7(2)&-3(2)&3(4)\\
\\
${R\over C_{\rm PS/V}}$
		&-&-		&1.0	&0.4(6)&-7(3)\\
\\
${r_0\Delta m\over C_{\rm spin}}$
		&-	&-	&0.0	&1.8(3)&1(1)\\[2ex]
\hline\\[-2ex]
\multicolumn{6}{c}{mass range $m_3\dots m_6$}\\
\\[-2ex]
\hline\\
$r_0^{3/2}{Y_{\rm  PS}}\over C_{\rm PS}$&1.7(1)&-2.4(7)&1.8(2)&-3(2)&4(8)\\
\\
${R\over C_{\rm PS/V}}$&1.0	&-0.8(2)	&1.0	&-&-\\
\\
${r_0\Delta m\over C_{\rm spin}}$&0.0	&1.99(8)	&0.0	&1.6(8)&2(4)\\[2ex]
\hline\hline
\end{tabular}
\caption{Coefficients and associated error of the polynomial fit $a_0+{a_1\over r_0m_{\rm PS}}+{a_2\over (r_0m_{\rm PS})^2}$ for the interpolation between the relativistic data and the static approximation for various observables.}\label{HQETcoeffs}
\end{table}


\subsection{Interpolation to the ${\rm B_{s}^{(\ast)}}$-meson}    
In the following, the determination of the observables
\be
\fbs,\,\fbs/\fbsstar,\,{\rm and}\,r_0(m_{\bsubstar}-m_{\bsub})
\ee
will be presented. 
One first extracts the value of the corresponding interpolation at the physical point \cite{PDBook}
\be\ba{ccc}
m_{\bsub}    =5.3696(24){\rm GeV} &\to& r_0m_{\bsub}=13.604(61),\\
\ea
\ee
which is indicated by the blue circles in the respective plots \ref{interF}, \ref{interfof} and \ref{intersplit}.  
Then, the conversion functions have to be eliminated. In the case of the decay constant and the ratio of the decay constants, also the square roots of the meson masses have to be eliminated (cf. section \ref{asymptotics}).
To this end, the conversion functions $C_{\rm X}(M_Q/\Lambda_{\MSbar})$ were taken at the value of the $b$-quark mass $r_0M_b=16.12(29)$ which has been determined non-perturbatively in quenched QCD \cite{Heitger:2003nj}.  
The results for all observables for the $\bsub$-meson are summarized in table \ref{Bsresults} for the various interpolations that have been carried out.

The quadratic interpolation over the mass range $m_{\rm {PS}}^3\dots m_{\rm PS}^6$ leads to very large errors and significant results cannot be obtained. The linear interpolation over the larger mass range $m_{\rm {PS}}^1\dots m_{\rm PS}^6$ on the other hand stretches across a domain, where sizeable higher order contributions are to be expected. Both interpolations were therefore discarded for the further analysis.
\begin{table}
\centering
\begin{tabular}{lrrrrr}
\hline\hline\\[-2ex]
Fit		&\multicolumn{1}{c}{linear}&\multicolumn{1}{c}{quadr.} &\multicolumn{1}{c}{linear} 	&\multicolumn{1}{c}{quadr.} \\
Mass range 	&		$m_{\rm PS}^1\dots m_{\rm PS}^6$&$m_{\rm PS}^1\dots m_{\rm PS}^6$&$m_{\rm {PS}}^3\dots m_{\rm PS}^6$&$m_{\rm PS}^3\dots m_{\rm PS}^6$\\[1ex]
\hline\\[-2ex]

$r_0\fbs$	&0.50(3)&0.51(3)&0.50(3)&0.53(3)\\[1ex]
$\fbs/\fbsstar$		&0.907(9)&0.99(3)&0.93(1)&\multicolumn{1}{c}{-}\\[1ex]
$r_0(m_{\bsub}-m_{\bsubstar})$	&0.171(5)&0.16(2)&0.167(7)&0.15(4)\\[1ex]
\hline\hline\\
\end{tabular}\caption{Results for $\bsub$ from the interpolation.}\label{Bsresults}
\end{table}
In order to determine the final results, the linear interpolation including the masses $m_{\rm {PS}}^3\dots m_{\rm PS}^6$ and the quadratic interpolation including the masses $m_{\rm {PS}}^1\dots m_{\rm PS}^6$ were then compared. First, the corresponding results from table \ref{Bsresults} translated to physical units are
\begin{center}
\begin{tabular}{lrC}
\hline\hline&&\\[-2ex]
Fit		&\multicolumn{1}{c}{linear}&\multicolumn{1}{C}{quadr.} \\
Mass range 	&$m_{\rm PS}^3\dots m_{\rm PS}^6$&$m_{\rm PS}^1\dots m_{\rm PS}^6$\\[1ex]
\hline&&\\[-2ex]
$\fbs$			&200(10)MeV&\multicolumn{1}{R}{198(9)MeV}\\[1ex]
$\fbs/\fbsstar$		&0.93(1)&\multicolumn{1}{R}{0.99(3)}\\[1ex]
$m_{\bsub}-m_{\bsubstar}$	&66(3)MeV&\multicolumn{1}{R}{63(6)MeV}\\[1ex]
\hline\hline\\
\end{tabular}
\end{center}

The results of both fits are compatible within 1$\sigma$. While the error for the ratio of the decay constants and for the mass splitting is twice as large in the case of the interpolation with the 2nd order polynomial, the error is approximately the same for the decay constant. Comparing the results from both fits, no preference becomes apparent. However, in order to arrive at a conservative error estimate, the values obtained from the quadratic interpolation were chosen as the final results.
All final results for the $\rm B_s^{(\ast)}$-meson are collected in table \ref{BSres}

Using QCD sum rules, $\fbs=236(30)$MeV \cite{Narison:2001pu} and $\fbs=244(21)$MeV \cite{Jamin:2001fw} have been determined.
Other groups have done lattice simulations in the quenched theory as well as with dynamical quarks with $N_f = 2$ and $N_f = 3$ flavors. The corresponding results have been summarized in table \ref{otherfbs}. Again it turns out, that unquenching yields decay constants that are about 10\% larger than in the quenched case. However, dynamical simulations still suffer from large systematic errors.

The experimental value of the mass splitting is 47.0(26)MeV \cite{PDBook}.
\begin{table}\centering
\begin{tabular}{l@{\hspace{1.5mm}}cl@{\hspace{1.5mm}}rl@{\hspace{1.5mm}}cc}
\hline\hline\\[-2ex]
\multicolumn{2}{c}{Reference	}			&\multicolumn{1}{c}{${\rm F}_{\rm B_s}^{N_f=0}/$MeV}		&$N_f$&\multicolumn{1}{c}{$\fbs/$MeV}&\begin{tabular}{c}scale\\setting\end{tabular}\\[1ex]
\hline\\[-2ex]
ALPHA		&\cite{Rolf:2003mn}	&206(10) &&&$r_0$\\[.3ex]
Becirevic \emph{et. al.}&\cite{Becirevic:1998ua}	&$204(16)(^{+28}_{-0})$ 	&&&$m_\rho$\\[.3ex]
Bowler \emph{et. al.}&\cite{Bowler:2000xw}	&$220(6)(^{+23}_{-28})$ 	&&&$f_\pi$\\[.3ex]
CP-PACS&\cite{AliKhan:2000eg}		&$219(10)$ 		&2&$250(10)(^{+15}_{-13})$&$m_\rho$ \\[.3ex]
CP-PACS&\cite{AliKhan:2001jg}		&$220(4)(31)$ 		&2&$242(9)(^{+51}_{-34})$&$m_\rho$ \\[.3ex]
de Divitiis&\cite{deDivitiis:2003wy}	&192(6)(4) &&&$r_0$\\[.3ex]
JLQCD		&\cite{Aoki:2003xb}		&		&2&$215(9)(^{+14}_{-13})$ &$m_\rho$\\[.3ex]

MILC&\cite{Bernard:2002pc}		&$199(5)(^{+23}_{-22})$ 	&2&$217(6)(^{+  37}_{-28})$ &$f_\pi$\\[.3ex]
UKQCD&\cite{Maynard:2001zd}		&$220(6)(^{+23}_{-28})$ &&&$f_\pi$\\[.3ex]

Wingate \emph{et. al.}&\cite{Wingate:2003gm}	&	&3&260(7)(28) 		&$\Upsilon$ \\
Ryan (\emph{world av.})&\cite{Ryan:2001ej}			&$200(20)$ 		&&$230(30) $\\[1ex]
\hline\hline\\
\end{tabular}
\caption{Results for the decay constant $\fbs$ from other groups with statistical and systematic errors.}\label{otherfbs}
\end{table}

\subsection{Quenched scale ambiguity}
As in the case of the $\dsub$- and $\dsubstar$-meson, the quenched scale ambiguity has been estimated. The size 
\begin{center}
\begin{tabular}{lcrcrc}
\hline\hline\\[-2ex]
&&&&ambiguity\\
\hline\\[-2ex]
$\fbs_{|r_0^{\prime}}  $&=&$182(10)$ MeV&$\to$&$-8\%$ &\\
\\
$ (m_{\bsubstar}-m_{\bsub})_{|r_0^{\prime}}  $&=&$ 52(6)$ MeV&$\to$&-$18\%$\\[1ex]
\hline\hline\\

\end{tabular}
\end{center}
is roughly the same as for the corresponding observables of the $\dsub$-meson.
\subsection{Discussion}
The interpolation between the results from quenched QCD around the charm sector and HQET indicates, that the coefficients of the linear and quadratic
term in the heavy quark expansion
\be
a_0+{a_1\over r_0m_{\rm PS}}+{a_2\over (r_0m_{\rm PS})^2}+\dots
\ee
are of order $O(1)$ and one can therefore expect, that HQET is a good approximation for ${\rm B_{(s)}}$-mesons.

The interpolation has also been used successfully to determine observables of the $\bsub$-meson with reasonable errors. All final results have been summarized in table \ref{BSres}, together with experimental data and the associated errors within the quenched approximation.

The result for the decay constant is compatible with most of the previous studies. The mass splitting that was determined from the simulations is not compatible with experiment when setting the scale with the Kaon decay constant. It agrees however, when setting the scale with the nucleon mass instead.

\begin{table}
\centering
\begin{tabular}{lrrc}
\hline\hline\\[-2ex]
observable		&experiment\cite{PDBook}&\multicolumn{1}{c}{lattice}&precision\\
			&			&	&(quenched)\\[1ex]
\hline\\[-2ex]
$\fbs$ 			&		& 198(9){\rm MeV}&5\%\\[.5ex]
$\fbsstar$		&		&194.664406(1)MeV&6\%\\
${\fbs/ \fbsstar}$ 	&		& 0.99(3)&3\%\\[1ex]
$m_{\rm B_s}$		&5.3696(24) GeV &input\\[.5ex]
$m_{\rm B_s^\ast}$	&5.4166(35) GeV	&\\
$m_{\bsubstar} - m_{\bsub}$ &47.0(26) MeV& 63(6){\rm MeV}&11\%\\[1ex]
\hline\hline\\

\end{tabular}\caption{Summary of results for the $\bsub$- and the $\bsubstar$-meson.}\label{BSres}
\end{table}

\chapter{Summary and outlook}
\vspace{-.5cm}
The work for this thesis was focused on precision measurements of heavy-light meson observables in quenched QCD. The systematic errors stemming from
\bi
\item discretization effects\\[-4ex]
\item contributions from excited states\\[-4ex]
\item the continuum extrapolation\\[-4ex]
\item finite volume effects\\[-4ex]
\item interpolation to the physical quark mass
\ei
have been controlled, estimated and considered in the analysis. Only the unknown error due to the quenched approximation remains. In particular, the meson decay constants and the mass splitting for the ${\rm D_s^{(\ast)}}$- and the ${\rm B_s^{(\ast)}}$-meson and the charm quark mass were studied. Moreover, the order of magnitude of the coefficient of the leading order contributions to the static approximation in the heavy quark expansion was estimated.

Starting from the MILC-collaboration's computer program for SU(3) lattice gauge theory, a platform independent tool was created and tested, that can accomplish all the necessary computations, for example on a PC-cluster, using MPI-based parallelism.

In a scaling study with five lattices of constant volume but decreasing lattice spacings $a\approx 1,0.8,0.7,0.5,0.3$ fm, the desired observables were extrapolated to the continuum.
At each lattice spacing, simulations were carried out for six heavy quark masses in the region of the charm quark mass, while keeping a seventh quark mass at the physical value of the strange quark mass.
Deviations from the expected linear scaling in $(a/r_0)^2$ were observed in all considered observables. They were stronger for the heavier meson masses. The expected scaling could be recovered by restricting the data that enters the continuum extrapolation with the upper bound $aM_Q<0.64$ for the heavy quark mass  \cite{Kurth:2001yr}. 

The results at the physical mass of the $\dsub$-meson show, that a final combined statistical and systematic error of 3\% on the pseudo scalar meson decay constant in the continuum can be achieved. This is at the level of the precision of currently running experiments. For example, CLEO-c claims, that the error on $\fds$ from experiment will be reduced below 2\% in the near future \cite{CLEO-c}.
Also the results for the charm quark mass in the continuum could be determined with a final error of 2\%.

The conversion functions, that are necessary to relate observables in QCD and their renormalization group invariant analog in HQET were computed and parameterized as functions of the renormalization group invariant quark mass. The simulation results of quenched QCD in the region of the charm quark mass, extrapolated to the continuum, were successfully combined with predictions from HQET by means of an interpolation in the inverse meson mass. Also the error due to the finite order in perturbation theory in the conversion functions entered the data analysis.

The results from the interpolation indicate, that the leading spin- and flavor symmetry breaking corrections in the heavy quark expansion have coefficients that are sufficiently small to expect HQET to be a good approximation for mesons containing a $b$-quark as the heavy quark.
These findings are compatible with a similar study of very heavy relativistic quark masses in small volume \cite{Heitger:2004gb}. Furthermore, by evaluating the interpolation at the physical point of the $\bsub$-meson, predictions for the pseudo scalar and the vector meson decay constant and the mass splitting with a combined statistical and systematic error of 5\%, 6\% and 10\%, respectively, were obtained. 

On a wish list of what has to be done next in order to improve the current status are:
\bi
\item the calculation of the $1/m$-corrections to the static limit\\[-4ex]
\item the inclusion of a more precise value for the decay constant in the static limit\\[-4ex]
\item the extension of the calculations to the B-mesons containing a $u$ or $d$ quark as the light quark\\[-4ex]
\item non-perturbative matching\\[-4ex]
\item the repetition of the simulations with dynamical fermions
\ei
The first two wishes are work in progress by the ALPHA collaboration \cite{Rolf:2003mn,ALPHAnew} and will allow to further constrain the interpolation and thus to reduce the error on the observables of the $\bsub$-meson. 

Including the light $u$- and $d$-quarks in order to simulate for the B-meson is very costly and involves a chiral extrapolation which introduces an additional source of systematic errors. 

A program to non-perturbatively match HQET and QCD which would reduce the systematic uncertainty due to perturbation theory in the conversion functions has been set up by the ALPHA collaboration \cite{Heitger:2003nj}. 

Finally, only results from simulations of full QCD can be used to reliably predict the physical observables, that are necessary for a precision analysis of the Standard Model. When keeping all sources of systematic errors under control, this is still a very costly task.

\begin{appendix}
\chapter{Notation}\label{notation}
\vspace{-1cm}
{\bf Pauli matrices}\\
The Pauli matrices are defined as
\be\ba{ccc}
\sigma_1 = \left(\ba{cc}0&1\\1&0\ea\right),\;
\sigma_2 = \left(\ba{cc}0&-i\\i&0\ea\right),\;
\sigma_3 = \left(\ba{cc}1&0\\0&-1\ea\right).
\ea\ee
They fulfill the Lie-Algebra
\be
[\sigma_i,\sigma_j]=2i\epsilon_{ijk}\sigma_k,
\ee
and are, in context with the iso-spin algebra, also often referred to as $\tau_i={\sigma_i}$.\\
\\
{\bf Dirac Matrices}\\
The Dirac Matrices in Euclidean space and in Minkowski space are connected via
\be\ba{c}
\gamma_{1,2,3}^{\rm Euclidean}\equiv-i\gamma_{1,2,3}^{\rm Minkowski}\\
\\
\gamma_{0}^{\rm Euclidean}\equiv\gamma_0^{\rm Minkowski}\\
\ea\ee
They obey the anti-commutation relation 
\be
\{\gamma_\mu,\gamma_\nu\}=2\delta_{\mu\nu}.
\ee
and can be constructed from the Pauli matrices. In the chiral representation, the Euclidean Dirac matrices $\gamma_\mu$ ($\mu=0,1,2,3$) read
\be\ba{ccc}
\gamma_{1,2,3} = \left(\ba{cc}0&-i\sigma_{1,2,3}\\i\sigma_{1,2,3}&0\ea\right).
\ea\ee
The matrices for $\mu=0$ and $\gamma_5 = \gamma_0\gamma_1\gamma_2\gamma_3$ in the chiral representation are
\be
\gamma_{0} = \left(\ba{cc}0&1\\1&0\ea\right),\;
\gamma_5 = \left(\ba{cc}1&0\\0&-1\ea\right).
\ee
$\gamma_5$ anti-commutes with the $\gamma_\mu$,
\be
\{\gamma_\mu,\gamma_5\}=0.
\ee
\chapter{The relation between the pole mass and the renormalization group invariant quark mass}\label{poleRGI}
The pole mass $m_Q$ is related to the renormalization group invariant quark mass $M_Q$ via
\be\label{reparam}
m_Q=\left({m_Q\over \overline{m}(\overline{m})}\right)\left({\overline{m}(\overline{m})\over M_Q}\right)M_Q.
\ee
$\overline{m}(\overline{m})$ is the renormalized mass in the $\overline{\rm MS}$-scheme of dimensional regularization. 
The ratio $m_Q/\overline{m}(\overline{m})$ has been determined to three loop precision in \cite{Melnikov:2000qh,Gray:1990yh,Broadhurst:1991fy} and is quoted here for the quenched approximation:
\be\ba{rcl}\label{mqovermbar}
{m_Q\over \overline{m}(\overline{m})} &=&  1 + {\bar g^2(\overline{m})\over 3\pi^2}+13.4434{\bar g^4(\overline{m})\over 16\pi^4} + 190.595{\bar g^6(\overline{m})\over 64 \pi^6}.\\
\ea
\ee
The ratio ${\overline{m}(\overline{m})/ M_Q}$ on the other hand 
can be determined from the renormalization group equations
\be\label{RGE1}
\mu {{d\overline{m}(\mu)\over d\mu}}=\tau^{\MSbar}(\bar g(\mu))\;\;{\rm and}\;\;
\mu {{d\bar{g}(\mu)\over d\mu}}=\beta^{\MSbar}(\bar g(\mu)),
\ee
using the 4-loop anomalous dimension of the renormalized coupling $\beta^{\MSbar}(g)$ \cite{vanRitbergen:1997va} and the 4-loop quark mass anomalous dimension $\tau^{\MSbar}(g)$ \cite{Vermaseren:1997fq}, which have the expansions 
\be\ba{rclc}
\beta^{\MSbar}(g)&=&-b_0g^3-b_1g^5-b_2g^7-b_3g^9-\dots&{\rm and}\\
\\
\tau^{\MSbar}(g)&=&-d_0g^2-d_1g^4-d_2g^6-d_3g^8+\dots\,.
\ea
\ee
The corresponding coefficients are collected in table \ref{betantau}.
Integrating the renormalization group equations (\ref{RGE1}), one obtains
\be\label{movermbar}
{M_Q\over \overline{m}(\overline{m})}  =  (2b_0\bar g^2(\overline m))^{-d_0/2b_0}  \exp\left\{-\int\limits_0^{\bar g{(\overline{m})}}dg\left[{\tau^{\MSbar}(g)\over \beta^{\MSbar}(g)}-{d_0\over b_0g}\right] \right\},
\ee
which can be evaluated using MAPLE. 
\begin{table}
\centering
\begin{tabular}{rcl|rcl}
\multicolumn{3}{c|}{$\beta^{\overline{\rm MS}}$ \cite{vanRitbergen:1997va}}&\multicolumn{3}{c}{$\tau^{\overline{\rm MS}}$ \cite{Vermaseren:1997fq}}\\[.5ex]
\hline
\vspace{-.4cm}&&&&\\
$b_0$&=&${11\over (4\pi)^2}$			&$d_0$&=& ${8\over(4\pi)^2}$\\[1.5ex]
$b_1$&=&${102\over(4\pi)^4}$			&$d_1$&=& ${404\over3 (4\pi)^4}$\\[1.5ex]
$b_2$&=&${2857\over 2(4\pi)^6}$			&$d_2$&=& ${2498\over(4\pi)^6}$\\[1.5ex]
$b_3$&=&${29243-5033/18\over(4\pi)^8}$		&$d_3$&=& ${50659\over(4\pi)^8}$\\
\end{tabular}
\caption{4-loop anomalous dimension of the coupling and the mass in the $\overline{\rm MS}$-scheme of dimensional regularization.}\label{betantau}
\end{table}

\chapter{Summary of improvement and renormalization constants}\label{parametrizationtables}
\begin{sidewaystable} 
\centering 
\begin{tabular}{ccrclccc} 
\hline\hline\\[-2ex]
non-pert.	&pert. 		&\multicolumn{3}{c}{parameterization}	&parameter	&error	&reference\\
		&\#-loops	&&&			&range		&&\\[1ex]
\hline
&&&&&&&\\
$\times$	&&$Z_A(g_0)$&$=$&$\frac{1-0.8496g_0^2+0.0610g_0^2}{1-0.7332g_0^2}$&$0\le g_0^2\le 1$&&\cite{Luscher:1997jn}\\
&&&&&&&\\
$\times$&	&$Z_V(g_0)$&$=$&$\frac{1-0.7663g_0^2+0.0488g_0^2}{1-0.6369g_0^2}			$&$0\le g_0^2\le1$&&\cite{Luscher:1997jn}\\
&&&&&&&\\
$\times$&	&$Z_P(\beta)$&$=$&$0.5233 - 0.0362  (\beta-6.0) + 0.0430  (\beta-6.0)^2			$&$6.0\le \beta\le 6.5$&&\cite{Capitani:1998mq}\\
&&&&&&&\\
$\times$&	&$Z(g_0)$&$=$&$(1+0.090514g_0^2)\frac{1-0.9678g_0^2+0.04284g_0^4			-0.04373g_0^6}{1-0.9678g_0^2}$	
			&$0\le g_0^2\le 1$	&$0.04\%$&\cite{Guagnelli:2000jw}\\
&&&&&&&\\
$\times$&	&$Z_M(\beta)$&$=$&$    1.754(19)+  0.27(10)(\beta-6) -0.10(10)(\beta-6)^2$		&$6.0\le\beta\le7.0$&	 &cf. (\ref{ZMnew})\\
&&&&&&&\\[1ex]
\hline\hline\\
\end{tabular}
\caption{Summary of renormalization constants.}\label{renormalizationtab} 
\end{sidewaystable}
\begin{sidewaystable} 
\centering 
\begin{tabular}{ccrclccc} 
\hline\hline\\[-2.3ex]
non-pert.	&pert. 		&\multicolumn{3}{c}{parameterization}	&parameter	&error	&reference\\
		&\#-loops	&&&			&range		&&\\[.7ex]	
\hline
&&&&&&&\\
$\times$&&	$c_{\rm SW}(g_0)$&$=$&$\frac{1-0.656g_0^2-0.152g_0^4-0.054g_0^6}			{1-0.922g_0^2}$	&$0\le g_0^2\le1$	&&\cite{Luscher:1997ug}\\
&&&&&&&\\
$\times$&&	$c_A$&$=$&$-0.00756g_0^2\frac{1-0.748g_0^2}{1-0.977g_0^2}$
			&$0\le g_0^2\le1$	&&\cite{Luscher:1997ug}\\
&&&&&&&\\
$\times$&&		$c_V$&$=$&$1-0.01633g_0^2\frac{1-0.257g_0^2}{1-0.963g_0^2}$ &	$0\le g_0^2\le1$&&\cite{Harada:2002jh}\\
&&&&&&&\\
$\times$&$$&		
\multicolumn{3}{c}{
\begin{tabular}{cccc}
$\beta$	&$6$		&$6.2$		&$6.4$		\\
\hline
$b_A$	&$1.28(3)(4)$ 	&$1.32(3)(4)$ 	&$1.31(2)(1)$	\\
\end{tabular}}

&&(stat.)(syst.)&\cite{Bhattacharya:2001ks}\\
&&&&&&&\\
$\times$&&	$b_A-b_P$&$=$&$ \frac{-0.00093g_0^2(1+23.3060g_0^2-27.371g_0^4}{1-0.9833g_0^2} $&$0.881\le g_0^2\le1$	&$<0.3\%$&\cite{Guagnelli:2000jw}\\
&&&&&&&\\
$\times$&&	$b_m$&$=$&$(-0.5-0.09623g_0^2)\frac{1-0.6905g_0^2+0.0584g_0^4}{ 1-0.6905g_0^2} $&$0\le g_0^2\le1$	&$<1.3\%$&\cite{Guagnelli:2000jw}\\
&&&&&&&\\
$\times$&&	$b_V$&$=$&$\frac{1-0.6518g_0^2-0.1226g_0^4}{1-0.8467g_0^2}$&$0\le g_0^2\le1$	&&\cite{Luscher:1997ug}\\
&&&&&&&\\
&$1$	&$\tilde{c}_t$&$=$&$1-\frac{4}{3}\,0.01346(1)g_0^2$&&&\cite{Sint:1997jx}\\
&&&&&&&\\
&$2$	&$c_t$&$=$&$1-0.08900(5)g_0^2-0.0294(3)g_0^4$&&&\cite{Bode:1999sm}\\[1ex]
\hline\hline
\end{tabular} 
\caption{Summary of improvement constants. 
}
\label{improvementtab}
\end{sidewaystable}
\chapter{Simulation results}
\begin{sidewaystable}
\centering
\begin{tabular}{ccc@{\hspace{3ex}}c@{\hspace{3ex}}c@{\hspace{3ex}}c@{\hspace{3ex}}c@{\hspace{3ex}}c@{\hspace{3ex}}c}
\hline\hline\\[-1ex]
&&$\kappa_1-\kappa_2$&$\kappa_1-\kappa_3$&$\kappa_1-\kappa_4$&$\kappa_1-\kappa_5$&$\kappa_1-\kappa_6$&$\kappa_1-\kappa_7$\\[1ex]

\multirow{5}{*}{$r_0m^{\rm eff}_{\rm PS}(x_0)$}&$\beta_1$&   3.07 -   4.38 &   3.26 -   4.56 &   3.26 -   4.75 &   3.26 -   4.75 &   3.26 -   4.75 &   3.26 -   4.75\\ 
&$\beta_2$&                                                  3.55 -   4.81 &   3.71 -   4.97 &   3.71 -   5.13 &   3.71 -   5.13 &   3.71 -   5.29 &   3.71 -   5.29\\ 
&$\beta_3$&                                                  3.32 -   4.81 &   3.32 -   5.08 &   3.45 -   5.21 &   3.45 -   5.21 &   3.32 -   5.21 &   3.32 -   5.21\\ 
&$\beta_4$&                                                  3.13 -   4.28 &   3.22 -   4.66 &   3.32 -   4.66 &   3.32 -   4.86 &   3.32 -   4.95 &   3.32 -   5.14\\ 
&$\beta_5$&                                                  3.34 -   4.22 &   3.22 -   4.59 &   3.09 -   4.78 &   3.09 -   4.84 &   3.03 -   4.97 &   2.72 -   3.53\\ 
\\                                                           
\multirow{5}{*}{$r_0m^{\rm eff}_{\rm V}(x_0)$}&$\beta_1$&    3.45 -   4.19 &   3.07 -   4.56 &   3.07 -   4.75 &   3.07 -   4.75 &   2.89 -   4.75 &   2.89 -   4.94\\ 
&$\beta_2$&                                                  3.55 -   4.66 &   3.71 -   4.97 &   3.55 -   5.13 &   3.39 -   5.13 &   3.39 -   5.29 &   3.39 -   5.44\\ 
&$\beta_3$&                                                  3.59 -   5.08 &   3.45 -   5.21 &   3.45 -   5.21 &   3.45 -   5.35 &   3.32 -   5.35 &   3.32 -   5.49\\ 
&$\beta_4$&                                                  3.61 -   4.66 &   3.51 -   4.66 &   3.22 -   4.86 &   3.22 -   4.86 &   3.22 -   4.86 &   3.13 -   5.05\\ 
&$\beta_5$&                                                  3.59 -   4.28 &   3.28 -   4.53 &   3.16 -   4.66 &   3.03 -   4.72 &   2.91 -   4.97 &   2.66 -   4.72\\ 
\\                                                           
\multirow{5}{*}{$r_0{\rm F_{PS}}(x_0)$}&$\beta_1$&           3.07 -   4.75 &   3.63 -   4.94 &   3.82 -   4.94 &   4.01 -   5.12 &   4.01 -   5.12 &   4.19 -   5.12\\ 
&$\beta_2 $&                                                 4.02 -   5.13 &   4.34 -   5.13 &   4.34 -   5.29 &   4.50 -   5.13 &   4.66 -   5.13 &   4.81 -   5.29\\ 
&$\beta_3 $&                                                 3.59 -   5.08 &   4.00 -   5.21 &   4.27 -   5.21 &   4.13 -   5.35 &   4.27 -   5.35 &   4.40 -   5.21\\ 
&$\beta_4 $&                                                 3.61 -   4.57 &   3.89 -   4.76 &   4.09 -   4.76 &   4.09 -   4.86 &   4.38 -   4.95 &   4.28 -   5.14\\ 
&$\beta_5 $&                                                 3.34 -   4.53 &   3.59 -   4.66 &   3.66 -   4.72 &   3.66 -   4.78 &   3.78 -   4.78 &   3.47 -   4.41\\ 
\\                                                         
\multirow{5}{*}{$r_0{\rm F_{V}}(x_0)$}&$\beta_1$&            3.07 -   4.56 &   3.07 -   4.94 &   3.07 -   4.94 &   2.89 -   5.12 &   2.89 -   5.12 &   2.70 -   5.12\\ 
&$\beta_2$&                                                  4.02 -   4.97 &   3.39 -   5.13 &   3.55 -   5.29 &   3.71 -   5.29 &   3.87 -   5.29 &   3.87 -   5.44\\ 
&$\beta_3$&                                                  3.18 -   5.21 &   4.00 -   5.35 &   4.54 -   5.35 &   4.54 -   5.49 &   4.54 -   5.35 &   4.54 -   5.49\\ 
&$\beta_4$&                                                  3.99 -   4.95 &   3.99 -   4.95 &   4.18 -   4.76 &   4.47 -   4.76 &   4.47 -   4.76 &   3.80 -   5.05\\ 
&$\beta_5$&                                                  3.34 -   4.03 &   3.28 -   4.22 &   3.28 -   4.28 &   3.34 -   4.28 &   3.28 -   4.47 &   3.28 -   4.53\\ [2ex]
\hline\hline
\end{tabular}
\caption{The plateau ranges in units of $r_0$.}\label{plateauranges}
\end{sidewaystable}
\begin{table}\begin{tabular}{cccccccccc}
\hline\hline\\[-2ex]
&&$\kappa_1-\kappa_2$&$\kappa_1-\kappa_3$&$\kappa_1-\kappa_4$&$\kappa_1-\kappa_5$&$\kappa_1-\kappa_6$&$\kappa_1-\kappa_7$\\[1ex]
\multirow{5}{*}{\begin{rotate}{90}$r_0\Delta_{\rm m_{PS}}$\end{rotate}}&$\beta_1$&1.556(64)& 1.408(85)& 1.355(98)& 1.32(11)& 1.30(13)& 1.29(15)&  \\ 
&$\beta_2$&1.529(56)& 1.396(81)& 1.349(96)& 1.32(12)& 1.29(13)& 1.28(13)&  \\ 
&$\beta_3$&1.571(56)& 1.449(71)& 1.397(87)& 1.36(10)& 1.35(11)& 1.32(13)&  \\ 
&$\beta_4$&1.732(84)& 1.549(99)& 1.48(12)& 1.42(13)& 1.38(15)& 1.37(17)&  \\ 
&$\beta_5$&1.552(92)& 1.50(11)& 1.50(18)& 1.44(17)& 1.37(19)& 1.36(56)&  \\ 

\\[-2ex]
\multirow{5}{*}{\begin{rotate}{90}$r_0\Delta_{\rm m_{V}}$\end{rotate}}&$\beta_1$&1.35(13)& 1.45(16)& 1.32(18)& 1.37(32)& 1.43(31)& 1.38(25)&  \\ 
&$\beta_2$&1.44(10)& 1.275(72)& 1.30(11)& 1.29(12)& 1.27(14)& 2.667(99)&  \\ 
&$\beta_3$&1.431(86)& 1.363(88)& 1.32(11)& 1.30(12)& 1.30(13)& 1.28(15)&  \\ 
&$\beta_4$&1.378(86)& 1.338(99)& 1.41(12)& 1.38(14)& 1.37(16)& 1.34(16)&  \\ 
&$\beta_5$&1.338(82)& 1.317(77)& 1.302(94)& 1.29(11)& 1.29(15)& 1.26(24)&  \\ 

\\[-2ex]
\multirow{5}{*}{\begin{rotate}{90}$r_0\Delta_{\rm F_{PS}}$\end{rotate}}&$\beta_1$&1.598(34)& 1.424(38)& 1.354(43)& 1.310(47)& 1.282(51)& 1.260(58)&  \\ 
&$\beta_2$&1.534(26)& 1.496(37)& 1.471(40)& 1.428(50)& 1.417(60)& 1.338(62)&  \\ 
&$\beta_3$&1.623(33)& 1.474(32)& 1.420(38)& 1.447(43)& 1.435(49)& 1.390(56)&  \\ 
&$\beta_4$&1.816(30)& 1.692(44)& 1.619(49)& 1.518(52)& 1.521(72)& 1.590(77)&  \\ 
&$\beta_5$&1.723(42)& 1.830(56)& 1.804(47)& 1.829(52)& 1.848(69)& 2.233(88)&  \\ 

\\[-2ex]
\multirow{5}{*}{\begin{rotate}{90}$r_0\Delta_{\rm F_{V}}$\end{rotate}}&$\beta_1$&-& -& -& -& -& -&  \\ 
&$\beta_2$&1.341(61)& 1.89(14)& 1.73(14)& 1.64(15)& 1.61(17)& 1.55(18)&  \\ 
&$\beta_3$&2.01(19)& 1.46(12)& 1.240(77)& 1.192(81)& 1.114(73)& 0.973(63)&  \\ 
&$\beta_4$&1.53(11)& 1.514(96)& 1.43(10)& 1.33(15)& 1.35(19)& 1.62(14)&  \\ 
&$\beta_5$&1.86(15)& 1.85(16)& 1.89(14)& 1.83(14)& 1.87(14)& 1.91(12)&  \\ 

\\[-2ex]
\hline\hline
\end{tabular}\caption{Estimates for the gap energy as obtained from fits to the effective masses and the decay constants. For $r_0\Delta_{\rm F_V}$ at $\beta_1$, the data was too noisy to allow for a sensible fit.}\label{gaps}
\end{table}

\begin{table}
\centering
\begin{tabular}{ccccccccc}
\hline\hline\\
&&$\kappa_1-\kappa_2$&$\kappa_1-\kappa_3$&$\kappa_1-\kappa_4$&$\kappa_1-\kappa_5$&$\kappa_1-\kappa_6$&$\kappa_1-\kappa_7$\\[1ex]
\multirow{5}{*}{\begin{rotate}{90} {{$r_0m_{\rm PS}$}} \end{rotate}}&$\beta_1$&3.147(15)&4.294(19)&4.971(22)&5.541(24)&6.005(26)&6.439(28)  \\ 
&$\beta_2$&3.069(14)&4.273(20)&4.988(23)&5.597(25)&6.099(28)&6.574(30)  \\ 
&$\beta_3$&3.093(15)&4.286(20)&5.012(23)&5.636(26)&6.153(28)&6.648(30)  \\ 
&$\beta_4$&3.095(18)&4.277(25)&5.020(29)&5.678(32)&6.229(35)&6.756(38)  \\ 
&$\beta_5$&3.742(28)&5.179(36)&6.250(42)&7.170(48)&8.818(58)&11.588(77) \\ 
[2ex]
\multirow{5}{*}{\begin{rotate}{90} {{$r_0m_{\rm V}$ }} \end{rotate}}&$\beta_1$&3.645(22)&4.637(26)&5.253(29)&5.766(33)&6.203(35)&6.609(38)  \\ 
&$\beta_2$&3.567(20)&4.612(23)&5.272(26)&5.843(29)&6.318(31)&6.769(35)  \\ 
&$\beta_3$&3.604(24)&4.653(26)&5.323(29)&5.908(31)&6.402(33)&6.884(36)  \\ 
&$\beta_4$&3.637(31)&4.656(34)&5.337(37)&5.954(40)&6.475(43)&6.967(45)  \\ 
&$\beta_5$&4.141(35)&5.461(41)&6.480(46)&7.370(51)&8.981(61)&11.698(77) \\ 
[2ex]
\multirow{5}{*}{\begin{rotate}{90} {{$r_0{\rm F_{PS}}$}}\end{rotate}}&$\beta_1$&0.493(12)&0.528(16)&0.539(19)&0.547(22)&0.551(24)&0.554(27)  \\ 
&$\beta_2$&0.524(10)&0.565(13)&0.577(16)&0.583(18)&0.589(20)&0.599(23)  \\ 
&$\beta_3$&0.539(11)&0.584(15)&0.599(18)&0.607(21)&0.606(24)&0.604(26)  \\ 
&$\beta_4$&0.5508(98)&0.594(13)&0.606(15)&0.616(17)&0.628(19)&0.646(22) \\ 
&$\beta_5$&0.555(11)&0.572(15)&0.571(18)&0.567(20)&0.558(26)&0.233(11)  \\ 
[2ex]
\multirow{5}{*}{\begin{rotate}{90} {{$r_0{\rm F_{V}}$}} \end{rotate}}&$\beta_1$&0.594(36)&0.565(36)&0.535(38)&0.498(42)&0.471(44)&0.441(47) \\ 
&$\beta_2$&0.572(26)&0.578(26)&0.570(28)&0.559(31)&0.551(33)&0.547(37) \\ 
&$\beta_3$&0.629(44)&0.663(44)&0.667(50)&0.661(53)&0.648(56)&0.651(59) \\ 
&$\beta_4$&0.732(49)&0.700(42)&0.689(42)&0.687(45)&0.687(48)&0.681(49) \\ 
&$\beta_5$&0.620(39)&0.600(38)&0.581(37)&0.568(39)&0.550(42)&0.248(19) \\ 
[2ex]
\multirow{5}{*}{\begin{rotate}{90} {{\hspace{-3mm}$\fds/\fdsstar$}}  \end{rotate}}&$\beta_1$&0.829(52)&0.935(61)&1.009(73)&1.097(92)&1.17(11)&1.26(13)   \\ 
&$\beta_2$&0.917(44)&0.978(45)&1.012(49)&1.043(55)&1.068(61)&1.095(68) \\ 
&$\beta_3$&0.857(58)&0.880(55)&0.897(61)&0.919(65)&0.936(71)&0.928(71) \\ 
&$\beta_4$&0.753(50)&0.848(47)&0.880(48)&0.897(51)&0.915(53)&0.949(55) \\ 
&$\beta_5$&0.895(52)&0.953(51)&0.982(53)&0.999(56)&1.015(63)&0.939(63) \\ 
[2ex]
\multirow{5}{*}{\begin{rotate}{90} {{\hspace{-3mm}$r_0M_{Q|_{m_{sc}}}$}}\end{rotate}}&$\beta_1$&1.657(33)&3.206(68)&4.332(96)&5.43(12)&6.43(15)&7.48(18)   \\ 
&$\beta_2$&1.578(30)&3.156(64)&4.280(89)&5.35(11)&6.33(14)&7.32(16)   \\ 
&$\beta_3$&1.657(28)&3.200(57)&4.293(79)&5.33(10)&6.26(12)&7.21(14)   \\ 
&$\beta_4$&1.691(25)&3.161(48)&4.204(65)&5.193(83)&6.072(99)&6.96(12) \\ 
&$\beta_5$&2.394(36)&4.278(65)&5.820(91)&7.23(11)&9.93(16)&15.04(26)  \\ 
[2ex]
\hline\hline
\end{tabular}\caption{Plateau averaged data for the effective masses, the decay constants and the renormalization group invariant quark mass.}\label{tabplatav1}
\end{table}

\begin{table}
\centering
\begin{tabular}{crrrrrrr}
\hline\hline\\
&&$\kappa_1-\kappa_2$&$\kappa_1-\kappa_3$&$\kappa_1-\kappa_4$&$\kappa_1-\kappa_5$&$\kappa_1-\kappa_6$&$\kappa_1-\kappa_7$\\[1ex]
\multirow{5}{*}{\begin{rotate}{90}$r_0^{3/2}{Y_{\rm PS}\over C_{\rm PS}}$\end{rotate}}&$\beta_1$&0.940(21)&1.053(30)&1.114(38)&1.163(45)&1.200(52)&1.232(59) \\ 
&$\beta_2$&0.996(19)&1.127(26)&1.196(32)&1.249(38)&1.295(44)&1.349(51) \\ 
&$\beta_3$&1.020(20)&1.165(29)&1.243(37)&1.306(45)&1.341(52)&1.370(59) \\ 
&$\beta_4$&1.038(19)&1.185(26)&1.264(31)&1.334(37)&1.402(42)&1.481(50) \\ 
&$\beta_5$&1.082(24)&1.208(33)&1.280(41)&1.335(49)&1.419(67)&0.660(32) \\ 
[2ex]
\multirow{5}{*}{\begin{rotate}{90}${R\over C_{\rm PS/V}}$\end{rotate}}&$\beta_1$&0.641(41)&0.804(54)&0.894(66)&0.991(84)&1.07(10)&1.16(12)   \\ 
&$\beta_2$&0.702(35)&0.840(39)&0.896(44)&0.939(50)&0.973(56)&1.005(63) \\ 
&$\beta_3$&0.661(46)&0.755(48)&0.793(55)&0.826(59)&0.850(65)&0.849(66) \\ 
&$\beta_4$&0.579(40)&0.726(42)&0.776(43)&0.805(46)&0.830(49)&0.869(51) \\ 
&$\beta_5$&0.741(44)&0.845(46)&0.891(48)&0.918(52)&0.946(59)&0.887(59) \\ 
[2ex]
\multirow{5}{*}{\begin{rotate}{90}$r_0{\Delta m\over C_{\rm spin}}$\end{rotate}}&$\beta_1$&0.395(15)&0.333(15)&0.264(16)&0.214(19)&0.181(21)&0.183(20) \\ 
&$\beta_2$&0.386(10)&0.3287(99)&0.2713(95)&0.238(10)&0.210(12)&0.193(13) \\ 
&$\beta_3$&0.410(15)&0.356(15)&0.300(16)&0.266(15)&0.242(16)&0.233(17) \\ 
&$\beta_4$&0.428(19)&0.367(18)&0.306(17)&0.272(18)&0.242(18)&0.219(18) \\ 
&$\beta_5$&0.389(17)&0.336(16)&0.282(16)&0.252(15)&0.227(15)&0.215(15) \\ 
[2ex]
\hline\hline
\end{tabular}\caption{Plateau averaged data for the decay constant, the ratio of the pseudo scalar to the vector meson decay constant and the mass splitting converted to HQET.}\label{tabplatav2}
\end{table}

\chapter{Running the code}\label{runit}
This appendix explains the structure of the PC-code, how it has to be compiled and how the input files that specify run parameters have to be designed. For the official MILC-Code documentation, please refer to \newline
\verb|http://www.physics.utah.edu/~detar/milc/milcv6.html| \newline or contact the author of this thesis in case of any questions.
\section{Directory structure}
After unpacking the code, the following directories will be created:

\begin{supertabular}{p{.28\linewidth}p{.68\linewidth}}
\\
	\verb|f_A|&\begin{minipage}{1\linewidth}	This is the project's main directory. It contains specific program code for the computations of correlation functions needed in this work.\end{minipage}\\ 
\\
	\verb|schroed_pg|&\begin{minipage}{1\linewidth} Gauge-update routines specific for Schr\"odinger functional boundary conditions.\end{minipage}\\ 
\\
	\verb|generic|&\begin{minipage}{1\linewidth}	Generic routines, like e.g. I/O or the layout for parallelization.\end{minipage}\\ 
\\
	\verb|generic_clover|&\begin{minipage}{1\linewidth} Inversion routine (BiCGstab). \end{minipage}\\
\\

	\verb|generic_pg|&\begin{minipage}{1\linewidth} Generic gauge update routines.\end{minipage}\\ 
\\
	\verb|generic_schroed|&\begin{minipage}{1\linewidth} Routines that are specific to Schr\"odinger Functional boundary conditions.\end{minipage}\\
\\
\verb|generic_wilson|&\begin{minipage}{1\linewidth} Routines for boundary sources and the Dirac operator.\end{minipage}\\
\\
\verb|include|	&\begin{minipage}{1\linewidth} Macros and declarations and definitions for structures like e.g. the site-structure.\end{minipage}\\
\\
\verb|library|&\begin{minipage}{1\linewidth} Linear algebra routines.\end{minipage}\\
\\
\end{supertabular}

\section{Job steps in a production run}

A production run is organised as follows: There is always one
binary for the gauge-updates. It starts from a cold gauge configuration for generating a thermalized field configuration or reloads the configuration from the previous Monte-Carlo step. The resulting field
configuration will be written to disc (In contrast to the experience 
with the APEMille computer at NIC/DESY Zeuthen, the I/O is not very time-consuming, even for very large lattices).
The binary for the calculation of the correlation functions reads
the gauge configurations from disc, computes the propagators
and evaluates the correlation functions. If desired, the computed propagators can be saved to disk.
All run-parameters are handed over to the programs by input files via STDIN. 

\section{Compiling the code}

Fist a list of important compiler flags:\\
{{\bf \verb|DOUBLE|:}}
\begin{flushright} \begin{minipage}[r]{.95\linewidth}
	If it is defined in the file
	\verb|include/config.h|, all arithmetics will be done with
	double precision arithmetics. If it is not defined, only some
	global sums will be done in double precision.
	Note however, that setting this flag will double
	the CPU-time and will also nearly double
	the amount of memory to be allocated.\\
\end{minipage} \end{flushright}
\vspace{-.3cm}
{\bf \verb|FIELD_MAJOR| and \verb|TMP_LINKS|:}
\begin{flushright} \begin{minipage}[r]{.95\linewidth}
	If defined in \verb|Make_template|, the code will be compiled 
	for use of field major (cf. section \ref{chapterperformance}).
	This improves performance. Additional memory for the allocation of
	temporary fields is needed. \\
\end{minipage} \end{flushright}
\vspace{-.3cm}
{\bf \verb|FORWARD| and \verb|BACKWARD|:}
\begin{flushright} \begin{minipage}[r]{.95\linewidth}
	Depending on which flag is defined, only the propagators in the 
	forward or in the backward direction will be computed. This is
	convenient for the simulation of large lattices with long run-times.
	At the HLRN, this allowed to submit the job-steps into job-queues
	with a shorter waiting time because of reduced wall-clock time.
	For smaller lattices just set both, the FORWARD and BACKWARD
	flag. Note, that saving propagators works only for one
	of the flags being defined.
\end{minipage}\\ \end{flushright}

Depending on the local installation of the MPICH library, one may have to change
the PATH to the correpsonding libraries in the \verb|Makte_linux_mpi| file.\\

Comilation of the Pure Gauge part:\\
\begin{verbatim}
# cd schroed_pg	
# make -f Make_linux_mpi su3_schr_ora
\end{verbatim}
Compilation of the Inversion routine
\begin{verbatim}
# cd f_A
# make -f Make_linux_mpi su3_schr_cl_bi
\end{verbatim}

The directories \verb|schroed_pg| and \verb|f_A| should now contain the executables \verb|su3_schr_ora| and \verb|su3_schr_cl_bi|.

\section{Job Scripts}

Sample job script (DESY-Zeuthen-Cluster):\\
\begin{verbatim}
##################################################################
#!/bin/csh
# QSUB -e multiple.err
# QSUB -r jobtest

# jump into working-directory
cd <your working dir.>

# start the thermalization procedure
mpirun -np <no. of cpus to use> -machinefile 
        <file containing machine names> 
        ./bin/su3_schr_ora_single \< ./<thermalize input file> >> 
        thermalize.out

# Keep thermalized configuration and copy it to working 
# configuration
cp thermalized checkpoint

# define a variable that counts job steps
set a=(1)
# start loop over job-steps (e.g. 100 measurements)
while ($a <= 100)
# start job
# Take care for seed, increase it before every update
awk '{if($1 ~ /iseed/)print $1,$2+123;else print $0}' 
        <update input file> > dummyfile
cp dummyfile <update input file>

mpirun -np <no. of cpus to use> -machinefile 
        <file containing machine names> 
        ./bin/su3_schr_ora \< ./<update input file> >>update.out

mpirun -np <no. of cpus to use> -machinefile 
        <file containing machine names>
        ./bin/su3_schr_cl_bi \< <measurement input file> 
        >>measure.out

# increase a by one
       set a=(`expr $a + 1`)
end
rm dummyfile
##################################################################
\end{verbatim}
The file \verb|<thermalize input file>| may be the following (It is important to remove all comments!):\\
\begin{verbatim}
##################################################################
prompt 0
nx 20                   # define lattice dimensions
ny 20
nz 20
nt 20
iseed 12318352          # seed for random number gen.

warms 0                 # warm ups
trajecs 500             # no. of trajectories
traj_between_meas 5     # output gauge-info every # steps
beta 7.8439             # beta
bc_flag 0               # boundary condition flag
steps_per_trajectory 10 # heatbath steps
qhb_steps 1             # OR steps
fresh                   # start with flat gauge config
save_serial             # save gauge config as binary after end
./thermalized           # name of gauge-config file
##################################################################
\end{verbatim}
The file \verb|<update input file>| may be the following (It is important to remove all comments!):\\
\begin{verbatim}
##################################################################
prompt 0
nx 20                   # define lattice dimensions
ny 20                                                                 
nz 20                                                                 
nt 20                                                                 
iseed 12318352          # seed for random number gen.
                                                                      
warms 0                 # warm ups
trajecs 25              # no. of trajectories
traj_between_meas 25    # output gauge-info every # steps
beta 7.8439             # beta
bc_flag 0               # boundary condition flag
steps_per_trajectory 10 # heatbath steps
qhb_steps 1             # OR steps
reload_serial           # start with flat gauge config
./checkpoint            # save gauge config as binary after end
save_serial             # name of gauge-config file
./checkpoint
##################################################################
\end{verbatim}
The file \verb|<measurement input file>| may be the followingi (It is important to remove all comments!):\\
\begin{verbatim}
##################################################################
prompt 0
nx 20                           # define lattice dimensions
ny 20
nz 20
nt 20

number_of_kappas 5              # total number of hopping params.
bc_flag 0                       # boundary condition flag
num_smear 0                     # this is redundant, 
                                # I will remove it soon
kappa 0.133373                  # values of the hopping params.
kappa 0.128989
kappa 0.128214
kappa 0.127656
kappa 0.125309
cttilde 0.9862                  # cttilde
clov_c 1.3066                   # csw
ferm_phases[0] 0.5              # theta
ferm_phases[1] 0.5
ferm_phases[2] 0.5
max_cg_iterations 100000	
max_cg_restarts 2
error_for_propagator 1e-14      # solver residuals
error_for_propagator 1e-14
error_for_propagator 1e-14
error_for_propagator 1e-14
error_for_propagator 1e-14
reload_serial                   # reload binary field config.
<filename>                      # name of binary field config.
num_prop_load aa                # number of propagators to reload
which_prop_load bb              # aa consecutive lines with 
				# numers of the hopping params. 
                                # of the propagator which you 
                                # want to save
reload_serial_prop <filename>   # conescutive lines with the names 
                                # of the propagators
num_prop_sav cc                 # number of propagators to save
which_prop_sav bb               # same as for load
save_serial_prop <filename>     # 
##################################################################
\end{verbatim}

If one of the lines \verb|num_prop_load| or \verb|num_prop_sav| has 0 input value, leave
out the \verb|which_prop_load/sav| and \verb|reload/save_serial_prop| lines\\

\section{Hints}
\bi
\item  For running the program on a single CPU, one can either use  the mpi-code and \verb|mpirun -np 1| or compile the code as "vanilla" version with \verb|Make_vanilla|.
\item Two test input files for the inverter are located in the directory \verb|f_A/| (\verb|in.csw1.7| und \verb|in.test|)
The output of the corresponding runs can be found in the files \verb|data.csw1.7| and \verb|data.test|.\\
\item There are some PERL scripts which the user might be interested in:\\
\bi
	\item \verb|singdoub|: converts a single precision arithmetics gauge config
		into a double precision. 
	\item \verb|alpha2milc|: a tool that converts binary or ASCII ALPHA-collaboration
		(APEMille for binary) gauge configurations into a format
		readable by the MILC code
	\item \verb|mouta|: extracts correlation function data from output files
		and produces output that is directly readable by the  
		data analysis program used here\\
\ei
\ei
\end{appendix}


\bibliographystyle{h-elsevier2}
\bibliography{outline}\cleardoublepage
\pagestyle{empty}
\noindent{\bf\Huge Erratum}
\vspace{1cm}\\
This addendum summarizes the changes and corrections to the thesis (apart from typos and corrected references), that were carried out as suggested by the referees.\\

\hspace{-.8cm}\begin{tabular}{p{1.2cm}p{12.5cm}}
Page	&Comment\\[.1cm]
\hline\\[-.3cm]
p. 1    &The photons of course couple to both, left- and right-handed fermions.\\
p. 2    &The branching ratio (\ref{branchingratio}) has been replaced by its tree-level expression.\\
p. 14	&In formula (\ref{erratum1}), the r.h.s. has been corrected. The factor $(2b_0 \bar g^2(m_Q))^{\gamma_0^{\rm X,\MSbar}/2b_0}$ and the wrong term in the exponent have been removed.\\
p.15 	&There was a wrong sign in formula (\ref{truegammamatch})\\
p. 17	&The definition of the vector meson decay constant (\ref{erratum2}) has been corrected.\\
p. 28ff	&The iso-structure of the currents has been formulated more consistently.\\
p. 47	&The integrated auto-correlation time $\tau_int$ is now given normalized with the total number of intermediate OR and update steps.\\
p. 52ff	&The arguments of the functions $Z_P$ have been corrected and rewritten in a less misleading way. The error estimates for $Z_P$ and $Z_M$ have been corrected in (\ref{erratum3}), (\ref{ZMnew}) and figure (\ref{ZPplot}).\\
p. 73, p. 85& The way, the scale has been set in the various lattice simulations for $\fds$ and $\fbs$ has been added to the tables (\ref{otherfds}) and (\ref{otherfbs}).\\
p. 80	&The values of the errors in table \ref{HQETcoeffs} have been corrected.\\
\end{tabular}
\vspace{.5cm}\\
In addition, reference \cite{AliKhan:2001jg} was added to table \ref{otherfbs} on page 85.

\appendix



\chapter*{Lebenslauf}

\begin{tabular}{ll}

Name: & \dcauthorname  \dcauthorsurname \\
1995			& Abitur am Martin-Behaim Gymnasium N\"urnberg\\
10/1996 - 10/2001 & Studium an der Friedrich-Alexander Universit\"at \\

 & Erlangen-N\"urnberg in der Fachrichtung Physik\\

4/2002 - 9/2004 & Wissenschaftlicher Mitarbeiter an \\

 & der Humboldt-Universit\"at zu Berlin,\\

 & Lehrstuhl Prof. Dr. U. Wolff, \\

 & Institut fuer Physik\\

\end{tabular}
\chapter*{Publications}
\bi
\item ALPHA collaboration (A. J\"uttner, J. Rolf), A precise determination of the decay constant of the ${\rm D_s}$ meson in quenched QCD, Published in Phys.Lett.\linebreak B560:59-63,2003, e-Print Archive: hep-lat/0302016
\item  J. Rolf, A. J\"uttner, Precision computation of the leptonic ${\rm D_s}$ meson decay constant in quenched QCD, Invited talk at the 2nd Workshop on the
CKM Unitarity Triangle, Durham, England, 5-9 April 2003, Published in eConf C0304052:WG508,2003, e-Print Archive: hep-ph/0306299
\vspace{-.2cm}
\item A. J\"uttner, J. Rolf, A precise determination of the decay constant of the ${\rm D_s}$ meson in quenched QCD, Presented at 21st International Symposium on Lattice Field Theory (LATTICE 2003), Tsukuba, Ibaraki, Japan, 15-19 Jul 2003, Nucl.Phys.Proc.Suppl.129:319-321,2004, e-Print Archive: hep-lat/0309069
\vspace{-.2cm}
\item ALPHA collaboration (J. Rolf, M. Della Morte, S. D\"urr, J. Heitger, A. J\"uttner, H. Molke, A. Shindler, R. Sommer), Towards a precision computation of $\rm F_{\rm B_s}$ in quenched QCD, Presented at 21st International Symposium on Lattice Field Theory (LATTICE 2003), Tsukuba, Ibaraki, Japan, 15-19 Jul 2003, Published in Nucl.Phys.Proc.Suppl.129:322-324,2004, e-Print Archive: hep-lat/0309072
\item ALPHA collaboration (J. Heitger, A. J\"uttner, R. Sommer, J. Wennekers), Non-perturbative tests of Heavy Quark Effective Theory, e-Print Archive: hep-ph/0407227
\ei
\chapter*{Acknowledgements}
\bi
\item I would like to thank Ulli Wolff and Juri Rolf for suggesting me the topic of this thesis and for their support throughout the whole time.
\item Special thanks to the members of the ALPHA collaboration. In particular to Jochen Heitger, Roland Hoffmann, Francesco Knechtli, Michele Della Morte,  and Rainer Sommer for the fruitful discussions during the whole time.
\item Also thanks to Hinnerk St\"uben from the Konrad-Zuse-Zentrum f\"ur Informationstechnik Berlin for his assistance. Thanks to the Norddeutscher Verbund f\"ur Hoch- und H\"ochstleistungsrechnen, for giving me the opportunity to use their computing resources.
\item Thanks to the DFG for the scholarship within the Graduiertenkolleg 271 and for the position within the Sonderforschungsbreich Transregio 9.
\item Thanks to the organizers of the Tuesday lattice seminar and to all participants of the discussion seminar.
\item Thanks AG-COM! And to my special friends Francesco, Michele, Anna and Rainer, with whom I did not only spend the working time - thanks guys!
\item I am indebted to my family, who gave me strong support and who always were there when I needed them.

\ei

\chapter*{Selbst\"andigkeitserkl\"arung}

\noindent Hiermit erkl\"are ich, die vorliegende Arbeit selbst\"andig ohne fremde Hilfe verfa{\ss}t und nur die angegebene Literatur und Hilfsmittel verwendet zu haben.\\

\vspace{5cm}
\noindent\dcauthorname \dcauthorsurname \\  
\dcdatesubmitted \\

\end{document}